
%
%
%
%
%
\def\unredoffs{\hoffset-.14truein\voffset-.2truein} 
 
%
%
\newbox\leftpage \newdimen\fullhsize \newdimen\hstitle \newdimen\hsbody
\tolerance=1000\hfuzz=2pt
\catcode`\@=11 
%
\magnification=1095\unredoffs\baselineskip=16pt plus 2pt minus 1pt
\hsbody=\hsize \hstitle=\hsize 
%
%
%
\newcount\yearltd\yearltd=\year\advance\yearltd by -1900

%
%

\def\draftmode{\message{ DRAFTMODE }\def\draftdate{{\rm preliminary draft:
\number\month/\number\day/\number\yearltd\ \ \hourmin}}%
\headline={\hfil\draftdate}\writelabels\baselineskip=16pt plus 2pt minus 2pt
 {\count255=\time\divide\count255 by 60 \xdef\hourmin{\number\count255}
  \multiply\count255 by-60\advance\count255 by\time
  \xdef\hourmin{\hourmin:\ifnum\count255<10 0\fi\the\count255}}}
\def\nolabels{\def\wrlabeL##1{}\def\eqlabeL##1{}\def\reflabeL##1{}}
\def\writelabels{\def\wrlabeL##1{\leavevmode\vadjust{\rlap{\smash%
{\line{{\escapechar=` \hfill\rlap{\sevenrm\hskip.03in\string##1}}}}}}}%
\def\eqlabeL##1{{\escapechar-1\rlap{\sevenrm\hskip.05in\string##1}}}%
\def\reflabeL##1{\noexpand\llap{\noexpand\sevenrm\string\string\string##1}}}
\nolabels
%
\global\newcount\secno \global\secno=0
\global\newcount\meqno \global\meqno=1
\def\newsec#1{\global\advance\secno by1\message{(\the\secno. #1)}
\global\subsecno=0\eqnres@t\noindent{\bf\the\secno. #1}
\writetoca{{\secsym} {#1}}\par\nobreak\medskip\nobreak}
\def\eqnres@t{\xdef\secsym{\the\secno.}\global\meqno=1\bigbreak\bigskip}
\def\sequentialequations{\def\eqnres@t{\bigbreak}}\xdef\secsym{}
\global\newcount\subsecno \global\subsecno=0
\def\subsec#1{\global\advance\subsecno by1\message{(\secsym\the\subsecno. #1)}
\ifnum\lastpenalty>9000\else\bigbreak\fi
\noindent{\bf\secsym\the\subsecno. #1}\writetoca{\string\quad 
{\secsym\the\subsecno.} {#1}}\par\nobreak\medskip\nobreak}
\def\appendix#1#2{\global\meqno=1\global\subsecno=0\xdef\secsym{\hbox{#1.}}
\bigbreak\bigskip\noindent{\bf Appendix #1. #2}\message{(#1. #2)}
\writetoca{Appendix {#1.} {#2}}\par\nobreak\medskip\nobreak}
%
%
\def\eqnn#1{\xdef #1{(\secsym\the\meqno)}\writedef{#1\leftbracket#1}%
\global\advance\meqno by1\wrlabeL#1}
\def\eqna#1{\xdef #1##1{\hbox{$(\secsym\the\meqno##1)$}}
\writedef{#1\numbersign1\leftbracket#1{\numbersign1}}%
\global\advance\meqno by1\wrlabeL{#1$\{\}$}}
\def\eqn#1#2{\xdef #1{(\secsym\the\meqno)}\writedef{#1\leftbracket#1}%
\global\advance\meqno by1$$#2\eqno#1\eqlabeL#1$$}
%
\newskip\footskip\footskip14pt plus 1pt minus 1pt 
\def\footnotefont{\ninepoint}\def\f@t#1{\footnotefont #1\@foot}
\def\f@@t{\baselineskip\footskip\bgroup\footnotefont\aftergroup\@foot\let\next}
\setbox\strutbox=\hbox{\vrule height9.5pt depth4.5pt width0pt}
\global\newcount\ftno \global\ftno=0
\def\foot{\global\advance\ftno by1\footnote{$^{\the\ftno}$}}
%
\newwrite\ftfile   
\def\footend{\def\foot{\global\advance\ftno by1\chardef\wfile=\ftfile
$^{\the\ftno}$\ifnum\ftno=1\immediate\openout\ftfile=foots.tmp\fi%
\immediate\write\ftfile{\noexpand\smallskip%
\noexpand\item{f\the\ftno:\ }\pctsign}\findarg}%
\def\footatend{\vfill\eject\immediate\closeout\ftfile{\parindent=20pt
\centerline{\bf Footnotes}\nobreak\bigskip\input foots.tmp }}}
\def\footatend{}
%
%
\global\newcount\refno \global\refno=1
\newwrite\rfile
\def\ref{[\the\refno]\nref}
\def\nref#1{\xdef#1{[\the\refno]}\writedef{#1\leftbracket#1}%
\ifnum\refno=1\immediate\openout\rfile=refs.tmp\fi
\global\advance\refno by1\chardef\wfile=\rfile\immediate
\write\rfile{\noexpand\item{#1\ }\reflabeL{#1\hskip.31in}\pctsign}\findarg}
\def\findarg#1#{\begingroup\obeylines\newlinechar=`\^^M\pass@rg}
{\obeylines\gdef\pass@rg#1{\writ@line\relax #1^^M\hbox{}^^M}%
\gdef\writ@line#1^^M{\expandafter\toks0\expandafter{\striprel@x #1}%
\edef\next{\the\toks0}\ifx\next\em@rk\let\next=\endgroup\else\ifx\next\empty%
\else\immediate\write\wfile{\the\toks0}\fi\let\next=\writ@line\fi\next\relax}}
\def\striprel@x#1{} \def\em@rk{\hbox{}} 
\def\lref{\begingroup\obeylines\lr@f}
\def\lr@f#1#2{\gdef#1{\ref#1{#2}}\endgroup\unskip}
\def\semi{;\hfil\break}
\def\addref#1{\immediate\write\rfile{\noexpand\item{}#1}} 
\def\startrefs#1{\immediate\openout\rfile=refs.tmp\refno=#1}
\def\xref{\expandafter\xr@f}\def\xr@f[#1]{#1}
\def\refs#1{\count255=1[\r@fs #1{\hbox{}}]}
\def\r@fs#1{\ifx\und@fined#1\message{reflabel \string#1 is undefined.}%
\nref#1{need to supply reference \string#1.}\fi%
\vphantom{\hphantom{#1}}\edef\next{#1}\ifx\next\em@rk\def\next{}%
\else\ifx\next#1\ifodd\count255\relax\xref#1\count255=0\fi%
\else#1\count255=1\fi\let\next=\r@fs\fi\next}
%

%
\newwrite\ffile\global\newcount\figno \global\figno=1
\def\fig{fig.~\the\figno\nfig}
\def\nfig#1{\xdef#1{fig.~\the\figno}%
\writedef{#1\leftbracket fig.\noexpand~\the\figno}%
\ifnum\figno=1\immediate\openout\ffile=figs.tmp\fi\chardef\wfile=\ffile%
\immediate\write\ffile{\noexpand\medskip\noexpand\item{Fig.\ \the\figno. }
\reflabeL{#1\hskip.55in}\pctsign}\global\advance\figno by1\findarg}
\def\xfig{\expandafter\xf@g}\def\xf@g fig.\penalty\@M\ {}
\def\figs#1{figs.~\f@gs #1{\hbox{}}}
\def\f@gs#1{\edef\next{#1}\ifx\next\em@rk\def\next{}\else
\ifx\next#1\xfig #1\else#1\fi\let\next=\f@gs\fi\next}
\newwrite\lfile
{\escapechar-1\xdef\pctsign{\string\%}\xdef\leftbracket{\string\{}
\xdef\rightbracket{\string\}}\xdef\numbersign{\string\#}}

\def\writestop{\def\writestoppt{\immediate\write\lfile{\string\pageno%
\the\pageno\string\startrefs\leftbracket\the\refno\rightbracket%
\string\def\string\secsym\leftbracket\secsym\rightbracket%
\string\secno\the\secno\string\meqno\the\meqno}\immediate\closeout\lfile}}
\def\writestoppt{}\def\writedef#1{}
\def\seclab#1{\xdef #1{\the\secno}\writedef{#1\leftbracket#1}\wrlabeL{#1=#1}}
\def\subseclab#1{\xdef #1{\secsym\the\subsecno}%
\writedef{#1\leftbracket#1}\wrlabeL{#1=#1}}
\newwrite\tfile \def\writetoca#1{}
\def\leaderfill{\leaders\hbox to 1em{\hss.\hss}\hfill}
\def\writetoc{\immediate\openout\tfile=toc.tmp 
   \def\writetoca##1{{\edef\next{\write\tfile{\noindent ##1 
   \string\leaderfill {\noexpand\number\pageno} \par}}\next}}}

%
\catcode`\@=12 
%
\edef\tfontsize{\ifx\answ\bigans scaled\magstep3\else scaled\magstep4\fi}
 \tfontsize  \tfontsize
 \tfontsize \font\titlei=cmmi10 \tfontsize
\font\titleis=cmmi7 \tfontsize \font\titleiss=cmmi5 \tfontsize
\font\titlesy=cmsy10 \tfontsize \font\titlesys=cmsy7 \tfontsize
\font\titlesyss=cmsy5 \tfontsize  \tfontsize
\skewchar\titlei='177 \skewchar\titleis='177 \skewchar\titleiss='177
\skewchar\titlesy='60 \skewchar\titlesys='60 \skewchar\titlesyss='60
 \ifx\answ\bigans\else scaled\magstep1\fi
\ifx\answ\bigans\else

 \font\absi=cmmi10 scaled\magstep1
\font\absis=cmmi7 scaled\magstep1 \font\absiss=cmmi5 scaled\magstep1
\font\abssy=cmsy10 scaled\magstep1 \font\abssys=cmsy7 scaled\magstep1
\font\abssyss=cmsy5 scaled\magstep1 
\skewchar\absi='177 \skewchar\absis='177 \skewchar\absiss='177
\skewchar\abssy='60 \skewchar\abssys='60 \skewchar\abssyss='60
\fi
\font\ninerm=cmr9 \font\sixrm=cmr6 \font\ninei=cmmi9 \font\sixi=cmmi6 
\font\ninesy=cmsy9 \font\sixsy=cmsy6 \font\ninebf=cmbx9 
\font\nineit=cmti9 \font\ninesl=cmsl9 \skewchar\ninei='177
\skewchar\sixi='177 \skewchar\ninesy='60 \skewchar\sixsy='60 
\def\ninepoint{\def\rm{\fam0\ninerm}
\textfont0=\ninerm \scriptfont0=\sixrm \scriptscriptfont0=\fiverm
\textfont1=\ninei \scriptfont1=\sixi \scriptscriptfont1=\fivei
\textfont2=\ninesy \scriptfont2=\sixsy \scriptscriptfont2=\fivesy
\textfont\itfam=\ninei \def\it{\fam\itfam\nineit}\def\sl{\fam\slfam\ninesl}%
\textfont\bffam=\ninebf \def\bf{\fam\bffam\ninebf}\rm} 
%
%

\hyphenation{anom-aly anom-alies coun-ter-term coun-ter-terms}
\def\inv{^{\raise.15ex\hbox{${\scriptscriptstyle -}$}\kern-.05em 1}}

\def\Dsl{\,\raise.15ex\hbox{/}\mkern-13.5mu D} 
\def\dsl{\raise.15ex\hbox{/}\kern-.57em\partial}

\def\lspace{\ifx\answ\bigans{}\else\qquad\fi}
\def\lbspace{\ifx\answ\bigans{}\else\hskip-.2in\fi} 
\def\boxeqn#1{\vcenter{\vbox{\hrule\hbox{\vrule\kern3pt\vbox{\kern3pt
	\hbox{${\displaystyle #1}$}\kern3pt}\kern3pt\vrule}\hrule}}}
\def\mbox#1#2{\vcenter{\hrule \hbox{\vrule height#2in
		\kern#1in \vrule} \hrule}}  
%

\def\darr#1{\raise1.5ex\hbox{$\leftrightarrow$}\mkern-16.5mu #1}

\def\half{{\textstyle{1\over2}}} 
\def\fourninths{{\textstyle{4\over9}}} 
\def\threeovertwo{{\textstyle{3\over2}}} 
\def\roughly#1{\raise.3ex\hbox{$#1$\kern-.75em\lower1ex\hbox{$\sim$}}}

\def\p2inf{\mathrel{\mathop{\sim}\limits_{\scriptscriptstyle
{p^2 \rightarrow \infty }}}}
\def\kap2inf{\mathrel{\mathop{\sim}\limits_{\scriptscriptstyle
{\kappa \rightarrow \infty }}}}
\def\x2inf{\mathrel{\mathop{\sim}\limits_{\scriptscriptstyle
{x \rightarrow \infty }}}}
\def\Lam2inf{\mathrel{\mathop{\sim}\limits_{\scriptscriptstyle
{\Lambda \rightarrow \infty }}}}
\def\frac#1#2{{{#1}\over {#2}}}
\def\half{\hbox{${1\over 2}$}}

\def\Mev{{\rm MeV}}\def\Gev{{\rm GeV}}

\def\eigplus{\exp\biggl[\int_{\alpha_s(Q^2)}^{\alpha_s(Q_0^2)}
\!\!{\tilde\gamma^0_{gg}
(\alpha_s(q^2)/N) \over\alpha^2_s(q^2)}d\,\alpha_s(q^2)\biggr]}

\def\eigplusq{\exp\biggl[\int_{\alpha_s(q^2)}^{\alpha_s(Q_0^2)}\!\!
{\tilde\gamma^0_{gg}
(\alpha_s(r^2)/N)\over \alpha^2_s(r^2)}d\,\alpha_s(r^2)\biggr]}

\def\eigvecplus{g_0(N)+\fourninths f^S_{0}(N)}

\def\eigpluscor{\exp\biggl[\int_{\alpha_s(Q^2)}^{\alpha_s(Q_0^2)}\!\!
{\tilde\gamma^0_{gg}
(\alpha_s(q^2)/N) \over\alpha^2_s(q^2)} + {\tilde
\gamma^1_{gg}(\alpha_s(q^2)/N)+ \fourninths \tilde
\gamma^1_{fg}(\alpha_s(q^2)/N)\over\alpha_s(q^2)}
d\,\alpha_s(q^2)\biggr]}

\def\eigpluscorp{\exp\biggl[\int_{\alpha_s(Q^2)}^{\alpha_s(Q_0^2)}\!\!
{\tilde\Gamma^0_{LL}
(\alpha_s(q^2)/N) \over\alpha^2_s(q^2)} + {\tilde
\Gamma^1_{LL}(\alpha_s(q^2)/N)-\biggl({36-8N_f\over 27}\biggr)
 \tilde\Gamma^1_{2L}(\alpha_s(q^2)/N)\over \alpha_s(q^2)}
d\,\alpha_s(q^2)\biggr]}

\def\eigcomiC{\exp\biggl[\int_{\alpha_s(Q^2)}^{\alpha_s(Q_0^2)}\!\!{\tilde
\gamma^1_{gg}(\alpha_s(q^2)/N)+ \fourninths \tilde
\gamma^1_{fg}(\alpha_s(q^2)/N)\over \alpha_s(q^2)}
-{d\over d\alpha_s(q^2)}\biggl(\ln\biggl({C^g_{L,1}(\alpha_s(q^2)/N)\over
C^g_{L,1,0}}\biggl)\biggl)d\,\alpha_s(q^2)\biggr]}

\def\lneigcomiC{\int_{\alpha_s(Q^2)}^{\alpha_s(Q_0^2)}\!\!{\tilde
\gamma^1_{gg}(\alpha_s(q^2)/N)+ \fourninths \tilde
\gamma^1_{fg}(\alpha_s(q^2)/N)\over \alpha_s(q^2)}
-{d\over d\alpha_s(q^2)}\biggl(\ln\biggl({C^g_{L,1}(\alpha_s(q^2)/N)\over
C^g_{L,1,0}}\biggl)\biggl)d\,\alpha_s(q^2)}

\def\eigveccor{\int_{\alpha_s(Q^2)}^{\alpha_s(Q_0^2)}
\!\!{(\tilde \gamma^1_{gg}(\alpha_s(q^2)/N)+ 
\fourninths\tilde\gamma^1_{fg}(\alpha_s(q^2)/N))
\over \alpha_s(q^2)}d\alpha_s(q^2)}

\def\lsim{\mathrel{mathpalette\@versim<}}
\def\gsim{\mathrel{mathpalette\@versim>}}
\def\higherorder{\hbox{\rm higher order in $\alpha_s$ and/or $N$}}

\catcode`@=11 
\def\slash#1{\mathord{\mathpalette\c@ncel#1}}
 \def\c@ncel#1#2{\ooalign{$\hfil#1\mkern1mu/\hfil$\crcr$#1#2$}}
\def\lsim{\mathrel{\mathpalette\@versim<}}
\def\gsim{\mathrel{\mathpalette\@versim>}}
 \def\@versim#1#2{\lower0.2ex\vbox{\baselineskip\z@skip\lineskip\z@skip
       \lineskiplimit\z@\ialign{$\m@th#1\hfil##$\crcr#2\crcr\sim\crcr}}}
\catcode`@=12 

\def\PR{{\it Phys.~Rev.~}}
\def\PRL{{\it Phys.~Rev.~Lett.~}}
\def\NP{{\it Nucl.~Phys.~}}
\def\PL{{\it Phys.~Lett.~}}

\def\SJNP{{\it Sov.~Jour.~Nucl.~Phys.~}}
\def\ZP{{\it Zeit.~Phys.~}}

\def\vyp#1#2#3{#1 (#2) #3}

\def\Asl{\raise.15ex\hbox{/}\mkern-11.5mu A}
\def\psl{\lower.12ex\hbox{/}\mkern-9.5mu p}
\def\qsl{\lower.12ex\hbox{/}\mkern-9.5mu q}
\def\rsl{\lower.03ex\hbox{/}\mkern-9.5mu r}
\def\ksl{\raise.06ex\hbox{/}\mkern-9.5mu k}
\def\Mev{\hbox{MeV}}


\pageno=0\nopagenumbers\tolerance=10000\hfuzz=5pt
\line{\hfill RAL-96-065}
\vskip 36pt
\centerline{\bf A Complete Leading--Order, Renormalization--Scheme--Consistent 
Calculation}
\vskip 6pt
\centerline{\bf of Small--$x$ Structure Functions, 
Including Leading--$\ln (1/x)$ Terms.}
\vskip 36pt
\centerline{Robert~S.~Thorne}
\vskip 12pt
\centerline{\it Rutherford Appleton Laboratory,}
\centerline{\it Chilton, Didcot, Oxon., OX11 0QX, U.K.}
\vskip 0.9in
{\narrower\baselineskip 10pt
\centerline{\bf Abstract}
\medskip
We present calculations of the structure
functions ${\cal F}_2(x,Q^2)$ and ${\cal F}_L(x,Q^2)$, concentrating on  
small $x$. After discussing the standard expansion of the structure 
functions in powers of $\alpha_s(Q^2)$
we consider a leading--order expansion in $\ln (1/x)$ and finally
an expansion which is leading order in both $\ln(1/x)$ and $\alpha_s(Q^2)$,
and which we argue is the only really correct expansion scheme. 
Ordering the calculation in a renormalization--scheme--consistent manner, 
there is no factorization scheme dependence, as there should not be
in calculations of physical quantities. The calculational method
naturally leads to the ``physical anomalous dimensions'' of Catani, but
imposes stronger constraints than just the use of these effective 
anomalous dimensions. In particular, a relationship between the small--$x$ 
forms of the inputs ${\cal F}_2(x,Q_0^2)$ and ${\cal F}_L(x,Q_0^2)$ 
is predicted. Analysis of a wide range of data for ${\cal F}_2(x,Q^2)$
is performed, and a very good global fit obtained, particularly for data at 
small $x$. The fit allows a prediction for ${\cal F}_L(x,Q^2)$ to be produced, 
which is smaller than those produced by the usual NLO--in--$\alpha_s(Q^2)$ 
fits to ${\cal F}_2(x,Q^2)$ and different in shape.}   

\vskip 0.7in
\line{RAL-96-065\hfill}
\line{January 1997\hfill}
\vfill\eject
\footline={\hss\tenrm\folio\hss}



\newsec{Introduction.}

The recent measurements of ${\cal F}_2(x,Q^2)$ at HERA have provided data on
a structure function at far lower values of $x$ than any previous
experiments, and show that there is a marked rise in ${\cal F}_2(x,Q^2)$ at
very small $x$ down to rather low values of $Q^2$ \ref\hone{H1 collaboration,
\NP \vyp{B470}{1996}{3}.}\ref\zeus{ZEUS collaboration: M. Derrick {\it et al},
\ZP \vyp{C69}{1996}{607}\semi preprint DESY 96--076 (1996), 
\ZP C, in print.}. Indeed, the most recent
measurements have demonstrated that the rise persists for values of
$Q^2$ as low as 1.5 $\Gev^2$.

\medskip

The qualitative result of a steep rise at small $x$ 
initially led to surprise in many quarters. The main
reason for this was that standard methods used to fit
the data were based on the solution of the Altarelli--Parisi evolution
equation \ref\apeqn{G. Altarelli and G. Parisi, \NP \vyp{B126}{1977}{298}\semi
Yu.L. Dokshitzer, {\it Sov. Jour. JETP} \vyp{46}{1977}{641}\semi
L.N.Lipatov, \SJNP \vyp{20}{1975}{95}\semi V.N. Gribov and L.N. Lipatov,
\SJNP \vyp{15}{1972}{438}.}
(along with convolution with coefficient functions) at the
two--loop level, using flat input parton distributions at starting
scales of $Q_0^2 \sim 4 \Gev^2$ (e.g. \ref\mrsi{A.D. Martin, R.G. Roberts and
W.J. Stirling, \PR \vyp{D47}{1993}{867}; \PL \vyp{B306}{1993}{145}.}).
This method
followed the reasoning that steep behaviour can only come about from
perturbative physics (the Donnachie--Landshoff pomeron used to
describe soft physics has behaviour $x^{-0.08}$
\ref\dlp{A. Donnachie and P.V. Landshoff, \NP \vyp{B244}{1984}{322};
\NP \vyp{B267}{1986}{690}.}, and we will take
steep to mean any powerlike behaviour steeper than this), and that a starting
scale for perturbative evolution should be high enough 
for $\alpha_s(Q^2)$ to be fairly small ($\lsim 0.3$) and to avoid any
significant corrections from ``higher twist'' ($\Lambda_{QCD}^2/Q^2$)
corrections. This procedure results in an
effectively steep\foot{We use the term effectively steep since the
slope at any $x$ and $Q^2$ is a function of these variables 
rather than a constant,
i.e. ${\cal F}_2(x,Q^2) \sim x^{-\lambda}$ where $\lambda \sim 0.7\ln\bigl(
{\ln(Q^2/\Lambda^2)\over\ln(Q_0^2/\Lambda^2)}\bigr)^{\half}
\ln^{-\half}(0.1/x)$, and is smaller than any power of $x$ as $x\to 0$.} 
behaviour at small $x$ \ref\wilc{A. de R\'ujula {\it et al}, \PR 
\vyp{D10}{1974}{1649}.}, but only after a long evolution length, 
and therefore at values of $Q^2 \gg 4 \Gev^2$.
 
Thus, the data  led to a degree of optimism amongst those
advocating an alternative description of small--$x$ structure
functions, i.e. using the BFKL equation \ref\bfkl{L.N. Lipatov, \SJNP 
\vyp{23}{1976}{338}\semi E.A. Kuraev, L.N. Lipatov and V.S. Fadin {\it Sov.
Jour. JETP} \vyp{45}{1977}{199}\semi Ya. Balitskii and L.N. Lipatov, \SJNP
\vyp{28}{1978}{6}.}. This equation provides the
unintegrated gluon Green's function which includes the leading power of
$\ln (1/x)$ for any power of $\alpha_s$ (where $\alpha_s$ is taken to
be fixed). It was traditionally solved analytically in the asymptotic
limit $x \to 0$, or numerically for finite $x$, and predicted a
powerlike behaviour of $x^{-1-\lambda}$ for the gluon distribution function,
where $\lambda = 4 \ln 2 \bar \alpha_s$ and $\bar \alpha_s =(3/\pi)
\alpha_s$, i.e. $\lambda \sim 0.5$ if $\alpha_s \sim 0.2$. This was assumed to
lead to ${\cal F}_2(x,Q^2)$ behaving like $x^{-\lambda}$ and could be  
claimed to be
in rough qualitative agreement with the data, even if $\lambda$ was somewhat
high. It could also be seen as some justification for choosing
powerlike inputs (with $\lambda \sim 0.2-0.3$) for the parton
distributions (e.g. \ref\mrsii{A.D Martin, R.G. Roberts and W.J. Stirling,
\PL \vyp{B354}{1995}{155}\semi H.L. Lai {\it et al}, \PR  
\vyp{D51}{1995}{4763}.}), which could then enable a good fit to the 
data using the Altarelli--Parisi equation. 

However, it was also convincingly demonstrated that it was possible to
generate the observed steep behaviour by being a little less conservative
concerning the region in which perturbative evolution could be
applied. Gl\"uck, Reya and Vogt had in fact predicted a sharp rise 
in $F_2(x,Q^2)$ at small $x$, even for $Q^2 \sim 1\Gev^2$, by using 
two--loop evolution from roughly valence--like parton distributions at a 
starting scale of $Q_0^2 = 0.34 \Gev^2$ \ref\grv{M. Gl\"uck, E. Reya and 
A. Vogt, \ZP \vyp{C48}{1990}{471}; \ZP \vyp{C53}{1992}{127}; \PL 
\vyp{B306}{1993}{391}; \ZP \vyp{C67}{1995}{433}.}.  
This attracted criticism not only on the grounds that higher twist 
corrections should be important in the region of
evolution, but also because $\alpha_s(Q^2)$ was as high as $\sim 0.5$ at
the lower end of the range, and therefore perturbation theory itself
should be questionable. Starting at a higher scale,
$Q_0^2 = 1 \Gev^2$, Ball and Forte were able to fit the small--$x$ data
using their double asymptotic scaling (DAS) formula \ref\bfi{R.D. Ball and
S. Forte, \PL \vyp{B335}{1994}{77}.}, which is a simple, but
very accurate, approximation to the solution of the
one--loop evolution equation with flat\foot{Parton inputs behaving like 
$x^{-1}$, as opposed to $x^{-1-\lambda}$, are referred to as flat, i.e. the 
parton density rescaled by $x$ is flat.} inputs and which is valid in the 
region of small $x$ ($x \lsim 0.01$). They also showed that
two--loop evolution with flat inputs starting at $Q_0^2 \approx 2
\Gev^2$ could fit the data available at that time very
well \ref\bfii{R.D. Ball and S. Forte, \PL \vyp{B336}{1994}{77}.}, 
and had a similar shape to the
DAS result (as indeed it must, since both fit the data). 

This state of affairs clearly left scope for argument about the real
underlying physics describing the small--$x$ behaviour of
${\cal F}_2(x,Q^2)$. Those who used Altarelli--Parisi evolution from small
scales could be accused firstly of working in regions where perturbation
theory was questionable and, perhaps more importantly, of ignoring
terms of higher order in $\alpha_s$ (but also higher order in $\ln
(1/x))$, which seemed from the BFKL equation to have very important
effects. Conversely, those who used the BFKL approach could be accused
of ignoring all but the leading--$\ln (1/x)$ terms (and hence ignoring
the large--$x$ data) and also of working
in a less well--defined theoretical framework than the renormalization
group approach based on the factorization of collinear
singularities \ref\collfac{J.C. Collins, D.E. Soper and G. Sterman, in: 
Perturbative Quantum Chromodynamics, ed. A.H. Mueller (World Scientific,
Singapore, 1989), and references therein.}. Starting from an input for 
the parton
distributions with $\lambda \sim 0.25$ at values of $Q_0^2 \sim
4\Gev^2$ was taking the best
of both worlds. However, this lacked a real justification 
for the choice
of input, which was significantly steeper than that expected from 
non--perturbative physics, but rather smaller than that from the BFKL equation,
and also ignored potentially important $\ln (1/x)$ terms in
the evolution. 

A significant step forward in the investigation of small--$x$ structure 
functions was the development of the
$k_T$--factorization theorem \ref\kti{S. Catani, M. Ciafaloni and 
F. Hautmann, \PL \vyp{B242}{1990}{97}; \NP \vyp{B366}{1991}{135}; \PL
\vyp{B307}{1993}{147}.}\ref\ktii{J.C. Collins and R.K. Ellis, \NP 
\vyp{B360}{1991}{3}.}. This is the prescription for the way in
which an off--shell photon--gluon scattering amplitude can be
convoluted with the unintegrated gluon Green's function calculated
using the BFKL
equation to provide the small--$x$ structure functions themselves
(once convoluted with a bare, off--shell gluon density). Hence, it
enables one to find effective moment--space coefficient functions (or
in the case of ${\cal F}_2(x,Q^2)$ and massless quarks, 
where $d\,{\cal F}_2(x,Q^2)/d \ln
Q^2$ is calculated, a mixture of coefficient function and anomalous
dimensions) within the BFKL framework. Numerical calculations performed
using this method were able to match the available data in a 
qualitative manner \ref\akms{A.J. Askew, J. Kwieci\'nski, A.D. Martin and 
P.J. Sutton, \PR \vyp{D49}{1994}{4402}\semi A.J. Askew {\it et al}, \PL 
\vyp{B325}{1994}{212}.}, as did similar calculations 
\ref\ccfmphen{J. Kwieci\'nski, 
A.D. Martin and P.J. Sutton, \PR \vyp{D53}{6094}{1996}; \ZP  
\vyp{C71}{1996}{585}.} using a modification of the
BFKL equation, i.e. the CCFM equation \ref\ccfm{M. Ciafaloni, \NP  
\vyp{B296}{1988}{49}\semi S. Catani, F. Fiorani and G. Marchesini, \PL 
\vyp{B234}{1990}{339}; \NP \vyp{B336}{1990}{18}.}. 

However, the $k_T$--factorization formula was also shown to be very important
if one insisted on working within the rigorous framework of the 
traditional renormalization group approach \ref\cathaut{S. Catani and 
F. Hautmann, \PL \vyp{B315}{1993}{157}; \NP \vyp{B427}{1994}{475}.}. In
this approach the infrared poles in the calculated coefficient
functions, or photon--parton scattering amplitudes, are removed order
by order in $\alpha_s$ and absorbed into
the essentially nonperturbative parton distribution functions. 
In order to guarantee the independence of physical quantities on the
factorization scale $\mu_F$, the
latter then evolve in $\ln (\mu_F^2)$ according to the renormalization group
equations governed by calculable perturbative anomalous dimensions. By
showing how $k_T$--factorization fits within the collinear
factorization framework, Catani and Hautmann were able to calculate
all the renormalization group anomalous dimensions to lowest
nontrivial order in $\alpha_s$ for each power of $\ln (1/x)$, and similarly
for a number of coefficient functions. 

In order to explain this properly we digress for a moment. We may
write the Altarelli--Parisi splitting functions as
\eqn\fullsp{P(x,\alpha_s(Q^2))= {1 \over x}
\sum_{n=1}^{\infty}\alpha_s^n(Q^2)\biggl(\sum_{m=0}^{n-1} a_{nm} {\ln^m(1/x)
\over m!}+
\hbox{\rm terms less singular as $x \to 0$}\biggr),}
or taking the Mellin transform of the splitting function (weighted by $x$)
\eqn\melltrans{\gamma(N,\alpha_s(Q^2)) = \int_0^1\,x^N P(x,\alpha_s(Q^2))dx,}
and working in moment space,
\eqn\fullad{\gamma(N,\alpha_s(Q^2)) = \sum_{n=1}^{\infty}
\alpha_s^n(Q^2)\Bigl(\sum_{m=1}^n a_{nm}  N^{-m} +
a_n(N)\Bigr),}
where $A(N)$ is regular as $N \to 0$. In the normal loop expansion
one solves the evolution equations
order by order in $\alpha_s$. We may however write 
\eqn\fullspalt{P(x,\alpha_s(Q^2))= {1\over x}
\sum_{n=0}^{\infty}\alpha_s^n(Q^2)\biggl(\sum_{m=1}^{\infty} a_{nm} 
\alpha_s^m(Q^2) {\ln^{m-1}(1/x)\over m!}\biggr)+
\hbox{\rm terms less singular as $x \to 0$},}
or in moment space, 
\eqn\fulladalt{\gamma(N,\alpha_s(Q^2)) = \sum_{n=0}^{\infty}
\alpha_s^n(Q^2)\sum_{m=1}^{\infty} \tilde a_{nm} 
\alpha_s^m(Q^2) N^{-m} +
\sum_{n=1}^{\infty}\alpha_s^n(Q^2) a_n(N),}
where the $a_n(N)$ are regular as $N \to 0$, but have singularities at
negative integer values of $N$. When one calculates the
structure function by converting back to $x$ space (i.e. performing
the inverse Mellin transform),
these singularities lead to contributions suppressed by powers of $x$, 
and so are
negligible at small $x$. Therefore, ignoring these contributions and expanding
the $a_n(N)$ about $N=0$ we get
\eqn\fulladalti{\gamma(N,\alpha_s(Q^2)) = \sum_{n=0}^{\infty}
\alpha_s^n(Q^2)\sum_{m=1-n}^{\infty} \tilde a_{nm} 
\alpha_s^m(Q^2) N^{-m}\equiv
\sum_{n=0}^{\infty}\alpha_s^n(Q^2)\gamma^n(\alpha_s(Q^2)/N) ,}
where the expressions are strictly convergent only for $\vert N \vert <1$.
Thus, in order to include the leading powers in $\ln (1/x)$ for the
splitting function, we have to take the $n=0$ part of the expression
\fulladalti\ for the anomalous dimension. Next to leading order (NLO) in
$\ln (1/x)$ is the $n=1$ part, and so on. 

Using the $k_T$--factorization formula Catani and Hautmann demonstrated
that within the renormalization group framework $\gamma^0_{gg}(N,Q^2)$
and $\gamma^0_{gf}(N,Q^2)$ were the same 
renormalization--scheme--independent expressions as the effective anomalous
dimensions \ref\jaro{T. Jaroszewicz, \PL \vyp{B116}{1982}{291}.} given
by the BFKL equation. i.e. 
\eqn\gammadef{\gamma^0_{gg}(\alpha_s(Q^2)/N) = 
\sum_{m=1}^{\infty} a_{0,m} \biggl(
{\bar \alpha_s(Q^2) \over N}\biggr)^m, \hskip 0.6in
\gamma^0_{gf}(\alpha_s(Q^2)/N)= {4 \over 9} \gamma^0_{gg}(\alpha_s(Q^2)/N),}
where $\gamma^0_{gg}(\alpha_s(Q^2)/N)$ is given by the iterative solution of 
\eqn\chieqn{1={\bar\alpha_s(Q^2) \over N} \chi(\gamma^0_{gg}),}
and,
\eqn\chifunc{\chi(\gamma) = 2\psi(1) -\psi(\gamma) -\psi(1-\gamma).}
This solution as a power series in $\bar \alpha_s(Q^2)/N$
exists only for $\vert N\vert \geq \lambda(\bar
\alpha_s(Q^2))$. The anomalous dimension develops a branch point at $N
= \lambda(\bar\alpha_s(Q^2))$, but the series in \gammadef\ is analytic
and convergent outside the circle $\vert N \vert =
\lambda(Q^2)$. In fact, this anomalous dimension leads to the
gluon distribution function having an asymptotic powerlike behaviour
of $x^{-\lambda(Q_0^2)}$, so it does not make sense to talk about its
Mellin transform for $\Re e N \leq \lambda(Q_0^2)$,
it simply does not exist in this region of moment space. However, the
inverse Mellin transform may be performed by analytic continuation
into the region $\Re e N < \lambda(Q_0^2)$ (but $\vert N\vert \geq
\lambda(Q_0^2)$) from $\Re e N  \geq \lambda(Q_0^2)$, if so desired. 

Catani and Hautmann also derived expressions for
$\gamma^1_{ff}(\alpha_s(Q^2)/N)$ and $\gamma^1_{fg}(\alpha_s(Q^2)/N)$ 
in certain
factorization schemes ($\gamma^0_{ff}(\alpha_s(Q^2)/N)$ and
$\gamma^0_{fg}(\alpha_s(Q^2)/N)$ being zero), and also for
the coefficient functions 
$C^g_{L,1}(\alpha_s(Q^2)/N)$, $C^f_{L,1}(\alpha_s(Q^2)/N)$, 
$C^g_{2,1}(\alpha_s(Q^2)/N)$ and $C^f_{2,1}(\alpha_s(Q^2)/N)$ (all
zeroth--order quantities being zero except $C^f_{2,0}$, which is
unity). This facilitated calculations of structure functions within
the normal renormalization group framework, but including much of what
is often called the BFKL physics (i.e. the leading--$\ln (1/x)$ terms),
and indeed, a number of calculations were performed \ref\bfresum{R.D. Ball 
and S. Forte, \PL \vyp{B351}{1995}{313};
\PL \vyp{B358}{1995}{365}.}\nref\ehw{R.K. Ellis, F. Hautmann and B.R. Webber,
\PL \vyp{B348}{1995}{582}.}\nref\frt{J.R. Forshaw, R.G. Roberts and 
R.S. Thorne, \PL \vyp{B356}{1995}{79}.}--\ref\brv{J. Bl\"umlein, S. 
Riemersma and A. Vogt, Proc. of the International Workshop {\sl QCD
and QED in higher orders}, Rheinsburg, Germany, April, 1996, 
eds. J. Bl\"umlein, F. Jegerlehner and T. Riemann, \NP B (Proc. Suppl.),
51C (1996) p. 30; {\tt hep-ph/9607329}, 
Proc. of the International Workshop on Deep Inelastic Scattering,
Rome, April, 1996, in print.}, and in most cases 
comparisons with data made. These calculations used 
different methods of solution,
made rather different assumptions and used different ans\"atze for unknown
terms. Consequently different results were obtained. The conclusions
which could be drawn regarding the inclusion of the leading--$\ln
(1/x)$ terms depended very much on which of the approaches was
taken. However, it seemed that by including these terms
it was not possible to
improve upon the best fits for the small--$x$ data using one-- or
two--loop evolution from soft inputs \bfresum\frt.
Indeed, many ways of including them made the
fits significantly worse, and this seemed to be universally true if
the fits were more global, i.e. constrained by large--$x$ data \bfresum.
Also, it seemed that there was a very strong
dependence on the factorization scheme used to perform the
calculations when including the leading--$\ln (1/x)$ terms \bfresum\frt
\ref\sdis{S. Catani, \ZP \vyp{C70}{1996}{263}.}\ref\qzero{M. Ciafaloni,
\PL \vyp{B356}{1995}{74}.}, and a
number of new factorization schemes were invented, e.g. the SDIS scheme \sdis\
and the $Q_0$ scheme \qzero. 

\medskip

The high precision of the most recent HERA data constrains theory far more 
than previously, and has changed the above picture somewhat. The best 
recent global fits seem to come from those intermediate approaches which 
use NLO perturbation theory with a quite steep (and completely unexplained) 
input for the singlet quark with 
$\lambda \sim 0.2$ and a similar form of small--$x$ input for the 
gluon \ref\mrsiii{A.D. Martin, W.J. Stirling and R.G. Roberts, 
\PL \vyp{B387}{1996}{419}.} (unless $Q_0^2$ is less than  
$\sim 4\Gev^2$, in which case the gluon must be flatter or even valence--like).
Fixed order perturbation theory using 
flat or valence--like inputs and low $Q_0^2$ fails
at the lowest $x$ values, and for fits to the small--$x$ data alone relatively
steep inputs for
the singlet quark, i.e. $\lambda \gsim 0.2$, seem to be absolutely necessary 
\ref\steep{R.D. Ball and S. Forte, {\tt hep-ph/9607289},
Proc. of the International 
Workshop on Deep Inelastic Scattering, Rome, April, 1996, in print.}.
Moreover, approaches including the leading--$\ln (1/x)$ terms now seem to 
fail \ref\faili{S. Forte and R.D. Ball, {\tt hep-ph/9607291},
Proc. of the International Workshop 
on Deep Inelastic Scattering, Rome, April, 1996, in print.}\ref\failii{I. 
Bojak and M. Ernst, {\tt hep-ph/9609378}, preprint DO--TH 96/18, 
September 1996.} in practically all factorization schemes.  

In this paper we will take issue with the above conclusions. 
In particular we will
demonstrate that the apparent failure of approaches
using the leading--$\ln (1/x)$ terms, and certainly the factorization scheme
dependence, is due to incorrect methods of incorporating these terms.
Indeed, Catani has already shown how to obtain
factorization--scheme--invariant results in the small--$x$ expansion by
writing evolution equations in terms of the physical quantities, the
structure functions and ``physical anomalous dimensions'', rather than 
in terms of parton densities and of the usual anomalous dimensions \ref\cat{S. 
Catani, talk at UK workshop on HERA physics, September 1995, unpublished;
{\tt hep-ph/9609263}, preprint DDF 248/4/96, April 1996; {\tt hep-ph/9608310},
Proc. of the International 
Workshop on Deep Inelastic Scattering, Rome, April, 1996, in print.}. 
In this paper we will go further and show that the
correct leading--order, renormalization--scheme--consistent (RSC) calculation 
of the structure functions must naturally include some leading--$\ln (1/x)$ 
terms in the form of these ``physical anomalous dimensions'' . It also 
provides limited predictive power at small $x$, giving justification 
for the input ${\cal F}_2(x,Q_0^2)$. We will 
discuss this method of calculation, then make detailed
comparisons to data, and demonstrate with the aid of the new HERA data 
that it leads to a very good global fit to all 
${\cal F}_2(x,Q^2)$ data. Indeed, the complete RSC calculation, including 
leading--$\ln (1/x)$ terms, is clearly preferred by the latest data, 
particularly that at small $x$. 

We note that a very brief presentation of the complete RSC calculation 
of structure functions has already appeared in \ref\mylet{R.S. Thorne,
{\tt hep-ph/9610334}, \PL B, in print.}.\foot{As in 
this shorter presentation, the present paper deals with the heavy
quark thresholds in a rather naive manner, i.e. the quarks are taken to be
massless, with a particular flavour becoming active only above a certain
$Q^2$.} However, this current paper gives a complete, and very detailed 
discussion of the correct calculation of structure functions, as well as 
examples of commonly encountered pitfalls,
whereas there is only the barest outline of the full calculation in \mylet.
Moreover, there is a far more comprehensive presentation of the comparison 
with experimental data, and with alternative approaches, in the current paper.

This paper will be structured as follows: after giving a brief 
outline of the different possible types of scheme
dependence in the calculation of structure functions, we 
also quickly review the work of Catani, 
illustrating that it is indeed very easy to 
guarantee factorization--scheme--invariant results. 
We then give a very detailed
description of the calculation, within moment space, 
of structure functions in various
expansion schemes using the normal parton language.
The first part of this, regarding the normal loop expansion, will be 
largely a review, but will highlight points usually not discussed in detail,
particularly the role played by the scale at which the parton distributions 
are input and the form of the inputs. The second part, 
discussing the leading--$\ln (1/x)$ expansion, will 
present the only correct way to perform this expansion. One finds that
by remedying the factorization scheme dependence one is forced towards
an alternative derivation of Catani's factorization--scheme--independent
anomalous dimensions. However, we will also show that, in order to make an
ordered calculation of quantities, more care is needed than simply working
with a factorization--scheme--independent set of
variables, and demonstrate that there is a certain degree of predictive power
for the form of the inputs for the structure functions at small $x$.
We also explain why the standard solutions using the small--$x$
expansions are strongly factorization scheme dependent.
To conclude this section we present the argument that there is a unique
renormalization--scheme--consistent calculation of structure functions, 
which applies to both large
and small $x$. We present this calculation for both the currently 
academic case of the nonsinglet
structure functions and for the phenomenologically important case of the 
singlet structure functions. We then discuss
how we move from moment space and
obtain our $x$--space solutions, and the qualitative form 
these solutions must take, i.e. our best attempt at predictions.
After this long theoretical presentation we consider the comparison with
experiment. We fit the data for ${\cal F}_2(x,Q^2)$ 
using the renormalization--scheme--consistent solutions,
and compare to global fits at NLO using the normal loop expansion. We also 
comment on alternative fits to the data made by other groups, and on
determinations of $\alpha_s$. We conclude 
that the full renormalization--scheme--consistent calculation gives the best 
global fit to structure function data, particularly at small $x$. 
Finally we investigate the phenomenological consequences for 
${\cal F}_L(x,Q^2)$, and also preliminary indications for the charm structure 
function. Good measurements of either (but preferably both)
of these quantities, particularly at small $x$, would help determine whether
the approach developed in this paper, and hence the inclusion of 
leading--$\ln(1/x)$ terms in the calculation of structure functions, 
is indeed correct.  

\newsec{Scheme and Scale Choices.}

For simplicity we work in moment--space for much of this paper, i.e. 
define the moment--space structure functions by the Mellin transform, i.e.
\eqn\melltranssf{F(N,Q^2)= \int_0^1\,x^{N-1}{\cal F}(x,Q^2)dx.}
The moment--space coefficient function is defined similarly but, as
with the definition of the anomalous dimension, we define the moment
space expression for the parton distribution as the Mellin transform
of a rescaled parton density i.e 
\eqn\melltranspd{f(N,Q^2)= \int_0^1\,x^{N}{\rm f}(x,Q^2)dx.}
Let us consider the most
general moment--space expression for a structure function, i.e. the sum of the
products of the expressions for hard scattering with a certain
parton (the coefficient functions) with the corresponding,
intrinsically nonperturbative parton distributions.
\eqn\strcfunc{F(N,Q^2)= \sum_a
C^a(N,\alpha_s(\mu_R^2),Q^2/\mu_F^2, \mu_R^2/\mu_F^2)
f_a(N,\alpha_s(\mu_R^2),\mu_R^2/\mu_F^2),}
$\mu_F$ is a factorization scale separating the ultraviolet physics
from the soft infrared physics, and on which the left--hand side of
\strcfunc\ is independent. $\mu_R$ is the renormalization scale, i.e. the 
appropriate scale at which to define the coupling constant, and on 
which $F(N,Q^2)$ is again independent. 

Despite being intrinsically nonperturbative,
the parton distributions evolve according to the perturbative renormalization
group equation
\eqn\apeqn{{d \, f_a(\alpha_s(\mu_R^2),\mu_R^2/\mu_F^2) 
\over d \ln \mu_F^2}= \sum_b
\gamma_{ab}(\alpha_s(\mu_R^2),\mu_R^2/\mu_F^2) 
f_b(\alpha_s(\mu_R^2),\mu_R^2/\mu_F^2).}

In principle there are many choices to be made when performing
a perturbative calculation of structure functions to a finite order. 
However, two of the choices left open in the above expressions are 
common to all perturbative calculations in quantum field theory:
the choice of renormalization scheme and subsequently the
choice of renormalization scale. When one
removes ultraviolet divergences from perturbative
calculations, there is always a freedom in what type of regularization is
used and/or how much of the finite part of a calculation is removed at
the same time as the divergent parts. Although the all orders
calculation of a physical quantity is independent of the convention,
the renormalization scheme used, the perturbative expansion of the
quantity is not. Hence, neither is the definition of the expansion
parameter, the running coupling constant. All
couplings in quantum field theory satisfy a renormalization group
equation,
\eqn\runcoup{{d \,\alpha (\mu^2) \over d 
\ln(\mu^2)}=-\sum_{n=0}^{\infty}b_n\alpha^{n+2}(\mu^2)\equiv 
-\beta(\alpha(\mu^2)),}
where $\mu$ is the scale at which the coupling is defined, and the
convention of the minus sign is introduced in order to make the 
$\beta$--function for QCD positive. The
solution to this equation depends on the coefficients $b_n$ and on a
scale $\Lambda$. This scale, and the value of the coefficients
beyond $n=1$ are renormalization scheme dependent ($b_0 =
(11-2N_f/3)/4\pi$ and $b_1 = (102-38N_f/3)/16\pi^2$ in all
renormalization schemes which use dimensional regularization, 
where $N_f$ is the number of quark flavours). 
Hence, a choice of
renormalization scheme amounts to a choice of which expansion
parameter is to be used, and consequently the form of the perturbative
expansion. One would hope the choice would be such as to make the
series converge as quickly as possible, but this is difficult to
guarantee. Conventionally the $\overline{\hbox{\rm MS}}$ renormalization
scheme is used.  
Also the choice of the appropriate scale to be used in the coupling for
a given process must be made. Sometimes this is relatively clear, but
in deep inelastic scattering it is not entirely obvious. Traditionally,
the simple choice $\mu_R = \mu_F$  is taken, and indeed, the data seem to
favour this choice \ref\mrsiv{A.D. Martin, R.G. Roberts and W.J. Stirling, \PL
\vyp{B266}{1991}{173}.}. Hence, we will assume this to be the case from now
on. 

Having settled for a particular choice of renormalization scheme and
renormalization scale the ambiguities due to ultraviolet regularization 
have been dealt with. Our fundamental equations become
\eqn\strcfunci{F(N,Q^2)= \sum_a
C^a(N,\alpha_s(\mu_F^2),Q^2/\mu_F^2)f_a(N,\mu_F^2),}
and
\eqn\apeqni{{d \, f_a(\mu_F^2) \over d \ln \mu_F^2}= \sum_b
\gamma_{ab}(\alpha_s(\mu_F^2)) f_b(\mu_F^2),}
where $\alpha_s$ is defined in a particular renormalization scheme. 
The remaining
ambiguities are due to the particular problems in calculating 
quantities in QCD, i.e separating the physical quantity into the
perturbative coefficient function and the intrinsically
nonperturbative parton distribution. To begin with we have the freedom
of choosing the factorization scale $\mu_F$. As with renormalization scheme
dependence this does not affect the all--orders calculation, but does
affect the form of the perturbative expansion, since it affects the
scale at which the coupling in the coefficient function is
evaluated. One might imagine that it is desirable to choose $\mu_F^2$
to be large in order to make the expansion parameter
$\alpha_s(\mu_F^2)$ as small as possible, and indeed $\mu_F^2$ is
nearly always chosen to be equal to the hard scattering scale $Q^2$. We 
shall also make this simple choice.

This leaves us with our defining equations\foot{Because we consider only 
massless quarks, once we set $Q^2=\mu_F^2$ the only scale dependence in 
the coefficient functions comes from $\alpha_s(Q^2)$.}
\eqn\strcfuncii{F(N,Q^2)= 
\sum_a C^a(N,\alpha_s(Q^2)) f_a(N,Q^2),}
and
\eqn\apeqni{{d\, f_a(Q^2) \over d \ln Q^2}=
 \sum_b\gamma_{ab}(\alpha_s(Q^2))  f_b(Q^2).}
In principle there are now choices to be made for the scale at which to 
begin the evolution of the parton densities, $Q_0^2$, and for the inputs for 
the partons at this scale. This question will be dealt with in detail later 
in this paper, so we will leave it for the moment.
This still leaves us one more freedom in our calculation,
i.e. how we choose to remove the infrared divergences from the bare 
coefficient functions and hence how we define our parton distributions. 
Starting from any particular choice for the definition of parton 
distributions one may always choose a new set of
parton distributions by an invertible transformation
\eqn\transpart{\breve f_a(N,Q^2) =
\sum_b U_{ab}(N,\alpha_s) f_b(N,Q^2),}
where $ U_{ab}(N,\alpha_s)$ has a power series expansion in
$\alpha_s$ such that $U_{ab}(N,\alpha_s)=\delta_{ab}+ {\cal
O}(\alpha_s)$. The structure functions will clearly be unchanged
as long as the coefficient functions obey the transformation rule
\eqn\coefftrans{\breve C^a(N,\alpha_s) =
(U^T)^{-1}_{ab}(N,\alpha_s)C^b(N,\alpha_s).}
By substituting $f_a(N,Q^2) =
\sum_b U^{-1}_{ab}(N,\alpha_s) \breve f_b(N,Q^2)$ into \apeqni\ we
easily find that 
the new parton densities evolve according to the standard evolution
equations but with the new anomalous dimensions
\eqn\transgamma{\breve \gamma_{ab}(N, \alpha_s) = \sum_c\sum_d 
U_{a,c}(N,\alpha_s)\gamma_{cd}(N,\alpha_s)U^{-1}_{db}(N,\alpha_s) +
\sum_c \beta(\alpha_s){\partial   U_{ac}(N,\alpha_s) \over \partial \alpha_s}
 U^{-1}_{cb}(N,\alpha_s).}
The matrix $U$ must obey a number of conditions in order that physical
requirements on the parton distributions are maintained, e.g. flavour
and charge conjugation invariance, fermion number conservation and
longitudinal momentum conservation (see for example the second of \cathaut).
However, none of these needs to be
satisfied simply in order to keep the structure functions unchanged. 

The transformation defined above is called a change of factorization scheme
and is constructed precisely so that the
physical structure functions are left invariant. However, 
it is important to realize that, unlike the changes in renormalization
scheme, a change in factorization scheme leaves the expression for the
structure functions unchanged not only to all orders, but 
order by order in $\alpha_s$, and
calculations performed carefully at a given well--defined order in
one scheme will lead to precisely the same results for the
structure functions as those in another scheme. Also, the coupling constant to 
be used depends only on the ultraviolet renormalization.\foot{It is perfectly
possible (if somewhat perverse) to choose the $\overline{\hbox{\rm MS}}$
scheme to remove ultraviolet divergences, but the ${\hbox{\rm MS}}$ scheme
to remove infrared divergences from the bare coefficient functions. In this
case it is the $\overline{\hbox{\rm MS}}$ scheme coupling constant that 
must be used.} We will illustrate
this point in \S 4, where we will calculate structure functions to a
well--defined order in a variety of expansion schemes and demonstrate
factorization scheme independence of our expressions. However, we will
first illustrate Catani's recent proposal for the construction of
factorization--scheme--independent structure functions. 

\newsec{Evolution Equations for Structure Functions.}

It is, as Catani noticed \cat, very simple to obtain 
factorization--scheme--independent expressions for the structure 
functions, or more
precisely, factorization--scheme--independent effective anomalous
dimensions governing the evolution of the structure functions
(reflecting the fact that we cannot make a concrete prediction of the
structure functions at a given scale using QCD, but only how they change
with $Q^2$). In order to obtain these effective  anomalous dimensions all one
has to do is eliminate the parton densities from the equations
\strcfuncii\ and \apeqni. 

In order to demonstrate this,
let us now be a little more careful in our definition of the relevant
structure functions. There are two independent structure functions 
$F_2(N,Q^2)$ and $F_L(N,Q^2)$. In general we may write
\eqn\srucdef{F_i(N,Q^2) =
{1\over N_f}\biggl(\sum_{j=1}^{N_f}e^2_j\biggr)F^S_i(N,Q^2) +
F^{NS}_{i}(N,Q^2) \hskip 0.5in (i=2,L),}
where the singlet and nonsinglet structure functions are defined by 
\eqn\fsing{F^S_i(N,Q^2) = C^f_i(N,\alpha_s)f^S(N,Q^2) +
C^g_i(N,\alpha_s)g(N,Q^2),}
and
\eqn\fnonsing{F^{NS}_i(N,Q^2) = C^{NS}_i(N,\alpha_s)\sum_{j=1}^{N_f}
e_j^2 f^{NS}_{q_j}(N,Q^2),}
where $N_f$ is the number of active quark flavours, $f^S(N,Q^2)$ and
$f^{NS}_{q_j}(N,Q^2)$ are the singlet and nonsinglet quark
distribution functions respectively, and $g(N,Q^2)$ is the gluon distribution.
In any factorization scheme obeying the requirements of  
flavour and charge conjugation invariance the renormalization group
equations for the singlet and nonsinglet sectors will be
decoupled. The equations for the nonsinglet distributions will be ordinary
differential equations,
\eqn\nonsingevol{{d \, f^{NS}_{q_j}(N,Q^2) \over d \ln Q^2} =
\gamma_{NS}(N,\alpha_s)f_{q_j}^{NS}(N,Q^2),}
while those for the singlet sector are coupled
\eqn\singevol{ {d\over d \ln Q^2} \pmatrix{f^S(N,Q^2)\cr g(N,Q^2)\cr} =
\pmatrix{\gamma_{ff}(N,\alpha_s) &\gamma_{fg}(N,\alpha_s)\cr 
\gamma_{gf}(N,\alpha_s) & \gamma_{gg}(N,\alpha_s)\cr}
\pmatrix{f^S(N,Q^2)\cr g(N,Q^2)\cr}.}
\medskip 

Let us first consider the simple case of the nonsinglet
structure function $F_2^{NS}(N,Q^2)$. 
Multiplying both sides of \nonsingevol\ by
$\sum_{j=1}^{N_f}e_j^2$ we can clearly write 
\eqn\nonsingfsi{{{d   \over 
d \ln Q^2} \Bigl({F_2^{NS}(N,Q^2)\over C^{NS}_2(N,\alpha_s)}\Bigr)
=\gamma_{NS}(N,\alpha_s){F_2^{NS}(N,Q^2) \over 
C^{NS}_2(N,\alpha_s)},}}
which becomes the factorization--scheme--independent equation
\eqn\nonsingfsi{{{d \, F_2^{NS}(N,Q^2) \over 
d \ln Q^2} =\Gamma_{2,NS}(N,\alpha_s)F_2^{NS}(N,Q^2),}}
where $\Gamma_{2,NS}(N,\alpha_s)=
\gamma_{NS}(N,\alpha_s)+d\,\ln(C^{NS}_2(N,\alpha_s))/d\ln Q^2$.
Therefore, we have
an effective anomalous dimension governing the evolution of each of
the nonsinglet structure functions, and clearly
$\Gamma_i^{NS}(N,\alpha_s)$ must be a factorization--scheme--independent
quantity (and is in principle measurable). The solution to this
equation is trivial:
\eqn\nonsingfsisol{{F_2^{NS}(N,Q^2)  
 =F_2^{NS}(N,Q_0^2)\exp\biggl[\int_{\ln Q_0^2}^{\ln Q^2}
\Gamma_{2,NS}(N,\alpha_s)d\ln q^2\biggr].}}

The situation for the singlet structure functions is more
complicated. As we see from \singevol\ the evolution equations for the
singlet quark density and gluon density are coupled. However, using
\fsing\ for $i=2,L$ we may solve for the parton densities in terms of
the two structure functions and the coefficient
functions. Substituting these into \singevol\ we then obtain the
coupled evolution equations
\eqn\singevolfsi{ {d\over d \ln Q^2} \pmatrix{F^S_2(N,Q^2)\cr 
F^S_L(N,Q^2)\cr} =
\pmatrix{\check\Gamma_{22}(N,\alpha_s) &\check\Gamma_{2L}(N,\alpha_s)\cr 
\check\Gamma_{L2}(N,\alpha_s) & \check\Gamma_{LL}(N,\alpha_s)\cr}
\pmatrix{F^S_2(N,Q^2)\cr F^S_L(N,Q^2)\cr}.}
The expressions for the physical anomalous dimensions,
$\check\Gamma_{22}(N,\alpha_s)$, $\check\Gamma_{2L}(N,\alpha_s)$,
etc., are straightforward to derive 
in terms of anomalous dimensions and coefficient functions in any
particular factorization scheme using
the above procedure, but result in rather cumbersome expressions.
It is simplest first to
define a factorization scheme such that $F^S_2(N,Q^2)= f^S(N,Q^2)$,
i.e. $C^f_2(N,\alpha_s)=1$, $C^g_2(N,\alpha_s)=0$ (this is generally
known as a DIS type scheme \ref\dis{G. Altarelli, R.K. Ellis and 
G. Martinelli, \NP \vyp{B157}{1979}{461}.}\foot{We call it a DIS 
``type'' scheme
because satisfying the above requirement still leaves freedom in how we
may define the gluon density, and thus we are still considering a
family of schemes.}). In terms of the coefficient functions and
anomalous dimensions in this type of scheme we have
\eqn\physanom{\eqalign{\check\Gamma_{22}(N,\alpha_s)&=
\gamma_{ff}(N,\alpha_s)
-{C^f_L(N,\alpha_s)\over
C^g_L(N,\alpha_s)}\gamma_{fg}(N,\alpha_s),\cr
\check\Gamma_{2L}(N,\alpha_s)&=
{\gamma_{fg}(N,\alpha_s)\over C^g_L(N,\alpha_s)},\cr
\check\Gamma_{L,2}(N,\alpha_s)&= C^g_L(N,\alpha_s)\gamma_{gf}(N,\alpha_s)-
C^f_2(N,\alpha_s)\gamma_{gg}(N,\alpha_s)+{d
C^f_L(N,\alpha_s)\over d \ln
Q^2}+C^f_L(N,\alpha_s)\gamma_{ff}(N,\alpha_s)\cr
&-C^f_L(N,\alpha_s){d \ln(C^g_L(N,\alpha_s))\over
d\ln Q^2} -{(C^f_L(N,\alpha_s))^2\over C^g_L(N,\alpha_s)}
\gamma_{fg}(N,\alpha_s),\cr
\check\Gamma_{LL}(N,\alpha_s)&=\gamma_{gg}(N,\alpha_s) + 
{d \ln(C^g_L(N,\alpha_s))\over d \ln
Q^2}+ {C^f_L(N,\alpha_s)\over C^g_L(N,\alpha_s)}\gamma_{fg}(N,\alpha_s).\cr}}

Thus, we have a factorization--scheme--invariant set of anomalous
dimensions governing the evolution of the structure functions. Before
going any further let us remark that we believe there is a 
(purely technical) problem
with the above expression. As is well known (and as we will discuss in the
next section), $F_L(N,Q^2)$ starts at an order of $\alpha_s$ higher than
$F_2(N,Q^2)$. Because of this there is an intrinsic asymmetry in the
above definitions, with $\check\Gamma_{2L}(N,\alpha_s)$ beginning at zeroth
order in $\alpha_s$, $\check\Gamma_{22}(N,\alpha_s)$ and 
$\check\Gamma_{LL}(N,\alpha_s)$
beginning at first order, and $\Gamma_{L2}(N,\alpha_s)$ beginning at
second order. The fact that one of the physical anomalous dimensions has a 
part at zeroth order in $\alpha_s$
seems against the spirit of the perturbative approach. In practice,
the result is that, if one solves the evolution equations order by
order in $\alpha_s$ the solutions are very different from those obtained
from an order--by--order solution using the parton densities, which we
know work very well for all but possibly the lowest values of $x$
yet probed. (We will discuss this point rather more in \S 4.)  

A trivial modification of Catani's approach is therefore to accept 
that $F_L(N,Q^2)$ contains an extra power of
$\alpha_s$, and to define the new structure function $\hat
F_L(N,Q^2)=F_L(N,Q^2)/(\alpha_s/(2\pi))$. The longitudinal coefficient 
functions are likewise changed to $\hat
C^a_L(N,\alpha_s)=C^a_L(N,\alpha_s)/(\alpha_s/(2\pi))$, and the
singlet evolution equations become 
\eqn\singevolfsimod{ {d\over d \ln Q^2} \pmatrix{F^S_2(N,Q^2)\cr 
\hat F^S_L(N,Q^2)\cr} =
\pmatrix{\Gamma_{22}(N,\alpha_s) &\Gamma_{2L}(N,\alpha_s)\cr 
\Gamma_{L2}(N,\alpha_s) & \Gamma_{LL}(N,\alpha_s)\cr}
\pmatrix{F^S_2(N,Q^2)\cr \hat F^S_L(N,Q^2)\cr},}
where the $\Gamma(N,\alpha_s)$'s are defined precisely as in \physanom, but
in terms of of $\hat C^a_L(N,\alpha_s)$ rather than $C^a_L(N,\alpha_s)$. 
This procedure
restores the symmetry between the physical anomalous dimensions, and
makes the order--by--order--in--$\alpha_s$ calculations 
essentially the same as when using
evolution of parton distributions. It is, of course, trivial to obtain the
physical $F_L(N,Q^2)$ from $\hat F_L(N,Q^2)$.\foot{Similarly, 
it is also 
best to work with the rescaled nonsinglet structure function $\hat
F^{NS}_L(N,Q^2)$. Doing this, the analogous procedure to
\nonsingfsi\ and \nonsingfsisol\ leads to $\hat F^{NS}_L(N,Q_0^2)$
beginning at zeroth order in $\alpha_s(Q_0^2)$, as does
$F^{NS}_2(N,Q_0^2)$.}  

Having made our redefinition of the quantities with which we work, we
now have a direct relationship between possible calculations using the
evolution equations for structure functions and the solutions using the parton
densities. At present the parton anomalous
dimensions and coefficient functions are known to order $\alpha_s^2$. It 
is easy to see that this allows us to derive each of the $\Gamma$'s to order
$\alpha_s^2$. For example, at first order in $\alpha_s$ they are 
\eqn\physanomone{\eqalign{\Gamma^{0,l}_{22}(N,\alpha_s)&=
\gamma^{0,l}_{ff}(N,\alpha_s)
-{\hat C^S_{L,0,l}(N,\alpha_s)\over
\hat C^g_{L,0,l}(N,\alpha_s)}\gamma_{fg}^{0,l}(N,\alpha_s),\cr
\Gamma_{2L}^{0,l}(N,\alpha_s)&={\gamma_{fg}^{0,l}(N,\alpha_s)
\over \hat C^g_{L,0,l}(N,\alpha_s)},\cr
\Gamma_{L,2}^{0,l}(N,\alpha_s)&= \hat C^g_{L,0,l}(N,\alpha_s)
\gamma_{gf}^{0,l}(N,\alpha_s)-
\hat C^S_{2,0,l}(N,\alpha_s)\gamma_{gg}^{0,l}(N,\alpha_s)
+\hat C^f_{L,0,l}(N,\alpha_s)
\gamma_{ff}^{0,l}(N,\alpha_s)\cr
&-{(\hat C^S_{L,0,l}(N,\alpha_s))^2\over \hat C^g_{L,0,l}(N,\alpha_s)}
\gamma_{fg}^{0,l}(N,\alpha_s),\cr
\Gamma_{LL}^{0,l}(N,\alpha_s)&=\gamma^{0,l}_{gg}(N,\alpha_s)  
+ {\hat C^S_{L,0,l}(N,\alpha_s)\over \hat C^g_{L,0,l}(N,\alpha_s)}
\gamma_{fg}^{0,l}(N,\alpha_s),\cr}}
where the super--subscript $n,l$ denotes the $(n+1)$--loop quantity. The
${\cal O}(\alpha_s^2)$ expressions for the physical anomalous
dimensions are straightforward to derive, but are rather complicated.
Similarly, from the known expansions of the parton
anomalous dimensions and coefficient functions in the form
$\alpha_s^n\sum_{m=1-n}^{\infty}a_m(\alpha_s/N)^m$, we can calculate 
$\Gamma_{LL}^0(N,\alpha_s)$ and $\Gamma^0_{L2}(N,\alpha_s)$,
$\Gamma_{2L}^0(N,\alpha_s)$ and $\Gamma_{22}^0(N,\alpha_s)$ 
(where both are trivially zero), and $\Gamma_{2L}^1(N,\alpha_s)$ 
and $\Gamma_{22}^1(N,\alpha_s)$.
This is the same order as
for the parton anomalous dimensions, with the longitudinal anomalous
dimensions having similar structure to the gluon anomalous dimensions and the 
$\Gamma_{2a}(N,\alpha_s)$'s having similar structure to the quark 
anomalous dimensions. The
exact form of these physical anomalous dimensions may be calculated using the
expressions for the parton anomalous dimensions and coefficient
functions in the standard DIS scheme in \cathaut. The resulting expressions are
relatively simple, being  
\eqn\physanomval{\eqalign{\Gamma^1_{22}(\alpha_s/N)&= 
-{1\over(2\pi)}(\hat C^{f}_{L,1,0}-
\fourninths \hat C^{g}_{L,1,0})
(\threeovertwo \gamma^0_{gg}(\alpha_s/N)+
\sum_{n=0}^{\infty}(\gamma^0_{gg}(\alpha_s/N))^n)\cr
&-\fourninths\gamma^{1,0}_{fg}(N,\alpha_s),\cr
\Gamma^1_{2L}(\alpha_s/N)&={1\over(2\pi)}(\threeovertwo 
\gamma^0_{gg}(\alpha_s/N)+
\sum_{n=0}^{\infty}(\gamma^0_{gg}(\alpha_s/N))^n),\cr
\Gamma^0_{L,2}(N,\alpha_s)&= -(\hat C^{f}_{L,1,0}-
\fourninths \hat C^{1,0}_{L,g})\gamma^0_{gg}(\alpha_s/N),\cr
\Gamma^0_{LL}(\alpha_s/N)&=\gamma^0_{gg}(\alpha_s/N),\cr}}
where $\gamma^{1,0}_{fg}$, $C^{f}_{L,1,0}$ and
$C^{g}_{L,1,0}$ are the one--loop contributions to 
$\gamma^{1}_{fg}(\alpha_s/N)$, $C^{f}_{L,1}(\alpha_s/N)$ and
$C^{g}_{L,1}(\alpha_s/N)$ respectively. We note that each of these
anomalous dimensions is renormalization scheme invariant as well as
factorization scheme invariant, as we would expect for leading--order 
physical quantities.   

Solving the evolution equations for the structure functions using any
subset of the currently known physical anomalous dimensions guarantees a
result which is factorization scheme independent. However, since the
evolution equations are coupled, a simple expression for the solutions,
such as \nonsingfsisol, is impossible. There is, in fact,
considerable freedom in how we may solve the
equations. We could, for example, simply put all of the anomalous
dimensions currently known into \singevolfsimod\ and then 
find the whole solution. 
Alternatively, we could solve using just the order $\alpha_s$ anomalous
dimensions and then try to perturb about this solution in an ordered
manner. These two approaches would lead to rather different answers, but 
both would be factorization scheme independent. The
problem of obtaining a correctly ordered solution for the structure
function will be discussed in detail in the next section. We will
initially use the familiar parton distributions and coefficient
functions, and show that, even when
using this approach, if we solve producing a well--defined expansion 
for the structure functions, we automatically avoid the problem of 
factorization scheme dependence.

\newsec{Ordered Calculations of Structure Functions.}

There are in principle many different expansion methods one may use when
obtaining solutions for the structure functions. The standard one is
simply solving order by order in
$\alpha_s$. But there is also the expansion in leading powers of $\ln
(1/x)$ for given powers in $\alpha_s$ (or equivalently in powers of 
$N^{-1}$ in moment space), as we
have already discussed. One can also combine the two expansion
methods, and indeed, we will later argue that this is the correct thing to do.
Nevertheless, we will begin by outlining the procedure for making
a well--ordered
calculation of structure functions using the standard loop
expansion. Although this is well known, we feel it is worth presenting
it pedagogically, and making some points which are not usually highlighted,
especially concerning the role of the starting scale. 
We may then discuss the more complicated cases
of the expansion in leading powers of $\ln (1/x)$ and the combined expansion. 

\subsec{Loop Expansion.}

We begin by introducing some new notation.
In order to solve the evolution equations for the parton densities
order by order in $\alpha_s$ and
hence obtain expressions for $F^{NS}_i(N,Q^2)$ and $F^S_i(N,Q^2)$ we
make use of equation \runcoup\ to rewrite the evolution equation for the
nonsinglet parton density as
\eqn\nonsingevolal{\alpha^2_s(Q^2) {d \, f^{NS}_{q_j}(N,Q^2) \over d 
\alpha_s(Q^2)} =- \tilde \gamma_{NS}(N,\alpha_s)f_{q_j}^{NS}(N,Q^2),}
where $\tilde \gamma_{NS}(N,\alpha_s) =
\alpha^2_s(Q^2)\gamma_{NS}(N,\alpha_s)/\beta(\alpha_s)$. Similarly
\eqn\singevolal{ \alpha^2_s(Q^2) {d\over d \alpha_s(Q^2)} 
\pmatrix{f^S(N,Q^2)\cr g(N,Q^2)\cr} =-
\pmatrix{\tilde \gamma_{ff}(N,\alpha_s) &\tilde \gamma_{fg}(N,\alpha_s)\cr 
\tilde\gamma_{gf}(N,\alpha_s) & \tilde\gamma_{gg}(N,\alpha_s)\cr}
\pmatrix{f^S(N,Q^2)\cr g(N,Q^2)\cr},}
with similar definitions for $\tilde\gamma_{ff}(N,\alpha_s)$ etc. as for
$\tilde \gamma_{NS}(N,\alpha_s)$. Each of the $\tilde \gamma$'s may now be
written as
\eqn\gamtilde{\tilde \gamma(N,\alpha_s) = \sum_{n=0}^{\infty}\tilde
\gamma^{n,l}(N) \, \alpha_s^{n+1},}
where we also have the analogous definition
\eqn\gamnorm{\gamma(N,\alpha_s) = \sum_{n=0}^{\infty}
\gamma^{n,l}(N) \, \alpha_s^{n+1},}
for the normal anomalous dimensions.\foot{We use the superscript $l$ in 
order to denote that
this is the standard $\alpha_s$ expansion of the anomalous dimensions rather
than the expansion in \fulladalti, i.e. the leading--$\ln (1/x)$
expansion, which will be used more frequently in this paper.}
Thus, at zeroth order the $\tilde \gamma(N)$'s only differ from the 
$\gamma(N)$'s by the normalization factor $b_0$. Beyond this order the 
differences are more complicated.
 
Using our definition, the evolution equations may be solved order by order in
$\alpha_s$. When doing this it is necessary to choose a starting
scale $Q_0^2$ for the perturbative evolution of the parton distributions 
(or equivalently a starting value of the coupling $\alpha_s(Q_0^2)$),
and specify input parton distributions at this scale.
Let us discuss the choice of this scale briefly.
$Q_0^2$ must clearly be chosen to be large enough that perturbative
evolution should be reliable and also such that higher twist 
corrections should be
very small. Traditionally evolution only takes place up from this starting 
scale. This has been both
for the simple reason of convenience, and also because some form of the 
inputs has been expected at low starting scales. Until a few years ago
the requirements described above led to a choice of 
$Q_0^2 \approx 4 \Gev^2$. In the past couple of years this value has 
often been chosen to be rather
lower, due to the apparent success of evolution from lower starting scales,
and also because much of the interesting small--$x$ data is now at $Q^2
\leq 4 \Gev^2$. These choices have been accompanied by guesses for the form of
the inputs at low starting scales, e.g valence--like \grv, or
flat \mrsi\bfi. However, with the quality of the most recent data, 
these guesses for the form of the inputs no 
longer seem to lead to good quantitative agreement with data.

We make no assumptions of the above sort about the value of $Q_0^2$. We
do require it to be high enough to be in the perturbative regime and to 
avoid higher twists, but acknowledge that  
there is no reason why $Q_0^2$ cannot be chosen to be
quite large, and evolution away from the starting scale 
performed both up and down in $Q^2$. Taking this open--minded approach we 
then simply assume that only perturbative effects can lead to deviations 
from soft behaviour of the structure functions, and also 
demand that the form of our well--ordered expressions for the structure 
functions is as insensitive as possible to this choice, thus making 
the choice of $Q_0^2$ as open as possible. We will see the 
consequences of this unusual
approach to the input scale and the inputs for the 
structure functions as we progress.   

\medskip

We begin by solving for the nonsinglet parton distributions, which are 
an easily understandable model. In this case the solution is
particularly simple. Integrating both sides of \nonsingevolal\ we
obtain
\eqn\nonsingsol{f^{NS}_{q_j}(N,Q^2)=\biggl[\sum_{k=0}^{\infty}\alpha_s^k(Q_0^2)
f^{NS}_{q_j,k}(N, Q_0^2)\biggr]
\biggl({\alpha_s(Q_0^2)\over\alpha_s(Q^2)}\biggr)^{\tilde \gamma^{0,l}_{NS}(N)}
\exp\biggl[\sum_{n=1}^{\infty}{(\alpha^n_s(Q_0^2)-\alpha_s^n(Q^2))\over
n}\tilde\gamma_{NS}^{n,l}(N)\biggr].}

Perhaps unconventionally, we explicitly express the input 
$f^{NS}_{q_j}(N,Q_0^2)$ as a power series in $\alpha_s(Q_0^2)$.
There are two reasons why this is necessary. Firstly, changes in the
starting scale $Q_0^2$ lead to $\alpha_s(Q_0^2)$--dependent changes in 
the expression for the evolution term, 
which must be compensated for by $\alpha_s(Q_0^2)$--dependent
changes in the starting distribution in order to leave 
the whole expression for
the parton distributions unchanged, as required.\foot{Since the structure 
function is obtained by multiplying the parton distributions by the 
$Q_0^2$--independent coefficient functions, the parton
distributions must be $Q_0^2$-independent to make the structure functions 
so.} Let us examine this briefly by looking at the change of the 
lowest--order piece of \nonsingsol, i.e.
\eqn\nonsingsolzero{f^{NS}_{q_j,0}(N,Q^2)=f^{NS}_{q_j,0}(N,Q_0^2)
\biggl({\alpha_s(Q_0^2)\over\alpha_s(Q^2)}\biggr)^{\tilde 
\gamma^{0,l}_{NS}(N)},}
under a change in starting scale, $Q_0^2 \to  (1+\delta) Q_0^2$,
where $\delta$ is some constant. Each of the terms in our expressions for
the structure functions would now be written in the forms shown above,
but as functions of $(1+\delta) Q_0^2$. We may regain expressions in
terms of $Q_0^2$ by expanding the coupling constant 
$\alpha_s((1+\delta) Q_0^2)$ in the form 
\eqn\inscalchange{\alpha_s((1+\delta)Q_0^2)= \alpha_s(Q_0^2)-
\delta b_0\alpha_s^2(Q_0^2) + {\cal O}(\alpha_s^3(Q_0^2)).}
Under this change in the input coupling constant, the evolution term in 
\nonsingsolzero\ undergoes a change
\eqn\zeroevolcha{\Delta \biggl({\alpha_s(Q_0^2)\over\alpha_s(Q^2)}
\biggr)^{\tilde \gamma^{0,l}_{NS}(N)} = -\alpha_s(Q_0^2)
\delta b_0 \tilde \gamma^{0,l}_{NS}(N)
\biggl({\alpha_s(Q_0^2)\over\alpha_s(Q^2)}
\biggr)^{\tilde \gamma^{0,l}_{NS}(N)} + \hbox{\rm higher order in  
$\alpha_s(Q_0^2)$}.} 
This change in the parton distribution due to the variation in the leading 
term can be countered, up to higher orders, by a change in the order 
$\alpha_s(Q_0^2)$ input of the form 
\eqn\chainput{\Delta f^{NS}_{q_j,1}(N,Q_0^2)=\delta b_0 \tilde 
\gamma^{0,l}_{NS}(N) f^{NS}_{q_j,0}(N,Q_0^2).}
Because changes in the evolution
term due to a change in $Q_0^2$ begin at first order in
$\alpha_s(Q_0^2)$, and are therefore absorbed by terms in the input 
at first order and beyond, the zeroth--order input must be insensitive
to such changes and is $Q_0^2$--independent: $f^{NS}_{q_j,0}(N,Q_0^2)\equiv
f^{NS}_{q_j,0}(N)$. It is not difficult to see that
higher--order changes in the parton distributions due to changes in  
$Q_0^2$ can all be accounted 
for by changes in the higher--order inputs and that in order to satisfy this 
constraint alone, it is consistent
for the higher--order inputs to be equal to functions of $N$ dependent 
only on the anomalous dimensions. Therefore, these higher--order inputs are
in some sense perturbative, consisting of perturbative parts
multiplying the fundamentally nonperturbative $f^{NS}_{q_j,0}(N)$. 
As we will soon discuss, there are other constraints to be satisfied, e.g.
factorization scheme independence of the input for the structure function, 
and this slightly complicates the above picture, but does not change the
main conclusions. 

Thus, we see that it is necessary to express the input as a power series in 
$\alpha_s(Q_0^2)$ in order to make the parton distribution 
$Q_0^2$--independent, but that only one intrinsically
nonperturbative input which is $Q_0^2$-independent is needed. 
Usually in analyses of structure functions the parton inputs are 
taken to be a single $\alpha_s(Q_0^2)$--independent function which is 
implicitly allowed to be $Q_0^2$--dependent. 
Phenomenologically, this is normally much the same,
but we stress the formally correct expression for the input here
since it is rather important when constructing properly ordered solutions,
and leads to some predictive power, especially in the small $x$ limit.  
 
Another reason for explicitly writing the input as a power series in 
$\alpha_s(Q_0^2)$ is that
it makes little sense to demand that the starting distribution should
not have a perturbative expansion, unless one insists that there is
something special about a particular factorization scheme. If there is
not, then any transformation from a scheme in which the starting parton
distribution is purely zeroth order in $\alpha_s(Q_0^2)$ will lead to a
starting distribution with a power series expansion in
$\alpha_s(Q_0^2)$, but again, where there seems to be some underlying 
nonperturbative input $f^{NS}_{q_j,0}(N)$ which is unaltered by the change in
factorization scheme.

Accepting that the parton inputs should be a power series in 
$\alpha_s(Q_0^2)$, and substituting our solution
for the parton distribution into \fnonsing, we obtain the general
expression for the nonsinglet structure functions,
\eqn\fnonsingsol{\eqalign{F^{NS}_i(N,Q^2)=
\biggl[\Bigl(\delta_{i,2} +\sum_{m=1}^{\infty}
C^{NS}_{i,m,l}(N)&\alpha_s^m(Q^2)\Bigr)\sum_{j=1}^{N_f}e_j^2
\sum_{k=0}^{\infty}\alpha_s^k(Q_0^2)
f^{NS}_{q_j,k}(N, Q_0^2)\biggr]\cr
&\biggl({\alpha_s(Q_0^2)\over\alpha_s(Q^2)}\biggr)^{\tilde 
\gamma_{NS}^{0,l}(N)}
\exp\biggl[\sum_{n=1}^{\infty}{(\alpha^n_s(Q_0^2)-\alpha_s^n(Q^2))\over
n}\tilde\gamma_{NS}^{n,l}(N)\biggr].\cr}}

It is clear that this may be written as
\eqn\fnonsingsolalt{F_i^{NS}(N,Q^2)=\biggl({\alpha_s(Q_0^2)\over\alpha_s(Q^2)}
\biggr)^{\tilde \gamma_{NS}^{0,l}(N)}
\sum_{n=0}^{\infty}\sum_{m=0}^nF^{NS}_{i,nm}(N,Q_0^2)\alpha^{n-m}_s(Q_0^2)
\alpha_s^m(Q^2),}
and that, once a choice of renormalization scheme and starting scale
have been made, each of the $F^{NS}_{nm}(N)$ must be invariant
quantities under changes of factorization scheme in order to guarantee
the scheme independence of the whole structure function. However, it is also
clear from \fnonsingsol\ that each of the $F^{NS}_{nm}(N)$ is
potentially a function of the starting distribution, the anomalous
dimension and the coefficient function. Any well--ordered calculation of the
structure function should include all complete terms in 
\fnonsingsolalt\ up to a given
order in $n$ and $m$, and no partial terms. In practice it is
possible to work to a given order in $n$ including all $m\leq n$, i.e. to
expand to a given order in powers of $\alpha_s(Q^2)$ plus powers of
$\alpha_s (Q_0^2)$, if
the $\tilde \gamma$'s and $C_i$'s are known to this order. 
(It is impossible to work to a given value of $m$
including higher values of $n-m$ without knowledge of $\tilde \gamma$ 
to order $n-m$, i.e. it is impossible to expand
just in powers of $\alpha_s(Q^2)$.)

We shall briefly describe how to construct this ordered solution for the
structure functions, working up from zeroth order, so that at each order
the solution is factorization scheme dependent. Consider first calculating
$F^{NS}_2(N,Q^2)$ by working
to zeroth order in $C^{NS}_2(N,\alpha_s)$, $\tilde
\gamma_{NS}(N,\alpha_s)$ and the starting
distribution (remembering that this is $Q_0^2$--independent). To this order
\eqn\nonsingsolzero{f^{NS}_{q_j,0}(N,Q^2)=f^{NS}_{q_j,0}(N)
\biggl({\alpha_s(Q_0^2)\over\alpha_s(Q^2)}\biggr)^{\tilde 
\gamma^{0,l}_{NS}(N)},}
and therefore
\eqn\fnonsingsolzero{F^{NS}_{2,0}(N,Q^2)=\sum_{j=1}^{N_f}e^2_j
f^{NS}_{q_j,0}(N)
\biggl({\alpha_s(Q_0^2)\over\alpha_s(Q^2)}\biggr)^{\tilde
\gamma^{0,l}_{NS}(N)}.}
Using the one--loop expression for the running coupling, as is
appropriate for a lowest--order calculation, each of the quantities 
in this expression is factorization scheme
independent and indeed, also renormalization scheme independent.
Therefore we have a consistent leading--order (LO) expression.

If we calculate $F_L^{NS}(N,Q^2)$, again by combining the coefficient
functions with the solutions for the parton evolution,
then working to the order $n=0$ in
all quantities leads to $F^{NS}_{L,0}(N,Q^2)=0$, since the zeroth--order
coefficient function is zero. However, looking at the expressions for
the general solutions, \fnonsingsol\ and \fnonsingsolalt, one sees
that the only contribution for $n=1$ in \fnonsingsolalt\ comes from
working to first order in $C^{NS}_{L}(N,\alpha_s)$ and to zeroth order
in $\tilde \gamma_{NS}(N,\alpha_s)$ and the starting distribution.
This leads to the
LO expression for $F_{L}^{NS}(N,Q^2)$ of
\eqn\fnonsingsolzerolong{F^{NS}_{L,1}(N,Q^2)=\alpha_s(Q^2)C^{NS}_{L,1,l}(N)
\sum_{j=1}^{N_f}e^2_j f^{NS}_{q_j,0}(N)
\biggl({\alpha_s(Q_0^2)\over\alpha_s(Q^2)}\biggr)^{\tilde
\gamma^{0,l}_{NS}(N)}.}
Again, using the one--loop expression for the running coupling, every
term in this expression is both factorization scheme and
renormalization scheme independent, giving a well--defined
LO expression.

We now consider the first correction to these expressions.
The first--order expression for the renormalization group equation is 
\eqn\firstorderrg{\alpha_s(Q^2) {d \, f^{NS}_{q_j,1}(N,Q^2) \over d
\alpha_s(Q^2)} = -\tilde \gamma_{NS}^{0,l}(N,\alpha,s)f^{NS}_{q_j,1}(N,Q^2)
- \alpha_s(Q^2)\tilde\gamma_{NS}^{1,l}(N)f^{NS}_{q_j,0}(N,Q^2),}
with solution
\eqn\firstordsol{f^{NS}_{q_j,1}(N,Q^2)=\Bigl[(\alpha_s(Q_0^2)-\alpha_s(Q^2))
\tilde \gamma_{NS}^{1,l}(N) f^{NS}_{q_j,0}(N)+\alpha_s(Q_0^2)
f^{NS}_{q_j,1}(N,Q_0^2)\Bigr]
\biggl({\alpha_s(Q_0^2)\over\alpha_s(Q^2)}\biggr)^{\tilde 
\gamma^{0,l}_{NS}(N)}.}
Multiplying $f^{NS}_{q_j,0}$ by $\alpha_s(Q^2)C^{NS}_{2,1,l}(N)$ and adding
to $f^{NS}_{q_j,1}$, we clearly obtain all terms in the expression 
\fnonsingsolalt\ for $F^{NS}_{2}(N,Q^2))$ at $n=1$. Adding this to 
\fnonsingsolzero\ we obtain the factorization--scheme--independent
expression for $F_2^{NS}(N,Q^2)$ up to $n=1$.

We note that this is not the same
as finding the complete solution to the renormalization group equation
including all terms in the anomalous dimension up to
first order in $\tilde \gamma$ and multiplying the solution by the
coefficient function up to first order. This procedure would involve
the exponentiation of the anomalous dimension, and thus would include
incomplete parts of the $F_{nm}^{NS}(N)$'s for $n\geq 1$, and would be
a factorization--scheme--dependent, and hence physically ambiguous quantity. 

Similarly to $F^{NS}_{2}(N,Q^2)$ we can obtain the NLO
factorization--scheme--independent expression for $F_L^{NS}(N,Q^2)$. The
expression at $n=2$ is obtained by
adding the product of the first--order coefficient function and
$f^{NS}_{q_j,1}(N,Q^2)$ to the product of the second--order coefficient
function and $f^{NS}_{q_j,0}(N,Q^2)$. The NLO
$F_L^{NS}(N,Q^2)$ is the sum of this and \fnonsingsolzerolong.
When working to NLO for either structure function 
we now have expressions which are
renormalization scheme dependent. This scheme dependence compensates
for the renormalization scheme dependence of the  
two--loop coupling constant (which has a 
renormalization--scheme--dependent value for $\Lambda_{QCD}$), and it is this
expression for the coupling that we should use at this level. Doing so
guarantees the renormalization scheme independence of the structure
functions up to corrections of higher order in $\alpha_s$, i.e.
${\cal O}(\alpha_s^2 F^{NS}_{2(L),0(1)})$.  

It is now simple to see how to construct 
factorization--scheme--independent 
structure functions order by order. Defining the 
$n_{\rm th}$ order renormalization group equation by  
\eqn\nthorderrg{\alpha_s(Q^2) {d \, f^{NS}_{q_j,n}(N,Q^2) \over d
\alpha_s(Q^2)} = -\sum_{m=0}^n\tilde \gamma_{NS}^{m,l}(N,\alpha,s)
f^{NS}_{q_j,n-m}(N,Q^2),}
it is easy to prove by induction that the solution contains all terms
in the full solution with a given sum of powers of $\alpha_s(Q^2)$ and
$\alpha_s(Q_0^2)$ multiplying the everpresent
$(\alpha_s(Q_0^2)/\alpha_s(Q^2))^{\tilde\gamma^{0,l}_{NS}(N)}$ factor. Thus,
defining part of \fnonsingsolalt\ by
\eqn\fnonsingsoln{F_{i,n}^{NS}(N,Q^2)=\biggl({\alpha_s(Q_0^2)
\over\alpha_s(Q^2)}\biggr)^{\tilde \gamma_{NS}^{0,l}(N)}
\sum_{m=0}^nF^{NS}_{i,nm}(N)
\alpha^{n-m}_s(Q_0^2)\alpha_s^m(Q^2),}
$F_{i,n}^{NS}(N,Q^2)$ is given by
\eqn\fnonsingnexp{F_{i,n}^{NS}(N,Q^2)=\sum_{j}^{Nf}e^2_j\sum_{m=0}^n 
C^{NS}_{i,m,l}\alpha_s^m(Q^2)f^{NS}_{q_j,n-m}(N,Q^2).}
The $n_{\rm th}$--order scheme--independent term in the expression for
the structure functions is given by the product of the $m_{\rm th}$--order
coefficient functions and the $(n-m)_{\rm th}$--order solutions to the
renormalization group equation summed over $m$. The $n_{\rm th}$--order
scheme--independent structure function is then the sum
of these terms up to order $n$. We must however
remember that including all $F_{2,m}^{NS}(N,Q^2)$ up to order $n$ is
working to $(n+1)_{\rm th}$ nontrivial order, and requires the $(n+1)$--loop
coupling in order to make the expression renormalization scheme
invariant up to higher orders in $\alpha_s$.
Similarly, including all $F_{L,m}^{NS}(N,Q^2)$ up to order $n$ is
working to $n_{\rm th}$ nontrivial order, and requires the $n$--loop
coupling.

\medskip

This procedure clearly provides factorization scheme independence and 
renormalization scheme independence for this method of expansion.
We can also see how it relates to the discussion of the
factorization--scheme--invariant evolution equations in terms of the
structure functions. In order to do this let us consider the solution
for the non--singlet structure function $F_2^{NS}(N,Q^2)$ again. We
may rewrite our general solution \fnonsingsol\ in the form 
\eqn\fnonsingsol{\eqalign{F^{NS}_2&(N,Q^2)=
\biggl[\Bigl(1 +\sum_{m=1}^{\infty}
C^{NS}_{2,m,l}(N)\alpha_s^m(Q_0^2)\Bigr)\sum_{j=1}^{N_f}e_j^2
\sum_{k=0}^{\infty}\alpha_s^k(Q_0^2)
f^{NS}_{q_j,k}(N, Q_0^2)\biggr]
\biggl({\alpha_s(Q_0^2)\over\alpha_s(Q^2)}
\biggr)^{\tilde \gamma_{NS}^{0,l}(N)}\times\cr
&\hskip -0.2in \exp\biggl[\sum_{n=1}^{\infty}
{(\alpha^n_s(Q_0^2)-\alpha_s^n(Q^2))\over n}\tilde\gamma_{NS}^{n,l}(N)
+\int_{\ln Q_0^2}^{\ln Q^2}{d \over d \ln q^2}
\ln\Bigl(1 +\sum_{m=1}^{\infty}
C^{NS}_{2,m,l}(N)\alpha_s^m(q^2)\Bigr)d\ln q^2\biggr].\cr}}
This way of writing $F^{NS}_2(N,Q^2)$ is particularly useful since it
separates the solution into the value of the structure function
at $Q_0^2$ (the term in square brackets), and the ratio of its value at
a different $Q^2$ with this initial value (the rest of the expression). 
Clearly these two quantities must be separately 
factorization scheme independent. Also, it is clear that this solution
is of exactly the same form as \nonsingfsisol, and we can express it
simply in terms of an input for $F^{NS}_2(N,Q^2)$ at $Q_0^2$ and a
physical anomalous dimension which governs the evolution, i.e. our
input and our prediction. Since these are both physical quantities they must
be separately renormalization scheme independent. 

The input and 
the evolution will mix with each other if we make a change in the starting 
scale, however, as with the parton distribution. Examining the effects of 
such a change for the full physical quantity gives us information about 
the form of the input. As we have already seen, at lowest order the 
nonsinglet structure function is just the sum of the charge weighted 
nonsinglet parton distributions. So the lowest--order input is just
\eqn\trivinp{F^{NS}_{2,0}(N)=\sum_{j=1}^{N_f}e_j^2f^{NS}_{q_j,0}(N).}
Making the change of starting scale and consequently of $\alpha_s(Q_0^2)$ 
already considered, the change in the lowest--order evolution is as in 
\zeroevolcha, and this leads to a change in the lowest--order structure 
function which is of higher order, and which can be absorbed by a change in 
the NLO input for the structure function of
\eqn\chainputsf{\Delta F^{NS}_{2,1}(N,Q_0^2)=\delta b_0 \tilde 
\gamma^{0,l}_{NS}(N) F^{NS}_{2,0}(N).}
In terms of parton distributions and coefficient functions 
\eqn\nonsinginone{F^{NS}_{2,1}(N,Q_0^2)=\sum_{j=1}^{N_f}e^2_j
(f_{q_j,1}^{NS}(N,Q_0^2)+C_{2,1,l}^{NS}(N)f_{q_j,0}^{NS}(N)).}
We chose the change in $f_{q_j,1}^{NS}(N,Q_0^2)$ in \chainput\ so that the 
structure function would be independent of $Q_0^2$, and it is clear that 
that is consistent with \chainputsf\ and \nonsinginone. However, we can now 
say more about the form of $f_{q_j,1}^{NS}(N,Q_0^2)$. Because it is 
the leading 
term in the expression for the input which depends on $\alpha_s(Q_0^2)$, 
$F_{2,1}^{NS}(N,Q_0^2)$ must be renormalization scheme independent. However,
$C_{2,1,l}^{NS}(N)$ is renormalization scheme dependent, so 
$f_{q_j,1}^{NS}(N,Q_0^2)$ must also be renormalization scheme dependent in a 
way such as to cancel this. 
Hence, $f_{q_j,1}^{NS}(N,Q_0^2)$ must not only have a
part like $\ln(Q_0^2)\gamma_{NS}^{0,l}(N)f_{q_j,0}^{NS}(N)$ in order to 
maintain $Q_0^2$--independence of the structure function, but also a 
part like $-C_{2,1,l}^{NS}(N)
f_{q_j,0}^{NS}(N)$ in order to maintain renormalization scheme 
independence, i.e.
\eqn\addon{f^{NS}_{q_j,1}(N,Q_0^2)=(\ln(Q_0^2/A_{NS})\gamma^{0,l}_{NS}(N)-
C^{NS}_{2,1,l}(N))f^{NS}_{q_j,0}(N),}
where $A_{NS}$ is some unknown scale parameter. So we see that
\eqn\addoni{F^{NS}_{2,1}(N,Q_0^2)=\ln(Q_0^2/A_{NS})\gamma^{0,l}_{NS}(N)
F^{NS}_{2,0}(N)\equiv \ln(Q_0^2/A_{NS})\Gamma^{0,l}_{NS}(N)
F^{NS}_{2,0}(N).}
It is 
clear that this does not spoil our argument that $f_{q_j,1}^{NS}(N,Q_0^2)$ 
consists of perturbatively calculable quantities multiplying
the fundamentally nonperturbative $f_{q_j,0}^{NS}(N)$. It is also clear
that all the higher--order inputs may be chosen to be perturbative functions 
multiplying this nonperturbative input, and therefore that the input for
the structure function is a perturbative power series 
(depending on the physical anomalous dimension) multiplying the single
nonperturbative factor $f_{q_j,0}^{NS}(N)$, which from \trivinp\ may be 
interpreted as a fundamentally nonperturbative input for the 
structure function $F_{2,0}^{NS}(N)$. Hence, demanding invariance of our 
expression for the structure function under changes in $Q_0^2$ leads us to
a power series expression for the input, but with only one (for each quark)
really nonperturbative factor for this input.  We also see that if
$Q_0^2=A_{NS}$, the first--order perturbative correction 
to $F_2^{NS}(N,Q_0^2)$ vanishes. Hence, we might expect $A_{NS}$
to be some scale typical of the transition between perturbative and 
nonperturbative physics, i.e $A_{NS} \lsim 1\Gev$.  

Separating the expression for the structure function into a definite
input and evolution part also enables us to view
the loop expansion in an alternative manner. We see that when
expanding out to $n_{\rm th}$ order in the loop expansion we are including
all terms where the order of the input part for the structure function
added to the order of the evolution part of the structure function
is less than or equal to $n$. This is clearly the same
as including all powers of $\alpha_s$ in the expression for the structure
function up to $n_{\rm th}$ order, but gives us some additional physical 
interpretation.  

Writing the solution as in \fnonsingsol\ does, however, also illustrate that
demanding factorization scheme invariance does not on its own force us
into the strictly defined loop expansion. 
It is clear that we could, if we wished,
expand the input and evolution term out to different orders in $\alpha_s$, 
still maintaining factorization scheme independence.  However, this is
not a sensible approach for reasons of renormalization scheme dependence. 
If we were to expand out the input and evolution terms to
different orders in $\alpha_s$ we should really use $\alpha_s$ itself
calculated to a different order in each case, surely a
perverse thing to do. Also, when using the resulting expression to
evaluate the structure function at some $Q^2$ away from $Q_0^2$ we
would only have a subset of the terms at some given power of
$\alpha_s$ (where $\alpha_s$ may represent $\alpha_s(Q_0^2)$ or
$\alpha_s(Q^2)$). Each of these terms is presumably of
similar magnitude, which is rather importantly equal to, or greater
than, the magnitude of the 
uncertainty due to renormalization scheme dependence.
This is even the case even if we simply use the
product of the input specified to $n_{\rm th}$ order and the evolution
calculated to $n_{\rm th}$ order. In this case there is no problem in what
definition of $\alpha_s$ to use: we simply use the $n$--loop expression
for the input and evolution and hence the $n$--loop coupling for both.
However, the resulting expression
contains, for example, a term like
$\alpha_s^{n-1}(Q_0^2)(\alpha_s^{n-1}(Q_0^2)-\alpha_s^{n-1}(Q^2))
F^{NS}_{2,0}(N,Q^2)$,
but none like $\alpha_s^{n-1-m}(Q_0^2)(\alpha_s^{n-1+m}(Q_0^2)-
\alpha_s^{n-1+m}(Q^2))F^{NS}_{2,0}(N,Q^2)$ which should be of the same size.
One only has a complete set of terms of order $\alpha_s^k$ up to $k=n-1$.
Moreover, this type of term is higher order 
than the overall uncertainty in the expression
due to the renormalization scheme dependence, which is of order
$\alpha_s^n F^{NS}_{2,0}(N,Q^2)$. Exponentiating the solution 
for the evolution part, once
it is found to a given order, is also redundant, since that
would introduce only a subset of the terms at higher orders in
$\alpha_s$ for the evolution, and these terms would be
renormalization scheme dependent.

Thus, all ways of obtaining factorization--scheme--invariant
expressions other than the loop expansion contain terms which are
essentially redundant using this expansion method, i.e. those beyond the
order where all complete parts of the loop expansion exist. These 
other terms in the expansion are the same order as uncertainties
due to renormalization scheme dependence. The loop expansion
gives the minimum expression which is totally factorization scheme
independent, and moreover, renormalization scheme consistent (forgetting
complications due to powers of $N^{-1}$ for the present) to a given
order. If solving using the factorization--scheme--invariant equations
in terms of physical variables it is useful to bear this in
mind.\foot{It is also true that, when solving using the parton model, 
other ways of obtaining
factorization--scheme--invariant expressions other than the loop
expansion would need, in comparison, more complicated prescriptions.} 

\medskip

The solution for the longitudinal structure function is much the same as for 
$F^{NS}_{2,0}(N,Q^2)$. We may write $F_L^{NS}(N,Q^2)$ as
\eqn\fnonsingsollong{\eqalign{F^{NS}_L&(N,Q^2)=
\biggl[\Bigl(\sum_{m=1}^{\infty}
C^{NS}_{L,m,l}(N)\alpha_s^m(Q_0^2)\Bigr)\sum_{j=1}^{N_f}e_j^2
\sum_{k=0}^{\infty}\alpha_s^k(Q_0^2)
f^{NS}_{q_j,k}(N, Q_0^2)\biggr]
\biggl({\alpha_s(Q_0^2)\over\alpha_s(Q^2)}
\biggr)^{\tilde \gamma_{NS}^{0,l}(N)-1}\times\cr
&\hskip -0.2in \exp\biggl[\sum_{n=1}^{\infty}
{(\alpha^n_s(Q_0^2)-\alpha_s^n(Q^2))\over n}\tilde\gamma_{NS}^{n,l}(N)
+\int_{\ln Q_0^2}^{\ln Q^2}{d \over d \ln q^2}
\ln\Bigl(\sum_{m=1}^{\infty}
C^{NS}_{L,m,l}(N)\alpha_s^{m-1}(q^2)\Bigr)d\ln q^2\biggr].\cr}}
So as with $F_2^{NS}(N,Q^2)$ the full solution in terms of parton densities,
anomalous dimensions and coefficient functions is easily separated into 
its input\foot{Comparing \fnonsingsollong\ with \fnonsingsol\ we see that 
there are no new fundamentally nonperturbative parts in the input. Hence, 
the complete expression may be interpreted as a perturbative part multiplying
the nonperturbative factor which is $F_{2,0}^{NS}(N)$. It is only the 
perturbative factor which is different in the two nonsinglet structure 
functions.} and evolution parts and therefore directly compared with the 
solution using the effective physical anomalous dimensions (where if we solve 
using the $\tilde \Gamma$s, the solution is expressed in terms of an input 
and evolution part once we write the overall power of $\alpha_s(Q^2)$ 
multiplying $\hat F_L^{NS}(N,Q^2)$ in the form $\alpha_s(Q_0^2)\times(
\alpha_s(Q^2)/\alpha_s(Q_0^2))$.) Once again, from \fnonsingsollong\ we 
see that the loop expansion multiplies orders in the input by orders in 
the evolution in a systematic manner.  

\bigskip

The situation for the singlet structure function is similar to that
for the nonsinglet, but is complicated by the fact that we now
have coupled evolution equations for the quark and gluon
distributions. This makes it impossible to write a closed form for the
solution to the renormalization group equations in the way we did for
the nonsinglet case in \fnonsingsol\ and \fnonsingsollong, but the 
equations may be solved order by order in the same way. 

The lowest--order solution to \singevolal\ is 
\eqn\evolsolsing{\eqalign{ f^{S}_{0}(N,Q^2) &=
f^{S,+}_{0}(N)\biggl({\alpha_s(Q_0^2) \over 
\alpha_s(Q^2)}\biggr)^{\tilde\gamma^{0,l}_+(N)}+
f^{S,-}_{0}(N)\biggl({\alpha_s(Q_0^2) \over 
\alpha_s(Q^2)}\biggr)^{\tilde\gamma^{0,l}_-(N)},\cr
g_{0}(N,Q^2) &=
g^+_{0}(N)\biggl({\alpha_s(Q_0^2) \over 
\alpha_s(Q^2)}\biggr)^{\tilde\gamma^{0,l}_+(N)}+
g^-_{0}(N)\biggl({\alpha_s(Q_0^2) \over 
\alpha_s(Q^2)}\biggr)^{\tilde\gamma^{0,l}_-(N)},\cr}}
where $\tilde \gamma^{0,l}_+(N)$ and $\tilde \gamma^{0,l}_-(N)$ are the
eigenvalues of the zeroth--order matrix for the $\tilde \gamma$'s, and
$f^{S,+}_{0}(N)+f^{S,-}_{0}(N)= f^S_{0}(N)$ and 
$g^+_{0}(N)+g^-_{0}(N)= g_{0}(N)$.  Similarly to the
nonsinglet case, the lowest--order factorization--scheme-- and
renormalization--scheme--independent expressions for the structure
functions are
\eqn\ftwosimp{F^S_2(N,Q^2)=f^S_{0}(N,Q^2),}
and 
\eqn\flsimp{F_L^S(N,Q^2)= 
\alpha_s(Q^2)(C^f_{L,1}(N)f^{S}_0(N,Q^2)+C^g_{L,1}(N)g_0(N,Q^2)).}
Under a change in starting scale, the change of the input for 
$F^S_2(N,Q^2)$ begins at order $\alpha_s(Q_0^2)$, and that for $F_L^S(N,Q^2)$
begins at order $\alpha_s^2(Q_0^2)$. Therefore, 
as with the nonsinglet quark distributions, the lowest--order 
inputs for the partons are $Q_0^2$--independent.
Also, if we expect any powerlike behaviour to come only from
perturbative effects, then these $Q_0^2$--independent
inputs for the quark and gluon are analytic for $N>0$.

Again in clear analogy to the nonsinglet case, the first--order
expression for the renormalization group equations is
\eqn\singevolalone{\eqalign{ \alpha_s(Q^2) {d\over d \alpha_s(Q^2)} 
\pmatrix{f^S_{1}(N,Q^2)\cr g_1(N,Q^2)\cr} &=-
\pmatrix{\tilde \gamma^{0,l}_{ff}(N,\alpha_s) &\tilde \gamma^{0,l}_{fg}
(N,\alpha_s)\cr 
\tilde \gamma^{0,l}_{gf}(N,\alpha_s) & \tilde \gamma^{0,l}_{gg}(N,\alpha_s)\cr}
\pmatrix{f^S_{1}(N,Q^2)\cr g_1(N,Q^2)\cr}\cr
&\hskip 0.6in -\alpha_s(Q^2)
\pmatrix{\tilde \gamma^{1,l}_{ff}(N,\alpha_s) &\tilde \gamma^{1,l}_{fg}
(N,\alpha_s)\cr 
\tilde \gamma^{1,l}_{gf}(N,\alpha_s) & \tilde \gamma^{1,l}_{gg}(N,\alpha_s)\cr}
\pmatrix{f^S_{0}(N,Q^2)\cr g_0(N,Q^2)\cr}\cr}}
and the solution is 
\eqn\singevolsolone{\eqalign{f^S_{1}(N,Q^2)&=
(\alpha_s(Q_0^2)-\alpha_s(Q^2))\Biggl( {\rm
e}^{S,+}_{1}(f^S_{0},g_0,\tilde \gamma^0, \tilde \gamma^1)\biggl(
{\alpha_s(Q_0^2) \over \alpha_s(Q^2)}\biggr)^{\tilde\gamma^{0,l}_+(N)}\cr
&+{\rm e}^{S,-}_{1}(f^S_{0},g_0,\tilde \gamma^0,\tilde\gamma^1)
\biggl({\alpha_s(Q_0^2) \over 
\alpha_s(Q^2)}\biggr)^{\tilde\gamma^{0,l}_-(N)}\Biggr) 
+\alpha_s(Q_0^2)\biggl(f^{S,+}_{1}(N,Q_0^2)\biggl({\alpha_s(Q_0^2) \over 
\alpha_s(Q^2)}\biggr)^{\tilde\gamma^{0,l}_+(N)}\cr
&+f^{S,-}_{1}(N,Q_0^2)\biggl({\alpha_s(Q_0^2) \over 
\alpha_s(Q^2)}\biggr)^{\tilde\gamma^{0,l}_-(N)}\biggr),\cr}}
and similarly for $g_1(N,Q^2)$, where
$f^{S,+}_{1}(N,Q_0^2)+f^{S,-}_{1}(N,Q_0^2)= f^S_{1}(N,Q_0^2)$.

This is clearly of the same form as \firstordsol\ and, as in the
nonsinglet case, the $n=1$ term in the factorization--scheme--independent
expression for $F^S_2(N,Q^2)$, including all single powers of
$\alpha_s(Q^2)$ or $\alpha_s(Q^2_0)$ multiplying
$(\alpha_s(Q_0^2)/\alpha_s(Q^2))^{\tilde \gamma^{0,l}_{+,-}(N)}$, is
\eqn\ftwonsimp{f^{S}_1(N,Q^2) +
\alpha_s(Q^2)(C^f_{2,1}(N)f_{0}^S(N,Q^2)+C^g_{2,1}(N)g_0(N,Q^2)).}
In the
same manner the $n=2$ term in the factorization--scheme--independent
expression for the longitudinal structure function is 
\eqn\flnsimp{\alpha_s(Q^2)
(C^f_{L,1}(N)f_{1}^S(N,Q^2)+C^g_{2,1}(N)g_1(N,Q^2))+\alpha_s^2(Q^2)
(C^f_{L,2}(N)f^{S}_0(N,Q^2)+C^g_{L,2}(N)g_0(N,Q^2)).}

As for the singlet structure function, there
is a simple prescription for calculating $n_{\rm th}$--order
factorization--scheme--independent structure functions. Defining the
$m_{\rm th}$--order solution to the renormalization group equation as the
solution to
\eqn\singevolm{ \alpha_s(Q^2) {d\over d \alpha_s(Q^2)} 
\pmatrix{f_{m}^S(N,Q^2)\cr g_m(N,Q^2)\cr} =-\sum_{k=0}^m
\alpha_s^k(Q^2)
\pmatrix{\tilde \gamma^k_{ff}(N,\alpha_s) &\tilde\gamma^k_{fg}(N,\alpha_s)\cr 
\tilde\gamma^k_{gf}(N,\alpha_s) & \tilde\gamma^k_{gg}(N,\alpha_s)\cr}
\pmatrix{f_{m-k}^S(N,Q^2)\cr g_{m-k}(N,Q^2)\cr},}
the $n_{\rm th}$--order term in the expression for the 
structure functions is given by
\eqn\nsolsf{F^S_{i,n}(N,Q^2)=\sum_{m=0}^n \alpha_s^m(Q^2)(C^f_{i,m}(N)
f^S_{n-m}(N,Q^2)+C^g_{i,m}(N)g_{n-m}(N,Q^2),)}
and the $n_{\rm th}$--order structure function is the sum of all such terms up
to $n$. The comments concerning the order of the renormalization
scheme independence and the order of the coupling constant
to be used for $F^S_2(N,Q^2)$ and $F^S_{L}(N,Q^2)$ for the singlet
case apply in exactly the same way as for the nonsinglet case. Once
again, the loop expansion is a well--ordered way to expand the
structure functions. 

\medskip

We may now discuss the relationship to the solutions using the evolution
equations for the structure functions. First, we can explain in more 
detail why we
feel it is appropriate to take out the power of $\alpha_s(Q^2)$ from
the longitudinal structure function when using this approach. Using
the parton model the lowest--order expression for $F_2^S(N,Q^2)$
consists of two $Q^2$--independent factors, the inputs for the
structure function, multiplying
$(\alpha_s(Q_0^2)/\alpha_s(Q^2))^{\tilde\gamma^{0,l}_{+,-}}$. Similarly,
we may think of the lowest--order expression for $F_L^S(N,Q^2)$ as
being two inputs for the structure function multiplying the
$Q^2$--dependent factors 
$(\alpha_s(Q_0^2)/\alpha_s(Q^2))^{\tilde\gamma^{0,l}_{+,-}-1}$, with the
extra power of $\alpha_s(Q^2)$ coming from the coefficient
function. So, at lowest non--trivial order the $Q^2$ evolution of 
the two structure
functions is different. If we were to solve the evolution equations
for the structure functions themselves at lowest order, then we would do it in
the same way as for the parton distributions: obtaining the
$Q^2$--dependence from the eigenvalues of the lowest--order anomalous
dimension matrix, and the inputs multiplying the two evolution
terms from $F^S_2(N,Q_0^2)$ and $F_L^S(N,Q_0^2)$ and the eigenvectors
of the matrix. Thus, the terms governing $Q^2$ evolution would have to
be the same for both structure functions, which is clearly in
contradiction with our lowest--order result using the parton model
(also, the lowest--order input for $F_L^S(N,Q^2)$ would be zero). This
leads to the solutions for the structure functions being built up in a
very different way using this method of solution from that using the
parton model.

Hence, we choose not to forget the success of the parton model, and
let the longitudinal structure function be multiplied by an overall
power of $\alpha_s(Q^2)$, as discussed earlier.
Working with the $\Gamma(N,\alpha_s)$'s in \physanomone\ and solving the 
lowest--order evolution equations,
one then obtains precisely the same lowest--order solution for the structure
functions as above. It is not difficult to see that, although the 
$\Gamma^{0,l}$'s are not identical to the $\gamma^{0,l}$'s, the
eigenvalues of the anomalous dimension matrix are the same.
Thus, we obtain the factors of
$(\alpha_s(Q_0^2)/\alpha_s(Q^2))^{\tilde\gamma^{0,l}_{+,-}}$ in these
solutions as we would hope, and get the extra factor of $(\alpha_s(Q^2)/
\alpha_s(Q_0^2))$ in the expression for $F_L^S(N,Q^2)$ from the overall 
power of $\alpha_s(Q^2)$ multiplying the solution (and simultaneously
get the extra power of $\alpha_s(Q_0^2)$ in the input). 
The eigenvectors are different from those in the parton model, of course,
but this takes account of the fact that the inputs are now those for the
structure functions themselves, rather than, in the case of
$F^{S}_L(N,Q^2)$, the parton densities weighted by coefficient functions.  
Working beyond this lowest order and 
finding the $n_{\rm th}$--order solutions to the evolution equations for the
structure functions, in the same way as those for the parton
densities, as outlined above, leads to all powers of $\alpha_s$ up to
power $n$ multiplying the factors of
$(\alpha_s(Q_0^2)/\alpha_s(Q^2))^{\tilde\gamma^{0,l}_{+,-}}$ for 
$F^S_2(N,Q^2)$, and powers of $\alpha_s$ up to power $n+1$ multiplying 
$(\alpha_s(Q_0^2)/\alpha_s(Q^2))^{\tilde\gamma^{0,l}_{+,-}-1}$ for 
$F^S_L(N,Q^2)$. Thus, it
produces the same result as the $n_{\rm th}$--order solution using the parton
distributions and the loop expansion. 

Unlike the nonsinglet case it is rather
difficult to see how to express the solution in terms of
factorized inputs and evolution parts; the evolution
parts will not be simple exponentials, as in \fnonsingsol. However,
the solution for the structure functions can be
written as a series of terms each of which can be 
factored into a part dependent only on $\alpha_s(Q_0^2)$ and one
depending on both $\alpha_s(Q_0^2)$ and $\alpha_s(Q^2)$, the latter either
vanishing at $Q^2=Q_0^2$, or being equal to $(\alpha_s(Q_0^2)/
\alpha_s(Q^2))^{\tilde\gamma^{0,l}_{+,-}(-1)}$ and thus equal to 
unity at $Q^2=Q_0^2$. For each of these terms the former part may
be interpreted as an input and the latter part may be interpreted as
the evolution. Within the loop expansion we then include all terms where
the order of the input part plus the order of the evolution part sums
to less than or equal to some integer $n$. 

By examining the form of the 
inputs under a change in starting scale, as with the nonsinglet 
structure functions, we find that the only fundamentally nonperturbative 
parts of the inputs are the zeroth--order parts, $F^S_{2,0}(N)$ and 
$F^S_{L,0}(N)$, with all other inputs being in
principle expressed as perturbative functions of the physical
anomalous dimensions
multiplying one of these nonperturbative components. If we wished to stay 
within the language of partons then we would equivalently have the two 
nonperturbative parton inputs $f^S_0(N)$ and $g_0(N)$, where the link between 
the two pairs of nonperturbative inputs is 
\eqn\linkf{F^S_{2,0}(N)=f^S_0(N),} 
and
\eqn\linkg{F^S_{L,0}(N)= C^f_{L,1}(N)f^S_0(N)+C^g_{L,1}(N)g_0(N).}
We choose to think of the expressions for the structure functions as the real
nonperturbative functions, since they are, of course, the physically relevant 
quantities.\foot{We could, if we wished, make a change in factorization 
scheme where the zeroth--order part of the matrix $U$ was not the unit matrix,
but a function of $N$. This changes no physical result, but does change the 
zeroth--order parton inputs.}  
This is consistent with the reasoning that $g_0(N)$
is $Q_0^2$--independent because 
the invariance of the physical quantity $F^S_L(N,Q^2)$ under
a change in starting scale leads to $F^S_{L,0}(N)$ being $Q_0^2$--independent,
which then, using \linkg, leads to $g_0(N)$ being $Q_0^2$--independent.

The extra complexity of the solution in the singlet case, when compared to
the nonsinglet case, also means it
is far from obvious how to express the physical anomalous dimensions in
terms of the parton anomalous dimensions and coefficient functions
simply by comparing the forms of the solutions in the partonic
language and purely in terms of structure functions. However, the
expressions for the physical anomalous dimensions can
be found at each order using Catani's expressions \physanom. 
One could then solve for the structure functions in a manner different
from the loop expansion. An example would be to calculate the solution 
to the whole evolution equation for
the structure functions using the physical anomalous dimensions up to a given
order. This would be by definition factorization scheme invariant, but
would only be renormalization scheme invariant up to the same order in
the solution as the order of the anomalous dimensions. The
rest of the solution would contain only a subset of the possible terms
of a given form obtained from the full calculation, and would therefore
still have no real significance. Hence, we conclude that the
evolution equations in terms of structure functions do nothing to alter the
strict ordering of the solution using this method of expansion, but do make 
finding this ordered solution somewhat easier.     

\bigskip

This whole discussion of the order--by--order--in--$\alpha_s$ 
expansion scheme
is perhaps a little academic since the errors invoked in performing the 
calculations
without paying strict heed to the formally correct procedures 
are rather small. For example, let us consider factorization scheme 
dependence.
When using this standard method of expansion only two different
factorization schemes are generally considered, the 
$\overline{\hbox{\rm MS}}$ scheme and the DIS scheme. The first of these simply
involves calculating the coefficient
functions using standard techniques and using the $\overline{\hbox{\rm
MS}}$ scheme to remove both the the infrared and ultraviolet
divergences and hence provide the 
anomalous dimensions. The second makes a change of parton
distributions, so that the singlet quark distribution is equal to the
singlet structure function. It is clear from the above discussion that,
if a calculation is made in one scheme to a well--defined order
(usually next to leading order), and then a transformation to the other
scheme made correctly, precisely the same result will be
obtained. In practice, small differences are 
sometimes noticed between calculations
using different schemes within 
the loop expansion to NLO, but these come 
from well--understood sources. One
common source is that the starting distributions in both schemes are
described by a simple functional form, e.g.
\eqn\start{xf(x,Q_0^2) = A(Q_0^2)x^{-\lambda(Q_0^2)}(1-x)^{\eta(Q_0^2)}
(1+\epsilon(Q_0^2) x^{1/2}+\gamma(Q_0^2) x),}
rather than the formally correct expression of a power series in 
$\alpha_s(Q_0^2)$ with essentially perturbative coefficients convoluted with
a $Q_0^2$--independent nonperturbative function of $x$. 
If the starting distribution is of the
form above in one scheme, then the starting distribution in  the other
scheme will not be modelled precisely by a function of the
same form. However, the error is in general small.
Alternatively, if the
calculations are not done in a well--ordered manner, e.g. simply solving
the whole evolution equation using the two--loop anomalous dimensions,
then differences between the calculation done in the two schemes, or
between this type of calculation and the correct NLO
calculation, will
be of NNLO. Again this usually results in
only small differences.\foot{For a comparison of calculations done 
at NLO using different methods see \ref\comparison{J. Bl\"umlein 
{\it et al}, Proc. of the 1996 HERA Physics Workshop, eds. G. Ingelman,
R. Klanner, and A. De Roeck, DESY, Hamburg, 1996, Vol. 1, p. 23.}. 
One sees here that starting with the same input parton distributions
the different calculations agree to better than $5\%$ except for very small
$x$ and large $Q^2$, where the discrepancy can approach $10\%$.
The differences observed can be compensated for almost
completely by small changes in inputs.}
Similarly, small differences would be
obtained by working in two different renormalization schemes, and of
course, there is nothing which can be done about this.

There are a couple of points we wish to make here. Firstly concerning the 
form of the input in \start\
we note that, if one considers the input to be consistent with the 
loop expansion, e.g. 
NLO--order evolution should be accompanied by a NLO input, 
the power of $\lambda(Q_0^2)$ should not correspond to parton 
distributions much 
steeper than flat for the singlet quark or gluon at this order. 
This is because the  first--order--in--$\alpha_s(Q_0^2)$ 
input should be accompanied by no more than a single power 
of $\ln(1/x)$, this being all that is required to absorb the change in the 
zeroth--order evolution under a change in starting scale at this order in 
$\alpha_s(Q_0^2)$. In practice 
restricting $\lambda(Q_0^2)$ in this way is rather important 
for the fits to the low $x$ data, and would mean that NLO fits to small--$x$ 
data would be very poor. The only way to avoid this is to let
$\lambda(Q_0^2)$ be an artificial 
free parameter, in which case, if \start\ describes the 
singlet quark density, it must be $\sim 0.2-0.3$ for practically any $Q_0^2$. 
This value is totally unjustified, and the need for this steepness in the 
input for the quark is a clear sign of the limited usefulness of the 
NLO--in--$\alpha_s$ calculation at small $x$. We will see that the situation 
is more more satisfactory when using the correct expansion method. 

Ignoring this problem with the inputs, there is another reason for being 
concerned about the validity of the loop expansion at small $x$.
The main reason for the smallness of the differences between inaccurately
performed NLO calculations noted above, even at small $x$, is
that the differences between these calculations do not contain terms
which are any more leading in $\ln (1/x)$ than the NLO
calculation itself. Hence they are genuinely an order of $\alpha_s$ down on
the NLO calculation, with no small--$x$
enhancement. This is also true for calculations done in different
renormalization schemes (or at different renormalization scales).
The real NNLO
contribution will be higher order in $\alpha_s$ but will also contain
terms at higher order in $\ln(1/x)$, and will therefore be potentially
large at small $x$. We therefore stress that the relative
insensitivity of structure functions to changes in
renormalization or factorization scheme for
calculations which are not carefully ordered 
is no guarantee that genuine higher--order
corrections will be small at small $x$ when using the loop
expansion. Indeed we would naively expect them to be large. 
This is a point to which we shall return. 

\medskip

In contrast to the insensitivity when using the loop expansion, when
using the leading--$\ln (1/x)$
expansion very large differences between calculations done in a large
number of different factorization schemes have been noted. This is an obvious 
sign that the calculations are not being done in a well--ordered
manner, and that the ambiguity introduced by lack of care in the 
calculations is
greater in this method of expansion than the standard loop expansion.   
We will now demonstrate that this is indeed the case.

\subsec{The Leading--$\ln (1/x)$ Expansion: Parton Distributions.}

It should clearly be possible to define a well--ordered, and hence, 
factorization scheme
independent expansion in leading powers of $\ln (1/x)$, or
equivalently, in leading powers of $1/N$ in moment space. We will now
demonstrate that this is indeed the case. As in the loop expansion, we
will first work in terms of the traditional parton distribution
functions and coefficient functions, and see how this results in
expressions containing the physical anomalous dimensions. Doing this
enables us to see how large factorization scheme dependence can arise
when calculating less carefully within this expansion scheme. 

First we must set up our
notation and qualify the statements made in the introduction
concerning the leading--$\ln (1/x)$ expansion. 
We stated that the anomalous dimensions could be written in the form
\fulladalti, and this means that using the form \nonsingevolal\ and
\singevolal\ for the evolution equations we may write,
\eqn\tildegamgen{\tilde\gamma(N,Q^2) =
\sum_{n=0}^{\infty}\alpha_s^n(Q^2)\sum_{m=1-n}^{\infty}\tilde
\gamma^{nm}\alpha_s^{m}(Q^2) N^{-m} \equiv
\sum_{n=0}^{\infty}\alpha_s^n(Q^2)\tilde\gamma^n(\alpha_s(Q^2)/N).}
So, the $\tilde\gamma^0$'s only differ from the $\gamma^0$'s by the 
normalization factor of $b_0$, but at higher orders the relationship
is more complicated. 
In particular, in the $\overline{\hbox{\rm MS}}$ renormalization and
factorization scheme we may write
\eqn\tildegamglu{\tilde \gamma_{gg}(N,\alpha_s(Q^2)) = \sum_{n=0}^{\infty}
\alpha_s^n(Q^2) \tilde\gamma_{gg}^n(\alpha_s(Q^2)/N),}
where the series expansion for $\tilde\gamma^0_{gg}(\alpha_s/N)$ is
known to all orders (all the coefficients being positive). This expression for
$\tilde\gamma^0_{gg}(\alpha_s/N)$ is renormalization scheme independent
since any change in renormalization scheme can only bring about a
change in the coupling at ${\cal O}(\alpha_s^2)$, and this does not
require a change in $\tilde\gamma^0_{gg}(\alpha_s/N)$ to keep physical
quantities invariant at leading order in this expansion
scheme, as we will soon see. $\tilde\gamma^1_{gg}(\alpha_s/N)$ is of 
course renormalization
scheme dependent, since it must change to absorb part of the effect of the
${\cal O}(\alpha_s^2)$ change in the coupling on
$\tilde\gamma^0_{gg}(\alpha_s/N)$. The renormalization scheme
independence is also true for $\tilde\gamma^0_{gf}(\alpha_s/N)$, which
obeys \gammadef, and also for $\tilde\gamma^0_{ff}(\alpha_s/N)$ and
$\tilde\gamma^0_{fg}(\alpha_s/N)$ which are both zero. There is also a
renormalization--scheme--independent relationship between
$\tilde\gamma^1_{ff}(\alpha_s/N)$ and $\tilde\gamma^1_{fg}(\alpha_s/N)$,
which tells us that 
\eqn\gamrel{\tilde\gamma^1_{ff}(\alpha_s/N)={4\over 9}
\biggl(\tilde\gamma^1_{fg}(\alpha_s/N)-{2N_f
\over 6\pi b_0}\biggr),}
where the
second term in the brackets is the one--loop contribution to
$\tilde\gamma^1_{fg}(\alpha_s/N)$. It is also known that the 
nonsinglet anomalous
dimension has no poles at $N=0$, and neither does the nonsinglet
coefficient function. Hence, the nonsinglet sector makes very little
contribution to the structure function at small $x$, and as such we
will ignore it for the remainder of this section. 

A general change in factorization scheme may be expressed by writing
an element of the transformation matrix $U$ as 
\eqn\Udef{U_{ab}(N,\alpha_s(Q^2)) = \sum_{n=-k}^{\infty} \alpha_s^n(Q^2)
\sum_{m=1-n}^{\infty}
U^{nm}_{ab}\alpha_s^m(Q^2) N^{-m}\equiv
\sum_{n=-k}^{\infty}\alpha_s^n(Q^2)U^n_{ab}(\alpha_s(Q^2)/N),}
with condition on the $U^{nm}_{ab}$ such that $U$ obeys
$U_{ab}=\delta_{ab} +{\cal O}(\alpha_s)$. This flexibility in the
factorization scheme means that all the above results on the low--order
$\tilde \gamma$'s are in principle factorization scheme dependent. In
particular, the expressions for the $\tilde \gamma$'s in \tildegamgen\ may
be insufficient in general, and should be replaced by
\eqn\hatgammoregen{\tilde\gamma(N,Q^2) =
\sum_{n=-k}^{\infty}\alpha_s^n(Q^2)\sum_{m=1-n}^{\infty}\tilde
a_{nm}\alpha_s^{m}(Q^2) N^{-m}.}
However, this is only necessary if one makes a change of scheme away from a
standard scheme, such as the $\overline {\hbox{\rm MS}}$ scheme, using a
matrix $U$ containing terms such as $\alpha_s^n(Q^2)N^{-m}$ where $m>
n$. For simplicity and because, of course, all physical results are
ultimately factorization scheme independent by definition using any 
expansion, we will forbid such singular changes of scheme and 
demand that $k=0$ in \Udef.

However, a change of factorization scheme with $k=0$ can still
introduce scheme dependence into the $\tilde
\gamma^0_{ab}(\alpha_s/N)$'s. Using \transgamma\ we see that 
\eqn\transgamzero{\breve{\tilde \gamma}^0_{ab}(\alpha_s/N) = 
\sum_c\sum_d U^0_{ac}(\alpha_s/N)\tilde\gamma^0_{cd}(\alpha_s/N)
(U^{-1})^0_{db}(\alpha_s/N).}
Insisting that $\tilde\gamma^0_{gg}(\alpha_s/N)$,
$\tilde\gamma^0_{gf}(\alpha_s/N)$,
$\tilde\gamma^0_{ff}(\alpha_s/N)$ and $\tilde\gamma^0_{fg}(\alpha_s/N)$
are unaltered by scheme changes leads to the requirement,
\eqn\urequire{\eqalign{&U^0_{fg}(N,\alpha_s)=0, \hskip 1in 
U^0_{ff}(N,\alpha_s)=1,\cr
&U^0_{gg}(\alpha_s/N) = 1
+\sum_{n=1}^{\infty}U^{0,n}_{gg}(\alpha_s(Q^2)/N)^n,
\hskip 0.5in U^0_{gf}(\alpha_s/N)=\fourninths(U^0_{gg}(\alpha_s/N)-1).\cr}}
This requirement also preserves the relationship \gamrel\ between
$\tilde\gamma^1_{ff}(\alpha_s/N)$ and
$\tilde\gamma^1_{fg}(\alpha_s/N)$. Again for simplicity, and due to
factorization scheme invariance of physical quantities, we will only
consider factorization scheme changes away from $\overline {\hbox{\rm
MS}}$ scheme of the type \urequire. The $U^n_{ab}(\alpha_s/N)$ for $n> 0$
will have no restrictions.\foot{To our knowledge, no--one has yet 
bothered considering the effects of
any transformations of the type we have forbidden (of course, there is no
point, since physical quantities are factorization scheme invariant).
A number of the type
described above with $U^0_{gg}(N,\alpha_s)\not= 0$ have been
considered (however, once again, physical quantities are invariant 
under such scheme
changes).}

Restricting ourselves to these schemes we may now write 
\eqn\defcstwo{C^S_2(N,\alpha_s) = 1 +
\sum_{n=1}^{\infty}\alpha_s^n(Q^2)\sum_{m=1-n}^{\infty}
C^{S}_{2,n,m}\alpha_s^m(Q^2)N^{-m} \equiv
1+\sum_{n=1}^{\infty}\alpha_s^n(Q^2) C^S_{2,n}(\alpha_s/N),}
and all other coefficient functions as
\eqn\defcoeffs{C^a_i(N,\alpha_s) = 
\sum_{n=1}^{\infty}\alpha_s^n(Q^2)\sum_{m=1-n}^{\infty}
C^{a}_{i,n,m}\alpha_s^m(Q^2)N^{-m} \equiv
\sum_{n=1}^{\infty}\alpha_s^n(Q^2) C^a_{i,n}(\alpha_s/N).}
All the $C^a_{i,n}(\alpha_s/N)$ are both renormalization--scheme-- and
factorization--scheme--dependent quantities. Indeed, all of the
$C^{a}_{i,n,m}$ are renormalization scheme and
factorization scheme dependent, except for the $C^a_{L,n,1-n}$, which
come from the one--loop longitudinal coefficient functions which, as
we saw in the previous subsection, are totally scheme independent. 
There are also two
renormalization and factorization scheme (with our restrictions)
independent relationships between coefficient
functions:
\eqn\coeffreltwo{C^S_{2,1}(\alpha_s/N) ={4\over 9}\Bigl(C^g_{2,1}(\alpha_s/N)
-C^g_{2,1,0}\Bigr),}
where the second term in brackets is the one--loop
contribution to $C^g_{2,1}(\alpha_s/N)$, which is itself
renormalization scheme and factorization scheme dependent (being equal
to $(N_f/6\pi)$ in $\overline{\hbox{\rm MS}}$ scheme); 
and 
\eqn\coeffrellong{\Bigl(C^S_{L,1}(\alpha_s/N)-{2\over 3\pi}\Bigr) 
={4\over 9}\Bigl(C^g_{2,1}(\alpha_s/N)-{2N_f\over 6\pi}\Bigr),}
where the second terms in the
brackets are the one--loop contributions to $C^S_{L,1}(\alpha_s/N)$
and $C^g_{L,1}(\alpha_s/N)$, both of which are renormalization and
factorization scheme independent.  

\bigskip

Working in an arbitrary factorization scheme 
(up to the above restrictions) and using the general
expressions for the $\tilde\gamma$'s and coefficient functions, we may
find expressions for the structure functions. The first step towards
this is solving the renormalization group equations for the parton 
distributions. The lowest--order part of the equation is,
\eqn\llevolzero{\alpha^2_s(Q^2) {d\over d \alpha_s(Q^2)} 
\pmatrix{f^S_{0}(N,Q^2)\cr g_0(N,Q^2)\cr} =-
\pmatrix{0 & 0\cr 
\fourninths\tilde \gamma^0_{gg}(\alpha_s/N) 
& \tilde \gamma^0_{gg}(\alpha_s/N)\cr}
\pmatrix{f^S_{0}(N,Q^2)\cr g_0(N,Q^2)\cr}.}
This may easily be solved to give
\eqn\solzero{\eqalign{f^S_0(N,Q^2)&=f^S_0(N,Q_0^2),\cr
g_0(N,Q^2)&= (g_0(N,Q_0^2)+\fourninths f_0^S(N,Q_0^2))\eigplus 
-\fourninths f_0^S(N,Q_0^2).\cr}}
This is analogous to the lowest--order solution within the loop
expansion and contains two factors, one of which must appear in all the
higher terms in the expansion: instead of
$(\alpha_s(Q_0^2)/\alpha_s(Q^2))^{\tilde\gamma^{0,l}_{+,-}(N)}$,
corresponding to the two eigenvalues of $\tilde\gamma^{0,l}(N)$ in the loop
expansion, we have $\eigplus$ and $1$, corresponding to the two
eigenvalues of $\tilde\gamma^0(\alpha_s/N)$ in the leading--$\ln (1/x)$
expansion, i.e. $\tilde\gamma^0_{gg}(\alpha_s/N)$ and $0$. 
In particular we notice that $f^S_0(N,Q^2)$ is $Q^2$--independent.

Let us briefly digress in order to discuss the form of the inputs
for the parton distributions. 
In particular, we examine how the terms in our expression change under the
change in input scale $Q_0^2\to (1+\delta)Q_0^2$, leading to the change in
coupling at the starting scale \inscalchange. 
Rather trivially, our expression for
$f_0^S(N,Q^2)\equiv f_0^S(N,Q_0^2)$ is unchanged by the change in the coupling
at the starting scale, and as such can be chosen to be independent 
of $Q_0^2$. Hence, as for the lowest order input in the loop expansion we may
write $f_0^S(N,Q^2)\equiv f_0^S(N,Q_0^2)\equiv f_0^S(N)$.
The expression for the gluon involves a little more work. The change
in the evolution term under a change in starting scale is 
\eqn\eigplusscalchange{\eqalign{\Delta &\eigplus \cr
&\hskip 0.4in =\exp\biggl[\int_{\alpha_s(Q_0^2)}^{\alpha_s((1+\delta)Q_0^2)}
{\tilde\gamma^0_{gg}(\alpha_s(q^2)/N)\over \alpha^2_s(q^2)}
d\alpha_s(q^2))-1\biggr]
\eigplus,\cr}}
and using \inscalchange\ we find that 
\eqn\eigplusscalchangeres{\eqalign{\Delta &\eigplus \cr
& \hskip 0.6in =(\exp[-\delta b_0
\tilde\gamma^0_{gg}(\alpha_s(Q_0^2)/N))-1\,]\eigplus \cr
&\hskip 0.6in + \higherorder.}}
We choose $g_0(N,Q_0^2)$ so that  
the change in $\eigplus$ can be compensated for by a change in
$g_0(N,Q_0^2)$ up to corrections of higher order. Hence, the gluon input
may be written as 
\eqn\gluin{g_0(N,Q_0^2)=g_0(N)+(\eigvecplus)\sum_{m=1}^{\infty}g_{0,m}(Q_0^2)
\biggl({
\alpha_s(Q_0^2)\over N}\biggr)^m\equiv g_0(N)+\tilde g_0(N,Q_0^2).}
The change in $g_0(N,Q_0^2)$ under a change in starting scale necessary to
make $g_0(N,Q^2)$ invariant under changes of $Q_0^2$ up to
higher orders may 
easily be calculated from \eigplusscalchangeres, and is equal to 
\eqn\changex{\eqalign{\Delta g_0(N,Q_0^2) &= \delta b_0 \tilde
\gamma_{gg}^0(\alpha_s(Q_0^2)/N) (\eigvecplus +\tilde g_0(N,Q_0^2))
+\hbox{\rm higher order}\cr
&\equiv \delta\gamma^0_{gg}(\alpha_s(Q_0^2)/N)(\eigvecplus 
+\tilde g_0(N,Q_0^2))+\hbox{\rm higher order},\cr}}
in the limit of small $\delta$. Thus, an appropriate expression for the input
which satisfies this condition is given by 
\eqn\inexp{\eqalign{(\eigvecplus+&\tilde g_0(N,Q_0^2))\cr
&\hskip -0.3in =(\eigvecplus)
\biggl(1+\sum_{m=1}^{\infty}\tilde g_{0,m} \biggl({\alpha_s(Q_0^2)\over N}
\biggr)^m\biggr)
\exp[\ln(Q_0^2/A_{gg})\gamma^0_{gg}(\alpha_s(Q_0^2)/N)],\cr}}
where $A_{gg}$ is an unknown scale. 
The series $\sum_{m=1}^{\infty}\tilde g_{0,m}(\alpha_s(Q_0^2)/ N)^m$ 
is at yet undetermined, but is potentially renormalization and 
factorization scheme dependent. It will only be determined when we come to 
construct the structure functions themselves. 

We also consider the form of the $N$--dependence of our inputs. If we take
the point of view that any steep behaviour in the parton distributions
only comes about due to perturbative effects, then we assume that
$f^S_0(N)$ and $g_0(N)$ are both soft, i.e. either flat or even
valence--like when the transform to $x$--space is performed (or at most
going like a finite, small power of $\ln(1/x)$). This
requires that they both be analytic for $N> 0$. We also note that the 
soft parts of the input are common for the whole of the gluon input, as shown 
in \gluin, i.e. $\tilde g_0(N,Q_0^2)$ is just the soft $(\eigvecplus)$ 
multiplied by a 
series of the form $\sum_{m=1}^{\infty}a_m(\alpha_s(Q_0^2)/N)^m$. Thus, as in 
the loop expansion, we may think of $g_0(N)$, and $f^S_0(N)$ as 
fundamentally soft, nonperturbative parts of the input. \foot{Of course, we 
should be trying to find fundamentally nonperturbative inputs for the 
structure functions rather than the partons, as discussed for the 
loop expansion. However, in this expansion scheme $f^S_0(N)$ and $g_0(N)$ are
trivially related to $F_{2,0}(N)$ and $\hat F_{L,0}(N)$ as we will see
in the next subsection.} The parts multiplying 
these are then really determined by perturbation theory.  
Since we are meant to be expanding our solution for the structure functions in
powers of both $\alpha_s$ and $N$, it might be argued that we should expand
$f^S_0(N)$ and $g_0(N)$ in powers of $N$. We feel this is not really 
appropriate since it is the perturbative part of the solution for which we 
are able to solve, and thus which we are able to order correctly, 
and in the expressions 
for the structure functions the whole of the nonperturbative 
inputs should multiply the well--ordered perturbative parts of the solution.

\medskip
 
Before using our solutions for the parton densities to 
construct expressions for the structure functions we will
solve higher--order renormalization group equations in order to
determine the general form of the solutions for the parton
distributions. This is made easier by the fact that
$\tilde\gamma^0_{ff}(\alpha_s/N)$ and $\tilde\gamma_{fg}^0(\alpha_s/N)$
are both zero, and hence,
\eqn\firstordfeq{\alpha_s(Q^2) {d\,f^S_1(N,Q^2)\over d\alpha_s(Q^2)} =
-(\tilde\gamma^1_{ff}(\alpha_s/N)f^S_0(N,Q^2)
+\tilde\gamma^1_{fg}(\alpha_s/N) g_0(N,Q^2)).}
All of the above quantities on the right--hand side are in principle
already known. Thus, we have an expression for
$(d\,f^S_1(N,Q^2)/d\alpha_s(Q^2))$ which can be written in the simple form
\eqn\firstordfeqi{\eqalign{\alpha_s^2(Q^2){d \,f^S_1(N,Q^2)\over
d\alpha_s(Q^2)}=&-\alpha_s(Q^2)\tilde\gamma^1_{fg}(\alpha_s/N)(\eigvecplus
+\tilde g_0(N,Q_0^2))\times\cr
&\hskip 0.5in \eigplus
+\alpha_s(Q^2){4\over 9}{2N_f\over 6\pi b_0} f_0^S(N),\cr}}
where \gamrel\ has been used. Thus, $\alpha^2_s(Q^2)d\,f^S_1(N,Q^2)/
d\alpha_s(Q^2)$ is simply a sum of two power series of the form
$\alpha_s\sum_{m=0}^{\infty}a_m(\alpha_s/N)^{m}$ multiplying input
densities and our eigenvalue determined evolution factors (the second series
rather trivially having $a_m=0, m\geq 0$). Thus, all terms in the expression
are of the same order in this expansion scheme. 

Integrating \firstordfeqi\ we get
\eqn\fisrtordf{\eqalign{f^S_1(N,Q^2) =
&-{4\over 9} {2N_f\over 6\pi b_0}\ln\biggl({\alpha_s(Q_0^2)\over
\alpha_s(Q^2)}\biggr) f^S_0(N)
+f^S_1(N,Q_0^2)+(\eigvecplus +\tilde g_0(N,Q_0^2))\times\cr
&\int_{\alpha_s(Q^2)}^{\alpha_s(Q_0^2)}\!\!
{\tilde\gamma^1_{fg}(\alpha_s(q^2)/N)\over\alpha_s(q^2)}
\eigplusq d\alpha_s(q^2).\cr}}
We may express the last term differently by integrating by
parts. This gives
\eqn\fisrtordfi{\eqalign{f^S_1(N,Q^2) =
&\biggl(\alpha_s(Q^2)
{\tilde\gamma^1_{fg}(\alpha_s(Q^2)/N)\over \tilde
\gamma^0_{gg}(\alpha_s(Q^2)/N)}\eigplus \cr
&\hskip 1.5in - \alpha_s(Q_0^2){\tilde\gamma^1_{fg}(\alpha_s(Q_0^2)/N)\over 
\tilde\gamma^0_{gg}(\alpha_s(Q_0^2)/N)}\biggr)(\eigvecplus+\tilde 
g_0(N,Q_0^2))\cr
&-\int_{\alpha_s(Q^2)}^{\alpha_s(Q_0^2)}\!\!
{d\over d\alpha_s(q^2)}
\biggl(\alpha_s(q^2){\tilde\gamma^1_{fg}(\alpha_s(q^2)/N)\over 
\tilde\gamma^0_{gg}(\alpha_s(q^2)/N)}\biggr)\times\cr
&\hskip 0.5in \eigplusq d\alpha_s(q^2)
(\eigvecplus+\tilde g_0(N,Q_0^2))\cr
&-{4\over 9}
{2N_f \over 6\pi b_0}\ln\biggl({\alpha_s(Q_0^2)\over\alpha_s(Q^2)}\biggr)
f^S_0(N)+ f^S_1(N,Q_0^2).\cr}}
$\alpha_s(\tilde\gamma^1_{fg}(\alpha_s/N)/
\tilde\gamma^0_{gg}(\alpha_s/N))$ is a power series of the form
$N\sum_{m=0}^{\infty} a_m(\alpha_s/N)^m$. The integration by parts
may be repeated indefinitely, producing power series which behave like
$N\alpha_s^k(Q^2) \sum_{m=-k}^{\infty}a_m(\alpha_s(Q^2)/N)^m$, with
$k$ increasing by one at each integration. Hence, the solution to the 
next--to--leading--order renormalization group equation for $f^S(N,Q^2)$ 
is not of 
a single consistent order in this expansion scheme. We may write it as
\eqn\firstordfii{\eqalign{f^S_1&(N,Q^2) =
(\eigvecplus+\tilde g_0(N,Q_0^2))\times \cr
&\biggl(\alpha_s(Q^2)
{\tilde\gamma^1_{fg}(\alpha_s(Q^2)/N)\over \tilde
\gamma^0_{gg}(\alpha_s(Q^2)/N)}
\eigplus - \alpha_s(Q_0^2){\tilde\gamma^1_{fg}(N,\alpha_s(Q_0^2))\over 
\tilde\gamma^0_{gg}(N,\alpha_s(Q_0^2))}\biggr)\cr
&+f_1^S(N,Q_0^2)-{4\over 9} {2N_f\over 6\pi b_0}
\ln\biggl({\alpha_s(Q_0^2)\over\alpha_s(Q^2)}\biggr) f^S_0(N)
+\higherorder.\cr}}
Let us examine the form of our solution for $f^S_1(N,Q^2)$. No part of
it is of the same form as the zeroth--order parton distributions,
i.e. either just a constant soft distribution or
such a constant soft distribution multiplied by powers of $\alpha_s(Q_0^2)/N$
and the evolution term $\eigplus$. Thus, the lowest--order corrections to this
zeroth--order solution are given by the lowest--order parts of \firstordfii.
 
In the same manner as for the zeroth--order gluon
distribution we can determine the general form of our input, in this
case $f_1^S(N,Q_0^2)$, by considering a change of starting
scale. Under the change of scale leading to the change of
$\alpha_s(Q_0^2)$ in \inscalchange\ each of the
$\alpha_s(Q_0^2)$--dependent terms in \firstordfii\ will change. The
change of $\tilde g_0(N,Q_0^2)$ has already been chosen to cancel (at 
leading order) the change due to $\eigplus$, and hence the first term in
\firstordfii\ is stable to the change in starting scale. The last
explicit term in \firstordfii\ is also stable since the change induced
in the $\ln(\alpha_s(Q_0^2))$ term is of higher order in
$\alpha_s(Q_0^2)$. The second term does however vary; the change due
to the variation of $\alpha_s(Q_0^2)\tilde\gamma^1_{fg}(\alpha_s(Q_0^2)/N)
/\tilde\gamma_{gg}^0(\alpha_s(Q_0^2)/N)$ is of higher
order in $\alpha_s(Q_0^2)$, but $\tilde g_0(N,Q_0^2)$ changes as 
prescribed in \changex, and hence the
second term changes by an amount of the same order as itself, i.e. by a
series of the type $\alpha_s(Q_0^2)\sum_n a_n
(\alpha_s(Q_0^2)/N)^n$. In order to keep the whole of \firstordfii\
unchanged at this order, $f_1^S(N,Q_0^2)$ must change under a change in
$Q_0^2$ in a manner which compensates this change in the second
term. Hence, we choose 
\eqn\findef{f_1^S(N,Q_0^2)= (\eigvecplus)N\sum_{m=0}^{\infty}f^S_{1,m}(Q_0^2)
\biggl({\alpha_s(Q_0^2)\over N}\biggr)^m.} 
The change of $f^S_{1,m}(Q_0^2)$ under a change in $Q_0^2$ may
easily be calculated using the known change in $\tilde g_0(N,Q_0^2)$, i.e.
\eqn\changexi{\Delta f^S_1(N,Q_0^2)=\delta \alpha_s(Q_0^2)
\gamma^1_{fg}(\alpha_s(Q_0^2)/N)(\eigvecplus+\tilde g_0(N, Q_0^2))
+\hbox{\rm higher order},}
in the limit of small $\delta$. Rather obviously, the simplest
choice which satisfies the requirement is
\eqn\inexpf{\eqalign{f^S_1(N,Q_0^2)&=(\eigvecplus)\alpha_s(Q_0^2)
{\gamma^1_{fg}(\alpha_s(Q_0^2)/N)\over\gamma^0_{gg}
(\alpha_s(Q_0^2)/N)}\biggl(1+\sum_{m=0}^{\infty}\tilde g_{0,n} 
\biggl({\alpha_s(Q_0^2)\over N}\biggr)^m\biggr)\times\cr
&\hskip 3in \exp[\ln(Q_0^2/A_{gg})\gamma^0_{gg}(\alpha_s(Q_0^2)/N)],\cr
&\equiv (\eigvecplus +\tilde g_0(N,Q_0^2))\alpha_s(Q_0^2)
{\gamma^1_{fg}(\alpha_s(Q_0^2)/N)\over\gamma^0_{gg}(\alpha_s(Q_0^2)/N)}.\cr}}
However, as when deriving the ${\cal O}(\alpha_s(Q_0^2))$ input for the 
nonsinglet structure function within the loop expansion \chainputsf, the 
invariance of the parton distribution under changes in $Q_0^2$ is not the
sole consideration. We also require that at lowest order the expression 
for the input for the structure function is renormalization scheme 
independent, and that at all orders it is factorization scheme independent. 
Also, we think of the scale $A_{gg}$ (or $A_{NS}$ {\it etc.}) as the 
value of $Q_0^2$ where the input for the structure function becomes just 
the nonperturbative input, so $f^S_1(N,Q_0^2)$ should be consistent with this.
If we add a series of the form 
\eqn\ftildedef{\tilde f^S_1(N,Q_0^2)=N\sum_{n=0}^{\infty}\tilde f^S_{1,m}
\biggl({\alpha_s(Q_0^2)\over N}\biggr)^m(\eigvecplus)}
to \inexpf\ then the resulting expression still
satisfies \changexi, but gives us the flexibility to satisfy the 
other requirements above. Thus, similarly to \addon\ we write
\eqn\inexpfc{f^S_1(N,Q_0^2)=(\eigvecplus +\tilde g_0(N,Q_0^2))\alpha_s(Q_0^2)
{\gamma^1_{fg}(\alpha_s(Q_0^2)/N)\over\gamma^0_{gg}(\alpha_s(Q_0^2)/N)}
+\tilde f^S_1(N,Q_0^2),}
as the appropriate input, where $\tilde f^S_1(N,Q_0^2)$ is potentially
renormalization and factorization scheme dependent, and is not yet
determined. 

Therefore, the leading correction to the zeroth--order input $f^S_0(N)$ 
is a series of the form
$N\sum_{m=0}^{\infty} (\alpha_s(Q_0^2)/N)^m$ multiplying $(\eigvecplus)$. The
lowest--order parts of the $Q^2$--dependent expression for $f^S(N,Q^2)$ are a
series of the form $N\sum_{m=0}^{\infty}
(\alpha_s(Q^2)/N)^m$ multiplying $\eigplus$ and a zeroth--order input,
and also a logarithm of
$(\alpha_s(Q_0^2)/\alpha_s(Q^2))$ multiplying the other common factor
of unity and a zeroth--order input. $f^S_1(N,Q^2)$ then contains other
terms which are subleading, i.e. of higher order in $\alpha_s$ and/or
$N$, to these. 

Hence, this solution for the next--to--leading evolution equation for the
singlet quark distribution clearly demonstrates that we will not 
obtain well--ordered,
factorization--scheme--independent expressions in the leading--$\ln
(1/x)$ expansion in as straightforward a manner as when using the loop
expansion.  

\medskip

In order to investigate this further, we now consider the NLO 
renormalization group
equation for $g_1(N,Q^2)$. This is a little more complicated that that
for $f^S_1(N,Q^2)$:
\eqn\gfirstordeq{\eqalign{\alpha^2_s(Q^2){d\,g_1(N,Q^2) \over
d\alpha_s(Q^2)}
&= -\Bigl(\alpha_s\tilde\gamma^1_{gf}(\alpha_s/N)f^S_0(N,Q^2)+\alpha_s\tilde
\gamma^1_{gg}(\alpha_s/N)g_0(N,Q^2)\cr
& \hskip 1.5in + \tilde \gamma^0_{gf}(\alpha_s/N)f^S_1(N,Q^2)+
\tilde\gamma^0_{gg}(\alpha_s/N)g_1(N,Q^2)\Bigr).\cr}}
This may be solved, and integrating some of the resulting
terms by parts, we obtain
\eqn\gfirstord{\eqalign{g_1(N,Q^2)=&\biggl(\eigveccor\times\cr
&(\eigvecplus+\tilde g_0(N,Q_0^2))+g_1(N,Q_0^2)\biggl)\eigplus\cr
& -\fourninths f^S_0(N){4\over9}{2N_f\over
6\pi b_0}\ln\biggl({\alpha_s(Q_0^2)\over\alpha_s(Q^2)}\biggr)
+\higherorder.\cr}}
The integral in the first term gives a series of the form
$\sum_{m=1}^{\infty} a_m(\alpha_s/N)^m$ plus a term of the sort
$\ln(\alpha_s(Q_0^2)/\alpha_s(Q^2))$. So those terms explicitly
shown above are the lowest--order terms in $g_1(N,Q^2)$. The 
input $g_1(N,Q_0^2)$ is present in order to 
compensate for the change in $\eigveccor$ under a change in $Q_0^2$. 
However, this means it 
is a series of the form $\alpha_s(Q_0^2)\sum_n a_n(\alpha_s(Q_0^2)/N)^n$, 
and as such is higher order than the other terms present, and may be dropped.  
We see that the  solution for $g_1(N,Q^2)$ 
contains terms of the same apparent order as the zeroth--order
solutions for the parton distributions. Therefore $g_0(N,Q^2)$ seems to be only
part of the lowest--order expression for the gluon distribution, and the terms
shown above appear to be on the same footing. These
are of a similar form to the leading parts of $f^S_1(N,Q^2)$,
with one term being a logarithm of $(\alpha_s(Q_0^2)/\alpha_s(Q^2))$
multiplying $f^S_0(N,Q_0^2)$, but the series multiplying $\eigplus$ is
a power of $N$, or equivalently of $\alpha_s$, 
lower than the corresponding series in the case
of $f^S_1(N,Q^2)$. 

\medskip

So we have the rather undesirable situation that solutions 
to the NLO renormalization group
equations for both the quark and gluon distribution functions result
in solutions which are of mixed order, but fortunately, ones from which we 
can clearly extract the leading--$\ln (1/x)$ behaviour. 
However, the situation gets worse, as may be seen by solving the NNLO
renormalization group equations. For the quark distribution function
this is
\eqn\fsecondordeq{\eqalign{\alpha_s(Q^2){d\,f^S_2(N,Q^2) \over
d\alpha_s(Q^2)}
&= -\Bigl(\alpha_s(Q^2)
\tilde\gamma^2_{ff}(\alpha_s/N)f^S_0(N,Q^2)+\alpha_s(Q^2)
\tilde\gamma^2_{fg}(\alpha_s/N)
g_0(N,Q^2)\cr
& \hskip 1.6in + \tilde \gamma^1_{ff}(\alpha_s/N)f^S_1(N,Q^2)+
\tilde\gamma^1_{fg}(\alpha_s/N)g_1(N,Q^2)\Bigr).\cr}}
As with the equation for $f^S_1(N,Q^2)$ all terms on the right are
already known in principle (it is clear that this will be true for the 
equations for all the $f^S_n(N,Q^2)$). Moreover, part of the
right--hand side of \fsecondordeq\ is of the same order as the
right--hand side of \firstordfeq, as is seen by inserting the leading
parts of $f^S_1(N,Q^2)$ and $g_1(N,Q^2)$, i.e.  
\eqn\fsecondordeqi{\eqalign{\alpha_s^2(Q^2){d\,f^S_2(N,Q^2) \over
d\alpha_s(Q^2)}
&= -\alpha_s(Q^2)\tilde\gamma^1_{fg}(\alpha_s/N)\biggl(\eigveccor\times\cr
&(\eigvecplus)+\tilde g_0(N,Q_0^2)\biggr)\eigplus\cr
& - f^S_0(N)
\alpha_s(Q^2)\biggl({4\over 9}{2N_f \over 6\pi b_0}\biggr)^2
\ln\biggl({\alpha_s(Q_0^2)\over\alpha_s(Q^2)}\biggr)
+\higherorder.\cr}}
Solving \fsecondordeq\ we find that $\tilde\gamma^2_{ff}(\alpha_s/N)$ and
$\tilde\gamma^2_{fg}(\alpha_s/N)$ play no role as far as the leading part 
of the solution is concerned, and
\eqn\fsecondord{\eqalign{f^S_2(N,Q^2)= 
&\alpha_s(Q^2)
{\tilde\gamma^1_{fg}(\alpha_s(Q^2)/N)\over
\tilde\gamma^0_{gg}(\alpha_s(Q^2)/N)}\biggl(\eigveccor\times\cr
&\hskip 0.3in (\eigvecplus)+\tilde g_0(N,Q_0^2)\biggr)\eigplus\cr
&+{1\over 2} \biggl({4\over 9} {2N_f\over 6\pi b_0}
\ln\biggl({\alpha_s(Q_0^2)\over\alpha_s(Q^2)}\biggr)\biggr)^2
f^S_0(N)+\higherorder.\cr}}
Thus, the leading part of $f^S_2(N,Q^2)$ has terms of the same order as the
leading part of $f^S_1(N,Q^2)$, i.e. a sum of the form
$N\sum_{m=0}^{\infty}a_m(\alpha_s/N)^m$ 
multiplying $\eigplus$ and soft inputs. We also explicitly include above the
parts of $f^S_2(N,Q^2)$ which have terms of the form
$\ln(\alpha_s(Q_0^2)/\alpha_s(Q^2))$ directly multiplying
$f^S_0(N)$, even though
these are not of the same form as in $f^S_1(N,Q^2)$. We will take these terms
to be part of the lowest--order parton distribution
since they are not a power of $\alpha_s$ or $N$ higher than any terms
already produced. However, we ignore terms of the form 
$\ln(\alpha_s(Q_0^2)/\alpha_s(Q^2))$ multiplying series of the form 
$N\sum_{m=0}^{\infty}a_m(\alpha_s/N)^m$, such as $f^S_1(N,Q_0^2)$, 
because they are clearly
of higher order in $N$ than the term consisting of logarithms
multiplying $f_0^S(N)$. 

Similarly, for the gluon the presence of lowest--order terms is not limited to
$g_1(N,Q^2)$, but is also seen in $g_2(N,Q^2)$:
\eqn\gsecondord{\eqalign{g_2(N,Q^2)=&{1\over
2}\biggl(\eigveccor\biggr)^2\times\cr
&(\eigvecplus+\tilde g_0(N,Q_0^2))\times\eigplus \cr
&+\fourninths f_0^S(N)
{1\over 2}\biggl({4\over 9} {2N_f \over 6\pi b_0}\ln
\biggl({\alpha_s(Q_0^2)\over\alpha_s(Q^2)}\biggr)\biggr)^2 +\higherorder.\cr}}
This phenomenon persists to all orders, but is independent of 
the $\tilde\gamma^n$'s for $n\geq 1$. It can be proved by induction
in a straightforward manner that the leading part of the solutions to
the $n_{\rm th}$--order renormalization group equations are
\eqn\fnthord{\eqalign{f^S_n(N,Q^2)= 
&\alpha_s(Q^2)
{\tilde\gamma^1_{fg}(\alpha_s(Q^2)/N)\over
\tilde\gamma^0_{gg}(\alpha_s(Q^2)/N)}\eigplus \times\cr
&\hskip 1in {1\over (n-1)!}\biggl(\eigveccor\biggr)^{n-1}\times\cr
&(\eigvecplus+\tilde g_0(N,Q_0^2))\eigplus\cr
&+{(-1)^n\over n!} \biggl({4\over 9} {2N_f\over 6\pi b_0}
\ln\biggl({\alpha_s(Q_0^2)\over\alpha_s(Q^2)}\biggr)\biggr)^n
f^S_0(N)+ \higherorder,\cr}}
for $n\geq 2$, and
\eqn\gnthord{\eqalign{g_n(N,Q^2)=&{1\over n!}\biggl(
\eigveccor\biggr)^{n}\times\cr
&\hskip 1in (\eigvecplus +\tilde g_0(N,Q_0^2))\eigplus\cr
&-\fourninths f_0^S(N)
{(-1)^n\over n!}\biggl({4\over 9} {2N_f \over 6\pi b_0}\ln
\biggl({\alpha_s(Q_0^2)\over\alpha_s(Q^2)}\biggr)\biggr)^n +\higherorder,\cr}}
for $n\geq 0$. One can therefore find well--ordered parts of the full
solutions for
$f^S(N,Q^2)$ and $g(N,Q^2)$ by summing all such terms, leading to
\eqn\lof{\eqalign{f^S(N,Q^2)=& \alpha_s(Q^2)
{\tilde\gamma^1_{fg}(\alpha_s(Q^2)/N)\over
\tilde\gamma^0_{gg}(\alpha_s(Q^2)/N)}(\eigvecplus+\tilde g_0(N,Q_0^2))\times\cr
&\hskip 0.5in \eigpluscor\cr
&+f^S_0(N)\biggl({\alpha_s(Q_0^2)\over
\alpha_s(Q^2)}\biggr)^{-\fourninths{2N_f\over 6\pi b_0}}+
\tilde f^S_1(N,Q_0^2)+\higherorder,\cr}}
and,
\eqn\loglu{\eqalign{g(N,Q^2)=& (\eigvecplus+\tilde g_0(N,Q_0^2))\times\cr
&\hskip 0.2in \eigpluscor\cr
&-\fourninths f^S_0(N)\biggl({\alpha_s(Q_0^2)\over
\alpha_s(Q^2)}\biggr)^{-\fourninths{2N_f\over 6\pi b_0}}+ \higherorder.\cr}}
We may also write the expression for the rate of change of the quark
distribution 
\eqn\loderivf{\eqalign{-\alpha_s^2(Q^2) {d\, f^S(N,Q^2)\over
d\alpha_s(Q^2)}
=& \alpha_s(Q^2)
\tilde\gamma^1_{fg}(\alpha_s(Q^2)/N)(\eigvecplus+\tilde g_0(N,Q_0^2))\times\cr
&\eigpluscor\cr
&-\fourninths \alpha_s(Q^2)
f^S_0(N){2N_f\over 6\pi b_0} \biggl({\alpha_s(Q_0^2)\over
\alpha_s(Q^2)}\biggr)^{-{4\over 9}{2N_f\over 6\pi b_0}}+\higherorder.\cr}}
This last expression will be important since at leading order
$\alpha_s^2(Q^2)(d\,F_2(N,Q^2)/d\alpha_s(Q^2))$ is directly related to 
$d\, F_2(N,Q^2)/d\,\ln Q^2$ which  we will wish to study as well as
$F_2(N,Q^2)$. 

\subsec{The Leading--$\ln(1/x)$ Expansion: Structure Functions.}

It is now possible to examine the form of the solutions for the
structure functions. We can construct the leading part of
the full solutions by combining our leading solutions for the
parton distributions with the zeroth-- and first--order
coefficient functions. This gives  
\eqn\fullfl{\eqalign{F_L(N,Q^2)&= \alpha_s(Q^2)C^g_{L,1}(\alpha_s(Q^2)/N)
(\eigvecplus+\tilde g_0(N,Q_0^2))\times \cr
&\hskip 0.2in \eigpluscor\cr
&+\alpha_s(Q^2) (C^S_{L,1,0}-\fourninths C^g_{L,1,0})f^S_0(N)
\biggl({\alpha_s(Q_0^2)\over
\alpha_s(Q^2)}\biggr)^{-\fourninths{2N_f\over 6\pi b_0}}+\higherorder,\cr}}
where \coeffrellong\ has been used. Also
\eqn\fullftwo{\eqalign{F_2(N,Q^2)&= 
\alpha_s(Q^2)
{\tilde\gamma^1_{fg}(\alpha_s(Q^2)/N)\over
\tilde\gamma^0_{gg}(\alpha_s(Q^2)/N)}+
C^g_{2,1}(\alpha_s(Q^2)/N))
(\eigvecplus +\tilde g_0(N,Q_0^2))\times\cr
&\eigpluscor\cr
&+f^S_0(N)\biggl({\alpha_s(Q_0^2)\over
\alpha_s(Q^2)}\biggr)^{-\fourninths{2N_f\over 6\pi b_0}}+\tilde f^S_1(N,Q_0^2)
+\higherorder.\cr}}
We can also write 
\eqn\fullftwoderiv{\eqalign{-\alpha_s^2(Q^2){d\,F_2(N,Q^2)\over
d\alpha_s(Q^2)}= &(\alpha_s(Q^2)
\tilde\gamma^1_{fg}(\alpha_s(Q^2)/N)+\cr
&\hskip -0.35in \alpha_s(Q^2)
C^g_{2,1}(\alpha_s(Q^2)/N)\tilde\gamma^0_{gg}(\alpha_s(Q^2)/N))
(\eigvecplus+\tilde g_0(N,Q_0^2))\times \cr
&\eigpluscor\cr
&\hskip -0.3in -\alpha_s(Q^2){4\over 9}
{2N_f\over 6\pi b_0}f^S_0(N)\biggl({\alpha_s(Q_0^2)\over
\alpha_s(Q^2)}\biggr)^{-{4\over 9}{2N_f\over 6\pi b_0}}+\higherorder.\cr}}
In each case, since we only consider the singlet structure functions in this
expansion scheme, we drop the superscript $S$ for the structure functions for
the rest of this subsection.

It is clear that each of these expressions will be factorization
scheme independent since each represents the expression for a physical
quantity up to corrections of a form different from the terms explicitly
appearing and which we have deemed to be higher order in our expansion scheme.
$(\alpha_s^2(Q^2)(d\,F_2(N,Q^2)/d\alpha_s(Q^2))$ has as much right to
be considered as a physical quantity as $F_2(N,Q^2)$, since 
at leading order it
is proportional to $(d \, F_2(N,Q^2)/d\ln Q^2)$. However, we would
like to find out if each of these expressions may itself be 
more rigorously ordered and, as a by--product, split into
factorization--scheme invariant pieces.  This would be desirable 
from a theoretical point of view, since the current expressions will soon 
be seen to be incompatible with renormalization scheme consistency.
Also, from the practical point of view
we do not know all of the terms appearing in the explicit parts of
these expressions in any factorization scheme, i.e. $\tilde
\gamma^1_{gg}(\alpha_s(Q^2)/N)$ is not at present known, and we would
clearly like to have some sort of scheme--independent expression
for the structure functions involving terms we already know. 

\medskip

In order to see if we can obtain scheme--independent and thus 
physically relevant subsets of the solutions
\fullfl--\fullftwoderiv, we shall examine in detail the form of these
solutions. Since it is the least complicated, and because we will find
it useful, we begin with the expression \fullfl\ for the longitudinal
structure function. Since this expression contains two terms of rather
different form, i.e. the first depending on the factor $\eigplus$ and
the second having an evolution unenhanced by leading--$\ln (1/x)$ terms, 
it would be surprising if these terms were not
separately factorization scheme invariant. It is simple to check that
this is indeed the case.

Beginning with the second term, it is clear that this is factorization scheme 
independent. The
power of $(\alpha_s(Q_0^2)/\alpha_s(Q^2))$ comes from the one--loop
contribution to $\tilde\gamma^1_{fg}(\alpha_s/N)$ which is both
factorization scheme and renormalization scheme invariant and which we
have already written in its actual numerical form. $\alpha_s(Q^2)$ and
$f^S_0(N)$ are both clearly renormalization and factorization
scheme dependent, and the one--loop contributions to the longitudinal
coefficient functions are also, i.e. $C^S_{L,1,0}=(2/3\pi)$ and
$C^g_{L,1,0}= (2N_f/6\pi)$ in all schemes. Hence the whole term is
both factorization scheme and renormalization scheme independent, as it
must be, depending only on leading--order quantities.

The first term is not as simple to deal with. 
It must be factorization scheme independent; however, many of the
pieces appearing in the term are clearly not, e.g. 
$\tilde\gamma^1_{fg}(\alpha_s/N)$ and
$C^g_{L,1}(\alpha_s/N)$. So, to a certain extent it must be the
interplay between the $\tilde\gamma$'s, the coefficient function and
the parton inputs which leads to a factorization--scheme--independent
result. Remembering our discussion of the solutions for the
structure function in the loop expansion, 
we see that there is at least one way of writing this term as the
product of two factorization--scheme--independent pieces, i.e.
\eqn\firstterm{\eqalign{&\alpha_s(Q^2)C^g_{L,1}(\alpha_s(Q^2)/N)
(\eigvecplus+\tilde g_0(N,Q_0^2))\times \cr
&\hskip 1in \eigpluscor\cr
&=\alpha_s(Q_0^2)C^g_{L,1,0}
(\eigvecplus)\biggl({C^g_{L,1}(\alpha_s(Q_0^2)/N)\over
C^g_{L,1,0}}.\biggl(1+{\tilde g_0(N,Q_0^2)\over \eigvecplus}\biggr)
\biggr)\times \cr
&\hskip 2in \eigplus\exp\biggl[-\ln\biggl({\alpha_s(Q_0^2)\over \alpha_s(Q^2)}
\biggr)\biggr]\times\cr
&\eigcomiC.\cr}}
The left--hand side is factored into its value at $Q_0^2$ and
a term which gives the evolution from this value.
Clearly these are physically distinguishable and as such must be
individually factorization scheme invariant. Also, we already know
that $\eigplus$ is factorization scheme invariant on its own, and 
$\exp[-\ln(\alpha_s(Q_0^2)/ \alpha_s(Q^2))]$ clearly is. Hence, 
we make the definition
\eqn\phioneplus{\Phi^+_1(Q^2,Q_0^2)=\!\!\lneigcomiC,}
where $\Phi^+_1(Q^2,Q_0^2)$ must now be factorization--scheme independent
(as can be checked using the rules \transpart--\transgamma)
and $\Phi^+_1(Q_0^2,Q_0^2)=1$. 

Having isolated the factorization--scheme--independent parts, we may
factorize \fullfl\  completely into factorization--scheme--invariant input
and evolution parts, i.e. 
\eqn\fullflalt{\eqalign{F_L(N,Q^2)&= \alpha_s(Q_0^2){2N_f\over 6\pi}
(\eigvecplus)\eigplus\times \cr
&\biggl({C^g_{L,1}(\alpha_s(Q_0^2)/N)\over
C^g_{L,1,0}}.\biggl(1+{\tilde g_0(N,Q_0^2)\over \eigvecplus}\biggr)\biggr)
\exp\biggl[\Phi^+_1(Q^2,Q_0^2)-\ln\biggl({\alpha_s(Q_0^2)
\over \alpha_s(Q^2)}\biggr)\biggr]\cr
&+\alpha_s(Q^2) \biggl({18-4N_f\over 27\pi}\biggr)f^S_0(N)
\biggl({\alpha_s(Q_0^2)\over
\alpha_s(Q_0^2)}\biggr)^{-\fourninths{2N_f\over 6\pi b_0}-1} + 
\higherorder.\cr}}
It is now possible to attach direct physical significance to each of the
factorization--scheme--independent pieces appearing in this
expression. 

We first consider the inputs. Going to \fullftwo\ for the 
moment we see that
$f^S_0(N)$ is the only term in $F_2(N,Q_0^2)$ which is zeroth--order 
in our expansion scheme. As such it is the
zeroth order input for $F_2(N,Q^2)$, and we may write 
\eqn\defftwozero{f^S_0(N)=F_{2,0}(N),}
and this is one of our two fundamentally nonperturbative inputs.
Also, from \fullflalt\ we see that
the total $\alpha_s(Q_0^2)$--independent input for $F_L(N,Q_0^2)$ (once
we have divided out a single power of $\alpha_s(Q_0^2)/(2\pi)$)
is
\eqn\flinput{{2N_f \over 3}(g_0(N)+\fourninths f^S_0(N))+
\biggl({36-8N_f\over 27}\biggr)f^S_0(N).}
This whole expression may therefore be written as $\hat F_{L,0}(N)$,
and we have 
\eqn\defflzero{\hat F_{L,0}(N)={2N_f \over 3}(g_0(N)+{2\over N_f} f^S_0(N)).}
$\hat F_{L,0}(N)$ is our other fundamentally nonperturbative physical input. 
We note that the expression for $F_{2,0}(N)$ in terms of the parton inputs is
the same in this case, i.e. \defftwozero, as in the loop expansion \linkf.
However, the definition of $F_{L,0}(N)$ in terms of the parton inputs, 
\defflzero, is not the same as in the loop expansion \linkg. Thus, 
the definition of the 
nonperturbative gluon, $g_0(N)$, is different in the two expansion schemes. 

We also make a similar definition for the part of the input 
for $\hat F_L(N,Q^2)$ which is of the form
$\sum_{n=1}^{\infty}a_n(Q_0^2)(\alpha_s(Q_0)/N)^n$ multiplying 
$(\eigvecplus)$, or more correctly multiplying $(\hat F_{L,0}(N)-((36-8N_f)/27)
F_{2,0}(N))$, to complete our
definition of the input in \fullflalt\ and write
\eqn\defflone{\eqalign{{2N_f\over 3}(g_0(N)+\fourninths f^S_0(N)+
\tilde g_0(N,Q_0^2)) &\biggl({C^g_{L,1}(\alpha_s(Q_0^2)/N)
\over C^g_{L,1,0}}\biggr)\cr
&=\hat F_{L,0}(N)-\biggl({36-8N_f\over 27}\biggr)F_{2,0}(N)
+\tilde {\hat F}_{L,0}(N,Q_0^2).\cr}}
This whole expression must be both factorization scheme and 
renormalization scheme independent, facts which reveal information 
about the form of the gluon 
input. $C^g_{L,1}(N,\alpha_s(Q_0^2))$ is both renormalization and 
factorization--scheme dependent, so the scheme dependence of $\tilde 
g_0(N,Q_0^2)$ must be precisely so as to cancel this out, i.e. 
\eqn\cancel{\eqalign{(\eigvecplus&+\tilde g_0(N,Q_0^2))\cr
&\hskip -0.6in\equiv (\eigvecplus)\biggl(1+\sum_{m=1}^{\infty}\tilde g_{0,m}
(\alpha_s(Q_0^2)/N)^m\biggr)\exp[\ln(Q_0^2/A_{LL})
\gamma^0_{gg}(\alpha_s(Q_0^2)/N)]\cr
&\hskip -0.6in =(\eigvecplus){\rm g}_{0}(N,Q_0^2)
(C^g_{L,1,0}/ C^g_{L,1}(\alpha_s(Q_0^2)/N))\exp[\ln(Q_0^2/A_{LL})
\gamma^0_{gg}(\alpha_s(Q_0^2)/N)],}} 
where ${\rm g}_0(N,Q_0^2)$ is scheme independent. Hence, there is no reason 
for $\tilde {\hat F}_{L,0}(\alpha_s(Q_0^2)/N)$ to depend on the 
leading--order longitudinal gluon coefficient function at all, 
and indeed, a natural 
choice seems to be that $(1+\sum_{m=1}^{\infty}\tilde g_{0,m}
(\alpha_s(Q_0^2)/N)^m)$ is chosen equal to 
$(C^g_{L,1,0}/C^g_{L,1}(N,\alpha_s(Q_0^2)))$ (it is difficult to see
what else it could be chosen equal to), i.e. ${\rm g}_{0}(N,Q_0^2)=1$,
and therefore that
\eqn\inexpfl{\eqalign{\hat F_{L,0}(N) -\biggl(&{36-8N_f\over 27}\biggr)
F_{2,0}(N)+\tilde {\hat F}_{L,0}(N,Q_0^2)\cr
&=\biggl(\hat F_{L,0}(N) 
-\biggl({36-8N_f\over 27}\biggr)F_{2,0}(N)\biggr)
\exp[\ln(Q_0^2/A_{LL})\gamma^0_{gg}(\alpha_s(Q_0^2)/N)].\cr}} 
Hence, we have a prediction for the input for the longitudinal structure 
function at small $x$ in terms of the nonperturbative inputs
and some scale $A_{LL}$ (where $A_{LL}=A_{gg}$ from the previous subsection).
As with $A_{NS}$ earlier, $A_{LL}$ is the scale at which the input is equal to 
the nonperturbative input alone, and hence we would expect it to be typical
of the scale where perturbation theory starts to break down. 
As we have already stressed, 
$Q_0^2$ is a completely free parameter. We have constructed the solution 
to be insensitive to $Q_0^2$ at leading order, but there is clearly some
residual $Q_0^2$--dependence. Hence, there 
will also be some optimum $Q_0^2$ to choose as the starting scale. 

We may now examine the terms governing the evolution, continuing the
convention started in
\phioneplus. $\eigplus$ is the factorization--scheme--independent factor
coming from the eigenvalue of the zeroth--order anomalous dimension
which will govern the small--$x$ growth with $Q^2$. As such we make the
definition 
\eqn\phizeroplus{\Phi^+_0(Q^2,Q_0^2) =
\int_{\alpha_s(Q^2)}^{\alpha_s(Q_0^2)}{\tilde
\gamma^0_{gg}(\alpha_s(q^2)/N)\over
\alpha^2_s(q^2)}d\alpha_s(q^2).}
Of course, the other evolution factor resulting from the eigenvalues
of the zeroth--order anomalous dimension was simply unity and as such
$\Phi^-_0(Q^2,Q^2_0)$ does exist, but
is implicitly zero. However, there is a
correction to this factor of unity in \fullfl--\fullftwoderiv, and we make
the definition 
\eqn\phioneminus{\Phi^-_1(Q^2,Q_0^2)=-{4\over 9}{2N_f\over
6\pi b_0}\ln\biggl({\alpha_s(Q_0^2)\over \alpha_s(Q^2)}\biggr).}

Having made these factorization--scheme--invariant definitions for the
inputs and evolution we may write the solution for $F_L(N,Q^2)$ as
\eqn\fullflalti{\eqalign{F_L(N,Q^2)&= {\alpha_s(Q_0^2)
\over 2\pi}(\hat F_{L,0}(N)+\tilde {\hat F}_{L,0}(N,Q_0^2)
-\biggl({36-8N_f\over 27}\biggr)F_{2,0}(N))\times\cr
&\exp\biggl[\Phi^+_0(Q^2,Q_0^2)+\Phi^+_1(Q^2,Q_0^2)-\ln\biggl(
{\alpha_s(Q_0^2)\over \alpha_s(Q^2)}\biggr)\biggr]\cr
&+{\alpha_s(Q_0^2) \over 2\pi} \biggl({36-8N_f\over 27}\biggr)F_{2,0}(N)
\exp\biggl[\Phi^-_1(Q^2,Q_0^2)-\biggl(\ln\biggl(
{\alpha_s(Q_0^2)\over \alpha_s(Q^2)}\biggr)\biggr)\biggr]\cr
&+\higherorder.\cr}}
It is now relatively obvious how we may separate out the ``leading
part'' from this expression. $[\Phi^+_1(Q^2,Q_0^2)-\ln(
\alpha_s(Q_0^2)/ \alpha_s(Q^2))]$ contains 
the same type of terms as $\Phi^+_0(Q^2,Q_0^2)$, but each is a power of
$N$ higher. Thus $[\Phi^+_1(Q^2,Q_0^2)-\ln(
\alpha_s(Q_0^2)/ \alpha_s(Q^2))]$ is subleading to
$\Phi^+_0(Q^2,Q_0^2)$, and indeed $\Phi^+_1(Q^2,Q_0^2)$ is a 
renormalization--scheme--dependent
quantity, as it must be in order to absorb the change in 
$\Phi^+_0(Q^2,Q_0^2)$ resulting from 
a change in the definition of $\alpha_s(Q^2)$ under a change 
in renormalization scheme when working beyond leading order. 
Hence, we should factor $\exp[\Phi^+_1(Q^2,Q_0^2)-\ln(
\alpha_s(Q_0^2)/ \alpha_s(Q^2))]$ out of the first term. Since
$[\Phi^-_1(Q^2,Q_0^2)-\ln(\alpha_s(Q_0^2)/ \alpha_s(Q^2))]$ is of the 
same form as the zeroth--order--in--$N$ 
part of $\Phi^+_1(Q^2,Q_0^2)$, then it 
should also be factored out of the leading--order expression for
$F_L(N,Q^2)$. Thus, we are left with 
\eqn\lofl{\eqalign{F^0_L(N,Q^2)= &{\alpha_s(Q_0^2)\over 2\pi}
(\hat F_{L,0}(N)+\tilde {\hat F}_{L,0}(N,Q_0^2)-
\biggl({36-8N_f\over 27}\biggr)F_{2,0}(N))
\exp[\Phi^+_0(Q^2,Q_0^2)]\cr
&+{\alpha_s(Q_0^2)\over 2\pi} 
\biggl({36-8N_f\over 27\pi}\biggr)F_{2,0}(N).\cr}}
Using the one--loop running coupling constant, as is appropriate for a
leading--order expression, the whole of \lofl\ is not only manifestly
factorization scheme independent, but also renormalization scheme
independent, as one would hope. Also, $\tilde {\hat F}_{L,0}(N,Q_0^2)$ is 
constructed precisely so as to make the expression unchanged,
at this order, under a change in
starting scale. Hence, within this expansion scheme \lofl\
is genuinely the leading--order expression for $F_L(N,Q^2)$.

\medskip

We now turn our attention to the more phenomenologically important
case of the structure function $F_2(N,Q^2)$. Here we immediately have
an ambiguity: should we consider the expression for the
structure function itself in this expansion scheme, or that for its
$\alpha_s$--derivative, $\alpha_s^2(Q^2)(d F_2(N,Q^2)/d
\alpha_s(Q^2))$ (which, using the definition of the running coupling
\runcoup, is directly related to $(d F_2(N,Q^2)/d\ln Q^2)$
at leading order)? It may be
argued that in certain senses the latter is more natural because it is
a ``real'' perturbative quantity, beginning at first order in
$\alpha_s$, as does $F_L(N,Q^2)$. 

There is in fact a distinction between $F_2(N,Q^2)$ and 
$(d\,F_2/d\,\ln Q^2)$ in the usual loop expansion. Differentiating a 
fixed--order expression for $F_2(N,Q^2)$ and using the $\beta$--function
evaluated to the appropriate order in $\alpha_s$ results
in the fixed--order expression for $(d\,F_2/d\,\ln Q^2)$\foot{$(d\, 
F_2(N,Q^2)/d\,\ln Q^2)$ 
is analogous to the longitudinal structure function, i.e. it has an 
overall power of $\alpha_s(Q^2)$ and the lowest--order expression
consists of first order--in--$\alpha_s(Q_0^2)$ 
inputs multiplying the eigenvalue--determined 
evolution terms $\bigl({\alpha_s(Q_0^2)\over \alpha_s(Q^2)}
\bigr)^{\tilde\gamma^{0,l}_{+,-}-1}$.} plus terms of higher
order in $\alpha_s(Q^2)$, which depend on the $\beta$--function beyond lowest 
order (and thus are absent when working to leading order only).
Hence, the size of these extra terms is of the same order as
the renormalization scheme uncertainty, and therefore the distinction 
between the fixed--order expressions for $F_2(N,Q^2)$ and 
$(d\,F_2/d\,\ln Q^2)$ is of similar magnitude to the 
distinction between renormalization schemes.
    
There is also a distinction when using the small--$x$ expansion, and it appears
more graphically, and is not only dependent on terms in the
$\beta$-function beyond lowest order. We distinguished between $F_2(N,Q^2)$ and
$\alpha_s^2(Q^2)(d\,F_2(N,Q^2)/d\alpha_s(Q^2))$ in \fullftwo\ and
\fullftwoderiv\ because the expression to a given order in $\alpha_s$
and $N$ even for $\alpha_s^2(Q^2)(d\,F_2(N,Q^2)/d\alpha_s(Q^2))$ is
no longer simply obtained by differentiating $F_2(N,Q^2)$ at given order,
and conversely given order in $F_2(N,Q^2)$ is not obtained just by integrating
$\alpha_s^2(Q^2)(d\,F_2(N,Q^2)/d\alpha_s(Q^2))$ with respect to
$\alpha_s$ at fixed order. This was clearly illustrated
when finding the leading part of $f^S_1(N,Q^2)$ by solving the
equation for $\alpha_s^2(Q^2)(d\,f^S_1(N,Q^2)/d\alpha_s(Q^2))$
earlier, and it leads to a distinction 
between even the LO $(d\,F_2/d\,\ln Q^2)$ and the 
derivative of the LO $F_2(N,Q^2)$ in this expansion
scheme. Thus, even at leading order we have to decide 
which of the two expressions to
use, although we would hope that there is not too much difference between
the choices in practice. We will postpone the decision for the 
present and will derive the form of the
leading--order expressions for both $F_2(N,Q^2)$
and its derivative. We can then examine these expressions  
and use them in order to help us decide. 

For technical simplicity we begin with the expression for
$\alpha^2_s(Q^2)(d\,F_2(N,Q^2)/d\alpha_s(Q^2))$. 
Using the definitions introduced in
our discussion for the longitudinal structure function we may write
\fullftwoderiv\  as
\eqn\fullftwoderivalt{\eqalign{-\alpha^2_s(Q^2){d \, F_2(N,Q^2)\over
d\alpha_s(Q^2)} =& (\alpha_s(Q^2)\tilde\gamma^1_{fg}(\alpha_s(Q^2)/N)+
\alpha_s(Q^2)C^g_{2,1}(\alpha_s(Q^2)/N)\tilde\gamma^0_{gg}(\alpha_s(Q^2)/N))
\times\cr 
&\hskip -1in {3\over 2N_f}\biggr({C^g_{L,1,0}\over
C^g_{L,1}(\alpha_s(Q^2)/N)}\biggl)(\hat F_{L,0}(N) +\tilde {\hat 
F}_{L,0}(N,Q_0^2)-\biggl({36-8N_f\over 27}\biggr)F_{2,0}(N))\times\cr
& \hskip -1in \exp[\Phi^+_0(Q^2,Q_0^2)+\Phi_1^+(Q_0^2,Q^2)] 
-\alpha_s(Q^2){4\over 9}{2N_f \over
6\pi b_0}F_{2,0}(N)\exp[\Phi_1^-(Q^2,Q_0^2)]\cr
& + \higherorder.\cr}}
The factorization scheme independence of this complete expression  
guarantees that the term $(\tilde\gamma^1_{fg}(\alpha_s(Q^2)/N)+
C^g_{2,1}(\alpha_s(Q^2)/N)\tilde\gamma^0_{gg}(\alpha_s(Q^2)/N)) 
(C^g_{L,1,0}/C^g_{L,1}(\alpha_s(Q^2)/N))$ is a 
factorization--scheme--invariant quantity, of the form
$\sum_{m=0}^{\infty}a_m(\alpha_s(Q^2)/N)^m$. 
It is straightforward to verify this. Indeed, it was
shown by Catani and Hautmann that  
$(\tilde\gamma^1_{fg}(\alpha_s(Q^2)/N)+
C^g_{2,1}(\alpha_s(Q^2)/N)\tilde\gamma^0_{gg}(\alpha_s(Q^2)/N))$ and
$ C^g_{L,1}(\alpha_s(Q^2)/N)$ could each always be expressed
in terms of the product of a factorization--scheme-- and 
renormalization--scheme--independent factor (which they calculated) 
and a scheme--dependent factor, where the
scheme--dependent part was the same for both \cathaut. Using their results for
the scheme--independent parts it is a trivial matter to find that 
\eqn\schindepratio{\eqalign{(\gamma^1_{fg}(\alpha_s(Q^2)/N)
+C^g_{2,1}(\alpha_s(Q^2)/N)\gamma^0_{gg}&(\alpha_s(Q^2)/N)))
{3\over 2N_f}\biggl({C^g_{L,1,0}\over C^g_{L,1}(\alpha_s(Q^2)/N)}\biggr)\cr
&= {1\over 2\pi}\biggl({3\over
2}\gamma^0_{gg}(\alpha_s(Q^2)/N)
+\sum_{n=0}^{\infty}(\gamma^0_{gg}(\alpha_s(Q^2)/N))^n\biggr),\cr}}
which is clearly both factorization scheme and renormalization scheme
independent. Thus, making the definition
\eqn\gamtwol{\alpha_s(Q^2)\tilde\gamma^1_{2L}(\alpha_s(Q^2)/N)=
{1\over 2\pi b_0}\biggl({3\over
2}+\sum_{n=0}^{\infty}(\alpha_s(Q^2)
\gamma^0_{gg}(\alpha_s(Q^2)/N))^n\biggr),}
our expression for the explicit part of \fullftwoderiv\ is entirely in
terms of factorization--scheme--invariant, and hence physically
meaningful quantities, i.e. 
\eqn\fullftwoderivalt{\eqalign{-\alpha^2_s(Q^2){d \, F_2(N,Q^2)\over
d\alpha_s(Q^2)} =& \alpha_s(Q^2)\tilde\gamma^1_{2L}(\alpha_s(Q^2)/N)
(\hat F_{L,0}(N) +\tilde {\hat F}_{L,0}(N,Q_0^2)-
\biggl({36-8N_f\over 27}\biggr)F_{2,0}(N)) \times\cr
& \exp[\Phi^+_0(Q^2,Q_0^2)+\Phi_1^+(Q_0^2,Q^2)] 
-\alpha_s(Q^2){4\over 9}{2N_f \over
6\pi b_0}F_{2,0}(N)\exp[\Phi_1^-(Q^2,Q_0^2)] \cr
&+ \higherorder.\cr}}

We now have an expression which is analogous to that for
$\hat F_L(N,Q^2)$ in \fullflalti, except that it has still 
not been explicitly 
separated into inputs and evolution terms. In order to do this we must 
rewrite \fullftwoderivalt\ as 
\eqn\fullftwoderivalti{\eqalign{-\alpha^2_s(Q^2){d \, F_2(N,Q^2)\over
d\alpha_s(Q^2)} =& \alpha_s(Q_0^2)\tilde\gamma^1_{2L}(\alpha_s(Q_0^2)/N)
(\hat F_{L,0}(N) +\tilde {\hat F}_{L,0}(N,Q_0^2)-
\biggl({36-8N_f\over 27}\biggr)F_{2,0}(N))\times\cr
& \hskip 1.4in \exp[\Phi^+_0(Q^2,Q_0^2)+\Phi_1^+(Q_0^2,Q^2)+
\tilde \Phi^+_1(Q_0^2,Q^2)] \cr
&\hskip -1in -{4\over 9}{2N_f \over
6\pi b_0}\alpha_s(Q_0^2)F_{2,0}(N)\exp\biggl[\Phi_1^-(Q^2,Q_0^2)-\ln\biggl(
{\alpha_s(Q_0^2)\over \alpha_s(Q^2)}\biggr)\biggr] + \higherorder,\cr}}
where $\tilde \Phi^+_1(Q^2,Q_0^2)$ is a series of the same form as 
$\Phi^+_1(Q^2,Q_0^2)$, defined as 
\eqn\deftwidphi{\tilde \Phi^+_1(Q^2,Q_0^2)=
-\int_{\alpha_s(Q^2)}^{\alpha_s(Q_0^2)} {d\over d\, \alpha_s(q^2)}\ln(
\alpha_s(q^2)\tilde \gamma^1_{2,L}(N,\alpha_s(q^2)))d\alpha_s(q^2).}

It is now clear how we obtain the ``leading part'' of this expression. 
All of the inputs are of leading order, but we must factor out the 
subleading parts of the evolution. $[\Phi^+_1(Q^2,Q_0^2)+\tilde 
\Phi^+_1(Q^2, Q_0^2)]$
is subleading to $\Phi^+_0(Q^2,Q_0^2)$ (we note that 
$\tilde \Phi^+_1(Q^2,Q_0^2)$ is  a renormalization--scheme--independent 
contribution to
this subleading evolution), and must be factored out. Since it is of 
the same order, so must $[\Phi^-_1(Q^2,Q_0^2)-\ln(\alpha_s(Q_0^2)/ 
\alpha_s(Q^2))]$. This leaves us with the leading--order expression
\eqn\loftwoderiv{\eqalign{-\biggl(\alpha^2_s(Q^2){d \, F_2(N,Q^2)\over
d\alpha_s(Q^2)}\biggr)_0 =&\alpha_s(Q_0^2)
\tilde\gamma^1_{2L}(\alpha_s(Q_0^2)/N)\times\cr
&(\hat F_{L,0}(N) +\tilde {\hat F}_{L,0}(N,Q_0^2)-
\biggl({36-8N_f\over 27}\biggr)F_{2,0}(N)) \exp
(\Phi^+_0(Q^2,Q_0^2))\cr
&-\alpha_s(Q_0^2){4\over 9}{2N_f \over
6\pi b_0}F_{2,0}(N).\cr}}
Again, this expression is insensitive to changes in starting scale, 
up to higher order, and using the one--loop coupling constant, is
renormalization scheme independent as well as factorization scheme independent.

\medskip

Finally we consider the expression for $F_2(N,Q^2)$ itself.
It is now a relatively simple matter to write this in terms of
factorization--scheme--independent quantities and in terms of inputs and 
evolution terms. Using the definitions we
have already made and defining
$\hat \Phi^+_1(Q^2,Q_0^2)$ by
\eqn\defhatphi{\hat \Phi^+_1(Q^2,Q_0^2)=
-\int_{\alpha_s(Q^2)}^{\alpha_s(Q_0^2)} {d\over d\, \alpha_s(q^2)}\ln
\biggl({\alpha_s(q^2)\gamma^1_{2,L}(\alpha_s(q^2)/N)\over 
\gamma^0_{gg}(N,\alpha_s(q^2))}\biggr)d\alpha_s(q^2).}
we may write \fullftwo\ as 
\eqn\fullftwoalt{\eqalign{F_2(N,Q^2)=& \alpha_s(Q_0^2)
{\tilde\gamma^1_{2L}(\alpha_s(Q_0^2)/N)\over \tilde
\gamma^0_{gg}(\alpha_s(Q_0^2)/N)}
(\hat F_{L,0}(N)
+ \tilde {\hat F}_{L,0}(N,Q_0^2) -\biggl({36-8N_f\over 27}\biggr)
F_{2,0}(N))\times\cr
&\exp[\Phi^+_0(Q^2,Q_0^2)+\Phi_1^+(Q_0^2,Q^2)+\hat\Phi^+_1(Q^2,Q_0^2)]
+F_{2,0}(N)\exp[\Phi_1^-(Q^2,Q_0^2)]\cr
&+\tilde f^S_1(N,Q_0^2)+ \higherorder.\cr}}
So the factorization--scheme--independent input $F_2(N,Q_0^2)$ is 
\eqn\defftwoonein{F_{2,0}(N)+\alpha_s(Q_0^2)
{\tilde\gamma^1_{2L}(\alpha_s(Q_0^2)/N)\over \tilde
\gamma^0_{gg}(\alpha_s(Q_0^2)/N)}
(\hat F_{L,0}(N)
+ \tilde {\hat F}_{L,0}(N,Q_0^2) -\biggl({36-8N_f\over 27}\biggr)
F_{2,0}(N))+\tilde f^S_1(N,Q_0^2).}
By construction it is guaranteed that
the change of \defftwoonein\ under a change in starting scale
will cancel the change in the evolution in the first term in
\fullftwoalt\ under a change in starting scale up to higher orders.
It is also clear that \defftwoonein\ is 
both renormalization scheme and factorization scheme 
independent, as we require, as long as $\tilde f^S_1(N,Q_0^2)$ is scheme
independent. In fact, the requirement that if $Q_0^2=A_{LL}$ the input 
reduces to the nonperturbative input $F_{2,0}(N)$ determines 
$\tilde f^S_1(N,Q_0^2)$ uniquely. It must be the scheme--independent quantity
$-\alpha_s(Q_0^2)(\tilde\gamma^1_{2L}(\alpha_s(Q_0^2)/N)/ \tilde
\gamma^0_{gg}(\alpha_s(Q_0^2)/N))
(\hat F_{L,0}(N)-\biggl({36-8N_f\over 27}\biggr)F_{2,0}(N))$. Hence, the 
input for $F_2(N,Q^2)$ is
\eqn\defftwooneinc{F_{2}(N,Q_0^2)=F_{2,0}(N)+\alpha_s(Q_0^2)
{\tilde\gamma^1_{2L}(\alpha_s(Q_0^2)/N)\over \tilde
\gamma^0_{gg}(\alpha_s(Q_0^2)/N)}
\tilde {\hat F}_{L,0}(N,Q_0^2)+\higherorder,}
and the expression for $F_2(N,Q^2)$ is
\eqn\fullftwoaltc{\eqalign{F_2(N,Q^2)=& \alpha_s(Q_0^2)
{\tilde\gamma^1_{2L}(N,\alpha_s(Q_0^2))\over \tilde
\gamma^0_{gg}(N,\alpha_s(Q_0^2))}
(\hat F_{L,0}(N)
+ \tilde {\hat F}_{L,0}(N,Q_0^2) -\biggl({36-8N_f\over 27}\biggr)
F_{2,0}(N))\times\cr
&\exp[\Phi^+_0(Q^2,Q_0^2)+\Phi_1^+(Q_0^2,Q^2)+\hat\Phi^+_1(Q^2,Q_0^2)]
+F_{2,0}(N)\exp[\Phi_1^-(Q^2,Q_0^2)]\cr
&-\alpha_s(Q_0^2)
{\tilde\gamma^1_{2L}(N,\alpha_s(Q_0^2))\over \tilde
\gamma^0_{gg}(N,\alpha_s(Q_0^2))}
(\hat F_{L,0}(N)-\biggl({36-8N_f\over 27}\biggr)
F_{2,0}(N))\cr
&+ \higherorder.\cr}}

By comparing \defftwooneinc\ with \lofl\ we now see that there is a very
direct relationship 
between the inputs for our two structure functions at small $x$, i.e. 
we have a definite prediction, up to additive nonperturbative parts
which are flat at small $x$, for one in terms of the other, as well  
as approximate predictions for the form of each. 
Also comparing with the form 
of \loftwoderiv, these two inputs for the structure functions are directly
related to the slope of $(d\,F_2/d\ln Q^2)$ for small $x$ at $Q_0^2$.  
As already mentioned,
we do not yet know at what $Q_0^2$ it is most appropriate to choose the 
inputs, but it is a nontrivial requirement that the inputs for the 
three expressions are of the correct form and related in the above manner at 
any $Q_0^2$.    
 
Our expression for $F_2(N,Q^2)$
cannot be split into leading and next--to--leading pieces
in quite such a clear and symmetric way as our previous two examples, 
essentially because it begins at zeroth order, not
at first order in $\alpha_s$. Looking at
\fullftwoaltc\ it is clear that there is only one LO (in our 
expansion scheme) input multiplied by a LO evolution, and that is 
$F_{2,0}(N)$ multiplying unity. Hence we take 
the leading--order expression to be given
simply by
\eqn\loftwo{\tilde F_{2,0}(N,Q^2)=F_{2,0}(N).}
This is obviously completely independent of $\alpha_s$, 
and is rather trivial. At next--to--leading order, or equivalently, at 
leading--$\alpha_s$--dependent order, we include the whole of \fullftwoaltc\
except that we factor $\exp[\Phi^+_1(q^2,Q_0^2)+ \hat
\Phi^+_1(Q^2,Q_0^2)]$, out of the first term, i.e. we have NLO 
inputs multiplying LO evolution, and vice versa, as well as
$\tilde F_{2,0}(N,Q^2)$. Hence, at 
leading--$\alpha_s$--dependent order we have
\eqn\firstftwo{\eqalign{F_{2,0}&(N,Q^2)= F_{2,0}(N)\exp[\Phi_1^-(Q^2,Q_0^2)]\cr
&+\alpha_s(Q_0^2)
{\gamma^1_{2L}(\alpha_s(Q_0^2)/N)\over \gamma^0_{gg}(\alpha_s(Q_0^2)/N)}
(\hat F_{L,0}(N)
+\tilde {\hat F}_{L,0}(N,Q_0^2) -\biggl({36-8N_f\over 27}\biggr)F_{2,0}(N))
\exp[\Phi^+_0(Q^2,Q_0^2)]\cr
&-\alpha_s(Q_0^2)
{\gamma^1_{2L}(\alpha_s(Q_0^2)/N)\over \gamma^0_{gg}(\alpha_s(Q_0^2)/N)}
(\hat F_{L,0}(N)-\biggl({36-8N_f\over 27}\biggr)F_{2,0}(N))\cr
&+ \higherorder.\cr}}
In this expression there are clearly no terms which mix if we were to make a
change in definition of the coupling $\alpha_s \to \alpha_s + \epsilon
\alpha_s^2$, and hence we can consider it as a leading--order 
expression. If the one--loop coupling is used, it is both
factorization scheme and renormalization scheme independent.

\medskip

We now have the full set of LO expressions in the leading--$\ln(1/x)$ 
expansion scheme. We could obtain the correct scheme--independent expressions
for the structure functions at higher orders within this expansion scheme. 
We choose to finish at 
leading order, however. The labour required to obtain higher--order 
expressions becomes progressively greater and we would obtain 
expressions requiring unknown anomalous dimensions and coefficient functions.
Working to NLO we would need to calculate all the NLO evolutions and all
the NLO inputs. This would require all the NLO expressions 
for the anomalous dimensions and coefficient functions, i.e. 
the $\tilde \gamma^1_{g,a}$'s, $\tilde \gamma^2_{f,a}$'s and $C^a_{i,2}$'s.
There is optimism that these NLO terms will soon be known
\ref\nlobfkl{L.N. Lipatov and V.S. Fadin, \SJNP \vyp{50}{1989}{712}\semi
V.S. Fadin, R. Fiore and A. Quartarolo, \PR \vyp{D53}{1996}{2729}\semi
V. Del Duca, \PR \vyp{D54}{1996}{989}\semi
G. Camici and M. Ciafaloni, \PL \vyp{B386}{1996}{341}.}, and once this
is so the full NLO scheme--independent expressions should be calculated.

\bigskip

We should make some comments
about our LO scheme--independent expressions for $F_L(N,Q^2)$, 
$F_2(N,Q^2)$ and $(d F_2(N,Q^2)/d\ln (Q^2))$. First we note that all depend on
factorization--scheme--independent combinations of the $\tilde
\gamma^0_{ga}$'s, $\tilde\gamma^1_{fa}$'s 
and $C^a_{i,1}$'s (along with the input
parton distributions, where $g_0(N,Q_0^2)$ is also factorization
scheme dependent). There is, however, no terribly simple prescription
for how one uses the anomalous dimensions and coefficient functions in order 
to arrive at the expressions. One
does certainly not take the known anomalous dimensions,
solve the full renormalization group equations, and combine with the
known coefficient functions. This introduces terms we do not
even include in \fullfl--\fullftwoderiv, let alone in \lofl, \loftwoderiv\ and
\firstftwo. The additional terms will depend
on the factorization scheme used, and can be very large. The method of
determining LO scheme--independent
expressions is also more complicated than
solving the renormalization group equations to a given order and
combining with coefficient functions to a given combined order, as in the 
loop expansion. The only way to obtain the correct expressions using the
parton model directly seems to be to calculate carefully and keep all
terms of a given type, as explained.       
   
\medskip

We should also make some mention of why factorization scheme dependence can 
be very large in this expansion scheme. In order to do this let us consider
\firstterm\ as an example. A representative example of the way in which 
factorization--scheme--dependent calculations are done is to consider 
this expression evaluated with $\tilde\gamma^0_{gg}(\alpha_s(Q^2)/N)$, 
$\tilde\gamma^1_{fg}(\alpha_s(Q^2)/N)$ 
and $C^g_{L,1}(\alpha_s(Q^2)/N)$
known in some particular factorization scheme, 
but $\tilde\gamma^1_{gg}(\alpha_s(Q^2)/N)$, which is 
unknown, either set equal to zero, or guessed by imposing some ansatz such 
as momentum conservation (which may well give completely the wrong answer).

As a first comment we consider the input. The input
in terms of parton distributions is multiplied by 
$C^g_{L,1}(\alpha_s(Q_0^2)/N)$, a series of the form 
$\sum_{m=0}^{\infty}a_m(\alpha_s(Q_0^2)/N)^m$ which, in general, becomes 
singular at $N=\lambda(Q_0^2)$. Under a transformation of the type 
\urequire\ this series will be multiplied by 
$(U^0_{gg}(\alpha_s(Q_0^2)/N))^{-1}$ in order to compensate for the change in
the input parton distributions. A number of scheme transformations that are
considered have the series $U^0_{gg}(\alpha_s(Q_0^2)/N)=1+\sum_{m=0}^{\infty}
U^{0,m}_{gg}(\alpha_s(Q_0^2)/N)^m$ also becoming nonsingular at $N=
\lambda(Q_0^2)$ (e.g. \sdis\qzero). This nonsingular behaviour leads to 
powerlike growth of the form $x^{-1-\lambda(Q_0^2)}$ as $x\to 0$, but the 
magnitude of the powerlike behaviour and the manner in which  
in which it is approached depends on the strength of the singularity and/or 
on the precise behaviour of the coefficients in the series. Hence, changes in
definition of the form in \urequire\ with this type of singularity can lead 
to very marked differences in the form of the gluon at small $x$, or if the 
gluon is kept roughly constant (e.g. it is attempted to predict the form
of the input structure function by assuming a form for the input gluon),
to significant changes in the form of 
$F_L(x, Q_0^2)$. Very similar considerations also hold for 
$F_2(x,Q^2)$ because the 
input depends strongly on $C^g_{2,1,0}(\alpha_s(Q_0^2)/N)$, which 
transforms in the same way as $C^g_{L,1,0}(\alpha_s(Q_0^2)/N)$ under
changes in factorization scheme; i.e. the parton input (or prediction
for the structure function if the parton input is assumed, e.g. to be flat) 
depends very strongly on scheme.
We also note that some of the scheme changes away from the 
standard $\overline{\hbox{\rm MS}}$ scheme (e.g. the SDIS scheme) 
involve $U^0_{gg}(\alpha_s(Q_0^2)/N)$ with larger coefficients 
in the power series than in 
$C^g_{L,1}(\alpha_s(Q_0^2)/N)$ defined in the $\overline{\hbox{MS}}$
scheme. This leads to $C^g_{L,1}(\alpha_s(Q_0^2)/N)$ defined in the new
scheme to have negative coefficients and therefore to  
$F_L(x,Q_0^2)$ developing a powerlike behaviour with negative 
magnitude unless either the input gluon or quark (or both) has a 
powerlike growth of the form $x^{-1-\lambda(Q_0^2)}$ itself, with a large 
enough multiplicative factor, 
in order to counter this effect. This must be borne in mind when using
such schemes.  

Examining \firstterm\ we can also see how the evolution may be strongly
factorization scheme dependent. As we have already mentioned, the whole
of $\Phi^+_1(Q^2,Q_0^2)$ must be used in order to obtain a 
factorization--scheme--independent 
expression. If $\gamma^1_{gg}(\alpha_s(Q^2)/N)$ is 
omitted, or the wrong $\gamma^1_{gg}(\alpha_s(Q^2)/N)$ used, then this can
give a completely misleading result for the evolution. The integrand in 
$\Phi^+_1(Q^2,Q_0^2)$ is a series which is a power of $\alpha_s(Q^2)$
down on the series $\gamma^0_{gg}(\alpha_s(Q^2)/N)$. However, in many popular 
factorization schemes the coefficients in the incomplete, or incorrect
series for this integrand are much larger than 
those in $\gamma^0_{gg}(\alpha_s(Q^2)/N)$ (helped by the fact that many of 
the early coefficients in $\gamma^0_{gg}(\alpha_s(Q^2)/N)$ are zero),
e.g. they commonly behave roughly like $(12\ln 2/\pi)^n n^{-3/4}$, whereas the 
coefficients in $\gamma^0_{gg}(\alpha_s(Q^2)/N)$ behave roughly like 
$(12\ln 2/\pi)^n n^{-3/2}$. Hence, the incorrect $\Phi^+_1(Q^2,Q_0^2)$ 
can have a dominant effect on the evolution. Under changes of 
factorization scheme the coefficients in the 
factorization--scheme--dependent series can change by amounts similar to 
their own magnitude,
i.e. $\gamma^1_{gg}(\alpha_s(Q^2)/N)$ is very unstable under factorization
scheme changes as can be seen from \transgamma. Therefore, the evolution
of the structure functions in terms of a given input can appear to have 
a very strong factorization scheme dependence. Once again, this is true for 
the evolution of $F_2(x,Q^2)$ as well as for $F_L(x,Q^2)$: the influence 
of the incorrect $\Phi^+_1(Q^2,Q_0^2)$ can be more important than that of 
$\Phi^+_0(Q^2,Q_0^2)$ and $\hat\gamma_{2L}(N,\alpha_s(Q_0^2))$ combined, 
where for the latter the coefficients in the series are again 
relatively small.

Simply using an incorrect calculational procedure, such as solving for 
the parton distribution using the renormalization group equations up to 
some order, and then combining with the coefficient functions to a given 
(combined) order leads to expressions which are not only similar to 
\fullfl--\fullftwoderiv\ with incorrect or missing $\gamma^1_{gg}
(\alpha_s(Q^2)/N)$, but which have additional factorization--scheme--dependent 
terms. (Solving by using the full known anomalous dimensions
and combining the resulting parton distributions with all known coefficient 
functions is even worse.)
These will be formally of higher order than the terms in 
\fullfl--\fullftwoderiv, but again can have very large coefficients in 
the series expansions. This can lead to even more dramatic effects than 
those outlined above. One clear example of such incorrect effects is found in
\frt.
    
Once $\gamma^1_{gg}(N,\alpha_s(Q^2))$ is known in a given 
scheme and \fullfl--\fullftwoderiv\ can be calculated correctly 
there is no guarantee that the correct $\Phi^+_1(Q^2,Q_0^2)$ is not larger
than $\Phi^+_0(Q^2,Q_0^2)$. If this is the case, $\Phi^+_1(Q^2,Q_0^2)$ 
will then have a large, but at least definite, effect. 
However, because it is a formally NLO correction
to the structure functions, if $\Phi^+_1(Q^2,Q_0^2)$ is introduced
then the full set of NLO expressions, both evolution
and input factors, must be calculated at the
same time. The correct 
calculational method respects renormalization scheme independence as well 
as factorization scheme dependence. As mentioned earlier,
this requires many more terms than just $\Phi^+_1(Q^2,Q_0^2)$.
Hopefully, the complete NLO
expression, as well as being factorization scheme independent, will also 
cause only fairly small changes to the LO expressions. 

\medskip

Having obtained our full set of leading--order expressions, 
we can also now examine the difference between 
$(d\, F_{2}(N,Q^2)/d\, \ln(Q^2))_0$ and $\alpha_s(Q^2) d\, F_{2,0}(N,Q^2)
/d\, \ln(Q^2)$, i.e. the LO  derivative of $F_2(N,Q^2)$
and the derivative of the LO $F_2(N,Q^2)$. If we differentiate  
\firstftwo\ with respect to $\alpha_s(Q^2)$ we obtain 
\eqn\derfirstftwo{\eqalign{-\alpha^2_s(Q^2){d \, F_{2,0}(N,Q^2)\over
d\alpha_s(Q^2)} =&\alpha_s(Q_0^2)
\tilde\gamma^1_{2L}(\alpha_s(Q_0^2)/N)\biggl({\gamma^0_{gg}(\alpha_s(Q^2)/N)
\over \gamma^0_{gg}(\alpha_s(Q_0^2)/N)}\biggr)
\times\cr
&(\hat F_{L,0}(N) +\tilde{\hat F}_{L,0}(N,Q_0^2)-
\biggl({36-8N_f\over 27}\biggr)F_{2,0}(N)) \exp
[\Phi^+_0(Q^2,Q_0^2)]\cr
&-\alpha_s(Q^2){4\over 9}{2N_f \over
6\pi b_0}F_{2,0}(N)\exp[\Phi^-_1(Q^2,Q_0^2].\cr}}
This is clearly not exactly the same as \loftwoderiv, the 
difference being due to three additional terms in the above expression as 
compared to \loftwoderiv. These are the factor 
$(\gamma^0_{gg}(\alpha_s(Q^2)/N)/ \gamma^0_{gg}(\alpha_s(Q_0^2)/N))$ 
in the first term, 
and the factors $(\alpha_s(Q^2)/\alpha_s(Q_0^2))$ and  
 $\exp[\Phi^-_1(Q^2,Q_0^2)]$ in the second term. All of these
factors are unity at the boundary of the evolution, and the two 
expressions are therefore identical in this limit, i.e the inputs are 
the same. Therefore, it is the evolution terms which are different when 
comparing \loftwoderiv\ and \derfirstftwo. Writing 
\eqn\addonii{\biggl({\gamma^0_{gg}(\alpha_s(Q^2)/N)\over 
\gamma^0_{gg}(\alpha_s(Q_0^2)/N)}\biggr)=\exp\biggl[
-\int_{\alpha_s(Q^2)}^{\alpha_s(Q_0^2)} {d\over d\, \alpha_s(q^2)}\ln
(\gamma^0_{gg}(\alpha_s(q^2)/N))d\alpha_s(q^2)\biggr],}
and
\eqn\addoniii{\alpha_s(Q^2)=\alpha_s(Q_0^2)\exp\biggl[-\ln\biggl(
{\alpha_s(Q_0^2)\over \alpha_s(Q^2)}\biggr)\biggr]}
we see that the terms present in \derfirstftwo\ but absent in \loftwoderiv\ 
are NLO evolution terms. Thus, as in the loop 
expansion, the difference between the fixed--order expression for $(d\,F_2/d
\ln Q^2)$ and the $\ln Q^2$ derivative of the fixed--order $F_2$ consists
of terms of higher order. However, in the leading--$\ln (1/x)$ expansion 
this difference exists between even LO expressions.   
Close to the boundary the effect of the additional evolution terms in 
\derfirstftwo\ is very small, but they clearly become more significant
for $Q^2>>Q_0^2$ or $Q^2<<Q_0^2$. So, as we might hope, despite the 
formal difference between the derivative of the LO
structure function and the LO expression for the derivative
of the structure function, within the 
region of expected applicability, i.e. $x$ small and $Q^2$ relatively near 
$Q_0^2$, the two are very similar in practice. As one 
goes away from the boundary, especially at large $x$, the 
terms begin to differ markedly.\foot{Similarly, in the loop expansion 
significant differences appear between the derivative of fixed order
$F_2(N,Q^2)$ and 
$(d\,F_2(N,Q^2)/d\ln Q^2)$ at the same fixed order when higher powers of 
$\alpha_s(Q^2)$ 
become more important, i.e. at low $Q^2$. Hence the differences are small in 
the expected region of validity of the loop expansion, i.e. large $Q^2$.}
This highlights the fact that the main region 
of applicability of the leading--$\ln (1/x)$ expansion is small $x$ and 
$Q^2\sim Q_0^2$.  

\medskip

Finally, we should also make some comment about longitudinal momentum
conservation. The first moment of the parton distributions is
interpreted as the fraction of momentum carried by that type of
parton. As such it is usually required that $f^S(1,Q^2)+g(1,Q^2)=1$,
i.e. the total momentum carried by the partons is that of the
proton. For this to be true for all $Q^2$ then, of course,
$(d\,(f^S(1,Q^2)+g(1,Q^2))/d\ln Q^2)=0$, and from the renormalization
group equations this is true if 
\eqn\momcons{\gamma_{ff}(1,\alpha_s(Q^2))+
\gamma_{gf}(1,\alpha_s(Q^2))=0, \hskip 0.5in \gamma_{fg}(1,\alpha_s(Q^2))+
\gamma_{gg}(1,\alpha_s(Q^2))=0.}  

This is assumed to be true for the all--orders anomalous
dimensions, and thus momentum will be conserved for the all--orders
parton distributions. When expanding the anomalous dimensions order by
order in $\alpha_s(Q^2)$, i.e. as in \fullad, it is easy to specify
that \momcons\ be true for the anomalous dimensions at each order, and
to define a wide variety of factorization schemes which maintain this. It is
not difficult to see from \S 2.2 that this guarantees that the fraction
of momentum carried by the $n_{\rm th}$--order parton distributions is
conserved at each $n$. However, it does not necessarily tell us
anything about the amount of momentum carried by the $n_{\rm th}$--order
inputs, for any particular $n$. Sometimes, all the momentum is designated 
to be carried by the zeroth order part of the solution, 
but this need not be the case, and indeed, we see no good reason why it 
should be. Most often the input is implicitly assumed to be the all--orders 
input. In this case it is 
true that it must carry all the momentum, but this method destroys the 
strict ordering of the solution.

As always seems to be the case, the situation is not as simple for the
leading--$\ln (1/x)$ expansion. As can be seen from the matrix in
\llevolzero, the leading--order $\gamma$ contains entirely positive
entries for $N=1$, and is clearly not consistent with
momentum conservation: $f^S_0(Q^2)$ carries a
constant amount of momentum while that carried by $g_0(Q^2)$ is
constantly increasing with $Q^2$. In a general factorization scheme 
we assume that there is no reason that working to a finite higher
order will restore the relationship \momcons.

Two general methods have been proposed to restore momentum conservation. 
The first multiplies the known $\gamma^0$'s and $\gamma^1$'s in some
factorization scheme by some finite
power series in $N$ which vanishes at $N=1$ \ehw, e.g. the simplest example is 
$(1-N)$. The evolution equations are then solved using the whole of these
anomalous dimensions and momentum conservation is guaranteed. However, this
prescription destroys any sense of ordering the solution correctly and 
is extremely scheme dependent, causing the sort of large
factorization scheme variations described above. Also, since the full
anomalous dimensions have singularities at $N=-1$ the power series 
expansion about $N=0$ for given order in $\alpha_s$ does not even 
converge at $N=1$. Thus, it seems 
inappropriate to demand that the first few terms in the expansion about 
$N=0$ should approximate the correct value at $N=1$.
An examination of the first few terms in the expansion of the two--loop
anomalous dimensions about $N=0$ shows that a truncation of each
after a few terms is not at all similar to the above prescription.

An alternative method \bfresum\ is to assume that the relationship  
\eqn\momconsll{\eqalign{&\alpha_s(Q^2)
\gamma_{ff}^1(1,\alpha_s(Q^2))+
\gamma^0_{gf}(1,\alpha_s(Q^2)) + \alpha_s(Q^2)\gamma_{gf}^1
(1,\alpha_s(Q^2))=0\cr
& \alpha_s(Q^2)\gamma_{fg}^1(1,\alpha_s(Q^2))+\gamma^0_{gg}
(1,\alpha_s(Q^2)) +\alpha_s(Q^2)\gamma_{gg}^1(1,\alpha_s(Q^2))=0,}}  
is satisfied, and to determine the unknown $\gamma^1_{ga}(\alpha_s/N)$
this way. In fact it has been proved that it is always possible to choose
a factorization scheme where this is true \ref\bfmomcon{R.D. Ball and 
S. Forte, \PL \vyp{B359}{1995}{362}.}, and suggested that this might 
be some sort of preferred scheme. This will guarantee
momentum conservation if one truncates the series for the
$\gamma$'s at the $\gamma^1$'s and solves the whole
renormalization group equation using this truncated $\gamma$. 
However, this does not lead to a well--ordered solution for the parton
distributions in any sense. If one were to believe that solving the
renormalization group equations order by order and combining with the 
coefficient functions up to a given combined order
led to an ordered solution for the structure functions in this expansion 
scheme, as in the loop expansion (and as is used in some 
calculations), then the zeroth--order 
solutions for the partons do not conserve momentum, and even if \momconsll\ is
true then adding \llevolzero, \firstordfeq\ and \gfirstordeq\ still
leaves us with 
\eqn\momviol{\sum_{i=0,1}-\alpha_s(Q^2){d(f^S_i(1,Q^2)+g_i(1,Q^2))
\over d \alpha_s(Q^2)}= \tilde
\gamma^0_{gf}(1,\alpha_s(Q^2))f^S_1(1,Q^2)+\tilde
\gamma^0_{gg}(1,\alpha_s(Q^2)) g_1(1,Q^2),}
as well as terms coming from the difference between $\tilde \gamma^1_{gg}$
and $\gamma^1_{gg}$. Thus, momentum violation 
will not be zero in general, and may be quite large. Thus, it seems that
in order to enforce momentum conservation strictly within this expansion 
scheme one must make a guess at the
full anomalous dimensions (and coefficient functions) in terms of some 
truncated form of them, and hope that this is a good approximation to the 
full solution. This sacrifices any possibility of making a well--ordered 
expansion for the structure functions. 

Of course, in this paper we advocate that the most 
sensible approach is to obtain a well--ordered solution 
for the structure functions, and thus definitely
eliminate any questions of scheme dependence; i.e. regard a correct treatment
of the physical quantities as of paramount importance. 
Doing this, there does not seem to be
any way to ensure that what we choose to define as the parton distributions
within our final expressions are such as to conserve momentum. 
Thus, we simply take the hint offered us
by the zeroth--order anomalous dimension, and accept the fact that
momentum is not conserved order by order in this method of expansion.   
How badly it appears to be violated, however, will depend very much on which
factorization scheme we claim to work in. 
It will always be possible to
choose one where momentum violation is very small and, if one wishes, 
one may choose to think of this as a ``physical scheme'' for the partons. 
However, the structure functions themselves will be completely unaffected 
by this choice, and the real physical relevance is therefore rather
questionable. Because momentum is not in general 
conserved order by order in this expansion scheme the amount
carried by a certain order will vary with $Q^2$. Hence, using an
arbitrary factorization scheme there
seems little reason to demand that the momentum carried by the zeroth--order 
inputs for the partons should sum to one; it seems more
sensible to share the momentum amongst the different orders in the
inputs. It is clear that the momentum carried by the zeroth--order
parton distributions, for example, will increase quickly with $Q^2$,
and thus it makes sense when choosing their inputs to choose
distributions which sum to less than unity. As higher order
corrections to the evolution come in, acting to curb this growth in
momentum of the parton distributions (hopefully countered by increased 
growth with $Q^2$ of the first moment of the structure functions coming 
from the effects of the 
coefficient functions, leading to the overall behaviour of 
the structure functions being largely unchanged)
they can bring inputs carrying positive momentum with them. 
If we could work to
all orders then the momentum carried by the inputs would eventually sum to 
unity, and stay at this value for all $Q^2$, but this is
of course not possible.    

\bigskip

We can also discuss the relationship between our scheme--independent
solutions and the ones which would be obtained using Catani's physical
anomalous dimensions. One can solve for the LO structure functions 
using these effective anomalous dimensions in exactly the same way as we 
solved for the parton distributions in the previous subsection. In the same 
way that we have the relationships between the LO anomalous
dimensions, \gammadef\ and \gamrel, we have relationships between the 
effective anomalous dimensions \physanomval, i.e
\eqn\rellong{\Gamma^0_{L2}(\alpha_s(Q^2)/N)=-\biggl({36-8N_f\over 27}
\biggr)\Gamma^0_{LL}(\alpha_s(Q^2)/N),}
and 
\eqn\reltwo{\Gamma^1_{22}(\alpha_s(Q^2)/N)=-\biggl({36-8N_f\over 27}
\biggr)\Gamma^1_{2L}(\alpha_s(Q^2)/N)-\fourninths\alpha_s(Q^2)
{2N_f\over 6\pi},}
where the second term in \reltwo\ is the one--loop contribution to 
$\Gamma^1_{2L}(\alpha_s(Q^2)/N)$. Using these relationships it is 
straightforward to
follow through the steps in \llevolzero--\loderivf\ to obtain the analogous
expressions to \lof, \loglu\ and \loderivf:
\eqn\plof{\eqalign{F_2(N,Q^2)=& \alpha_s(Q^2)
{\Gamma^1_{2L}(\alpha_s(Q^2)/N)\over
\Gamma^0_{LL}(\alpha_s(Q^2)/N)}(\hat F_{L,0}(N)+\tilde 
{\hat F}_{L,0}(N,Q_0^2)
-\biggl({36-8N_f \over 27}\biggr)F_{2,0}(N))\times\cr
& \hskip -0.2in \eigpluscorp\cr
&\hskip -0.2in -\tilde F_{2,1}(N,Q_0^2)+F_{2,0}(N)\biggl({\alpha_s(Q_0^2)\over
\alpha_s(Q^2)}\biggr)^{-\fourninths{2N_f\over 6\pi b_0}}+
\higherorder,\cr}}
\eqn\ploglu{\eqalign{\hat F_L(N,Q^2)=& (\hat F_{L,0}(N)+
\tilde{\hat F}_{L,0}(N,Q_0^2)
-\biggl({36-8N_f \over 27}\biggr)F_{2,0}(N))\times\cr
&\hskip -0.2in \eigpluscorp\cr
&+ \biggl({36-8N_f \over 27}\biggr)F_{2,0}(N)\biggl({\alpha_s(Q_0^2)\over
\alpha_s(Q^2)}\biggr)^{-\fourninths{2N_f\over 6\pi b_0}}+ \higherorder,\cr}}
and 
\eqn\loderivfp{\eqalign{-\alpha_s^2(Q^2) {d\, F_2(N,Q^2)\over
d\alpha_s(Q^2)}
=& \alpha_s(Q^2)
\tilde\Gamma^1_{2L}(\alpha_s(Q^2)/N)(\hat F_{L,0}(N)+
\tilde {\hat F}_{L,0}(N,Q_0^2)
-\biggl({36-8N_f \over 27}\biggr)F_{2,0}(N))\times\cr
&\hskip -1in \eigpluscorp\cr
&-\fourninths \alpha_s(Q^2) 
F_{2,0}(N){2N_f\over 6\pi b_0} \biggl({\alpha_s(Q_0^2)\over
\alpha_s(Q^2)}\biggr)^{-{4\over 9}{2N_f\over 6\pi b_0}}+\higherorder,\cr}}
where $\tilde {\hat F}_{L,0}(N,Q_0^2)$ is a function of $\Gamma^0_{LL}$ rather
than $\gamma^0_{gg}$.

We see that, once we make the identifications 
\eqn\identi{\gamma^0_{gg}(\alpha_s(Q^2)/N)=\Gamma^0_{LL}
(\alpha_s(Q^2)/N),}
\eqn\identiii{\gamma^1_{2L}(\alpha_s(Q^2)/N)=
\Gamma^1_{2L}(\alpha_s(Q^2)/N),}
and
\eqn\identii{\eqalign{\tilde \gamma^1_{gg}(\alpha_s(Q^2)/N)+\fourninths \tilde 
\gamma^1_{fg}(\alpha_s(Q^2)/N)-&{d\over d\,\alpha_s(Q^2)}\biggl(\ln
\biggl({C^g_{L,1}(\alpha_s(Q^2)/N)\over C^g_{L,1,0}}\biggr)\biggr)\cr
&=\tilde\Gamma^1_{LL}(\alpha_s(Q^2)/N)-
\biggl({36-8N_f\over 27}\biggr)\tilde\Gamma^1_{2L}(\alpha_s(Q^2)/N),\cr}}
\ploglu\ is identical to \fullflalti, 
\loderivfp\ is identical to \fullftwoderivalt\ and  \plof\ is identical to
\fullftwoaltc. Of course, the identifications \identi--\identiii\ are 
exactly what we obtain from the definitions of the physical anomalous 
dimensions in \S 3. 

Thus, we are able to reach these expressions 
for the structure functions somewhat more directly by 
using the physical anomalous dimensions, and do not have to worry about 
problems with factorization scheme dependence (though we do have 
to calculate the physical anomalous dimensions in terms of known 
coefficient functions and anomalous dimensions of course). 
Once we have obtained these expressions using the 
physical anomalous dimensions we may then separate each of the terms into 
input parts and evolution parts (where more of this has already 
been done automatically when using the physical anomalous dimensions) and
keep the most leading parts, obtaining once again the LO expressions
\lofl, \loftwoderiv\ and \firstftwo.
So, using the physical anomalous dimensions leads us
in a rather more direct manner to the correct leading--order expressions. 
If we were to work to higher orders, the amount of simplification obtained by
using the physical anomalous dimensions rather than working in terms of 
parton densities would increase significantly.
However, we stress that one will always
automatically obtain factorization--scheme--independent answers by 
working to well--defined orders in physical quantities even when working 
in terms of partons. Also, even if one uses the physical anomalous dimensions, 
care is still needed to obtain expressions which are consistent with 
renormalization scheme dependence. 

\bigskip

In this section, we have derived well--ordered, 
factorization--scheme--independent
expressions for structure functions in the leading--$\ln (1/x)$
expansion (which should be useful at small $x$ and moderate $Q^2$) 
up to the order which is useful at present. This expansion does, however,
sacrifice any attempt to describe the structure functions at large $x$
at any reasonable distance from $Q_0^2$
(we will discuss later what large $x$ and small $x$ turn out to
be), in the same way that the loop expansion should show signs of failing 
at very low values of $x$. We would hope there is some expansion scheme
which will be useful at all values of $x$. In the next section we will
show that there is indeed an expansion scheme which satisfies this 
criterion, and argue that it is the only really correct expansion scheme.

\subsec{The Renormalization--Scheme--Consistent Expansion.}

In order to devise an expansion scheme which is useful at both large
and small $x$ we would {\it a priori} expect that we would need to
use the known anomalous dimensions and coefficient
functions at low orders in both $\alpha_s$ and in the leading--$\ln (1/x)$
expansion. There have already been various methods along these lines, and
the phrase ``double leading expansion'' was coined in \bfresum. 
However, these methods have all suffered from scheme dependence.
As when deriving our
expressions for the structure functions when using the leading--$\ln (1/x)$
expansion we will ensure that we obtain results which are
invariant under changes of factorization scheme 
and, as a stronger constraint, demand complete consistency of our 
expressions for physical quantities with renormalization scheme
invariance (which in itself automatically guarantees factorization scheme 
independence). Consequently, our approach will be rather different 
from those used previously, and the results and conclusions will also be 
somewhat different. 

\medskip  

To begin, let us consider what we have meant by 
``consistency with renormalization scheme dependence'' so far in this paper.
In both the loop expansion and the leading--$\ln (1/x)$ 
expansion we demanded that once we had chosen a particular renormalization
scheme and chosen to work to a particular order in this renormalization
scheme then we would include all terms in our expressions for the 
structure function which were of greater magnitude than the uncertainty
due to the freedom of choice of renormalization scheme (i.e. the uncertainty 
in the definition of the coupling constant), and no others. In both cases the 
leading--order term consisted of the lowest--order inputs multiplying the 
lowest--order evolution terms. If working with the $n$--loop coupling
constant in a particular renormalization scheme the
uncertainty in its definition is of order $\alpha_s^{n+1}$. 
Thus, the uncertainty of the 
input or evolution when working to $n_{\rm th}$--order is the change in the 
leading--order input or evolution if the coupling changes by $\delta
\alpha_s=\epsilon \alpha_s^{n+1}$, i.e $\alpha_s \to \alpha_s(1+\epsilon
\alpha_s^n)$. Hence, the uncertainty in the whole structure function
is of the order of the change of the leading--order part under such 
a change in the coupling. Therefore the $n_{\rm th}$--order 
renormalization--scheme--independent expression includes all complete terms
smaller than this change. 

This definition does give us a well--defined
way of building up an ordered solution to the structure functions, 
but relies upon the definition of a given expansion scheme.
It leaves an ambiguity about how we define the leading--order 
expressions and in how we define the order of terms compared to this 
leading--order term. Our two examples, i.e. the loop expansion, where the size
of a term is determined simply by its order in $\alpha_s$, and the 
leading--$\ln(1/x)$ expansion, where $\ln(1/x)$ is put on an equal footing 
to $\alpha_s$, are just the two most commonly used examples of expansion 
schemes (even though the $\ln(1/x)$ expansion has not previously
been presented in the quite same way as in this paper). 
Both have potential problems: in the former one does not 
worry about the fact that the large--$\ln(1/x)$ terms can cause enhancement 
at small $x$ of terms which are higher order in $\alpha_s$, 
and in the latter one does not worry about the fact that 
at large $x$, especially as $Q^2$ increases, it is the terms which are of 
lowest order in $\alpha_s$ that are most important. Hence, one would think
that both have limited regions of validity.

The shortcomings of these two expansion schemes 
come about because, even though any given order contains no terms which are 
inconsistent with working to the same given order in a particular
renormalization scheme, in neither case does it include every one of the terms 
which are consistent to working to a given order in the  
renormalization scheme. In each expansion scheme some of the terms 
appearing at what we call higher orders are not actually
subleading in $\alpha_s$ to any terms which have already appeared. Thus,
despite the fact that for a given expansion method these terms are 
formally of the same order as 
uncertainties due to the choice of renormalization scheme, they are not terms 
which can actually be generated by a change in renormalization 
scheme.\foot{Similarly, they cannot be generated by a change in 
renormalization scale.} 

\medskip

In order to demonstrate this point more clearly we consider a simple toy 
model. Let us imagine some hypothetical physical quantity which can be
expressed in the form
\eqn\hypothet{H(N,\alpha_s(Q^2))=\sum_{m=1}^{\infty}\alpha_s(Q^2)
\sum_{n=-m}^{\infty}a_{mn}N^n\equiv \sum_{i=0}^{\infty}\alpha^i_s(Q^2)
\sum_{j=1-i}^{\infty}b_{ij}\biggl({\alpha_s(Q^2)\over N}\biggr)^j,}
where the expansion in powers of $N$ about $N=0$ is convergent for all $N$. 
The first way of writing $H(N,\alpha_s(Q^2))$ as a power series corresponds to
the loop expansion, where we work order by order in $m$, out to $m=k$,  
and use the $k$--loop
coupling. The second corresponds to the leading--$\ln(1/x)$ expansion 
where we work order by order in $i$, out to $i=l$,
and use the $(l+1)$--loop coupling.
Let us, for a moment, consider the LO expression in the loop expansion,
$\alpha_s(Q^2) \sum_{n=-1}^{\infty}a_{1n}N^n$.
The coupling is uncertain by ${\cal O}(\alpha^2_s(Q^2))$ and hence the 
uncertainty of the leading--order expression (i.e. the change due to 
a change of the coupling) is $\sim\alpha_s^2(Q^2)\sum_{n=-1}^{\infty}
b_{1n}N^n$. There is no change with powers of $N$ less than $-1$,
and hence any such term is not really subleading.
Similarly, the uncertainty of the leading--order expression in the leading
$\ln (1/x)$ expansion contains no terms at first order in $\alpha_s$ (or
with positive powers of $N$), and such terms are not really 
subleading either. The full set of terms contained within the combination
of both leading--order expressions is genuinely leading order, and 
is therefore renormalization scheme independent by definition. 

Perhaps the best way in which to write our expression for 
$H(N,\alpha_s(Q^2))$ in order to appreciate these points is 
\eqn\hypotheti{H(N,\alpha_s(Q^2))=\sum_{m=-1}^{\infty}N^m
\sum_{n=1}^{\infty}c_{mn}\alpha_s^n(Q^2)+ \sum_{m=2}^{\infty}N^{-m}
\sum_{n=m}^{\infty}c_{mn}\alpha^n_s(Q^2),}
i.e. as an infinite number of power series in $\alpha_s(Q^2)$, one 
for each power on $N$. Each of these series in $\alpha_s(Q^2)$ is 
independent of the others, and the lowest order in $\alpha_s(Q^2)$ of each 
is therefore renormalization scheme independent and part of
the complete LO expression for $H(N,\alpha_s(Q^2))$.  
The full LO expression for $H(N,\alpha_s(Q^2))$ is therefore
\eqn\hypothetlo{\eqalign{H_0(N,\alpha_s(Q^2))&=\sum_{m=-1}^{\infty}N^m
c_{m1}\alpha_s(Q^2)+ \sum_{m=2}^{\infty}c_{mm}N^{-m}
\alpha^{m}_s(Q^2)\cr
&\equiv \alpha_s(Q^2)
\sum_{n=-1}^{\infty}a_{0n}N^n+ 
\sum_{j=2}^{\infty}b_{0j}\biggl({\alpha_s(Q^2)\over N}\biggr)^j.}}
Hence, the combined set of terms considered LO in both 
our expansion schemes comprise the full set of renormalization scheme
invariant, and thus truly leading--order, terms.
By considering $H(N,\alpha_s(Q^2))$ written in the form \hypotheti, 
and considering
how the coefficients in the expression must change in order to make the 
whole expression invariant under a redefinition of the coupling constant,
$\alpha_s(Q^2)\to \alpha_s(Q^2)+{\cal O}(\alpha^m_s(Q^2))$, we see that the 
$n_{\rm th}$--order expression for $H(N,\alpha_s(Q^2))$, which should be 
used with the $n$--loop coupling constant, consists of the sum of the first 
$n$ terms in each of the power series in $\alpha_s(Q^2)$. Thus, the full 
$n_{\rm th}$--order expression always consists of the $n_{\rm th}$--order 
expression in the loop expansion plus additional terms with inverse powers 
of $N$ greater than $n$.

Similar arguments have already been applied to the anomalous dimensions 
and coefficient functions, for example  \cathaut\ and particularly \bfresum.
The latter claims that one may, but need not, use expansions of the above 
form for the anomalous dimensions and coefficient functions, and moreover, 
in practice expresses the terms beyond fixed order in $\alpha_s$ as functions 
of $x/x_0$ for $x<x_0$ and sets them to zero otherwise ($x_0\leq 1$, and is 
in general $sim 0.1$), thus reducing their effect (see also \faili). 
Here we take a strong, inflexible viewpoint and insist that the complete 
renormalization--scheme--consistent expressions, with no artificial 
suppression of leading--$\ln (1/x)$ terms, must be used. Furthermore, 
and very importantly, the 
expressions used must be those for the physical structure functions, not for 
the factorization--scheme-- and renormalization--scheme--dependent 
coefficient functions and anomalous dimensions.

\medskip

When considering the real structure functions the situation is  
technically a great deal more  complicated than our toy model, 
but the principle is exactly the same. This can be seen by examining the 
the LO expressions for the structure functions in the 
two expansion schemes already considered. There is some overlap between 
the LO expressions for the structure functions when using 
the loop expansion and when using the leading--$\ln(1/x)$ expansion, 
but each contains an infinite number of terms not present in the other. 
However, we were previously happy to use the one 
loop coupling for both expressions. The uncertainty in the definition of 
this coupling is ${\cal O}(\alpha_s^2)$. Considering the change of each 
of our leading--order expressions under a change of coupling of  
${\cal O}(\alpha_s^2)$, the changes in the expressions are rather 
complicated. However, it is not too difficult to see that, as with our 
toy model, the change in the LO 
structure functions in the loop expansion contains no terms in the 
LO expressions in the leading--$\ln (1/x)$ expansion, and
{\it vice versa}. Thus, none of the terms contained within each of the LO
expressions are generated by uncertainties at higher order
in the opposing expansion scheme. Therefore, they  
should really all be regarded as genuinely LO,
and be included in the full expressions for the structure functions 
which use the one--loop coupling constant. 

Hence, as with the toy model,
there should be some combined expansion--scheme--independent 
expressions for the structure functions which we can 
genuinely call the ``leading order'' expressions. Since these expressions
will contain all the parts of the one--loop expressions, and also contain
leading--$\ln(1/x)$ terms as well, they should be able to 
describe the data over the full range of parameter space
(except very low $Q^2$, of course), as we would like 
from our correct LO expressions. 
We shall now demonstrate how we obtain these expressions. 

\medskip

There are two main complications when considering structure functions in 
comparison to our simple toy model. One is that the structure functions 
are combinations of perturbative evolution parts and input parts 
(which are viewed as partly perturbative with nonperturbative 
factors), rather than one simple power series in $\alpha_s(Q^2)$. The other
is that in general the physical anomalous dimensions, out of   
which the perturbative parts are constructed, are nonanalytic 
functions which cannot be expressed as power series about a particular
value $N_0$ for all $N$. 
The physical anomalous dimensions have singularities at $N=0$ (in the case
of the singlet structure function only), and also at
negative powers of $N$ (as well as possible $\alpha_s(Q^2)$--dependent 
nonanalyticities due to 
resummation effects, e.g. the branch point in $\Gamma^0_{LL}
(N,\alpha_s(Q^2))$ at $N=\lambda(Q^2)$). We will deal with this second 
complication first. 

Let us consider the perturbative parts of the expressions for the 
structure functions. The singularities at negative integer values of $N$
mean that we cannot write any physically meaningful quantity as just a power
series about $N=0$ (or about $(N+1)$ for the nonsinglet case).
Any such power series expansion will have a radius of 
convergence of unity, whereas the physical moments of any $x$-space 
quantity will exist for all real $N$ above some minimum $N_{min}$, which 
depends on the asymptotic form of the structure functions as $x \to 0$. 
A series 
expansion which applies over this whole range of $N$ does not exist: the valid
expression must include the nonanalytic functions explicitly. This seems to 
make it impossible to order the moment--space solution in such a simple
way as in \hypothet. 

In order to overcome this problem let us consider making the 
inverse transformation of some 
physically relevant perturbative function $A(N,\alpha_s(Q^2))$ to $x$--space.
The inverse of the Mellin transformation \melltranssf\ is 
\eqn\inverstran{{\cal A}(x,\alpha_s(Q^2))={1\over 2\pi i}
\int_{c-i\infty}^{c+i\infty}x^{-N}A(N,\alpha_s(Q^2))dN,}
where the line of integration is to the right of all nonanalyticities.
Making the substitution $\xi =\ln(1/x)$ this becomes
\eqn\inverstran{{\cal A}(x,\alpha_s(Q^2))={1\over 2\pi i}
\int_{c-i\infty}^{c+i\infty} \exp[\xi N]A(N,\alpha_s(Q^2))dN.}
Since $A(N,\alpha_s(Q^2))$ has, in general, singularities for all 
nonpositive integers, this whole integral may be evaluated by performing an 
infinite series of integrals, each with a contour centred on a given 
singularity, and not extending as far as unity from this singularity, i.e.
not reaching any of the other singularities. Within each of these contours 
the function $A(N,\alpha_s(Q^2))$ may be expanded as a power series about 
the singularity, i.e. we may write
\eqn\inverstranps{{\cal A}(x,\alpha_s(Q^2))={1\over 2\pi i}\sum_{n=0}^{\infty}
x^{n}\int_{c_n} \exp[\xi (N+n)]A_n((N+n),\alpha_s(Q^2))d(N+n),}
where $A_{n}((N+n),\alpha_s(Q^2))$ denotes $A(N,\alpha_s(Q^2))$ expanded as 
a power series about $N=-n$. The integrals will produce functions of $\xi$,
which do not sum to integer powers of $x$, and hence each of the integrals 
in \inverstranps\ will be independent and physically relevant in its own 
right. 

So this is the solution to our problem of how to order the moment
space expressions for physical expressions which are related to the real 
structure 
functions as powers series in $\alpha_s$ and $N$. We must consider the 
complete moment--space expression as an infinite number of expressions of the
form \hypotheti, each one having power series expansions in terms of $(N+n)$,
where $n=0
\to \infty$. The expression for each $n$ is then related to the part of 
the $x$--space expression behaving $\sim x^{n}$. 
Of course, in practice, unless we 
want to examine the details of the perturbative calculation of the 
structure function for $x$ very close to $1$, we can ignore all $n$ 
greater than a finite, relatively small constant. 

Thus, when we calculate the expressions for the perturbative part of the 
singlet structure functions, we will only 
be concerned about any LO terms beyond lowest order in $\alpha_s$ 
for the specific case of $n=0$. For $n>0$ we take the 
whole LO expression to be the one--loop expression. The 
terms we ignore by making this necessary decision are those which are 
LO in $\ln (1/x)$ at first order in $x$. Although these terms 
grow like $\alpha_s^m\ln^{2m-1}(1/x)$, as opposed to $\alpha^m_s\ln^m(1/x)$ at 
zeroth order in $x$, there is no evidence that their coefficients are any
larger than those for the zeroth--order--in--$x$ logarithms. Since the 
resummed terms at zeroth order in $x$ only begin to make an impact as $x$ 
falls to $\sim 0.1$ (as we will see), and only become dominant 
for $x$ much smaller than this,
the effect of terms like  $x\alpha_s^m\ln^{2m-1}(1/x)$ should be 
extremely small in comparison. Indeed, the effect of those terms of
the form $x\alpha_s^m\ln^{2m-1}(1/x)$ which are actually known, 
i.e. for $m=2$, can indeed 
be shown to be negligible. In a similar manner, we will only consider 
the one--loop expressions for the nonsinglet structure functions in 
practice: the other LO parts of the expressions again lead to small--$x$ 
enhancement of the form $x\alpha_s^m\ln^{2m-1}(1/x)$, which is very small
compared to the leading singlet small--$x$ enhancement, and there is only 
detailed data at very small $x$ for the total structure function.\foot{Also,
the full leading--order--in--$\ln (1/x)$ physical anomalous dimension is not
yet known for the nonsinglet structure functions, as will be discussed 
below.}
    
Hence, we only really need to consider calculating a full LO 
renormalization--scheme--consistent (RSC) expression 
for the perturbative contributions to the structure functions expressed as 
power series in $\alpha_s$ and in $N$ about $N=0$, as for our toy model. 
However, we now have to return to our first problem, i.e. the fact that the 
structure functions are expressed in terms of both inputs and evolution parts.
Using the results we 
have already obtained in the earlier subsections it is not too difficult to 
construct the full LORSC expressions for the inputs and for the evolution 
parts of structure functions. For the case of the nonsinglet structure 
functions the construction of the LORSC expression is then just
the product of these two terms. Let us discuss this as an example first.      

\subsec{The Renormalization--Scheme--Consistent 
Nonsinglet Structure Functions.}

We consider a nonsinglet longitudinal structure function. For the nonsinglet 
structure functions the physical anomalous dimensions contain no singularities
at $N=0$, so the leading--$\ln (1/x)$ behaviour comes from singularities at 
$N=-1$. Expanding about $N=-1$, the full LO physical 
anomalous dimension can be written in the form,
\eqn\lonslpan{\eqalign{\Gamma^{NS}_{L,0}(N+1, \alpha_s(Q^2))=&
\alpha_s(Q^2)\biggl[
\sum_{m=-1}^{\infty}a_m (N+1)^m \cr
&\hskip 0.3in +\sum_{m=1}^{\infty}b_m
(N+1)^{-1}\biggl({\alpha_s(Q^2)
\over (N+1)^2}\biggr)^m+\sum_{m=1}^{\infty}c_m\biggl({\alpha_s(Q^2)
\over (N+1)^2}\biggr)^m\biggr].\cr}}
The first sum is just $\Gamma_{L,0,l}^{NS}(N+1)$, the one--loop 
anomalous dimension expanded in powers of $(N+1)$. The second sum 
contains the leading singularities in $(N+1)$ for all other orders
in $\alpha_s(Q^2)$.
The final sum is included because, despite the obvious fact that it is a 
power of $(N+1)^{-1}$ down on the second sum, 
a series of this form cannot be created from the second sum by
a change in the definition of the coupling of ${\cal O}(\alpha^2_s(Q^2))$.
Therefore, the third sum is not subleading in $\alpha_s(Q^2)$ to 
the second sum, and 
must be renormalization scheme independent. Integrating  
\lonslpan\ between $Q_0^2$ and $Q^2$, and including the overall 
power of $\alpha_s(Q^2)$ for the 
longitudinal structure function we obtain the leading--order evolution 
\eqn\loevolnslsf{\biggl({\alpha_s(Q_0^2)\over \alpha_s(Q^2)}
\biggr)^{\tilde\Gamma_{L,0,l}^{NS}(N+1)-1}
\exp\biggl[
\int_{\alpha_s(Q^2)}^{\alpha_s(Q_0^2)}
\biggl(\sum_{m=1}^{\infty} \biggl({1\over N+1}b_m+c_m\biggr)
\biggl({\alpha_s(q^2)
\over (N+1)^2}\biggr)^m\biggr){d\alpha_s(q^2)
\over b_0\alpha^2_s(q^2)}\biggr].}

In the loop expansion the lowest--order input for $F^{NS}_{L}(N,Q^2)$
was $\alpha_s(Q_0^2)C^{NS}_{L,1,l}(N)F^{NS}_{2,0}(N)$, where 
$F^{NS}_{2,0}(N)$ is a nonperturbative factor and $C^{NS}_{L,1,l}(N)$ has
an expansion in powers of $(N+1)$ beginning at zeroth order. Factoring 
out the nonperturbative part, our lowest--order input is $\alpha(Q_0^2)$
multiplying a power series in $(N+1)$ which starts at zeroth order,
i.e $C^{NS}_{L,1,l}(N+1)$. In order 
to construct the full LO input we must consider how the evolution 
term \loevolnslsf\ changes under a change in starting scale $Q_0^2 \to
(1+\delta)Q_0^2$, and therefore how the input must change in order to 
compensate for this. The change in the first term is $
\propto \alpha_s(Q_0^2)(\tilde\Gamma^{NS}_{L, 0,l}(N+1)-1)$. This can be 
absorbed into a change of the input at order $\alpha_s^2(Q_0^2)$, and hence 
the second--order--in--$\alpha_s(Q_0^2)$ input has a part $\propto
(\tilde \Gamma^{NS}_{L, 0,l}(N+1)-1)C^{NS}_{L,1,l}(N+1)$. The power expansion
of this expression in terms of $(N+1)$ begins at order $-1$, 
and a term of this type 
cannot be generated by a change of the order--$\alpha_s(Q_0^2)$ input under
renormalization scheme changes. Hence, this part of the $\alpha^2_s(Q_0^2)$ 
input, i.e $ \propto\alpha^2_s(Q_0^2)a_{-1}C^{NS}_{L,1,l}(N=-1)(N+1)^{-1}$,
belongs to the LORSC input. The rest of the ${\cal O}(\alpha^2_s(Q_0^2))$
input is genuinely subleading to the ${\cal O}(\alpha_s(Q_0^2))$ input
and is renormalization scheme dependent.
Extending this argument, and considering the form of the input 
required to compensate for the change of the whole of \loevolnslsf\ under a 
change in starting scale (as with $g_0(N,Q_0^2)$ in subsection 4.2), 
we can see that the full LORSC input is    
\eqn\loinputnslsf{\eqalign{\alpha_s&(Q_0^2)F^{NS}_{2,0}(N)\biggl(
C^{NS}_{L,1,l}(N+1) + C^{NS}_{L,1,l}(N=-1)
\biggl(\ln(Q_0^2/A^{NS}_L)\alpha_s(Q_0^2)\times\cr
&\biggl({1\over N+1}
\sum_{m=0}^{\infty} b_m\biggl({\alpha_s(Q_0^2)
\over (N+1)^2}\biggr)^m
+ \sum_{m=0}^{\infty}\bigl(c_m
+\half \sum_{n=0}^{m}b_nb_{m-n}\ln(Q_0^2/A^{NS}_L)
\bigr)\biggl({\alpha_s(Q_0^2)
\over (N+1)^2}\biggr)^m\biggr)\biggr)\biggr),\cr}}
where $b_0\equiv a_{-1}$ and $c_0=0$.
The first term is just the lowest--order input in the loop expansion, 
while the second includes all the leading--$\ln (1/x)$ terms in a simple form 
which is compatible with making the full expression invariant 
under changes in starting scale $Q_0^2$. This second term only 
depends on the part of the one--loop coefficient function at zeroth order 
in $(N+1)$: the higher--order--in--$(N+1)$ 
parts multiply the part consisting of the leading--$\ln(1/x)$ terms
to give the type of terms which can be generated from \loinputnslsf\
by changes in the renormalization scheme.  
 
Hence, the full LORSC expression
for the nonsinglet longitudinal structure function, expanded about 
$N=-1$, is 
\eqn\rscflns{\eqalign{F^{NS}_{L,RSC,0}&(N,Q^2)=
\alpha_s(Q_0^2)F^{NS}_{2,0}(N)\biggl(C^{NS}_{L,1,l}(N+1) +
C^{NS}_{L,1,l}(N=-1)
\biggl(\ln(Q_0^2/A^{NS}_L)\alpha_s(Q_0^2)\times\cr
&\hskip -0.2in \biggl({1\over N+1}
\sum_{m=0}^{\infty} b_m\biggl({\alpha_s(Q_0^2)
\over (N+1)^2}\biggr)^m
+ \sum_{m=0}^{\infty}\bigl(c_m+
\half \sum_{n=0}^{m}b_nb_{m-n}\ln(Q_0^2/A^{NS}_L)\bigr)
\biggl({\alpha_s(Q_0^2)
\over (N+1)^2}\biggr)^m\biggr)\biggr)\biggr)\times\cr
&\biggl({\alpha_s(Q_0^2)\over \alpha_s(Q^2)}
\biggr)^{\tilde\Gamma_{L,0,l}^{NS}(N+1)-1}\exp\biggl[
\int_{\alpha_s(Q^2)}^{\alpha_s(Q_0^2)}
\biggl(\sum_{m=1}^{\infty} \biggl({1\over N+1}b_m+c_m\biggr)
\biggl({\alpha_s(q^2)
\over (N+1)^2}\biggr)^m\biggr){d\alpha_s(q^2)\over 
b_0\alpha^2_s(q^2)}\biggr].\cr}}
As we have already argued for the singlet case, as far as the expansion 
about the singularities at $N$ to the left of the rightmost singularity
are concerned, we may as well just take the one--loop expression.
In practice, we will only use the one-loop expression for the 
nonsinglet structure functions for all the singularities in the anomalous 
dimensions. This is because of the phenomenological 
reasons given at the end of the last subsection,
and also because of lack of knowledge of the full
physical anomalous dimensions. In the $\overline{\hbox{\rm MS}}$ scheme
the terms $\sim \alpha_s^m(N+1)^{2m-1}$ in the parton anomalous dimension are
all known \ref\nsresum{R. Kirschner and L.N. Lipatov, \NP  
\vyp{B213}{1983}{122}.}. It does not 
appear as though the coefficient functions 
contribute to these sort of terms in the physical anomalous dimension
(though there is no formal proof of this), and hence, it is believed all of 
these terms are known. However, there is little knowledge yet of the 
terms of the sort $\sim \alpha_s^m(N+1)^{2m-2}$. We have argued that these 
are an intrinsic part of the LORSC
expression for the nonsinglet structure function, and they should be 
calculated and included in order to give a true indication of the effect
of leading--$\ln (1/x)$ terms. Hence, we believe that calculations of the 
nonsinglet \brv\ref\nsresumphen{B.I. Ermolaev, S.I. Manaenkov and M.G. Ryskin,
\ZP\vyp{C69}{1996}{259} \semi J. Bl\"umlein and A. Vogt, 
{\it Acta Phys. Polon.} \vyp{B27}{1996}{1309}; \PL \vyp{B370}{1996}{149}.}  
(and polarized \ref\resumpol{J. Bartels, B.I. Ermolaev and M.G. Ryskin, 
{\tt hep-ph/9603204},
preprint DESY--96--025, March 1996; \ZP \vyp{C70}{1996}{273}\semi
J. Bl\"umlein and A. Vogt, \PL \vyp{B386}{1996}{350};
{\tt hep-ph/9610203}, Proc. of the 1996 HERA Physics Workshop, 
eds. G. Ingelman,
R. Klanner, and A. De Roeck, DESY, Hamburg, 1996, Vol. 2, p. 799;
Proc. of the International Conference SPIN'96, Amsterdam, September 1996,
eds. K. de Jager, P. Mulders {\it et al}, (World Scientific, Singapore, 1996),
in print.}) structure functions which claim to 
include leading--$\ln(1/x)$ corrections are incomplete, even at 
leading order,  
until the terms of the form $\alpha_s^m(N+1)^{2m-2}$ are known. 

\bigskip

We now consider the nonsinglet structure function $F^{NS}_2(N,Q^2)$. This 
leads us back to our previous question of whether we should use the
full RSC expression for $F_2(N,Q^2)$ or that
for $(d\, F_2(N,Q^2)/d\,\ln Q^2)$. In order to illustrate the
difference between the two, and help 
us make our choice, we consider the simpler nonsinglet case
before the singlet case. We also pretend for the moment 
that the inverse powers of $(N+1)$ in the expressions do not increase as 
the power of $\alpha_s$ increases, i.e. there is no small--$x$ enhancement
at higher orders in $\alpha_s$. This being the case, the LO term
in the evolution is just
\eqn\loevolnssfpret{\biggl({\alpha_s(Q_0^2)\over \alpha_s(Q^2)}
\biggr)^{\tilde\Gamma_{2,0,l}^{NS}(N+1)}.}
Of course, the input may be written as a power series in $\alpha_s(Q_0^2)$, 
as we saw in \S 4.1, and is of the form 
\eqn\nsinputftwo{F^{NS}_{2,0}(N,Q_0^2)=F^{NS}_{2,0}(N)[1+\alpha_s(Q_0^2)
\ln(Q_0^2/A^{NS}_2)\Gamma^{2,0,l}_{NS}(N+1)+ \hbox{\rm higher order in 
$\alpha_s(Q_0^2)$} ].}
Therefore, we have the problem that as well as the lowest order 
$\alpha_s$--, and hence $Q_0^2$--dependent part of the input there is also 
the ``sub--lowest--order'', $Q_0^2$--independent part. These two terms are 
clearly of different order, but
under a change in renormalization scheme, and hence in the definition of 
the coupling, both remain unchanged and both
should therefore appear in the LO definition of the structure function.
This mixing of orders seems rather unsatisfactory, and comes about 
because for 
$F^{NS}_2(N,Q^2)$ the structure function still exists in the formal limit of 
$\alpha_s \to 0$, i.e. the parton model limit,
being equal to the simple $Q_0^2$--independent function
$F^{NS}_{2,0}(N)$. Hence, it is not a
perturbative quantity in quite the sense way as $F_L^{NS}(N,Q^2)$
or $(d\, F_2^{NS}(N,Q^2)/d\ln Q^2)$, both of which vanish in this limit.

So in our simplified model the LORSC expression for $F_2^{NS}(N,Q^2)$, 
obtained by combining the LO input and evolution, is
\eqn\loexpnssfpret{F^{NS}_{2,RSC,0}(N,Q^2)=F^{NS}_{2,0}(N)[1+\alpha_s(Q_0^2)
\ln(Q_0^2/A^{NS}_2)\Gamma^{2,0,l}_{NS}(N+1)]
\biggl({\alpha_s(Q_0^2)\over \alpha_s(Q^2)}
\biggr)^{\tilde\Gamma_{2,0,l}^{NS}(N+1)}.}
This consists of two parts which are clearly of different magnitude, i.e. one
is a power of $\alpha_s$ down on the other with no small--$x$ enhancement.
This seems against the spirit of a well--ordered calculation. Indeed, the 
second part of this LO expression is of the same order of 
magnitude as part of the LO input multiplying the NLO evolution, i.e.
\eqn\samemag{F^{NS}_{2,0}(N)(\alpha_s(Q_0^2)-\alpha_s(Q^2))\hat 
\Gamma^{2,1,l}_{NS}(N+1)\biggl({\alpha_s(Q_0^2)\over \alpha_s(Q^2)}
\biggr)^{\tilde\Gamma_{2,0,l}^{NS}(N+1)}.}
Even when we take into account the higher inverse powers of $(N+1)$ at 
higher powers of $\alpha_s$, \loexpnssfpret\ will be part of the LO
expression for $F^{NS}_{2}(N,Q^2)$, and this unsatisfactory 
behaviour remains. It is also clear that the same effect will be 
seen for the singlet structure function.

If we instead consider $(d\,F_2^{NS}(N,Q^2)/d\,\ln Q^2)$ (again ignoring
small--$x$ enhancement for the moment) the situation improves. 
In this case the full expression is 
\eqn\ftwoderivnsexample{\eqalign{{d\,F_2^{NS}(N,Q^2)\over d\,\ln Q^2}  
 =&\Gamma_{2,NS}(N+1,\alpha_s(Q_0^2))F_2^{NS}(N,Q_0^2)\times\cr
&\exp\biggl[\int_{\ln Q_0^2}^{\ln Q^2}
\biggl(\Gamma_{2,NS}(N+1,\alpha_s(q^2))
-{d\over d\,\ln q^2}\ln(\Gamma_{2,NS}(N+1,\alpha_s(q^2)))\biggr)
d\ln q^2\biggr].\cr}}
Hence, the input may be written as
\eqn\nsinputderivftwo{\eqalign{\biggl({d\,F^{NS}_{2,0}(N,Q^2)\over d\,\ln Q^2}
\biggr)_{Q_0^2}=F^{NS}_{2,0}(N)\bigl[
\alpha_s(Q_0^2)\Gamma^{2,0,l}(N+1)
+\alpha^2_s(Q_0^2)&
\bigl(\ln(Q_0^2/A^{NS}_2)(\Gamma^{2,0,l}_{NS}(N+1))^2\cr
&+\Gamma^{2,1,l}(N+1)\bigr)+{\cal O} (\alpha^3_s(Q_0^2)) \bigr].\cr}}
Of course, the ${\cal O}(\alpha_s(Q_0^2))$ piece is renormalization scheme
independent by definition. The ${\cal O}(\alpha^2_s(Q_0^2))$ piece is 
renormalization scheme dependent ($\Gamma^{2,1,L}(N)$ is renormalization
scheme dependent) in order to absorb changes in the 
${\cal O}(\alpha_s(Q_0^2))$ piece under a change in the coupling of ${\cal O}
(\alpha_s^2(Q_0^2))$. So this time we have a LORSC input which is of a 
given order in $\alpha_s(Q_0^2)$. The full LO expression for 
$(dF_2(N,Q^2)/d\ln Q^2)$ is then 
\eqn\nsderivftwoex{\biggl({d\,F^{NS}_{2,0}(N,Q^2)\over d\,\ln Q^2}
\biggr)_{0}=F^{NS}_{2,0}(N)
\alpha_s(Q_0^2)\Gamma^{NS}_{2,0,l}(N+1)\biggl({\alpha_s(Q_0^2)\over 
\alpha_s(Q^2)}\biggr)^{\tilde\Gamma_{2,0,l}^{NS}(N+1)-1}.}
This is rather more satisfactory than the 
renormalization--scheme--independent expression for $F_2^{NS}(N,Q^2)$ itself,
and hence we choose $(dF_2(N,Q^2)/d\ln Q^2)$ to be the perturbative
quantity we calculate, in both the nonsinglet and singlet case. 

If we wish to calculate the structure function $F_2(N,Q^2)$ itself to a 
given order we will do this by integrating the given order expression for 
$(dF_2(N,Q^2)/d\ln Q^2)$ between $Q_0^2$ and $Q^2$, 
and adding it to $F_2(N,Q_0^2)$
evaluated to the same order. For example, in our simplified nonsinglet model 
we would integrate \nsderivftwoex\ and add this to the explicitly 
written part of \nsinputftwo. This results in the effective LO expression 
\eqn\loexpnssfpretalt{F^{NS}_{2,RSC,0}(N,Q^2)=F^{NS}_{2,0}(N)\biggl[
\biggl({\alpha_s(Q_0^2)\over \alpha_s(Q^2)}
\biggr)^{\tilde\Gamma_{2,0,l}^{NS}(N+1)} +\alpha_s(Q_0^2)
\ln(Q_0^2/A^{NS}_2)\Gamma^{2,0,l}_{NS}(N+1)].}
This is perhaps more sensible than \loexpnssfpret, since the two terms 
are now of a more comparable size. Moreover, this expression is more stable 
under changes in $Q_0^2$.

Of course, this whole discussion of $F^{NS}_{2}(N,Q^2)$ has been rather 
simplified by the assumption that the higher--order--in--$\alpha_s$ 
terms in the
physical anomalous dimension do not contain higher singularities in $(N+1)$. 
Recognizing that they do, we obtain expressions which are more complicated, 
as in the case of $F^{NS}_L(N,Q^2)$. With a little work it is possible to 
see that the full leading--order expressions are 
\eqn\rscftwons{\eqalign{\biggl(&{d\,F^{NS}_{2,RSC,0}(N,Q^2)\over d\,\ln Q^2}
\biggr)_0=\alpha_s(Q_0^2)F^{NS}_{2,0}(N)
\biggl(\Gamma^{NS}_{2,0,l}(N+1) +
{1\over N+1}\times\cr
&\biggl(\sum_{m=1}^{\infty}\bigl( \tilde b_m
+ (N+1)\tilde c_m\bigr)\biggl({\alpha_s(Q_0^2)
\over (N+1)^2}\biggr)^m+{\tilde a_{-1}\alpha_s(Q_0^2)\over (N+1)}
\ln(Q_0^2/A^{NS}_2) \sum_{m=0}^{\infty}
\tilde b_m\biggl({\alpha_s(Q_0^2)\over (N+1)^2}\biggr)^m
\biggr)\times\cr
&\biggl({\alpha_s(Q_0^2)\over \alpha_s(Q^2)}
\biggr)^{\tilde\Gamma_{2,0,l}^{NS}(N+1)-1}
\exp\biggl[
\int_{\alpha_s(Q^2)}^{\alpha_s(Q_0^2)}
\biggl(\sum_{m=1}^{\infty} \biggl({1\over N+1}\tilde b_m+\tilde c_m\biggr)
\biggl({\alpha_s(q^2)\over (N+1)^2}\biggr)^m\biggr){d\alpha_s(q^2)\over 
b_0\alpha^2_s(q^2)}\biggr]\times\cr
&\biggl({1+\sum_{m=1}^{\infty} \tilde b_m(\alpha^{m+1}_s(Q^2)/ (N+1)^{2m})
\over 1+\sum_{m=1}^{\infty} \tilde b_m(\alpha^{m+1}_s(Q_0^2)/ (N+1)^{2m})}
\biggr)\biggr)\cr}}
and
\eqn\loinputnstwosf{\eqalign{F^{NS}_{2,RSC,0}&(N,Q_0^2)=F^{NS}_{2,0}(N)\biggl(
1 +\ln(Q_0^2/A^{NS}_2)\alpha_s(Q_0^2)\biggl(\Gamma^{NS}_{2,0,l}(N+1)+\cr
&{1\over N+1}
\sum_{m=1}^{\infty} \tilde b_m\biggl({\alpha_s(Q_0^2)
\over (N+1)^2}\biggr)^m
+ \sum_{m=0}^{\infty}\bigl(\tilde c_m+
\half\sum_{n=0}^{m}\tilde b_n\tilde b_{m-n}\ln(Q_0^2/A^{NS}_2)
\bigr)\biggl({\alpha_s(Q_0^2)
\over (N+1)^2}\biggr)^m\biggr)\biggr),\cr}}
where for $F^{NS}_{2}(N,Q^2)$ the coefficients in the series in 
$(N+1)^{-1}$are not necessarily 
the same as for the longitudinal structure function, hence the slightly 
different notation. However, $a_{-1}=\tilde a_{-1}$ and, if the hypothesis 
that in the $\overline{\hbox{\rm MS}}$ renormalization and factorization 
scheme all the contribution to the $\alpha_s^m/(N+1)^{2m-1}$ singularities 
in the physical anomalous dimensions is due to the parton anomalous dimension,
$b_m =\tilde b_m$ as well. The $c_m$ are not equal to the 
$\tilde c_m$ in general though, so there is no guarantee that 
$F^{NS}_{2,RSC,0}(x,Q^2)$ and $F^{NS}_{L,RSC,0}(x,Q^2)$ will behave in the 
same way in the small--$x$ limit. 

\medskip

When all the $b_m$ and $c_m$ are known they can be used to 
present an argument 
for the form of the small--$x$ behaviour of nonsinglet structure functions, 
e.g. to explain any discrepancy between the input power at small $x$
for these structure functions and predictions from Regge physics.
This seems to us to be an interesting project, and we look forward to
the calculation of these coefficients. Until this happens 
our discussion of the LORSC calculation of the nonsinglet structure
functions is rather academic. 
However, it has enabled us to discuss many of the issues in a simpler 
framework than if we had gone directly to the singlet structure functions.
We will discuss these singlet structure functions next.

\subsec{The Renormalization--Scheme--Consistent Singlet Structure Functions.}

When calculating the singlet structure functions
we cannot just construct the complete LO
evolution and input and combine these to obtain the LO 
expression because the evolution mixes the two different structure functions.
Each of the component parts of the LO expressions for $F_2(N,Q^2)$ and 
$F_L(N,Q^2)$ (we omit the superscript $S$ in this section) must
consist of LO input parts and evolution parts, but 
it is not obvious what these are. In order to find the full LORSC 
expressions for the singlet structure functions we will have to work in 
steps. We will consider only the full LO expression with the perturbative 
factors expanded about the particular value of $N=0$ (the nonperturbative 
inputs are the full nonanalytic expressions for $\hat F_L(N)$ and $F_2(N)$), 
and the simplest way to proceed is to work directly with the physical 
quantities, solving the evolution equations in terms of physical anomalous 
dimensions and structure functions. We have already proved that in the loop
expansion the LO expressions only depend on the one--loop physical 
anomalous dimensions, and in the leading--$\ln (1/x)$ expansion the  
LO expressions depend only on $\Gamma^{0}_{LL}(\alpha_s(Q^2)/N)$,
$\Gamma^{0}_{L2}(\alpha_s(Q^2)/N)$, $\Gamma^{1}_{2L}
(\alpha_s(Q^2)/N)$ and $\Gamma^{1}_{22}(\alpha_s(Q^2)/N)$. Hence,  
it is only the combination of these anomalous dimensions which is 
considered in our solution. 

We cannot simply write the physical anomalous dimension matrix
\eqn\inclmatrix{\pmatrix{\alpha_s(Q^2)\Gamma^{0,l}_{LL}(N)+
\Gamma^{\tilde 0}_{LL}(\alpha_s(Q^2)/N)& \alpha_s(Q^2) 
\Gamma^{0,l}_{L2}(N)+\Gamma^{\tilde 0}_{L2}(\alpha_s(Q^2)/N)\cr
\alpha_s(Q^2)\Gamma^{0,l}_{2L}(N)+\alpha_s(Q^2)
\Gamma^{\tilde 1}_{2L}(\alpha_s(Q^2)/N)&\alpha_s(Q^2)\Gamma^{0,l}_{22}(N) +
\alpha_s(Q^2)\Gamma^{\tilde 1}_{22}(\alpha_s(Q^2)/N)\cr}}
(where $\Gamma^{\tilde 0}_{LL}(\alpha_s(Q^2)/N)=
\Gamma^{0}_{LL}(\alpha_s(Q^2)/N)$ with the one--loop component subtracted 
out, etc.), and solve the renormalization group equations. With this 
anomalous dimension matrix there is no simple closed form for the solution of
these equations, and the full solution contains terms 
which are not properly of leading order. We must choose some way of 
solving for the structure functions systematically which enables us to 
extract the true LO behaviour in as simple a manner as possible. 

In order to do this we take account of the fact that the one--loop 
solutions for $F_L(N,Q^2)$ and $(dF_2(N,Q^2)/d\ln Q^2)$ must be part of the 
complete LORSC solutions. Hence, we split our anomalous dimension 
matrix up into the form 
\eqn\inclmatrixsplit{\alpha_s(Q^2)\pmatrix{\Gamma^{0,l}_{LL}(N)+
& \Gamma^{0,l}_{L2}(N)\cr
\Gamma^{0,l}_{2L}(N)& 
\Gamma^{0,l}_{22}(N)\cr}
+\pmatrix{\Gamma^{\tilde 0}_{LL}(\alpha_s(Q^2)/N)& 
\Gamma^{\tilde 0}_{L2}(\alpha_s(Q^2)/N)\cr
\alpha_s(Q^2)\Gamma^{\tilde 1}_{2L}(\alpha_s(Q^2)/N)&
\alpha_s(Q^2)\Gamma^{\tilde 1}_{22}(\alpha_s(Q^2)/N)\cr},}
and solve by treating the second matrix as a perturbation to the first.
Doing this we
obtain the one--loop solutions as the lowest--order solutions and can
systematically calculate corrections to this, extracting the parts of these 
``corrections'' which are leading order. 

So, first let us consider the solution to the renormalization group equation
\eqn\fullsoli{{d\over d\, \ln Q^2}\pmatrix{\hat F^{0,l}_L(N,Q^2)\cr 
F^{0,1}_2(N,Q^2)\cr}=\alpha_s(Q^2)\pmatrix{\Gamma^{0,l}_{LL}(N)
& \Gamma^{0,l}_{L2}(N)\cr
\Gamma^{0,l}_{2L}(N)&
\Gamma^{0,l}_{22}(N)\cr}\pmatrix{\hat F^{0,l}_L(N,Q^2)\cr 
F^{0,1}_2(N,Q^2)\cr},}
with boundary conditions $\hat F^{0,l}_L(N,Q_0^2)=\hat F_L(N)$ and
$F^{0,l}_2(N,Q_0^2)=F_2(N)$. We may write the solution for the longitudinal
structure function as
\eqn\fullsolii{\hat F^{0,l}_L(N,Q^2)=\hat F^{0,l,+}_L(N)\biggl({\alpha_s(Q_0^2)
\over \alpha_s(Q^2)}\biggr)^{\tilde \Gamma^{0,l,+}(N)}+\hat F^{0,l,-}_L(N)
\biggl({\alpha_s(Q_0^2)
\over \alpha_s(Q^2)}\biggr)^{\tilde \Gamma^{0,l,-}(N)},}
where $\tilde \Gamma^{0,l,+,-}(N)$ are the two eigenvalues of the 
zeroth--order 
physical anomalous dimension matrix (which are the same as the eigenvalues 
of the zeroth--order parton anomalous dimension matrix), and $\hat 
F^{0,l,+}_L(N)
+\hat F^{0,l,-}_L(N)=\hat F_L(N)$. Having chosen to write the lowest--order 
solution for the longitudinal structure function in this way we may then 
write the lowest--order solution for $F_2(N,Q^2)$ as
\eqn\fullsoliii{\hat F^{0,l}_2(N,Q^2)=e^+(N)\hat F^{0,l,+}_L(N)
\biggl({\alpha_s(Q_0^2)
\over \alpha_s(Q^2)}\biggr)^{\tilde\Gamma^{0,l,+}(N)}+e^-(N)\hat F^{0,l,-}_L(N)
\biggl({\alpha_s(Q_0^2)
\over \alpha_s(Q^2)}\biggr)^{\tilde \Gamma^{0,l,-}(N)},}
where 
\eqn\fullsola{e^+(N)= \biggl({\Gamma^{0,l,+}(N)-
\Gamma^{0,l}_{LL}(N)\over
\Gamma^{0,l}_{L2}(N)}\biggr),\hskip 0.5in 
e^-(N)= \biggl({\Gamma^{0,l,-}(N)-\Gamma^{0,l}_{LL}
(N)\over \Gamma^{0,l}_{L2}(N)}\biggr),}
and $e^+(N)\hat F^{0,l,+}_L(N)+
e^-(N)\hat F^{0,l,-}_L(N)=F_2(N)$. In practice  
\eqn\fullsolb{e^+(N)= N/6+{\cal O}
(N^2), \hskip 0.5in e^-(N)=\biggl({27\over 36-8N_f}\biggr)+{\cal O}(N),}
and
\eqn\fullsolc{\hat F^{0,l,+}_L(N)=\hat F_L(N)-
{36-8N_f\over 27}F_2(N)+{\cal O}(N),\hskip 0.2in \hat F^{0,l,-}_L(N)={36-8N_f
\over 27} F_2(N)+{\cal O}(N).} 
It is then simple to see that 
\eqn\fullsold{\eqalign{\hat F^{0,l,+}_2(N)&\equiv e^+(N)F^{0,l,+}_L(N)
={N\over 6}\biggl(\hat F_L(N)-\biggl({36-8N_f\over 27}\biggr)\biggr)F_2(N)
+{\cal O}(N^2),\cr  
\hat F^{0,l,-}_2(N)&\equiv e^-(N)F^{0,l,-}_L(N)=F_2(N)+{\cal O}(N).\cr}} 

The first correction to the one--loop solution may be obtained by solving 
the equation,
\eqn\fullsoliv{\eqalign{{d\over d\, \ln Q^2}\pmatrix{\hat F^{c1}_L(N,Q^2)\cr 
F^{c1}_2(N,Q^2)\cr}&=\alpha_s(Q^2)\pmatrix{\Gamma^{0,l}_{LL}(N)
&\Gamma^{0,l}_{L2}(N)\cr
\Gamma^{0,l}_{2L}(N)& 
\Gamma^{0,l}_{22}(N)\cr}\pmatrix{\hat F^{c1}_L(N,Q^2)\cr 
F^{c1}_2(N,Q^2)\cr}\cr
&\hskip -0.6in+\pmatrix{\Gamma^{\tilde 0}_{LL}(\alpha_s(Q^2)/N)& 
-{(36-8N_f)\over 27}\Gamma^{\tilde 0}_{LL}(\alpha_s(Q^2)/N)\cr
\alpha_s(Q^2)\Gamma^{\tilde 1}_{2L}(\alpha_s(Q^2)/N)&
-{(36-8N_f)\over 27}\alpha_s(Q^2)\Gamma^{\tilde 1}_{2L}(\alpha_s(Q^2)/N)\cr}
\pmatrix{\hat F^{0,l}_L(N,Q^2)\cr F^{0,l}_2(N,Q^2)\cr},\cr}}
where we have used the relationships in \rellong\ and \reltwo\ in order to 
simplify the second matrix. We proceed as follows. First we define 
the vectors
\eqn\fullsolv{\underline e^+(N)=\pmatrix{1\cr e^+(N)\cr}, \hskip 0.5in
\underline e^-(N)=\pmatrix{1\cr e^-(N)\cr},\hskip 0.5in \underline 
F^{c1}(N,Q^2)=
\pmatrix{F^{c1}_L(N,Q^2)\cr F^{c1}_2(N,Q^2)\cr},}
and write 
\eqn\neeqni{\underline F^{c1}(N,Q^2)= \underline e^+(N)F^{c1,+}(N,Q^2)+
\underline e^-(N)F^{c1,-}(N,Q^2).}
We also define projection operators $\underline p^+(N)$ and 
$\underline p^-(N)$ by
\eqn\fullsolvi{\eqalign{&\underline p^+(N)\cdot\underline e^+(N)=1,\hskip 1in
\underline p^+(N)\cdot\underline e^-(N)=0,\cr
&\underline p^-(N)\cdot\underline e^+(N)=0, \hskip 1in \underline 
p^-(N)\cdot\underline e^-(N)=1,\cr}} 
which in practice gives
\eqn\fulsolvii{\underline p^+(N)=\pmatrix{1\cr{8N_f-36\over 27}\cr}
+ {\cal O}(N), \hskip 1in \underline p^-(N)=\pmatrix{0\cr{27 \over 8N_f-36}\cr}
+ {\cal O}(N).}

Multiplying \fullsoliv\ by $\underline p^+(N)$ now leads to the 
straightforward first--order differential equation 
\eqn\fullsoliix{\eqalign{{d\, F^{c1,+}(N,Q^2)\over d\, \ln Q^2}&=
\alpha_s(Q^2)\Gamma^{0,l,+}(N)
F^{c1,+}(N,Q^2) \, +\cr
&\hskip -0.3in \underline p^+(N)\cdot
\pmatrix{\Gamma^{\tilde 0}_{LL}(\alpha_s(Q^2)/N)& 
-{(36-8N_f)\over 27}\Gamma^{\tilde 0}_{LL}(\alpha_s(Q^2)/N)\cr
\alpha_s(Q^2)\Gamma^{\tilde 1}_{2L}(\alpha_s(Q^2)/N)&
-{(36-8N_f)\over 27}\alpha_s(Q^2)\Gamma^{\tilde 1}_{2L}(\alpha_s(Q^2)/N)\cr}
\pmatrix{\hat F^{0,l}_L(N,Q^2)\cr F^{0,1}_2(N,Q^2)\cr}.\cr}}
We also write the zeroth--order solution $\underline F^{0,l}(N,Q^2)$, 
in the form 
\eqn\neweqnii{\underline F^{0,l}(N,Q^2)=
\underline e^+(N)F^{0,l,+}(N,Q^2)+\underline e^-(N)F^{0,l,-}(N,Q^2),}
where $F^{0,l,+}(N,Q^2)=F_L^{0,l,+}(N)(\alpha_s(Q_0^2)/
\alpha_s(Q^2))^{\tilde \Gamma^{0,l,+}(N)}$ and similarly for 
$F^{0,l,-}(N,Q^2)$. Doing this \fullsoliv\ becomes
\eqn\fullsolix{\eqalign{{d\, F^{c1,+}(N,Q^2)\over d\, \ln Q^2}&=
\alpha_s(Q^2)\Gamma^{0,l,+}(N)F^{c1,+}(N,Q^2) +\cr
&\underline p^+(N)\cdot
\pmatrix{\Gamma^{\tilde 0}_{LL}(\alpha_s(Q^2)/N)& 
-{(36-8N_f)\over 27}\Gamma^{\tilde 0}_{LL}(\alpha_s(Q^2)/N)\cr
\alpha_s(Q^2)\Gamma^{\tilde 1}_{2L}(\alpha_s(Q^2)/N)&
-{(36-8N_f)\over 27}\alpha_s(Q^2)
\Gamma^{\tilde 1}_{2L}(\alpha_s(Q^2)/N)\cr}\times\cr
&\hskip -0.3in \biggl(\underline e^+(N)F^{0,l,+}_L(N)
\biggl({\alpha_s(Q_0^2)\over
\alpha_s(Q^2)}\biggr)^{\tilde \Gamma^{0,l,+}(N)}+
\underline e^-(N)F^{0,l,-}_L(N)\biggl({\alpha_s(Q_0^2)\over
\alpha_s(Q^2)}\biggr)^{\tilde \Gamma^{0,l,-}(N)}\biggr).\cr}}
This equation can now be solved by using the power series expansions of 
$\underline e^{+(-)}(N)$ and $\underline p^{+(-)}(N)$ in terms of $N$. 
Only a small part of the overall solution contributes at LO.
Considering the contraction of the matrices
between the two vectors in \fullsolix\ we obtain
\eqn\fullsolx{\eqalign{\underline p^+(N)&\cdot
\pmatrix{\Gamma^{\tilde 0}_{LL}(\alpha_s(Q^2)/N)& 
-{(36-8N_f)\over 27}\Gamma^{\tilde 0}_{LL}(\alpha_s(Q^2)/N)\cr
\alpha_s(Q^2)\Gamma^{\tilde 1}_{2L}(\alpha_s(Q^2)/N)&
-{(36-8N_f)\over 27}\alpha_s(Q^2)\Gamma^{\tilde 1}_{2L}(\alpha_s(Q^2)/N)\cr}
\underline e^+(N)\cr &\hskip 2.3in =\Gamma^{\tilde 0}_{LL}(\alpha_s(Q^2)/N)
+\hbox{higher order,}\cr}}
and 
\eqn\fullsolxi{\eqalign{\underline p^+(N)\cdot
\pmatrix{\Gamma^{\tilde 0}_{LL}(\alpha_s(Q^2)/N)& 
-{(36-8N_f)\over 27}\Gamma^{\tilde 0}_{LL}(\alpha_s(Q^2)/N)\cr
\alpha_s(Q^2) \Gamma^{\tilde 1}_{2L}(\alpha_s(Q^2)/N)&
-{(36-8N_f)\over 27}\alpha_s(Q^2)\Gamma^{\tilde 1}_{2L}(\alpha_s(Q^2)/N)\cr}
&\underline e^-(N)\cr
\hskip -0.6in= {\cal O}\biggl(\alpha_s(Q^2)\biggl({\alpha_s(Q^2)\over N}
\biggr)^m\biggr).\cr}}
Therefore \fullsolix\ becomes
\eqn\fullsolxii{\eqalign{{d\, F^{c1,+}(N,Q^2)\over d\, \ln Q^2}&
=\alpha_s(Q^2)\Gamma^{0,l,+}(N)F^{c1,+}(N,Q^2) +\cr
&\Gamma^{\tilde 0}_{LL}(\alpha_s(Q^2)/N)
\hat F_L^{0,l,+}(N)\biggl({\alpha_s(Q_0^2)\over
\alpha_s(Q^2)}\biggr)^{\tilde \Gamma^{0,l,+}(N)}+\hbox{higher order}.\cr}}
Thus
\eqn\fullsolxiii{\eqalign{F^{c1,+}(N,Q^2)&=
\int_{\alpha_s(Q^2)}^{\alpha_s(Q_0^2)}
{\tilde\Gamma^{\tilde 0}_{LL}(\alpha_s(q^2)/N)\over \alpha^2_s(q^2)}d
\alpha_s(q^2)\hat F^{0,l,+}_L(N)\biggl({\alpha_s(Q_0^2)\over
\alpha_s(Q^2)}\biggr)^{\tilde \Gamma^{0,l,+}(N)}\cr
&+\hbox{higher order}.\cr}}
This method of solution demonstrates the particularly nice feature that the 
leading--order ``corrections'' to the one--loop solution which are 
proportional to $\underline e^+(N)$ are completely independent of 
$F^{0,l,-}(N,Q^2)$. Since this is due to the form of the ``correction''
matrix contracted between $\underline p^+(N)$ and $\underline e^-(N)$, we 
can write 
the equation for the $n_{\rm th}$--order correction to $F^{0,l,+}(N,Q^2)$ as 
\eqn\fullsolxiii{\eqalign{{d\, F^{cn,+}(N,Q^2)\over d\, \ln Q^2}&=
\alpha_s(Q^2)\Gamma^{0,l,+}(N)F^{cn,+}(N,Q^2)\cr
&\hskip -0.7in+\underline p^+(N)\cdot
\pmatrix{\Gamma^{\tilde 0}_{LL}(\alpha_s(Q^2)/N)& 
-{(36-8N_f)\over 27}\Gamma^{\tilde 0}_{LL}(\alpha_s(Q^2)/N)\cr
\alpha_s(Q^2)\Gamma^{\tilde 1}_{2L}(\alpha_s(Q^2)/N)&
-{(36-8N_f)\over 27}\alpha_s(Q^2)\Gamma^{\tilde 1}_{2L}(\alpha_s(Q^2)/N)\cr}
\pmatrix{\hat F^{c(n-1)}_L(N,Q^2)\cr F^{c(n-1)}_2(N,Q^2)\cr},\cr}}
and precisely the same argument goes through each time. Therefore,
\eqn\fullsolxiv{{d\, F^{cn,+}(N,Q^2)\over d\,\ln Q^2}
=\alpha_s(Q^2)\Gamma^{0,l,+}(N)F^{cn,+}(N,Q^2)+
\Gamma^{\tilde 0}_{LL}(\alpha_s(Q^2)/N)
F^{c(n-1),+}(N,Q^2)+\hbox{higher order}.}
Thus, it is easy to prove by induction that 
\eqn\fullsolxv{\eqalign{F^{cn,+}(N,Q^2)&={1\over n!}\biggl(
\int_{\alpha_s(Q^2)}^{\alpha_s(Q_0^2)}
{\tilde \Gamma^{\tilde 0}_{LL}(\alpha_s(q^2)/N)\over \alpha^2_s(q^2)}d
\alpha_s(q^2)\biggl)^n \hat F^{0,l,+}_L(N)
\biggl({\alpha_s(Q_0^2)\over
\alpha_s(Q^2)}\biggr)^{\tilde \Gamma^{0,l,+}(N)}\cr
&\hskip 3in+\hbox{higher order}.\cr}}
This leads to the straightforward expression for the whole of 
the leading--order part of $F^+(N,Q^2)$,
\eqn\fullsolxvi{F^+_{RSC,0}(N,Q^2)=F_L^{0,l,+}(N)
\biggl({\alpha_s(Q_0^2)\over
\alpha_s(Q^2)}\biggr)^{\tilde \Gamma^{0,l,+}(N)}
\exp\biggl[\int_{\alpha_s(Q^2)}^{\alpha_s(Q_0^2)}
{\tilde\Gamma^{\tilde 0}_{LL}(N,\alpha_s(q^2))\over \alpha_s(q^2)}d
\alpha^2_s(q^2)\biggl].}
Of course, we have not yet considered the corrections to the one--loop
input. We could have built up these 
corrections to the input at the same time as we built up the 
evolution. However, it is easier to simply examine the change of the whole of 
\fullsolxvi\ under a change in starting scale, and choose the 
input necessary to make the expression insensitive to such changes. 
In order to do this it is simplest to rewrite \fullsolxvi\ in the form,
\eqn\fullsolxvii{\eqalign{F^+_{RSC,0}(N,Q^2)=\biggl({\alpha_s(Q_0^2)\over
\alpha_s(Q^2)}\biggr)^{\tilde \Gamma^{0,l,+}(N)}
&\biggl(\hat F_L(N)-\biggl({36-8N_f\over 27}\biggr)F_2(N)\biggr)\times\cr
&\exp\biggl[\int_{\alpha_s(Q^2)}^{\alpha_s(Q_0^2)}
{\tilde \Gamma^{\tilde 0}_{LL}(\alpha_s(q^2)/N)\over \alpha^2_s(q^2)}
d\alpha_s(q^2)\biggr] + \hbox{higher order in $N$},\cr}}
where the second term contains those parts of the one--loop input 
which are higher order in $N$. The factor $(\hat F_L(N)
-\biggl({36-8N_f\over 27}\biggr)F_2(N))$ 
must be multiplied by $\exp[\ln(Q_0^2/A_{LL})
\Gamma^0_{LL}(\alpha_s(Q_0^2)/N)]$ in order to absorb changes in the 
evolution term under a change in starting scale $Q_0^2\to (1+\delta )Q_0^2$.
This does not absorb the whole of the change of the evolution 
term, but all other changes required in the input are subleading to this. 
Similarly the change in the input of the ``higher--order--in--$N$'' terms 
needed to absorb the change in the evolution is entirely subleading 
in $\alpha_s(Q_0^2)$ to the one--loop input, or to $(\hat F_L(N)
-{(36-8N_f)\over 27}F_2(N))\exp[\ln(Q_0^2/A_{LL})
\Gamma^0_{LL}(\alpha_s(Q_0^2)/N)]$.
Hence, the full LO expression for $F^+(N,Q^2)$ is   
\eqn\fullsolxvii{\eqalign{F^+_{RSC,0}(N,Q^2)=&\biggl[
\biggl(\hat F_L(N)-\biggl({36-8N_f\over 27}\biggr)F_2(N)\biggr)
(\exp[\ln(Q_0^2/A_{LL})
\Gamma^0_{LL}(\alpha_s(Q_0^2)/N)]-1)\cr
+\hat F^{0,l,+}_L(N)\biggr]
&\biggl({\alpha_s(Q_0^2)\over
\alpha_s(Q^2)}\biggr)^{\tilde \Gamma^{0,l,+}(N)}
\exp\biggl[\int_{\alpha_s(Q^2)}^{\alpha_s(Q_0^2)}
{\tilde \Gamma^{\tilde 0}_{LL}(\alpha_s(q^2)/N)\over \alpha^2_s(q^2)}
d\alpha_s(q^2)\biggr].\cr}}

\medskip 

We can solve for the corrections which are proportional to $\underline 
e^-(N)$ in exactly the same manner as above. Multiplying \fullsoliv\ by 
$\underline p^-(N)$, instead of $\underline p^+(N)$ leads to 
\eqn\fullsoliixx{\eqalign{{d\, F^{c1,-}(N,Q^2)\over d\, \ln Q^2}&=
\alpha_s(Q^2)\Gamma^{0,l,-}(N)F^{c1,-}(N,Q^2)\,+ \cr
&\hskip -0.6in\underline p^-(N)\cdot
\pmatrix{\Gamma^{\tilde 0}_{LL}(\alpha_s(Q^2)/N)& 
-\biggl({36-8N_f\over 27}\biggr)\Gamma^{\tilde 0}_{LL}(\alpha_s(Q^2)/N)\cr
\alpha_s(Q^2)\Gamma^{\tilde 1}_{2L}(\alpha_s(Q^2)/N)&
-\biggl({36-8N_f\over 27}\biggr)\alpha_s(Q^2) 
\Gamma^{\tilde 1}_{2L}(\alpha_s(Q^2)/N)\cr}
\pmatrix{\hat F^{0,l}_L(N,Q^2)\cr F^{0,l}_2(N,Q^2)\cr}.\cr}}
Again considering the contraction of the matrix
between the two vectors in the last term we obtain
\eqn\fullsolixx{\eqalign{\underline p^-(N)&\cdot
\pmatrix{\Gamma^{\tilde 0}_{LL}(\alpha_s(Q^2)/N)& 
-\biggl({36-8N_f\over 27}\biggr)\Gamma^{\tilde 0}_{LL}(\alpha_s(Q^2)/N)\cr
\alpha_s(Q^2)\Gamma^{\tilde 1}_{2L}(\alpha_s(Q^2)/N)&
-\biggl({36-8N_f\over 27}\biggr)\alpha_s(Q^2)\Gamma^{\tilde 1}_{2L}
(\alpha_s(Q^2)/N)\cr}
\underline e^+(N)\cr
&\hskip 0.8in = \biggl({36-8N_f\over 27}\biggr)
(\alpha_s(Q^2)\Gamma^{\tilde 1}_{2L}(\alpha_s(Q^2)/N)-
{N\over 6}\Gamma^{\tilde 0}_{LL}(\alpha_s(Q^2)/N))
+\hbox{higher order,}\cr}}
and 
\eqn\fullsolxx{\eqalign{\underline p^-(N)\cdot
\pmatrix{\Gamma^{\tilde 0}_{LL}(\alpha_s(Q^2)/N)& 
-\biggl({36-8N_f\over 27}\biggr)\Gamma^{\tilde 0}_{LL}(\alpha_s(Q^2)/N)\cr
\alpha_s(Q^2)\Gamma^{\tilde 1}_{2L}(\alpha_s(Q^2)/N)&
-\biggl({36-8N_f\over 27}\biggr)\alpha_s(Q^2)\Gamma^{\tilde 1}_{2L}
(\alpha_s(Q^2)/N)\cr}&
\underline e^-(N)\cr
\hskip 0.3in= {\cal O}\biggl(\alpha^2_s(Q^2)\biggl({\alpha_s(Q^2)\over N}
\biggr)^m\biggr).\cr}}
Therefore we can write \fullsoliixx\ in the form
\eqn\fullsolxxi{\eqalign{{d\, F^{c1,-}(N,Q^2)\over d\, \ln Q^2}&
=\alpha_s(Q^2)\Gamma^{0,l,-}(N)F^{c1,-}(N,Q^2) +
\biggl({36-8N_f\over 27}\biggr)
\biggl({\alpha_s(Q_0^2)\over\alpha_s(Q^2)}\biggr)^{\tilde \Gamma^{0,l,+}(N)}
\times\cr
&(\alpha_s(Q^2)
\Gamma^{\tilde 1}_{2L}(\alpha_s(Q^2)/N)
-{N\over 6}\Gamma^{\tilde 0}_{LL}(\alpha_s(Q^2)/N))
\biggl(\hat F_L(N)-\biggl({36-8N_f\over 27}\biggr)F_2(N)\biggr)\cr
&+\hbox{higher order}.\cr}}
Rather than solving this equation, 
it is easier to solve for the whole of the correction to 
$F^{0,l,-}_L(N,Q^2)$ in one go. Using \fullsolixx\ and \fullsolxx\ we easily
obtain
\eqn\fullsolxxii{\eqalign{\sum_{n=1}^{\infty}&
\biggl[\biggl({d\, F^{cn,-}(N,Q^2)\over d\, \ln Q^2}\biggr)
-\alpha_s(Q^2)\Gamma^{0,l,-}(N)F^{cn,-}(N,Q^2)\biggr] =
\biggl({36-8N_f\over 27}\biggr)\biggl({\alpha_s(Q_0^2)\over
\alpha_s(Q^2)}\biggr)^{\tilde\Gamma^{0,l,+}(N)}\cr
& (\alpha_s(Q^2)\Gamma^{\tilde 1}_{2L}(\alpha_s(Q^2)/N)
-{N\over 6}\Gamma^{\tilde 0}_{LL}(\alpha_s(Q^2)/N))
\biggl(\hat F_L(N)-\biggl({36-8N_f\over 27}\biggr)F_2(N)\biggr)
\times\cr
&\exp[\ln(Q_0^2/A_{LL})
\hat\Gamma^0_{LL}(\alpha_s(Q_0^2)/N)]\exp\biggl[
\int_{\alpha_s(Q^2)}^{\alpha_s(Q_0^2)}
{\tilde \Gamma^{\tilde 0}_{LL}(\alpha_s(q^2)/N)\over \alpha^2_s(q^2)}
d\alpha_s(q^2)\biggr]\cr
&+\hbox{higher order}.\cr}}
The solution to this is relatively simple if we wish to keep only
the LO parts. Letting $F^{cfull,-}(N,Q^2)=\sum_{n=1}^{\infty}
F^{cn,-}(N,Q^2)$ we obtain, 
\eqn\fullsolxxiii{\eqalign{F^{cfull,-}(N,Q^2)&=F^{cfull,-}(N,Q_0^2)
\biggl({\alpha_s(Q_0^2)\over
\alpha_s(Q^2)}\biggr)^{\tilde \Gamma^{0,l,-}(N)}\cr
&\hskip -0.9in +\biggl({\alpha_s(Q_0^2)
\Gamma^{\tilde 1}_{2L}(\alpha_s(Q_0^2)/N)-{N\over 6}\Gamma^{\tilde 0}_{LL}
(\alpha_s(Q_0^2)/N)\over 
\Gamma^{0}_{LL}(\alpha_s(Q_0^2)/N)}\biggr)\biggl({36-8N_f\over 27}\biggr)
\biggl(\hat F_L(N)-\biggl({36-8N_f\over 27}\biggr)F_2(N)\biggr)\times\cr
&\hskip -0.5in \exp[\ln(Q_0^2/A_{LL})
\Gamma^0_{LL}(\alpha_s(Q_0^2)/N)]\biggl({\alpha_s(Q_0^2)\over
\alpha_s(Q^2)}\biggr)^{\tilde \Gamma^{0,l,+}(N)}\exp\biggl[
\int_{\alpha_s(Q^2)}^{\alpha_s(Q_0^2)}
{\tilde \Gamma^{\tilde 0}_{LL}(\alpha_s(q^2)/N)\over \alpha^2_s(q^2)}
d\alpha_s(q^2)\biggr]\cr
&\hskip -0.9in-\biggl({\alpha_s(Q_0^2) 
\Gamma^{\tilde 1}_{2L}(\alpha_s(Q_0^2)/N)
-{N\over 6}\Gamma^{\tilde 0}_{LL}(\alpha_s(Q_0^2)/N)\over
\Gamma^{0}_{LL}(\alpha_s(Q_0^2)/N)}\biggr)\biggl({36-8N_f\over 27}\biggr)
\biggl(\hat F_L(N)-
\biggl({36-8N_f\over 27}\biggr)F_2(N)\biggr)\times\cr
&\hskip -0.5in \exp[\ln(Q_0^2/A_{LL})
\Gamma^0_{LL}(\alpha_s(Q_0^2)/N)]\biggl({\alpha_s(Q_0^2)\over
\alpha_s(Q^2)}\biggr)^{\tilde \Gamma^{0,l,-}(N)}
+\hbox{higher order}.\cr}}
Adding to the solution at one--loop, the whole of the LO part of 
$F^-(N,Q^2)$ is
\eqn\fullsolxxv{\eqalign{F^{-}_{RSC,0}(N,Q^2)&=
\biggl(F^{0,l,-}_L(N) -\biggl({36-8N_f\over 27}\biggr)
\biggl(\hat F_L(N)-
\biggl({36-8N_f\over 27}\biggr)F_2(N)\biggr)\times\cr
&\biggl({\alpha_s(Q_0^2)
\Gamma^{\tilde 1}_{2L}(\alpha_s(Q_0^2)/N)
-{N\over 6}\Gamma^{\tilde 0}_{LL}(\alpha_s(Q_0^2)/N)\over 
\Gamma^{0}_{LL}(\alpha_s(Q_0^2)/N)}\biggr)\biggr)
\biggl({\alpha_s(Q_0^2)\over
\alpha_s(Q^2)}\biggr)^{\tilde \Gamma^{0,l,-}(N)}+\cr
&\hskip -0.7in\biggl({36-8N_f\over 27}\biggr)
\biggl({\alpha_s(Q_0^2)
\Gamma^{\tilde 1}_{2L}(\alpha_s(Q_0^2)/N)
-{N\over 6}\Gamma^{\tilde 0}_{LL}(\alpha_s(Q_0^2)/N)\over 
\Gamma^{0}_{LL}(\alpha_s(Q_0^2)/N)}\biggr)
\biggl(\hat F_L(N)-
\biggl({36-8N_f\over 27}\biggr)F_2(N)\biggr)\times\cr
&\hskip -0.5in \exp[\ln(Q_0^2/A_{LL})
\Gamma^0_{LL}(\alpha_s(Q_0^2)/N)]\biggl({\alpha_s(Q_0^2)\over
\alpha_s(Q^2)}\biggr)^{\tilde \Gamma^{0,l,+}(N)}\exp\biggl[
\int_{\alpha_s(Q^2)}^{\alpha_s(Q_0^2)}
{\tilde \Gamma^{\tilde 0}_{LL}(\alpha_s(q^2)/N)\over \alpha^2_s(q^2)}
d\alpha_s(q^2)\biggr]\cr
&+\hbox{higher order},\cr}}
where, in order to make the expression invariant under changes in starting 
scale, and also ensure that for $Q_0^2$ =$A_{LL}$ we have $F^{-}(N,Q_0^2)=
\hat F_L^{0,l,-}(N)$, we make the choice
\eqn\deffmininput{\eqalign{F^{cfull,-}(N,Q_0^2)=\biggl({36-8N_f\over 27}\biggr)
&\biggl({\alpha_s(Q_0^2)
\Gamma^{\tilde 1}_{2L}(\alpha_s(Q_0^2)/N)
-{N\over 6}\Gamma^{\tilde 0}_{LL}(\alpha_s(Q_0^2)/N)\over 
\Gamma^{0}_{LL}(\alpha_s(Q_0^2)/N)}\biggr)\times\cr
&\hskip -0.6in \biggl(\hat F_L(N)-
\biggl({36-8N_f\over 27}\biggr)F_2(N)\biggr)(\exp[\ln(Q_0^2/A_{LL})
\Gamma^0_{LL}(\alpha_s(Q_0^2)/N)]-1).\cr}}

\medskip

We have been able to obtain, without too much difficulty, 
the leading parts of $F^+(N,Q^2)$ and $F^{-}(N,Q^2)$. We must now use these 
in order to obtain LORSC expressions for the structure functions. The way 
in which we have set up the calculation 
makes this very straightforward for the longitudinal structure function: we  
multiply $F^+_{RSC,0}(N,Q^2)$ and $F^{-}_{RSC,0}(N,Q^2)$ by $(\alpha_s(Q^2)/
2\pi)$. However, we notice that all the parts of $F^-_{RSC,0}(N,Q^2)$, 
except the one--loop expression, are subleading
to $F^{+}_{RSC,0}(N,Q^2)$: all the 
terms in the inputs in the former are a power of $\alpha_s(Q_0^2)$ higher
for the same power of $N$. Thus, the only part of $F^{-}_{RSC,0}(N,Q^2)$
which contributes to the LORSC expression for $F_L(N,Q^2)$ is 
the one--loop part. Adding this to \fullsolxvii, and multiplying
by $(\alpha_s(Q^2)/2\pi)$, we obtain
\eqn\fullsolfl{\eqalign{F_{L,RSC,0}(N,Q^2)&={\alpha_s(Q_0^2)\over 2\pi}
\Biggl[\biggl({\alpha_s(Q_0^2)\over
\alpha_s(Q^2)}\biggr)^{\tilde\Gamma^{0,l,+}(N)-1}
\exp\biggl[\int_{\alpha_s(Q^2)}^{\alpha_s(Q_0^2)}
{\tilde \Gamma^{\tilde 0}_{LL}(\alpha_s(q^2)/N)\over \alpha_s(q^2)}
d\alpha_s(q^2)\biggr]\times\cr
&\hskip -0.3in\biggl(F^{0,l,+}_L(N)+
\biggl(\hat F_L(N)-\biggl({36-8N_f\over 27}\biggr)F_2(N)\biggr)
(\exp[\ln(Q_0^2/A_{LL})\Gamma^0_{LL}(\alpha_s(Q_0^2)/N)]-1)\biggr)\cr
&\hskip -0.3in +F^{0,l,-}_L(N)\biggl({\alpha_s(Q_0^2)\over
\alpha_s(Q^2)}\biggr)^{\tilde \Gamma^{0,l,-}(N)-1}\Biggr].\cr}}

\medskip

It is slightly more involved to find the LORSC expressions for 
$F_2(N,Q^2)$. In this case we wish to find the LORSC expressions for 
the derivative $(d\,F_2(N,Q^2)/d\,\ln Q^2)$ and for the input $F_2(N,Q_0^2)$.
We consider the former first. Using the form of $e^+(N)$ and $e^-(N)$ in 
\fullsolb\ it is clear that, besides for the one--loop contributions, 
the LORSC expression   
will come from $(27/(36-8N_f))\times(d\,F^-_2(N,Q^2)/d\,\ln Q^2)$
and from $N/6 \times(d\,F^+_2(N,Q^2)/d\,\ln Q^2)$. Explicitly we obtain 
\eqn\fullsolfderiv{\eqalign{\biggl({d\,F_{2}(N,Q^2)\over d\,\ln Q^2}
\biggr)_{RSC,0}&
=\alpha_s(Q_0^2)\Biggl[e^-(N)\Gamma^{0,l,-}(N)
\hat F^{0,l,-}_L(N)\biggl({\alpha_s(Q_0^2)\over
\alpha_s(Q^2)}\biggr)^{\tilde \Gamma^{0,l,-}(N)-1}\cr
&+\biggl(e^+(N)\Gamma^{0,l,+}(N)\hat F^{0,l,+}_L(N)- \Gamma^{1,0}_{2,L}(N)
\biggl(\hat F_L(N)-\biggl({36-8N_f\over 27}\biggr)F_2(N)\biggr)\cr
&\hskip-0.6in+\Gamma^{1}_{2L}(\alpha_s(Q_0^2)/N)
\biggl(\hat F_L(N)-\biggl({36-8N_f\over 27}\biggr)F_2(N)\biggr)
\exp[\ln(Q_0^2/A_{LL})
\Gamma^0_{LL}(\alpha_s(Q_0^2)/N)]\biggr)\times\cr
&\exp\biggl[
\int_{\alpha_s(Q^2)}^{\alpha_s(Q_0^2)}
{\tilde \Gamma^{\tilde 0}_{LL}(\alpha_s(q^2)/N)\over \alpha^2_s(q^2)}
d\alpha_s(q^2)\biggr]\biggl({\alpha_s(Q_0^2)\over
\alpha_s(Q^2)}\biggr)^{\tilde \Gamma^{0,l,+}(N)-1}\Biggr],\cr}}
where we have used the relationship 
\eqn\fullsolxxiix{{N\over 6}(\Gamma^0_{LL}(\alpha_s(Q^2)/N)
-\Gamma^{\tilde 0}_{LL}(\alpha_s(Q^2)/N))=\alpha_s(Q^2)(\Gamma^1_{2L}
(\alpha_s(Q^2)/N)-\Gamma^{\tilde 1}_{2L}(\alpha_s(Q^2)/N)),}
and $\Gamma^{1,0}_{2,L}(N)\biggl(
\hat F_L(N)-\biggl({36-8N_f\over27}\biggr)F_2(N)\biggr)$ is the part of the 
input which is common to both the one--loop and the leading--$\ln (1/x)$ 
input, and is subtracted from the former in order to avoid double counting. 
Similarly the input for $F_2(N,Q^2)$ is given by the relatively simple form
\eqn\fullsolfin{\eqalign{F_{2,RSC,0}(N,Q_0^2) &=F_2(N)\cr 
&\hskip -0.8in + \alpha_s(Q_0^2)
{\Gamma^1_{2L}(\alpha_s(Q_0^2)/N)\over\Gamma^0_{LL}(\alpha_s(Q_0^2)/N)}
\biggl(\hat F_L(N)-{(36-8N_f)\over 27}
F_2(N)\biggr)(\exp[\ln(Q_0^2/A_{LL})\Gamma^0_{LL}(\alpha_s(Q_0^2)/N)]-1)\cr
&\hskip -0.8in+\ln(Q_0^2/A_{LL})
\alpha_s(Q_0^2)\biggl(e^+(N)\Gamma^{0,l,+}(N)\hat 
F_L^{0,l,+}(N,Q_0^2) + e^-(N)\Gamma^{0,l,-}(N)\hat F_L^{0,l,-}(N,Q_0^2) \cr
&\hskip 1.2in -\Gamma^{1,0}_{2,L}(N)\biggl(
\hat F_L(N)-\biggl({36-8N_f\over27}\biggr)F_2(N)\biggr)\biggr).}}
The third term is the renormalization--scheme--invariant 
order--$\alpha_s(Q_0^2)$ input, which must compensate for changes in the 
one--loop evolution under changes in $Q_0^2$. Again we explicitly
extract a term $\propto\ln(Q_0^2/A)\Gamma^{1,0}_{2,L}(N)\biggl(
\hat F_L(N)-\biggl({36-8N_f\over27}\biggr)F_2(N)\biggr)$ in order to avoid
double counting. The fact that this term can be thought 
of as appearing from two different sources leads us to choose both our 
unknown scale constants in the input equal to the same value $A_{LL}$. 
We note that our choice of inputs not only ensures $Q_0^2$--invariance 
up to higher orders, but are also such that our expressions for
$\hat F_{L,RSC,0}(N,Q_0^2)$ and $F_{2,RSC,0}(N,Q_0^2)$ reduce to the 
nonperturbative inputs if $Q_0^2=A_{LL}$.
Having obtained our LORSC expressions for $(d\,F_2(N,Q^2)/d\,\ln Q^2)$ and 
$F_2(N,Q_0^2)$, then as already argued, in order to obtain our expression 
for $F_2(N,Q^2)$ we integrate $(d\,F_{2}(N,Q^2)/d\,\ln Q^2)_{RSC,0}$ from 
$Q_0^2$ to $Q^2$ and add to the input $F_{2,RSC,0}(N,Q_0^2)$.

\medskip

Thus, we have our complete leading--order, 
including leading--$\ln (1/x)$ terms, 
renormalization--scheme--consistent expressions for the structure functions. 
These are significantly different from both the one--loop expressions 
and the leading--$\ln(1/x)$ expressions, although they do reduce to them 
in the appropriate limits; i.e. to the one--loop expressions for very large 
$x$, and to the leading--$\ln (1/x)$ expressions in the limit of very small
$x$ and near the boundary of evolution, $Q^2 \approx Q_0^2$. Indeed, once 
we include the ${\cal O}(\alpha_s(Q_0^2))$ inputs for $F_2(N,Q^2)$ in 
the definition of the leading--order inputs in the loop--expansion
(as we should), each of 
the terms in the inputs and evolution terms in \fullsolfl, \fullsolfderiv\ 
and \fullsolfin\ contains a part which appears in both the LO
expression in the loop expansion and the LO expression in the leading
$\ln(1/x)$ expansion. Our full LORSC expressions are obtained rather 
more easily than above by simply letting each of the input and evolution terms 
become the combination of the terms in the two expansion schemes. In a sense 
this result is obvious, but it is necessary to verify this by deriving the
expressions as above.

Let us comment on the form of our final LORSC expressions.
We note that as for the leading--$\ln (1/x)$ expansion, 
the LORSC expansion still leads to 
predictions for all the small--$x$ inputs: predictions of each in terms
of the nonperturbative inputs (which we imagine should be quite flat) and the 
nonperturbative scale $A_{LL}$, and also stronger predictions for the 
relationships between the inputs (although the scale $Q_0^2$ at which they
should be chosen is not determined). 
 
Finally, we notice that each of the terms appearing in our expressions is
manifestly renormalization scheme invariant, and it is clear that 
no terms are subleading in $\alpha_s$ to any other terms, either in the
input or in the evolution. If we had simply solved the renormalization group
equations using the whole of the anomalous dimension matrix \inclmatrix\
then we would have obtained many terms which do not appear in our full 
leading--order expressions \fullsolfl, \fullsolfderiv\ and \fullsolfin. 
These would still be 
renormalization scheme independent (and trivially factorization scheme
independent of course), since our input anomalous dimensions 
are renormalization scheme independent. However, 
all of these extra terms would be of the same form as terms which must be
renormalization scheme dependent in order to absorb the changes of the 
leading--order  expressions under a change of the coupling coming from a
renormalization scheme change, i.e. $\alpha_s(Q^2)
\to \alpha_s(Q^2)+{\cal O}(\alpha^2_s(Q^2))$ (e.g. we saw 
in 4.2 that the subleading--in--$\ln (1/x)$ evolution 
$\Phi^+_1(Q^2,Q_0^2)$ has a manifestly 
renormalization--scheme--independent part depending on $\Gamma^1_{2L}
(\alpha_s(Q^2)/N)$ as well as a renormalization--scheme--dependent part 
depending on $\Gamma^1_{LL}(\alpha_s(Q^2)/N)$). Thus, these terms should 
be dropped, and \fullsolfl, \fullsolfderiv\ and \fullsolfin\ are the 
correct expressions for the structure functions to be used 
with the one--loop coupling constant. Finally,
as already stated, when considering the 
expressions for the structure functions expanded about negative integer
values of $N$ we will simply use the one--loop expressions, due both to lack
of knowledge of the explicit resummed anomalous dimensions which would 
occur, and also because it almost certainly makes practically no difference 
to do this as far as phenomenology is concerned.

Now that we have our desired expressions, \fullsolfl,
\fullsolfderiv\ and \fullsolfin, we can see how they compare to the data. 
The first step towards this is to 
consider the solution in $x$ space. After briefly doing 
this we will consider detailed fits to the data. 

\newsec{$x$--Space Solutions.}

We shall now discuss how we use the expressions \fullsolfl, \fullsolfderiv\ 
and \fullsolfin\ in order to obtain our expressions for 
the $x$--space structure 
functions and ultimately compare with data. The data on 
${\cal F}_2(x,Q^2)$ exist
over a range of $Q^2$ from $\sim 0.2\Gev^2$ to $5000 \Gev^2$, so it is clear
that we will need expressions for the structure functions which cross 
quark thresholds. Thus, before presenting details of the $x$--space solutions
we must consider how we treat these quark thresholds.

In practice we will impose a lower cut on the data of $Q^2=2\Gev^2$, 
except for the HERA data where we choose $Q^2=1.5\Gev^2$ simply in order not
to lose some of the very low $x$ data.  The threshold of heavy quark
production is $W^2 = 4m_H^2$, where
$W$ is the invariant mass of the hadronic system created from the struck
proton (or neutron) and in the limit of zero proton (and/or neutron) mass is
given by $W^2=Q^2(x^{-1}-1)$. $m_H$ is the mass of the heavy quark. 
Thus, we clearly work in the limit where the up, down and strange
quarks are effectively massless. However, we cross the $b$--quark
threshold of $W^2 \approx 20 \Gev^2$, and at the lower end of our range are 
in the region of the $c$--quark threshold of $W^2\approx 2\Gev^2$.
The correct treatment of these heavy quark thresholds is less well established
than the treatment of effectively massless quarks, and is certainly more
complicated (especially when considering leading--$\ln(1/x)$ terms). Hence, we
will delay a more sophisticated treatment to a future article, 
and in this paper use a relatively simple treatment. 

We use the prescription for treating heavy quarks outlined in 
\ref\coltung{J.C. Collins and W.K. Tung, \NP \vyp{B278}{1986}{934}.}. This
involves treating all the quarks as massless, but only allowing the
heavy quarks to become active above the simple threshold $Q^2=m_H^2$. Hence,
the value of $N_f$ appearing explicitly in any expressions
changes discontinuously at this threshold. The running
coupling constant, which itself depends on $N_f$,
is defined to be continuous at the thresholds. It is 
determined by the relationship 
\eqn\couplink{\alpha_{s,n}(Q^2)=\alpha_{s,n+1}(Q^2)\biggl(1+
{\alpha_{s,n+1}(Q^2)\over 6\pi}\ln(m^2_{n+1}/Q^2)\biggr),}
where the central $\alpha_s(Q^2)$ with $N_f=4$ is defined by 
\eqn\couplinkfour{\alpha_{s,4}(Q^2)={12\pi \over (33-2\cdot(N_f=4))
\ln(Q^2/\Lambda^2_{QCD,4})}.}
This leads to a kink in $\alpha_s(Q^2)$ at the thresholds, 
with the coupling used below a 
threshold being larger than the continuation downwards of
the coupling used above the threshold. 

This complete prescription for treatment of
the heavy quarks is consistent with the decoupling theorem, as it is 
guaranteed to provide the correct expressions far above or below any 
threshold, as discussed in \coltung: the increase in the coupling below a 
threshold compensating for the absence of virtual heavy quarks in calculations 
below this threshold. It is clearly going to be rather 
unsatisfactory in the region of
the threshold, with heavy quark structure functions having an abrupt
threshold in $Q^2$ rather than the physically correct smooth threshold 
in $W^2$. Work to rectify
this is in progress, and will certainly involve the use of the heavy 
quark coefficient functions at leading order in $\ln (1/x)$ already 
calculated \kti\cat. We note, however, that the treatment of quark thresholds 
in this paper is no less rigorous than in most of the other calculations 
of structure functions currently performed, and is perhaps better than some. 

The number of active quark flavours, $N_f$, appears in \fullsolfl, 
\fullsolfderiv\ and \fullsolfin, in a number of places other than the 
implicit dependence in $\alpha_s(Q^2)$. It appears in the one--loop physical
anomalous dimension eigenvalues $\tilde \Gamma^{0,l,+(-)}(N)$, in the 
eigenvector factors $e^{+(-)}(N)$, and in the definition of $\hat F_L(N)$; 
i.e. $C^g_{L,1,l}(N) \propto N_f$ and hence, from the definitions \linkf\ and
\linkg\ which define the nonperturbative inputs for the structure functions, 
we have 
\eqn\flonnf{\hat F_L(N) -\hat C^f_{L,1,l}(N) F_2(N) =
\hat F_L(N) -{8 \over 3(N+2)}F_2(N) \propto N_f.}
The first two forms of dependence have very little impact in practice, the 
$N_f$--dependence being relatively weak. The last dependence, the 
proportionality of what is in practice almost all of $\hat F_L(N)$ to $N_f$, 
has a large impact. It means that the expressions for the longitudinal 
structure function and for the $\ln Q^2$ derivative of $F_2(N,Q^2)$ are 
discontinuous at $Q^2=m_c^2$ and at $Q^2=m_b^2$. Also the definition of the 
input $F_2(N,Q_0^2)$ is sensitive to where $Q_0^2$ is chosen with respect 
to the thresholds. 

This dependence of $\hat F_L(N)$ on $N_f$ is obviously rather unsatisfactory.
When treating the heavy quark thresholds more rigorously we should  
subtract out of the full structure functions those contributions where there
are particles with heavy quarks in the final state. Doing this, we could 
define intrinsic light quark structure functions where the final state 
particles do not contain heavy quarks, and which have a fixed number of
active flavours.  Our definition of the nonperturbative inputs 
could come from these light quark structure functions alone. 
There would then also be heavy quark structure functions which 
could be calculated separately from and entirely 
in terms of the light quark structure 
functions. However, using our somewhat simplistic
prescription for turning on the heavy quark contributions,
even if we were to define heavy quark 
structure functions completely separately from the light ones, the 
expressions for $F_{L,H}(N,Q^2)$ and 
$(d\, F_{2,H}(N,Q^2)/d\ln (Q^2))$ would be discontinuous 
at the thresholds. Thus,
in this paper we feel that the treatment of quark thresholds makes 
it more appropriate to define the whole structure function in terms of a 
discontinuous number of active quark flavours. The contribution to this 
complete
structure function from the production of heavy quarks can then be extracted 
straightforwardly. Hence, we treat the problem of quark thresholds as
described above. We will discuss the effect of the discontinuities at 
the thresholds on fits to data in detail in the next section. 
We note, however, that 
the expressions for the whole structure functions in terms of the singlet
and nonsinglet components in \srucdef\ depend on the average value of the 
squared charge of the active quarks, and this will also change 
discontinuously at the thresholds. This means that 
overall the effects of heavy quark thresholds on a best fit to 
${\cal }F_2(x,Q^2)$ data are in practice rather small. 
The effects of the $b$--quark contribute only like the
charge squared, i.e. $1/9$, and hence have little impact. 
The effects due to the $c$--quark threshold are proportionally
much larger, but because they are at $Q^2\approx 2\Gev^2$ they
only affect a very small proportion of the complete set
of data.  

\medskip

Having defined our treatment of quark thresholds we can now discuss the form 
of the $x$--space solutions for the structure functions. As already discussed
it is the perturbative part of the moment--space structure functions for 
which we can produce well--ordered, RSC expressions and 
the nonperturbative parts of the expressions for these structure functions, 
$F_2(N)$ and $\hat F_L(N)$, will be nonanalytic, complicated functions of $N$. 
Our complete moment--space expression for a general structure 
function  (or derivative of a structure function) will be
\eqn\compstrN{ F_i(N,Q^2) = P_{i,2}(N, \alpha_s)F_2(N) + P_{i,L}(N,\alpha_s) 
\hat F_L(N),}
where the $P_{ij}(N,\alpha_s)$'s are the calculable perturbative components  
of the complete expressions which can be expanded as power series in 
both $\alpha_s$ and $N$. Hence, the $P_{ij}(N,\alpha_s)$'s 
are an example of the physically relevant 
perturbative functions we discussed in \S 4.4.
Taking the inverse Mellin transformation of \compstrN\ back to $x$--space
we obtain 
\eqn\compstrx{x{\cal F}_i(x,Q^2) ={\cal P}_{i,2}(x,\alpha_s)\otimes
{\cal F}_2(x)+{\cal P}_{i,L}(x,\alpha_s)\otimes
{\cal F}_L(x),}
where $\otimes$ denotes the convolution of two quantities, i.e. 
\eqn\defconv{{\cal A}(x)\otimes{\cal B}(x) = x\int_{x}^1 {d\, z\over z}
{\cal A}(x/z){\cal B}(z).}
Making an ordered calculation of the structure function is then equivalent to 
making an ordered calculation of the ${\cal P}_{ij}(x,\alpha_s)$'s.  
As proved in \S 4.4, this is equivalent to calculating these 
perturbative parts of the expressions for structure functions
as RSC expansions about each of their nonanalyticities in 
$N$, and then taking the inverse Mellin transformation of each of these 
expansions by integrating around curves enclosing each nonanalyticity. 

Our moment--space expressions \fullsolfl, \fullsolfderiv\ and 
\fullsolfin, are part of the full expressions for the structure 
functions in the form in \compstrN. They contain the complete
nonanalytic expressions for $F_2(N)$ and 
$\hat F_L(N)$, but the factors multiplying these are LORSC expansions of the 
perturbative parts of the expressions about $N=0$. Thus, in a sense the 
expressions \fullsolfl, \fullsolfderiv\ and \fullsolfin\ are misleading.  
They need to be interpreted correctly. We cannot simply take an inverse Mellin
transformation of these expressions because the result would depend on 
all the singularities in the nonperturbative functions, whereas the 
perturbative factors in these expressions are defined as power series 
expansions in $N$ about $N=0$ with  
radii of convergence of only unity. The expressions do not have a direct
meaning for general values of $N$. Instead, we must interpret them as 
providing us with the perturbative factors multiplying $F_2(N)$ and 
$\hat F_L(N)$ which are LORSC when expanded about $N=0$. 

Hence, we use \fullsolfl, 
\fullsolfderiv\ and \fullsolfin\ to obtain the $x$--space structure 
functions as follows. We can factor out 
the perturbative part of each of these expressions which is proportional to 
either $\hat F_L(N)$ or $F_2(N)$, and take the inverse Mellin transformation
of this perturbative function
by integrating around a contour encircling $N=0$. 
This provides us with the leading--in--$\ln(1/x)$, at 
lowest--order--in--$x$ parts of the 
perturbative ${\cal P}(x,\alpha_s)$'s as well as all the 
lowest--order--in--$\alpha_s$ parts at lowest--order--in--$x$.
In order to obtain the leading--order--in--$\ln(1/x)$--and--$\alpha_s$,
at lowest--order--in--$x$, part of the perturbatively calculated 
structure functions we then convolute with the whole of the nonperturbative
$\hat F_L(x)$ and $F_2(x)$ which are obtained by a complete inverse Mellin
transformation of $\hat F_L(N)$ and $F_2(N)$, including contributions 
from singularities at many values of $N$.  
In principle, the full LORSC $x$--space structure functions are calculated by
repeating this procedure for the LORSC moment--space structure functions 
where the perturbative parts are expanded about $N$ equal to all 
negative integers. This would systematically include all 
leading--order--in--$\ln(1/x)$--and--$\alpha_s$ 
terms at progressively higher orders in $x$. 
As already explained, in practice one only really needs
include the one--loop, or leading--in--$\alpha_s$ part of the full 
LORSC perturbative parts for all 
negative integers, and also only need to work to a relatively small negative
integer unless one is concerned with values of $x$ very close to one. Thus, 
it is this type of calculation we aim to make in order to compare our 
expressions with data. 

\medskip

In practice the calculation of the structure functions is performed in a 
rather different manner. The structure functions are calculated using a 
modification of the computer program that is used by Martin, 
Roberts and Stirling in 
their global fits to structure function data. This works in terms of parton
densities and also directly in $x$--space. Input parton densities are 
specified at some scale $\tilde Q_0^2$ and the $Q^2$ evolution is calculated 
on a grid in $x$ and $Q^2$. This $Q^2$ evolution is obtained by
integrating up the complete renormalization group equations involving the 
complete parton distributions and full specified splitting functions. 
The structure functions are calculated by numerically 
performing the convolutions of the resulting $Q^2$--dependent 
parton distributions with coefficient functions. 
Thus, in fact we obtain the factorization--scheme--independent LORSC 
structure functions by working in terms of parton densities obtained from 
the full solution of the renormalization group equations
and choosing coefficient functions and splitting functions (i.e. an effective
factorization scheme) which will reproduce the correct expressions
as closely as possible. 

In order to do this we choose to work in a DIS--type scheme for simplicity, 
and take the longitudinal coefficient functions to be just the one--loop 
expressions. By breaking the expressions for the LORSC structure 
functions in terms of the structure functions themselves into
expressions involving partons with the above choice of coefficient 
functions, and also using the power series expansions of the one--loop 
anomalous dimensions and eigenvalues in terms of $N$ it is possible to choose  
effective 
splitting functions which when used in the computer program will reproduce the 
correct expressions for $(d\,{\cal F}_2(x,Q^2)/d\ln(Q^2))$, 
${\cal F}_2(x,Q_0^2)$ and ${\cal F}_L(x,Q^2)$ to very good 
accuracy. We note that this choice
has to be different when calculating ${\cal F}_2(x,Q^2)$ from when calculating 
${\cal F}_L(x,Q^2)$.
Due to the method of calculation used by the MRS program, it is extremely 
difficult to obtain the exact behaviour one desires of the 
structure functions, i.e. a given evolution away from a 
particular input at some fixed $Q_0^2$, with any choice of 
splitting functions. However, we aim to get as close as possible to this exact
behaviour with relatively simple choices.

We will briefly describe our choice of effective splitting functions
as follows. At first
order in $\alpha_s$ we choose the normal parton splitting functions. We then 
add corrections to these in order to reproduce our desired results. 
Denoting these corrections by $\Delta \gamma_{ab}(N,\alpha_s(Q^2))$,
for a given parton to parton splitting function, and expressing these 
in moment space and in terms of previously discussed quantities 
for simplicity we get for ${\cal F}_L(x,Q^2)$ 
\eqn\anomcorrl{\eqalign{ \Delta \gamma_{fg}(N,\alpha_s(Q^2))&=0\cr
\Delta \gamma_{ff}(N,\alpha_s(Q^2))&=0\cr
\Delta \gamma_{gg}(N,\alpha_s(Q^2)) &= 
\Gamma^{\tilde 0}_{LL}(\alpha_s(Q^2)/N)\cr
\Delta \gamma_{gf}(N,\alpha_s(Q^2)) &= 
\fourninths \Gamma^{\tilde 0}_{LL}(\alpha_s(Q^2)/N).\cr}}
This form is not too difficult to understand by 
looking at  e.g. \fullsolfl: the only leading--$\ln (1/x)$
enhancement of the one-loop expression comes from the evolution in the 
longitudinal sector which is directly related to the 
enhanced gluon evolution and hence corrections to the gluon anomalous 
dimensions.
For ${\cal F}_2(x,Q^2)$ the corrections are a little more involved: 
\eqn\anomcorrtwo{\eqalign{ \Delta \gamma_{fg}(N,\alpha_s(Q^2))&=
\alpha_s(Q^2)\hat C^g_{2,1,l}(N) \Gamma^{\tilde 1}_{2L}(\alpha_s(Q_0^2)/N)\cr
\Delta \gamma_{ff}(N,\alpha_s(Q^2))&=\alpha_s(Q^2)\Biggl[\biggl(
{8N_f\over 27}-{4\over 3}\biggr)+{2\over N_f}\hat C^g_{2,1,l}(N)\Biggr]
\Gamma^{\tilde 1}_{2L}(\alpha_s(Q_0^2)/N)\cr
\Delta \gamma_{gg}(N,\alpha_s(Q^2)) &= 
\Gamma^{\tilde 0}_{LL}(\alpha_s(Q^2)/N)-\fourninths \Delta 
\gamma_{fg}(N,\alpha_s(Q^2))\cr
\Delta \gamma_{gf}(N,\alpha_s(Q^2)) &= 
\fourninths \Gamma^{\tilde 0}_{LL}(\alpha_s(Q^2)/N)-\fourninths
\Delta \gamma_{ff}(N,\alpha_s(Q^2)).\cr}}
The corrections to the quark anomalous dimensions are functions of $Q_0^2$ 
rather than $Q^2$, except for the single power of $\alpha_s(Q^2)$, because 
$\Gamma^1_{2L}(\alpha_s/N)$ appears only in the input terms in 
\fullsolfderiv. In the small--$x$, or small--$N$ limit we have the 
simplifications that $\Delta \gamma_{fg}(N,\alpha_s(Q^2))\to
\alpha_s(Q^2)(2N_f/3)\Gamma^{\tilde 1}_{2L}(\alpha_s(Q_0^2)/N)$ and 
$\Delta \gamma_{ff}(N,\alpha_s(Q^2))\to \fourninths \Delta 
\gamma_{fg}(N,\alpha_s(Q^2))$.
Thus, in this limit these corrections to the one--loop quark 
anomalous dimensions take on the standard form of 
$\alpha_s(Q^2)\gamma^1_{fg}(\alpha_s(Q_0^2)/N)$ and 
$\alpha_s(Q^2)\gamma^1_{ff}(\alpha_s(Q_0^2)/N)$ 
minus their one--loop components. The prefactors multiplying these terms, 
which depend on the one--loop longitudinal coefficient functions, 
ensure that the leading--$\ln(1/x)$ enhancement of the rate of growth of 
${\cal F}_2(x,Q^2)$ is directly coupled to the longitudinal structure, 
not to the gluon. This delays the 
enhancement to slightly smaller values of $x$, and reduces it a little. The 
contributions to the gluon anomalous dimensions beyond LO in $\ln (1/x)$ are
present to counter the exponentiation of the corrections to the 
quark anomalous dimensions, i.e. this is similar to choosing 
$\gamma^1_{gg}(\alpha_s/N)$  equal to $-\fourninths \gamma^1_{fg}
(\alpha_s/N)$ in \fullfl--\fullftwoderiv\ 
in order to ensure that the exponentiation of
$\gamma^1_{fg}(\alpha_s/N)$ has no effect in these expressions. The effect of
these terms is in practice rather small.
  
We may now perform checks to see if the particular
choice above does indeed lead to an accurate representation of the correct 
expressions. In order to do this  we obtain  
solutions for the structure functions which are the same as those 
obtained by the MRS program for our given choice of splitting functions,
by analytically performing the same calculation as performed by this program,
but in moment space, where the calculation is possible. 
The resulting expressions can 
be compared with the correct LORSC moment--space expressions in the region of 
relatively small $x$ ($x\lsim 0.1$) by expanding both 
expressions in powers of $N$, and 
an accurate approximation of the differences between the two found.\foot{The 
ease of obtaining an accurate expression at small $x$ is aided 
immensely by the fact that the expansions
about $N=0$ of the one--loop parton anomalous dimensions, and subsequently 
their eigenvalues, eigenvectors and projection operators   
have coefficients which stay roughly constant as the power of $N$
increases, or 
even where the lowest powers have the largest coefficients. For example,
the expansion of $\tilde \gamma^{0,l,+}(N)$ about $N=0$ goes like 
$(6/N -5.64+0.24N-2.37N^2+3.38N^3+ \cdots)$. This enables one to truncate 
these series after a very small number of terms when working at small $x$, 
or conversely work to 
quite large $x$ without keeping too many terms in the series. 
It is also the reason why
the ``double asymptotic scaling'' formula, which effectively truncates the 
above series after the second term, is such a good approximation to the
full one--loop solution (with flat parton inputs) up to quite high values of 
$x$. In the $\overline{\hbox{\rm MS}}$ renormalization and factorization
scheme, or the DIS factorization scheme, the two--loop anomalous dimensions, 
and in the former case coefficient functions, do not exhibit nearly such 
a nice behaviour. e.g. the $\overline{\hbox{\rm MS}}$ scheme
$\tilde \gamma^{1,l}_{qg}(N)$ expanded about $N=0$ 
goes like $(26.7/N-57.7+104N-176N^2+\cdots)$, and the expansion of the 
coefficient function is even worse, $C^g_{2,1,l}(N) \propto (1.33-6.16N+9.2N^2
-12N^3+\cdots)$. Hence, truncations of the two--loop anomalous dimensions and
coefficient functions expanded as series about $N=0$ give a far worse 
approximation to the effects of the complete anomalous dimensions and 
coefficient functions at small $x$ than in the one--loop case.
This has been explicitly checked, and indeed, evolution and convolution of
a given set of parton distributions using the series expansions for the
two--loop 
anomalous dimensions and coefficient functions truncated at lowest order in 
$N$ lead to structure functions up to a factor of two larger than those 
obtained from calculations performed using the full anomalous dimensions and 
coefficient functions.} 
These differences between the expressions
can then be estimated by transforming back to $x$--space. With the  
choice of the splitting functions above  the differences 
between the expressions are very small. 

One may also take a more direct approach.
For a simple, but physically reasonable, 
choice of input parton densities, and hence input structure functions,
one can use the types of techniques 
outlined in \frt\ to begin with the exact expressions \fullsolfl, 
\fullsolfderiv\ and \fullsolfin\ in moment space and then transform to 
$x$--space obtaining analytic solutions for the structure functions. These 
expressions are very complicated, but are almost exact for very low values
of $x$. They begin to become approximate for larger values of 
$x$, particularly at high $Q^2$, due to the effects of 
truncating power series in $N$. The range of $x$ for which the expressions 
are reliable depends on the severity of the truncation but it is not too 
difficult to have accurate expressions for $x \lsim 0.1$. These enable one 
to compare the results obtained from the inverse Mellin
transformation of our correct moment--space expressions with the 
calculations performed using the computer program in $x$-space. For 
$x \lsim 0.1$ our choice of the anomalous dimensions leads to
agreement with the analytic expressions to an accuracy much better than the
errors on the data in any appropriate range of parameter space. There is no 
reason to believe that input distributions which are very similar in form, 
but which need to be described by more complicated expressions, should lead to
deviations much different from our particular choice. Hence, using this 
direct test and the more general, if slightly more approximate check above,
we are confident that
our choices do lead to the correct $x$--space expressions for $x\lsim 0.1$ up 
to very small corrections.

At higher $x$ the check is simpler, in fact almost trivial. 
Simply setting to
zero all terms in the MRS program other than those coming from one loop, we 
observe that the expressions for the structure functions for $x > 0.1$,
particularly at high $Q^2$, are 
very close to the one--loop expressions alone (becoming essentially 
identical for $x>0.3$). This result agrees with calculations taking the 
inverse Mellin transformation of the LORSC moment--space expressions. 
Hence, the direct $x$--space calculation must be 
the same as our desired expression in this relatively high $x$ range. 
Therefore, the correct choice of effective
splitting functions leads to 
the MRS program producing a very accurate approximation to our correct 
LORSC expressions for the structure functions over the complete range of 
parameter space.\foot{In fact the agreement between the result obtained 
from the computer program and the ``correct'' result is certainly
better than the agreement between NLO calculations performed using different
programs and different prescriptions for truncating the solution at NLO 
\comparison.} 

\medskip

We note that with our choice of splitting functions and 
definitions of parton densities then momentum is not conserved by the 
evolution: in the best fit to ${\cal F}_2(x,Q^2)$ discussed in the next 
section the total momentum carried by the partons at $Q^2 =2\Gev^2$
is $87\%$ and at $Q^2=5000\Gev^2$ it is about $94\%$. Hence, the amount of 
momentum violation is at the level of a few percent. 
We have already defended this 
violation of momentum conservation in subsection 4.3. We now also point 
out that starting with our definition of the partons we could define 
new parton densities and splitting functions
by defining non--zero $C^{f(g)}_{2,1}(\alpha_s/N)$
and $C^{f(g)}_{L,1}(\alpha_s/N)$ beyond one--loop, and use the 
transformation rules in \S 2 to keep the structure functions unchanged.
If these coefficient functions had negative coefficients then, compared to 
our prescription above, the low $Q^2$ parton distributions would need to 
be larger, and would hence carry more momentum. As $Q^2$ increased the 
effect of these new coefficient functions would decrease, and the extra
amount of momentum carried by the new parton distributions compared to the 
original ones would decrease. Thus, the effect of such a redefinition of 
parton distributions would be to increase the amount of momentum carried by 
the partons at low $Q^2$ and also to 
slow the growth of momentum with $Q^2$ (or even 
to turn it into a fall). With a judicious redefinition of coefficient
functions of this sort (with perhaps some dependence on $\alpha_s(Q_0^2)$ 
as well as $\alpha_s(Q^2)$) it should clearly be possible to find an 
effective factorization scheme where the structure functions are 
identical to those given by 
our prescription, but where the momentum violation will be extremely small. 
As already mentioned, one may think of this as a ``physical scheme''. We 
have not seriously investigated this redefinition of parton densities in any
quantitative manner since it will not affect any physical quantities. 
  
Now that it has been determined that our choice of splitting functions is 
correct, one can vary the input parameters in the standard way to
obtain the best fit. In practice the starting scale for the 
numerical evolution is 
chosen to be $\tilde Q_0^2=m_c^2$ so that the charm contribution to the 
structure functions is guaranteed to turn on at this scale. Evolving 
upwards, the bottom quark contribution turns on at $Q^2=m_b^2$.
Thus, the parton distributions and, via convolutions with coefficient 
functions, the structure functions are input at $\tilde Q_0^2$.      
However, this does not mean that
the true starting scale, $Q_0^2$, has to be the same as 
$\tilde Q_0^2$. This true starting scale is the scale at which some of the 
couplings in the 
anomalous dimensions in our full LORSC expressions have to be fixed. When
performing the numerical integration upwards form a lower  
scale we simply interpret the values of ${\cal F}_L(x,Q^2)$, 
$(d\, {\cal F}_2(x,Q^2)/d\ln Q^2)$ and ${\cal F}_2(x,Q^2)$ at the fixed
scale $Q_0^2$ as our input values for these structure functions. We then 
demand that at $Q_0^2$ the relationship between these quantities is the same 
(up to an accuracy much better than the errors on the data)
as that demanded by the expressions \fullsolfl, \fullsolfderiv\ and 
\fullsolfin. We
also demand that the value of each of these quantities at $Q_0^2$ is 
compatible with them being written as the convolution of our completely 
determined, up to the scale $A_{LL}$, perturbative parts and our 
nonperturbative soft inputs. In practice we demand that these nonperturbative 
inputs must be flat at small $x$, i.e. they must be described well by
a function of the form 
\eqn\softinput{{\cal F}_i(x) =F_i(1-x)^{\eta_i}(1+\epsilon_i x^{0.5}+
\gamma_i x).} 
It is by satisfying these requirements on the inputs, while 
simultaneously choosing $Q_0^2$ as
the fixed scale appearing in the anomalous dimension or splitting function
dependent input terms in our expressions, which enables us to identify $Q_0^2$
as the true starting scale. The evolution which takes place upwards from 
$\tilde Q_0^2(=m_c^2)$ to $Q_0^2$ would be identical if it were to take place 
backwards from $Q_0^2$ to $\tilde Q_0^2$. The former manner of 
performing the evolution is simply more convenient in practice.\foot{As 
already mentioned, a more sophisticated treatment of the heavy quarks
would involve calculating heavy quark structure functions separately from 
the light quark structure functions. More sophisticated treatments within the 
usual loop expansion, e.g \ref\grs{M. Gl\"uck, E. Reya and M. Stratmann, \NP
\vyp{B422}{1994}{37}.}, have the thresholds for heavy quark production 
built into the coefficient functions, and it is not important to begin 
numerical evolution at, or below any particular value of $Q^2$ in this case.}

We should perhaps comment more on this starting scale $Q_0^2$, and on its
relationship to $A_{LL}$. $A_{LL}$ has been defined as the particular choice 
of $Q_0^2$ for which the inputs for the structure functions reduce to 
the purely nonperturbative inputs. Hence, we imagine it is a scale
typical of nonperturbative physics, i.e. $A_{LL} \lsim 1\Gev^2$. If 
we choose $Q_0^2$ different from (in practice higher than) $A_{LL}$, 
then $A_{LL}$ 
is not precisely the scale at which the full expressions for the structure 
functions reduce to the nonperturbative values, although it will remain
fairly close to this. Hence, $A_{LL}$ is simply some phenomenological 
parameter which we will fine tune in order to get the best fit, though it 
does maintain its interpretation as a scale typical of the transition 
between perturbative and nonperturbative physics, and it would be 
surprizing if it were much greater than say $1\Gev^2$.
As already mentioned, the choice of $Q_0^2$ is undetermined. However, it 
should be such that it does make our full expressions relatively 
insensitive to its value. (We also reduce this sensitivity by letting 
$A_{LL}$ be a free parameter for each choice of $Q_0^2$, though it should 
not vary a great deal.) This insensitivity to $Q_0^2$ imposes two 
restrictions on its value. Firstly, $Q_0^2$
should not be too small, otherwise the value of $\alpha_s(Q_0^2)$ will be 
very sensitive to our choice of $Q_0^2$, making it difficult to make
our expressions $Q_0^2$ insensitive. However, we cannot simply choose 
very high $Q_0^2$. This is because terms beyond LO in the full RSC 
expressions will contain terms of higher order in $\ln(Q_0^2/A_{LL})$, albeit
along with higher orders in $\alpha_s(Q_0^2)$. Hence, we do not wish to 
choose $Q_0^2$ to be a great deal larger than the scale $A_{LL}$. Thus, 
$Q_0^2$ is probably some intermediate value. It will, of course be 
determined in practice by the quality of the fits to data using the 
expressions \fullsolfl, \fullsolfin\ and \fullsolfderiv\ evaluated at 
particular values of $Q_0^2$.

\bigskip

Now that we know precisely how we will perform our calculations of 
structure functions we can make some comment on the general form 
the structure functions have to take. 
The expressions for ${\cal F}_L(x,Q^2)$ and ${\cal F}_2(x,Q^2)$
depend on $\Gamma^0_{LL}(\alpha_s/N)$ and $\Gamma^1_{2L}(\alpha_s/N)$. 
Both of these series have coefficients which behave like 
$n^{-3/2}(12\ln 2)/\pi$ for very large $n$. This leads to a cut in 
the $N$--plane
at $N=\bar \alpha_s 4\ln 2$ in both cases, and to the structure functions 
having asymptotic behaviour ${\cal F}(x,Q^2) \sim (\ln x)^{-3/2}
x^{-4\ln 2 \bar \alpha_s}$ as $x \to 0$, where $\alpha_s=\alpha_s(Q_0^2)$ 
and $Q^2\geq Q_0^2$.
   
However, this is only for the strict asymptotic limit $x\to 0$. It was 
convincingly demonstrated in \bfresum\ (and discussed from a different 
point of 
view in \frt) that one only need keep a finite number of terms in the 
leading--$\ln (1/x)$ series if working at finite $\ln (1/x)$. In practice,
if one works down to $x\approx 10^{-5}$ and the nonperturbative inputs behave 
roughly like $(1-x)^5$, then keeping terms up to $10_{\rm th}$ order in 
the series for 
$\Gamma^0_{LL}(\alpha_s/N)$ and $\Gamma^1_{2L}(\alpha_s/N)$ is more than 
sufficient (truncating at $8_{\rm th}$ order leads to only tiny errors). 
The series up to this order have the explicit form
\eqn\serieslong{\eqalign{\Gamma^0_{LL}(\alpha_s/N)&=
\biggl({\bar\alpha_s\over N}\biggr)
+2.40\biggl({\bar\alpha_s\over N}\biggr)^4+
2.07\biggl({\bar\alpha_s\over N}\biggr)^6+
17.3\biggl({\bar\alpha_s\over N}\biggr)^7
+2.01\biggl({\bar\alpha_s\over N}\biggr)^8\cr
&\hskip 2.7in +39.8\biggl({\bar\alpha_s\over N}\biggr)^9+
168.5\biggl({\bar\alpha_s\over N}\biggr)^{10} +\cdots,\cr}}
and
\eqn\seriestwo{\eqalign{2\pi\Gamma^1_{2L}(\alpha_s/N)&=1+
2.5\biggl({\bar\alpha_s\over N}\biggr)+
\biggl({\bar\alpha_s\over N}\biggr)^2+
\biggl({\bar\alpha_s\over N}\biggr)^3+
7.01\biggl({\bar\alpha_s\over N}\biggr)^4+
5.81\biggl({\bar\alpha_s\over N}\biggr)^5
+13.4\biggl({\bar\alpha_s\over N}\biggr)^6\cr
&\hskip 1in +58.1\biggl({\bar\alpha_s\over N}\biggr)^7
+64.7\biggl({\bar\alpha_s\over N}\biggr)^8+
196.8\biggl({\bar\alpha_s\over N}\biggr)^9
+650\biggl({\bar\alpha_s\over N}\biggr)^{10}+\cdots.\cr}}
It is clear that the low--order terms in both these series have coefficients 
which generally grow far less quickly than the asymptotic relationship $a_{n+1}
=4\ln 2 a_n$; some fall, and in the case of $\Gamma^0_{LL}(\alpha_s/N)$
are even zero. Thus, in the range of parameter space we are considering we
will have rather less steep behaviour than the asymptotic limit of 
${\cal F}(x,Q^2) \sim (\ln x)^{-3/2}x^{-4\ln 2 \bar \alpha_s}$. Using the 
techniques in \frt\ to derive analytic expressions for the structure 
functions in the small--$x$ limit, it becomes clear that the extreme 
smallness of the first few coefficients in the series for 
$\Gamma^0_{LL}(\alpha_s/N)$ leads to the small--$x$ form of ${\cal F}_L(x,Q^2)$
being driven largely by the first term, with the $4_{\rm th}$--order term 
having a reasonable effect at the smallest $x$ values. Similarly, the 
small--$x$ form of ${\cal F}_2(x,Q^2)$ is determined largely by the series 
$\Gamma^1_{2L}(\alpha_s/N)$ and the first term in 
$\Gamma^0_{LL}(\alpha_s/N)$, again with the $4_{\rm th}$--order term in 
$\Gamma^0_{LL}(\alpha_s/N)$ playing some role at the smallest $x$ values. 

One can be a little more quantitative. Choosing $Q_0^2= 25\Gev^2$ and 
making a guess that $A_{LL}=0.8\Gev^2$ we obtain $\ln(Q_0^2/A_{LL})\approx
3.4$. We also choose nonperturbative inputs which behave like
$F_i(1-x)^5$ for $i=L,2$. As argued in \frt, for values of $x\lsim 0.01$ 
it is a good approximation to replace the $(1-x)^5$ behaviour with the 
behaviour $\Theta(0.1-x)$.\foot{The argument of the $\Theta$--function 
depends on the power of $(1-x)$. Higher powers would require the step to 
occur at lower $x$ and {\it vice versa}.} Doing this we can derive simple 
expressions for the form of the inputs for $x\lsim 0.01$ which will be 
accurate up to a few percent error. Using the series \serieslong\ and 
\seriestwo\ and the expressions \fullsolfl\ and \fullsolfin\ we can take the 
inverse Mellin transformations (using the techniques of \frt), obtaining
\eqn\anflinput{\eqalign{{\cal F}_L(x,Q_0^2) &\approx 
{\alpha_s(Q_0^2)/2\pi}\biggl[\biggl(F_L+{4\over 27}F_2\biggr)
\biggl(1+3.4\bar\alpha_s(Q_0^2)\xi +
5.8{(\bar\alpha_s(Q_0^2)\xi)^2\over 2!}+6.6{(\bar\alpha_s(Q_0^2)\xi)^3
\over 3!}\cr&+13.7{(\bar\alpha_s(Q_0^2)\xi)^4\over 4!}+31.5{
(\bar\alpha_s(Q_0^2)\xi)^5\over 5!}
+56{(\bar\alpha_s(Q_0^2)\xi)^6\over 6!}+136{(\bar\alpha_s(Q_0^2)\xi)^7
\over 7!}\biggr)-{4\over 27}F_2\biggr],\cr}}
and 
\eqn\anftwoinput{\eqalign{{\cal F}_2(x,Q_0^2) &\approx 3.4
{\alpha_s(Q_0^2)/2\pi}\biggl(F_L+{4\over 27}F_2\biggr)\biggl(1+
4.2\bar\alpha_s(Q_0^2)\xi +
7.1{(\bar\alpha_s(Q_0^2)\xi)^2\over 2!}+9.1{(\bar\alpha_s(Q_0^2)\xi)^3\over 
3!}\cr&+19.7{(\bar\alpha_s(Q_0^2)\xi)^4\over 4!}+44.1{(\bar\alpha_s(Q_0^2)
\xi)^5\over 5!}+84{(\bar\alpha_s(Q_0^2)\xi)^6\over 6!}+206{
(\bar\alpha_s(Q_0^2)\xi)^7\over 7!}\biggr)+F_2,}}
where $\xi=\ln(0.1/x)$, and $\alpha_s(Q_0^2)\approx 0.2$ if $\Lambda_{QCD,4}
\approx 100\Mev$. Putting in values of $F_L=2.5$ and $F_2=1$, choices which, as
with the $(1-x)^5$ behaviour, are roughly compatible with the high $x$ data,
we have a rough estimate of the form of the input structure functions at 
$Q_0^2=25\Gev^2$. We see that the coefficients in the series in 
$\alpha_s(Q_0^2)\xi$ multiplying $(F_L+4/27 F_2)$ grow a little more quickly 
for ${\cal F}_2(x,Q_0^2)$ than for ${\cal F}_L(x,Q_0^2)$. However, 
in the former case this contribution which rises with falling $x$ is 
accompanied by the flat $F_2$, whereas in the latter it is accompanied only
by the nearly insignificant $-4\alpha_s(Q_0^2)/(54\pi)F_2$. Thus, we find   
that for $x\approx 10^{-4}$ ${\cal F}_L(x,25)$
behaves approximately like $0.1x^{-0.3}$ but that 
${\cal F}_2(x,25)$ is slightly flatter,
behaving approximately like $0.6 x^{-0.28}$. 
These powers of $x$ are clearly somewhat 
less steep than the asymptotic $x^{-0.5}$.
Comparison of the estimate of ${\cal F}_2(x,25)$ with the data in \hone\ 
and \zeus\ shows a very reasonable qualitative agreement. Of course, 
there is no data for ${\cal F}_L(x,25)$
for values of $x$ anything like this low. Being rather more general, we find
that for any $Q_0^2$ between $10\Gev^2$ and $100\Gev^2$, and with any 
sensible choice of $A_{LL}$ (e.g $0.2\Gev^2
\geq A_{LL} \geq 2\Gev^2$), then both $F_2(x,Q_0^2)$ and $F_L(x,Q_0^2)$
behave roughly like $x^{-0.3}$ for $0.01 \geq x \geq 0.00001$. Hence, this type
of behaviour can be taken as a prediction of the theory. 

One can also use the same 
techniques to make an estimate of the structure functions at values of $Q^2$
away from $25\Gev^2$. The expressions involved are relatively 
straightforward to derive but become rather less concise than \anflinput\ and
\anftwoinput, and we will not write them explicitly here. With the particular 
inputs above they do lead to good qualitative agreement with the data 
below $x=0.01$. 
Therefore we have every reason to feel encouraged by our results
and believe that we really are proceeding correctly. However, the 
real test of our approach will be a complete global fit to the available data
for ${\cal F}_2(x,Q^2)$ using the rather more accurate calculations,
especially at large $x$, 
of the MRS program. We will therefore discuss these detailed fits next. 

\newsec{Fits to the Data and Predictions.}

As explained in detail in the previous section we indirectly
use the expressions \fullsolfl--\fullsolfin\ to calculate the 
lowest--order--in--$x$, RSC part of the perturbative part of the solution 
for the  $x$--space
singlet structure functions, and convolute this with the nonperturbative parts.
The--higher--order--in--$x$ parts of the perturbative solution for the singlet 
structure function, along with the nonsinglet structure functions are 
calculated using the normal one--loop prescription. Phenomenologically this 
procedure is practically identical to using the full LORSC solution 
for the structure 
functions. Let us now explain how we use the procedure to obtain 
fits to structure function data.

Once we have the general LORSC expression then by combining the
singlet and nonsinglet components and varying all the free parameters 
($A_{LL}$, the soft inputs for ${\cal F}^S_L(x,Q^2)$ and 
${\cal F}^S_2(x,Q^2)$ and the 
soft nonsinglet inputs), we obtain the  
best fit for the available ${\cal F}_2$ structure function data using a 
particular 
starting scale $Q_0^2$. We note that at this starting scale 
the input ${\cal F}^S_2(x,Q_0^2)$ and the evolution 
$(d\,{\cal F}_2(x,Q^2)/d\ln Q^2)_{Q_0^2}$
are forced by \fullsolfderiv\ and \fullsolfin\ to be trivially related
at small $x$. This is not the case when working at fixed order in $\alpha_s$,
and indeed, is a new feature of the approach to the small--$x$ structure 
functions in this paper, and in particular of the manner of determining 
the inputs that is used.
Choosing the renormalization scale to be $Q^2$,
the one--loop value for $\Lambda_{N_f=4}$ is fixed at $100 \hbox{\rm MeV}$,
thus giving $\alpha_s(M_Z^2)=0.115$.\foot{Of course, since we are using a 
genuinely leading--order expression, any change in renormalization scale 
$\mu_R^2
\to kQ^2$ is exactly countered by a change in $\Lambda_{N_f=4}$ of the
form $\Lambda_{N_f=4}\to k^{-1}\Lambda_{N_f=4}$. However, it is encouraging
that making the simple choice $\mu_R^2=Q^2$ leads to a value of 
$\alpha_s(M_Z^2)$ which is nicely compatible with the usual value.}
This precise value is not determined by a best fit, but a value near
to this is certainly favoured. 

There are some further details we should mention. Firstly,
when obtaining a fit using the above approach, the values of 
${\cal F}_2(x,Q^2)$ used are 
not precisely those published in \hone\ and \zeus. This is because it 
is not ${\cal F}_2(x,Q^2)$ that is measured directly at HERA, but 
the differential cross--section. This is related to the structure functions as 
follows
\eqn\crosstosf{{d^2\sigma\over d x d Q^2}={2\pi\alpha^2\over Q^4 x}
\biggl(2-2y+{y^2\over 1+R}\biggr){\cal F}_2(x,Q^2),}
where $y=Q^2/xs\approx Q^2/90000x$ for the HERA experiment
and $R={\cal F}_L(x,Q^2)/({\cal F}_2(x,Q^2)-{\cal F}_L(x,Q^2))$.
Over most of the range of parameter space $y$ is very small and 
${\cal F}_L(x,Q^2)$ is likely to be small. Hence, the measurement is 
essentially directly that of ${\cal F}_2(x,Q^2)$. However, at the lowest 
$x$ values, especially as $Q^2$ increases, the value of ${\cal F}_2(x,Q^2)$ 
must be extracted using some prescription for the value of 
${\cal F}_L(x,Q^2)$. Both the H1 and ZEUS collaborations, roughly
speaking, obtain their values of ${\cal F}_L(x,Q^2)$ from predictions coming 
from NLO calculations of ${\cal F}_2(x,Q^2)$; H1 use the GRV structure 
function parameterization while ZEUS perform their own fit to the data
on ${\cal F}_2(x,Q^2)$.\foot{We note that both 
NLO calculations of ${\cal F}_2(x,Q^2)$ use a smooth treatment of the 
charm threshold 
while the predictions for ${\cal F}_L(x,Q^2)$ use a LO formula which uses  
NLO parton distributions and assumes 4 massless quarks in the 
expression for the longitudinal coefficient functions. 
Thus, there seems to be an 
internal inconsistency in this method of determining ${\cal F}_2(x,Q^2)$.
Fortunately the errors due to using the LO formula in general cancel with 
those from using 4 massless quarks. The net effect
is an overestimate of at most about $2-3 \%$ on the extracted
value of ${\cal F}_2(x,Q^2)$ at the largest values of $y$. For most of the 
parameter space the effect is negligible.} 
Using these values of ${\cal F}_L(x,Q^2)$ can 
lead to an increase in ${ \cal F}_2(x,Q^2)$ of about $12\%$ 
for the highest values of $y$ compared to the values of 
${\cal F}_2(x,Q^2)$ obtained using the assumption that ${\cal F}_L(x,Q^2)=0$. 
Since the approach in this paper leads in practice to a somewhat lower 
prediction of ${\cal F}_L(x,Q^2)$ than the standard NLO--in--$\alpha_s$ 
approach, the values of ${\cal F}_2(x,Q^2)$ used in the fit must be altered 
to take account of the fact that our predictions for ${\cal F}_L(x,Q^2)$ 
are not the same as those used by H1 and ZEUS.
Thus, the ${\cal F}_2(x,Q^2)$ values are a little (at most about $6\%$) 
lower for the largest values of $Q^2/x$ than in those in \hone\ and \zeus.
In practice, the best fit is first obtained using the published values of 
${\cal F}_2(x,Q^2)$, a prediction obtained for ${\cal F}_L(x,Q^2)$, 
the values of ${\cal F}_2(x,Q^2)$ altered accordingly and the best fit 
obtained once again. The values of ${\cal F}_2(x,Q^2)$ are not altered 
a second time using the corrected 
prediction for ${\cal F}_L(x,Q^2)$ since this changes by only a small amount, 
leading to further changes in ${\cal F}_2(x,Q^2)$ of much less than $1\%$.  

We should also comment on our choice of $m_c^2$ for the best fit. This is 
chosen to be equal to $4\Gev^2$ in order to obtain a reasonable description 
of the
available data on the charm structure function coming from EMC 
\ref\emccharm{EMC collaboration: J.J. Aubert {\it et al}, \NP 
\vyp{B213}{1983}{31}.} and also from preliminary measurements at HERA 
\ref\heracharm{H1 collaboration, preprint DESY 96--138 (1996), to be 
published in \ZP C.}. The quality of the fit is shown in 
\fig\loxcharm{The description of the EMC and preliminary H1 data for 
${\cal F}^c_2(x,Q^2)$ using the LO(x) fit.}. As one can see 
the fit is of a fair quality, with the predicted ${\cal F}^c_2(x,Q^2)$ 
perhaps being a little large in general at large $x$, and a little small 
at small $x$. This result is qualitatively
consistent with the fact that we have used a 
threshold at $Q^2=m_c^2$, rather than at $W^2=Q^2(x^{-1}-1)=4m_c^2$, since 
in the latter case at
very low $x$ the threshold will effectively occur at lower values of $Q^2$, 
and hence ${\cal F}^c_2(x,Q^2)$ should increase relatively, while at large 
$x$ the threshold will be at higher values of $Q^2$, and 
${\cal F}^c_2(x,Q^2)$ will decrease relatively. 
Hopefully, a correct treatment of the charm quark threshold will support these
qualitative conclusions. Since our treatment of the charm 
threshold is so primitive we do not regard our value of $m_c^2$ as
a determination of the charm mass. Rather, $m_c$ is used simply as a 
phenomenological parameter, ensuring that the charm contribution to the 
full structure function is qualitatively correct. Of course, $m_c^2=4\Gev^2$ 
is somewhat high compared with the values obtained from reliable 
determinations. We will comment on this value later.

The strange quark is treated as being massless in our calculations, which is
presumably a good approximation for the values of $Q^2$ considered. However,
we insist that the strange contribution to the structure function 
is $0.2$ of the singlet structure function minus the valence
contributions (i.e. the ``sea structure function'') at $Q^2=m_c^2$. This 
ensures compatibility with the data on neutrino--induced deep inelastic 
di--muon production obtained by the CCFR collaboration \ref\strange{CCFR 
collaboration: A.O. Bazarko {\it et al}, \ZP \vyp{C65}{1995}{189}.}. 

Finally we consider the form of the gluon at large $x$. Within our effective
factorization scheme we would expect the gluon distribution to be quite
similar to that in the $\overline{\hbox{\rm MS}}$ scheme at 
NLO--in--$\alpha_s$ for very large $x$.
Thus, we demand that our gluon is qualitatively similar to that obtained 
from the WA70 prompt photon data \ref\gluon{WA70 collaboration: M. Bonesini
{\it et al}, \ZP \vyp{C38}{1988}{371}.} at $x\geq 0.3$. This means that 
the gluon distribution must be roughly of the form $2.5(1-x)^6$ at 
$Q^2=20\Gev^2$ for values of $x$ this large.
Encouragingly, this is the type of large $x$ behaviour that the best fit 
chooses for the gluon, and no strong constraint is needed.      

\medskip

The fit is performed for a wide variety of data: the H1 \hone\ and Zeus 
\zeus\ data on ${\cal F}^{ep}_2(x,Q^2)$ with $0.000032\geq x \geq 
0.32$ and $1.5\Gev^2 \geq Q^2\geq 5000 \Gev^2$; The BCDMS data 
\ref\bcdms{BCDMS collaboration: A.C. Benvenuti {\it et al}, \PL 
\vyp{B223}{1989}{485}.} on 
${\cal F}^{\mu p}_2(x,Q^2)$ with $0.07 \geq x\geq 0.75$ and 
$ 7.5 \Gev^2 \geq Q^2 
\geq 230 \Gev^2$; the new NMC data \ref\nmc{NMC 
collaboration, {\tt hep-ph/9610231}, \NP B, in print.} on 
${\cal F}^{\mu p}_2(x,Q^2)$ and ${\cal F}^{\mu d}_2(x,Q^2)$ with 
$0.008\geq x\geq 0.5$ and 
$2.5\Gev^2 \geq Q^2 \geq 65 \Gev^2$; NMC data on the ratio of 
${\cal F}^{\mu n}_2(x,Q^2)$ to ${\cal F}^{\mu p}_2(x,Q^2)$ 
\ref\nmcrat{NMC collaboration: 
M. Arneodo {\it et al}, \PL \vyp{B364}{1995}{107}.} with $0.015 \geq
x \geq 0.7$ and $5.5 \Gev^2 \geq Q^2 \geq 160 \Gev^2$, CCFR data 
\ref\ccfr{CCFR collaboration: P.Z. Quintas {\it et al}, \PRL 
\vyp{71}{1993}{1307}.} on ${\cal F}^{\nu N}_2(x,Q^2)$ and 
${\cal F}^{\nu N}_3(x,Q^2)$
with $0.125\geq x\geq0.65$ and $5\Gev^2 \geq Q^2\geq 501.2\Gev^2$; and the 
E665 \ref\esixsixfive{E665 collaboration: M.R. Adams {\it et al}, 
\PR\vyp{D54}{1996}{3006}.} data on ${\cal F}^{\mu p}_2(x,Q^2)$
with $0.0037 \geq x\geq 0.387$
and $2.05\Gev^2 \geq 64.3\Gev^2$.\foot{It has recently been realized that 
the MRS fit program has been treating the errors for the E665 data incorrectly;
the errors as a fraction of the value of ${\cal F}_2(x,Q^2)$ have been 
interpreted as the absolute errors, and the errors used have therefore been 
on average about 3 times too big. This has resulted in the very low 
$\chi^2$ in, for example, \mrsiii\ and \mylet.  In this paper the correct 
errors are used, though the MRSR fits are not performed again. As will be seen,
including the correct errors raises the $\chi^2$ for the fit to the E665 data 
by a factor of about $8$, as we would expect. However, it does not really 
affect any comparison of the best fits since all the fits we perform give 
descriptions of this data which are almost identical in quality.}  
For each data set the lowest 
$x$ bins cover a range of $Q^2$ from the minimum up to somewhat less than the
maximum, and similarly the highest $x$ bins cover a range of $Q^2$ from 
somewhat higher than the minimum up to the maximum. We use the NMC data 
on the ratio ${\cal F}^{\mu n}_2(x,Q^2)/{\cal F}^{\mu p}_2(x,Q^2)$ 
as well as on 
${\cal F}^{\mu p}_2(x,Q^2)$ and ${\cal F}^{\mu d}_2(x,Q^2)$ since the 
parameter space
covered by the two types of data sets is by no means identical. One can see 
that the full range of data used in the fit covers an extremely 
wide range of both $x$ 
and $Q^2$ and thus provides a very stringent test of any approach used to 
describe it.\foot{We have not included ZEUS data for $Q^2\geq 2000\Gev^2$
since this was not included in the MRSR fits \mrsiii\ against which we 
compare our results. The fit to this high $x$, high $Q^2$ data is practically
identical for all the fits discussed in this paper, giving a $\chi^2$ of 
about $17$ for the $7$ data points. This fairly large value of $\chi^2$ is due
to the large scatter of these points.}
     
The result of the best fit to this data using the leading--order, 
including leading--$\ln (1/x)$ terms, RSC expressions 
(henceforth referred to as the LO(x) fit) with 
$Q_0^2$ chosen to be equal to $40\Gev^2$, $m_c^2=4\Gev^2$
and $m_b^2=20\Gev^2$ is shown in table 1. 
Also the result of the fit to the 
small--$x$ data is shown in \fig\data{The curves correspond to
the value of the proton structure function ${\cal F}_2(x,Q^2)$ obtained 
from the 
leading--order, renormalization--scheme--consistent (LO(x)) calculation at 12
values of $x$ appropriate for the most recent HERA data. For clarity of 
display we add $0.5(12-i)$ to the value of ${\cal F}_2(x,Q^2)$ each time 
the value 
of $x$ is decreased, where $i=1\to12$. The data are assigned to the $x$
value which is closest to the experimental $x$ bin (for more details see
the similar figure displaying the two--loop fits
in \mrsiii). E665 data are also shown on the curves with 
the five largest $x$ values. The H1 and ZEUS data are normalized by $1.00$ 
and $1.015$ respectively in order to produce the best fit.}.   
As one can see there is a very 
good quality fit to the whole selection of data, and thus over the whole 
range of $x$ and $Q^2$. Only the NMC 
${\cal F}^{\mu n}_2(x,Q^2)/{\cal F}^{\mu p}_2(x,Q^2)$
data give a $\chi^2$ of much more than one per point, but this is
due to the scatter of the data points and is   
true for any global fit. Overall, the fit gives a $\chi^2$ of $1105$ for
$1099$ data points.\foot{The LO(x) fit is a little different from that in 
\mylet\ since $m_c^2=4\Gev^2$ is chosen here in order to give a rather better
description of the charm structure function data. In fact this value of 
$m_c^2$ leads to the best overall fit for the data also: lower $m_c^2$ leads
to a slightly worse fit to the HERA data with the fit to the other data 
remaining more or less unchanged, while higher $m_c^2$ leads to a 
slightly worse fit to the large $x$ data. We also point out that there was
a problem in the computer program when at $Q^2 <m_c^2$ in \mylet; the slope
of $d\,{\cal F}_2(x,Q^2)/d\ln Q^2$ was a little too small in this region. 
Thus, in fig. 1 in \mylet\ this slope should be a little larger below 
$Q^2=3\Gev^2$, and hence the kink at this value of $Q^2$ 
becomes less obvious. As far
as the fit is concerned, the effect of this error is minute since the error
is small, and because there are so few data points below $Q^2=3\Gev^2$. In 
practice the overall $\chi^2$ should be $\sim 2$ better than quoted in 
\mylet.}      

The fit shown is for the particular starting scale $Q_0^2=40\Gev^2$,
but the quality of the fit is extremely insensitive to changes in this scale, 
as we expect from the method of construction of the solutions. The fit is
essentially unchanged over the range $20-80 \Gev^2$, 
and we choose $40\Gev^2$ as the geometric mean. 
When $Q_0^2$ drops below $20\Gev^2$ the fit immediately gets markedly worse.
This effect is due to the bottom quark threshold, and we can explain it 
briefly. Both $d\,{\cal F}^S_2(x,Q^2)/d\ln Q^2$ and ${\cal F}^S_2(x,Q_0^2)$
are nearly proportional to $F^S_L(x,Q^2_{(0)})$ and hence nearly 
proportional to $N_f$ 
at small $x$, as can be seen from \fullsolfin\ and \fullsolfderiv\ and from 
\linkf\ and \linkg. Thus, when starting at $Q_0^2$ above $Q^2=20\Gev^2$ 
they are both 
$\propto N_f=5$. As we go below the bottom quark threshold, $Q^2=20\Gev^2$,
$d\,{\cal F}^S_2(x,Q^2)/d\ln Q^2$ becomes smaller, and is 
$\propto 4/5{\cal F}^S_2(x,Q_0^2)$. 
However, if $Q_0^2<20\Gev^2$, then $d\,{\cal F}^S_2(x,Q^2)/d\ln Q^2$ is
proportional to ${\cal F}^S_2(x,Q_0^2)$, with the same constant of 
proportionality as above, when below $Q^2=20\Gev^2$, and is 
$\propto 5/4{\cal F}^S_2(x,Q_0^2)$ above $Q^2=20\Gev^2$. In the latter case,
choosing the correct value of ${\cal F}^S_2(x,Q_0^2)$ to fit the data 
leads to $d\,{\cal F}^S_2(x,Q^2)/d\ln Q^2$ being too large at very small 
$x$. The quality of the fit then gets continuously
worse as $Q_0^2$ lowers further, becoming completely uncompetitive long
before reaching $m_c^2$.  We expect that a correct treatment of 
quark thresholds would lead to similar results to above for $Q_0^2\approx 
40\Gev^2$ ($d\,{\cal F}^S_2(x,Q^2)/d\ln Q^2\propto 
{\cal F}^S_2(x,Q^2)$ at the input here, but 
$d\,{\cal F}^S_2(x,Q^2)/d\ln Q^2$ falls off
towards 3/4 of this as $Q^2$ falls to $\sim 4\Gev^2$ and the charm quark 
effects die away) and then a smooth falling off of the quality of the 
fit as $Q_0^2$ was lowered, with it again beginning to deteriorate somewhere
in the region of $20\Gev^2$, due to $\alpha_s(Q_0^2)$ becoming too 
large below this value. 

There are 18 free parameters used in the fit: $\Lambda_{QCD,4}$
(which we choose to describe as a parameter since, although it is fixed at 
$100 \hbox{\rm MeV}$, and small variations would no doubt improve the fit 
slightly, it can certainly not vary much); four parameters for each of the 
valence quark contributions, which are described by functions of the form
\start\ where the normalization is set by the number of valence quarks; 
four parameters for the nonperturbative inputs for the two singlet 
structure functions, which are of the form \softinput; and the unknown 
scale $A_{LL}$, where we allow $A_{LL}$ to be a free parameter for each 
$Q_0^2$. We do not consider $Q_0^2$ as a free parameter, since it can take 
a very wide range of values.
The parameter $A_{LL}$, which we have argued should be a scale typical of soft 
physics, turns out to be $0.55\Gev^2$ for the fit starting at $Q_0^2=40
\Gev^2$. This decreases a little as $Q_0^2$ increases and {\it vice versa}.
For $Q_0^2=40\Gev^2$ the soft inputs for the fit are roughly
\eqn\softinputs{\hat F^S_L(x)\approx 3(1-x)^5(1+0.1x^{0.5}-0.2x)
, \hskip 0.5in F^S_2(x)\approx
(1-x)^{3.4}(1-0.65x^{0.5}+4.5x),}
where they have been forced to be flat as $x\to 0$. These nonperturbative 
inputs lead to complete inputs of $\hat F^S_L(x,Q_0^2)\approx 3x^{-0.33}$
and $F^S_2(x,Q_0^2)\approx 0.65 x^{-0.3}$ for $0.01 \geq x\geq 0.0001$, 
with the effective $\lambda$ increasing in both cases for even smaller $x$. 
Instead of forcing the nonperturbative inputs to be flat as $x\to 0$ we
could allow an asymptotic behaviour $x^{-\lambda}$, where 
$\lambda \lsim 0.08$. This leads to an
equally good fit as for the flat nonperturbative inputs.

\medskip

Thus, the LO(x) fit seems to be a success. However, in order to gauge its true 
quality it is helpful to have some points of comparison. 
Hence, we will discuss some alternative fits.
As in \mylet, the most recent MRS fits R$_1$ and R$_2$ are shown. These
are obtained using the standard two--loop method, where R$_1$ allows 
$\Lambda^{N_f=4}_{{\overline{\rm MS}}}$ to be free (giving 
$\Lambda^{N_f=4}_{{\overline{\rm MS}}}=241{\hbox{\rm MeV}}$) and 
R$_2$ fixes $\Lambda^{N_f=4}_{{\overline{\rm MS}}}=344{\hbox{\rm MeV}}$
to force a better fit to the HERA data. The new NMC data for 
${\cal F}^{\mu p}_2(x,Q^2)$
and ${\cal F}^{\mu d}_2(x,Q^2)$ are used rather than the previous sets 
which were used in the MRS fits. This new data is not 
used directly in the MRS fits, i.e they are not performed
again using these new data, the $\chi^2$ values for the new NMC data sets are 
simply calculated using the same input parameters as in \mrsiii. 
However, there is little 
indication that these would be changed much by the new data. The MRS fits 
are useful for purposes of comparison for a number of reasons.
Firstly the treatment of the errors in
this paper is identical to that in the MRS fits, so consistency in this 
respect is guaranteed. Also, any systematic differences due to differences
in computer programs is guaranteed to be absent. Finally, the cuts in $Q^2$ 
for each data set are chosen to be the same in this paper as for the 
MRS fits.   

The number of free parameters in the NLO--in--$\alpha_s$ 
fit is the same as in the LO(x) 
fit. There is one more parameter in the NLO--in--$\alpha_s$ 
fit due to the powerlike forms
of the input gluon and singlet quark at small $x$ being independent\foot{We 
claim that the allowed powerlike form of the inputs at small $x$ is against 
the spirit of a well--ordered perturbative expansion, and that the small--$x$ 
form of the inputs should be determined in terms of the nonperturbative inputs 
and the requirement that the expressions are insensitive to the choice of 
$Q_0^2$, as already discussed in \S 4.2. However, enforcing this rule would
mean that the quality of the fit to the small--$x$ data was very poor. Thus, we
allow the more usual, unjustified choice for the small--$x$ form of the 
inputs. In fact, at $Q_0^2=1\Gev^2$ the gluon is strongly valence--like, while
the sea--quark distribution $\sim x^{-1-0.15}$at small $x$.} whereas the 
small--$x$ shape of the inputs for both structure functions is completely
determined by the one parameter $A_{LL}$ in the LO(x) fit.
However, there is one less parameter in the
NLO fit due to the normalization of the gluon being determined by 
momentum conservation. Essentially, the LO(x) fit is a little less 
constrained at large $x$ than the NLO fit, but rather more predetermined at
small $x$ than the NLO fit. Indeed, the LO(x) fit has a certain amount of 
predictive power in the low $x$ region, as seen at the end of the last 
section, whereas the NLO has none.   

Comparing the LO(x) fit and the MRS fits, it is clear that
the LO(x) scheme--independent fit is much better for the HERA data (even
when compared to R$_2$), much better for the BCDMS data (even when compared
to R$_1$) and similar in standard for the rest of the data. The overall fit is 
$\sim 200$ better for the whole data set; clearly a lot better. 
However, this direct comparison with the MRS fits is rather unfair, since 
there are a number of ways in which it differs from the LO(x) fit. 
First, and not terribly importantly, the normalizations of the data sets are 
allowed to vary more in the LO(x) fits than in the MRS fits, as described 
in the caption to table 1. (There is a small
systematic difference in the normalization required by the H1 data and ZEUS 
data.) Rather more 
importantly the MRS fits use the SLAC data \ref\slac{L.W. Whitlow {\it et al},
\PL \vyp{B282}{1992}{475}\semi L.W. Whitlow {\it et al}, preprint SLAC--357
(1990).} on ${\cal F}^{ep}_2(x,Q^2)$ in the fits as well as the data sets 
mentioned in table 1. This covers the range $0.07\geq x\geq 0.65$ and 
$2\Gev^2\geq Q^2\geq 22.2\Gev^2$, i.e. the same sort of high $x$ range as the 
BCDMS data, but rather lower $Q^2$ in general. Hence the SLAC data is not 
included in the LO(x) fit due to much greater sensitivity of this 
data to potential higher twist effects than any of the other data sets.   
If it is included in a fit, then in certain regions of parameter space it is 
incompatible with the BCDMS data, as can be seen in fig. 4 of \mrsiii.  
Hence a best fit to both the BCDMS data and the SLAC data must, to a certain 
extent, be a compromise between the two. Therefore the fit to the BCDMS data
in the MRS fits is worsened somewhat by the need to fit the SLAC data. If
the SLAC data were ignored the fits to the BCDMS data would improve by 
about 40. The LO(x) fit also differs from the MRS fits in the starting
scale of the evolution and in the treatment of the quark thresholds. The 
charm contribution to the structure function in the MRS fits turns on at
$Q_0^2=1\Gev^2$ but is suppressed by a phenomenological factor of a
function of $(3.5\Gev^2/Q^2)$ in order to give a reasonable description of 
the charm data. The bottom contribution to the structure function, which is 
very small, is not included at all. Finally, the fact that the quark and gluon 
distributions are input at $1\Gev^2$ in the MRS fits is not helpful to the
quality of the fit. If the best fit is obtained for the HERA data using the 
NLO calculation with massless quarks with inputs at $Q_0^2\approx 4\Gev^2$, 
as in \steep, and evolution performed downwards in $Q^2$ then the gluon 
becomes negative at very small $x$ by the time $Q^2=1\Gev^2$ is reached. 
Hence, the valence--like input gluon distributions in MRSR$_1$ and MRSR$_2$ 
are not the 
ideal distributions for fitting the HERA data. A parameterization which would
allow the gluon distribution to turn over and become negative at very small 
$x$ would be better, as would simply starting at larger $Q^2$ where the gluon 
distribution is happy to be positive everywhere.\foot{Starting the evolution 
at $Q_0^2 < 1\Gev^2$ quickly leads to much worse fits to the data. Also 
starting at fairly high $Q_0^2$, i.e. $Q_0^2 \gsim 20\Gev^2$, leads to a 
worsening of the fit due to the given form of parameterization not being 
adequate to describe the parton distributions at these scales, 
see e.g. \faili. Thus, $Q_0^2$ is no less a parameter in a NLO fit as in 
the LO(x) fit.}

Thus, in order to make a more meaningful comparison of the LO(x) fit with 
a NLO--in--$\alpha_s$ fit we have ourselves performed a NLO--in--$\alpha_s$
fit, called NLO$_1$, allowing 
the normalizations to vary in the same way as in the LO(x) fit, 
with exactly the same treatment of 
quark thresholds as the LO(x) fit, with the input parton distributions
chosen at $Q_0^2=m_c^2$ and with the fit to exactly the same data, i.e. to 
those sets shown in table 1, and with the constraint on the gluon from the 
WA70 prompt photon data.\foot{Once again we allow the small--$x$ form of the 
inputs to be unjustified powerlike behaviours. At $Q_0^2=2.75\Gev^2$ the 
gluon is quite flat while the sea--quark distribution $\sim x^{-1-0.22}$.}
As in the MRS fits the values of ${\cal F}_2(x,Q^2)$ at small $x$ are those 
quoted in \zeus\ and \hone. Strictly speaking they should be altered to 
take account of fact that the predicted values of ${\cal F}_L(x,Q^2)$ from 
these fits are not precisely the same as those used in \hone\ and \zeus\
to extract ${\cal F}_2(x,Q^2)$ from the measurement of the differential 
cross-section, i.e. an iterative procedure should be used as with the LO(x)
fit. However, the differences between the values of ${\cal F}_L(x,Q^2)$ in 
\hone\ and \zeus\ and from these NLO--in--$\alpha_s$ fits are far smaller 
than the differences between those in \hone\ and \zeus\ and the LO(x) 
fit. Hence, in practice the errors introduced by not altering the values of 
${\cal F}_L(x,Q^2)$ are very small (the $\chi^2$ for the fit to the 
HERA data might improve by a couple of points if the formally correct 
procedure was used). The value of $m_c^2$ is chosen to provide a 
good description of the data on the charm structure function. The value 
needed is $m_c^2=2.75\Gev^2$, somewhat lower than in the LO(x) fit, and 
the fit to the charm structure function data is shown in 
\fig\nloxcharm{The description of the EMC and preliminary H1 data for 
${\cal F}^c_2(x,Q^2)$ using the NLO$_1$ fit.}.
\foot{As in the LO(x) fit 
the choice of $m_c^2$ which gives a good description of the charm data is 
also the choice which gives the best global fit. Raising the value of $m_c^2$
leads to the value of $\Lambda^{N_f=4}_{{\overline{\rm MS}}}$ chosen going 
up, which improves the fit to the HERA data, but the fit to the rest 
of the data deteriorates by more than this improvement. Lowering the value of 
$m_c^2$ leads to the value of $\Lambda^{N_f=4}_{{\overline{\rm MS}}}$ chosen 
going down, improving the fit to the large $x$ data, but again leading to 
overall deterioration.} The value of $\Lambda^{N_f=4}_{{\overline{\rm MS}}}$ 
determined by the fit is intermediate to those chosen by the MRS fits, 
being equal to $299 \hbox{\rm MeV}$. The quality of the fit is shown in 
table 1. 

As one can see, the different treatment of the quark thresholds and the 
higher starting scale for evolution has produced a better fit to the HERA 
data than the MRS fits, even though the value of $\alpha_s(Q^2)$ is lower
than in the R$_2$ fit. However, the best fit comes from allowing 
the normalization
of the H1 data to be at the lower limit allowed by the error on the 
normalization\foot{The fit to the H1 data continues to improve slightly for
a normalization going down to $0.96$ if this is allowed.}, and is still not as
good as the LO(x) fit to the HERA data. Not including the SLAC data, and in 
the case of the R$_2$ fit the lowering of 
$\Lambda^{N_f=4}_{{\overline{\rm MS}}}$, leads to a much better fit to the 
BCDMS data than the MRS fits, but again this is clearly not as good as the 
LO(x) fit. The fit to the NMC data is much the same for the LO(x) fit as for 
the NLO$_1$ fit, and the NLO$_1$ fit is a little better for the CCFR data. 
The overall quality of the NLO$_1$ fit is $1169$ for the $1099$ points, and 
thus is $64$ worse than the LO(x) fit. Therefore, the LO(x) fit is still 
clearly better than the NLO$_1$ fit, but not nearly as convincingly as 
appeared to be the 
case when compared to the MRS fits. Nevertheless, it is encouraging that, 
while the overall fits to the relatively high $x$ data, i.e. the BCDMS, 
NMC, E665 and CCFR data are similar in the NLO$_1$ fit and the LO(x) fit, 
it is the fit to the small--$x$ HERA data that is 
definitely better for the LO(x)
fit, as we would expect. This can be seen even more clearly if we simply 
examine the quality of the fit by separating the data into two sets: one 
where $x< 0.1$ and one where $x\geq 0.1$. This is shown in table 2, and 
demonstrates that the LO(x) fit is superior at small $x$, while not
quite as good as the NLO$_1$ fit at large $x$. 

This qualitative result above is exactly what we would expect. 
The importance of the leading--$\ln(1/x)$ terms in the LORSC calculation 
can be quantitatively judged
by how they affect the fit. If, after obtaining the best LO(x) fit, all terms
other than those in the one--loop expressions are set to zero, the quality 
of the fit is unchanged above $x=0.3$, begins to 
alter slightly below this, and is clearly worse by the time we reach 
$x=0.1$. Thus, the leading--$\ln(1/x)$ terms are important by this 
value of $x$. However, much of this effect 
is due to the terms at ${\cal O}(\alpha_s^2)$, so the NLO
expression at fixed order in $\alpha_s$ should be insensitive to 
higher--order--in--$\alpha_s$ leading--$\ln(1/x)$ terms down to $x$ somewhat 
lower than $0.1$. Therefore, above $x=0.1$ the NLO fit should in principle 
be better than the LO(x) fit since it contains terms at NLO in $\alpha_s$
which are important at large $x$. However, the NLO fit should be 
considerably worse at small $x$ since it does not contain many important
leading--$\ln (1/x)$ terms. This is qualitatively in agreement with the 
comparison of the NLO$_1$ fit and the LO(x) fit. 

However, the above comparison is 
somewhat incorrect because in the process of obtaining the best fit for all
the data the NLO$_1$ fit may choose some parameters, particularly   
$\Lambda^{N_f=4}_{{\overline{\rm MS}}}$, such that they mimic the effects 
of the leading--$\ln(1/x)$ terms, and a decent fit for the small
$x$ data is obtained to the detriment of the fit to the large $x$ data. 
In order to check this hypothesis we have also performed a 
NLO--in--$\alpha_s$ fit with 
$\Lambda^{N_f=4}_{{\overline{\rm MS}}}$ fixed at $250 \hbox{\rm MeV}$, 
which we will denote by NLO$_2$. The results of this fit in terms of the 
different data sets is shown in table 1. The fit clearly improves compared 
to the NLO$_1$ fit for the BCDMS and CCFR data, and gives the best overall fit
for the high $x$ data sets. It worsens somewhat for the NMC data, and also
for the HERA data, and overall is slightly worse than the NLO$_1$ fit, 
having a $\chi^2$ of $1184$ for the $1099$ data points. Nevertheless, it is
perhaps a truer representation of a real NLO--in--$\alpha_s$ 
fit than NLO$_1$ since, as can 
be seen in table 2, it gives a better fit to high $x$ data, but not such a 
good fit to the small--$x$ data  which presumably require the 
leading--$\ln(1/x)$ terms. Whether one believes this argument or not, 
it is certainly
clear that the LO(x) fit does provide a better description of the data than 
any standard NLO--in--$\alpha_s$ fit.  

We can be even more general in our NLO--in--$\alpha_s$ type fit. 
We have already argued that,
when working at fixed order in $\alpha_s$, despite the fact that the amount
of momentum carried by the partons can be constant, it does not have to 
sum to unity, i.e. higher--order inputs can carry some fraction of the 
momentum. Thus, we have also performed a fit where the total momentum of the 
partons is not constrained. When doing this we have again allowed the 
small--$x$ form of the parton inputs to be $\propto x^{-1-\lambda}$ 
which, as we have already stressed, we do not believe is correct 
procedure for a NLO--in--$\alpha_s$ calculation. However, we will allow 
as much freedom in this fit as possible. Doing so we obtain a fit which 
is nearly, but still not as good as the LO(x) fit. The fit to the data 
for $x\geq 0.1$ is 
about 25 better than the LO(x) fit, i.e. similar to the NLO$_2$ fit, but
still about 45 worse for $x< 0.1$. The value of 
$\Lambda^{N_f=4}_{{\overline{\rm MS}}}$ chosen by the fit is $230 \hbox{\rm
MeV}$. The amounts of momentum carried by the quarks are similar to the 
other NLO--in--$\alpha_s$ fits; these are constrained by 
fitting the data near to the 
input scale. The momentum carried by the gluon is $57\%$, far higher than 
in the other NLO--in--$\alpha_s$ fits. 
This large amount of gluon momentum is necessary in
order make the gluon large at small $x$ and hence
provide a good fit to the small--$x$ data for ${\cal F}_2(x,Q^2)$ even
with the small value of $\Lambda^{N_f=4}_{{\overline{\rm MS}}}$.  The large 
amount of gluon momentum leads to the total momentum carried by the 
NLO partons being $114\%$ of the proton's momentum. We do not believe the 
validity of this fit, having already argued that the LORSC expressions are
the theoretically correct expressions, and that leading--$\ln(1/x)$ terms are
necessary for the correct description of small--$x$ data. Indeed, this fit 
leads to very different predictions for other quantities from those 
obtained using the LO(x) fit,
as we will soon see. However, we do 
believe that the lack of the momentum constraint on the inputs allows this
NLO--in--$\alpha_s$ fit to be perhaps the most realistic at high $x$: 
the fact that the NLO--in--$\alpha_s$ fit is not designed to work at 
small $x$ due to the lack of leading--$\ln(1/x)$ terms can be largely 
compensated by an overly large input gluon distribution at small $x$, and the 
high $x$ part of the fit can be as good as possible without having to 
compromise itself to match the low $x$ data. In fact, at  $x\geq 0.1$ the 
details
of this fit are very similar to the NLO$_2$ fit, but the gluon is much larger
at smaller $x$ in this fit. 

\medskip

Let us comment briefly on other possible NLO--in--$\alpha_s$ fits. 
As already mentioned it is difficult to compare directly with global fits 
from groups other than MRS since there may be differences in treatment of 
errors or systematic differences in the outputs of the computer program. 
However, we can make some comments. The latest CTEQ fits 
\ref\cteqlat{H.L. Lai {\it et al}, {\tt hep-ph/9606399},
preprint MSUHEP--60426 or CTEQ--604,
June 1996.}, e.g. CTEQ4M, are performed in a very similar manner to the 
NLO$_1$ fit, though with different cuts on data, and the results also appear 
to be very similar.\foot{There is a clear improvement to the fit to the BCDMS
data compared to the NLO$_1$ fit, and clear worsening of the fit to the
CCFR ${\cal F}_2(x,Q^2)$ data.
These seem to be typical systematic differences between all CTEQ fits and 
MRS type fits and are presumably due to systematic differences in the 
treatment of errors or in the  numerical calculation. The comparison of the 
codes in \comparison\ certainly supports the latter conclusion.} Hence,
there seems little more to say than that the NLO$_1$ fit is similar to the 
CTEQ results. 

We also feel we must comment on 
the fit performed by H1 in \hone. This takes account of the charm quark 
threshold rather more correctly than in the present paper, i.e. there is a 
smooth threshold at $W^2=4m_c^2$. However, it is a fit to only the H1, 
NMC and BCDMS data. It is difficult to make a direct comparison with this 
fit due to the completely different way of performing the calculation.
It is however possible to perform a NLO--in--$\alpha_s$ fit to the 
same data as the H! fit using the same 
prescription as for NLO$_1$ and NLO$_2$. 
This has been done, and many similarities to features of the H1 fit noticed. 
It is possible to obtain a NLO--in--$\alpha_s$ fit for the H1, NMC and 
BCDMS data of a comparable quality to that in the LO(x) fit. 
However, if one simply inserts the 
values of the ZEUS and CCFR data after finding the best fit one finds that 
the $\chi^2$ for the ZEUS fit is over $300$, i.e. there is a certain degree of
incompatibility between finding the best fit to the H1 data and the ZEUS data,
and the fit to the CCFR data is very poor indeed. Perhaps more importantly,
since the fit does not put any direct constraint on the gluon, then as in 
\hone\ one obtains a gluon which at $Q^2=5\Gev^2$ behaves like $2x^{-0.2}$
at small $x$, which helps the fit to the H1 data, but behaves like 
$2(1-x)^8$ at large $x$. This leads to a gluon like $2(1-x)^{8.5}$ 
at large $x$ at $Q^2=20\Gev^2$ rather than the $2.5(1-x)^6$ required by
the WA70 prompt photon data. For values of $x\approx 0.4$, $2(1-x)^{8.5}$ is
clearly a great deal smaller than $2.5(1-x)^6$, and so the gluon 
obtained from this type of approach is hopelessly incompatible with the
prompt photon data. 

We feel that similar, if less dramatic, 
considerations are probably true of the type of approach 
adopted in \bfresum, \steep\ and \faili, where fits are performed to 
small--$x$ data alone. Despite the fact that the input parton distributions
are constrained to be very similar to standard parton distributions at 
large $x$, e.g. the MRSD0 distributions, it must be remembered that these 
parton distributions provide good descriptions of the large $x$ data when
evolved at NLO in $\alpha_s$ only when using a particular value of 
$\Lambda^{N_f=4}_{{\overline{\rm MS}}}$.
Evolving the MRSD0 parton distributions, which should have 
$\Lambda^{N_f=4}_{{\overline{\rm MS}}}=230\hbox{\rm MeV}$, using instead 
$\Lambda^{N_f=4}_{{\overline{\rm MS}}}=360\hbox{\rm MeV}$ will produce 
very different results. Indeed, as some sort of comparison
we are able to obtain a fit to the H1 data and NMC data (thus constraining
the high $x$ parton distributions) alone to a quality rather better 
than in the full LO(x) fit by using a NLO--in--$\alpha_s$ fit. 
However, the fit does indeed choose 
$\Lambda^{N_f=4}_{{\overline{\rm MS}}}=360\hbox{\rm MeV}$. The 
resulting fit to virtually all the other data is very poor. 

Thus, we have a number of clear demonstrations of the dangers of 
performing fits to only a restricted set of data. In order to avoid very 
obvious inconsistencies with some particular data it is necessary to 
perform proper global fits. One may investigate the possibility that 
the data in some region of parameter space is not really expected to be 
fit well using a particular expression. Indeed,  this is what we have done 
in the NLO$_2$ fit, and believe that we have sound theoretical arguments 
for doing so, and moreover that the results of the fits back these up.       

This leads us to the question of the determination of 
$\alpha_s(M_Z^2)$ using global fits to structure function data. The complete 
RSC expression for structure functions only exists at leading order, and, 
as already mentioned, this leaves the renormalization scale completely
undetermined. Hence, a determination of $\alpha_s(M_Z^2)$ does not really 
take place. (Note, however, the previous comments on this question.) 
It is not yet possible to extend the RSC
calculation beyond the leading order due to lack of knowledge of 
NLO--in--$\ln (1/x)$ terms, but hopefully these will shortly become available 
\nlobfkl, and when they do the NLO versions of \fullsolfl--\fullsolfin\ 
can be derived and
put to use. Only then should the NLO coupling constant really be used in 
any global fit.  Until the full renormalization--scheme--consistent 
NLO expressions become available, we believe that it is incorrect to use 
NLO--in--$\alpha_s$ fits to small--$x$ structure function data in order to 
determine the NLO coupling 
constant (unless, of course, direct measurements of ${\cal F}_L(x,Q^2)$ 
and other less inclusive quantities at very small $x$ turn out to 
verify standard two--loop predictions).
However, as already argued, the fixed--order--in--$\alpha_s$ expressions 
should be accurate 
for CCFR, BCDMS and NMC data (except perhaps at the lowest $x$ values), 
which after all are still much more 
precise than HERA data, and fits to these data alone will provide the 
best determination of the NLO $\alpha_s(Q^2)$. 
However, let us be wary here. One might think that the above comments mean 
that we believe that the value of $\Lambda^{N_f=4}_{{\overline{\rm MS}}}$
at NLO is about $250\hbox{\rm MeV}$, since our fits to large $x$ data at 
NLO--in--$\alpha_s$ support this. To a certain extent this is true, 
but we also believe that 
our naive treatment of the heavy quark thresholds can introduce an 
error in the determination of $\alpha_s(M_Z^2)$ which could easily be as
large as $0.005$. Hence, until this naive treatment is improved, we have
nothing concrete to say about the value of $\alpha_s(Q^2)$. 

\bigskip

We have given what we hope are convincing arguments for our 
advocated approach for calculating structure functions. 
Not only do we claim theoretical correctness, and 
limited predictive power, but we also have good quality, very 
comprehensive fits to data on ${\cal F}_2(x,Q^2)$.
However, we are well aware that only further experimental tests can prove us 
right (or wrong). Hence, we now discuss our predictions for 
${\cal F}_L(x,Q^2)$. 
   
So far we have only probed ${\cal F}_L(x,Q^2)$ indirectly, 
i.e. it is simply related to the $\ln Q^2$ derivative of ${\cal F}_2(x,Q^2)$
(as well as to the input ${\cal F}_2(x,Q_0^2)$). 
Having tied down the nonperturbative inputs and $A_{LL}$ and $Q_0^2$ from
our fit to ${\cal F}_2(x,Q^2)$, we have a prediction for 
${\cal F}_L(x,Q^2)$. The result of this prediction for the fit with 
$Q_0^2=40\Gev^2$ is shown in \fig\fl{Comparison of predictions for 
${\cal F}_L(x,Q^2)$ using the full leading--order,
renormalization--scheme--consistent (LO(x)) fit and the two--loop NLO$_1$ 
fit. For both sets of curves ${\cal F}_L(x,Q^2)$ 
increases with increasing $Q^2$ at the lowest $x$ values.}, 
where it is compared to the prediction using the NLO--in--$\alpha_s$ approach, 
in particular the NLO$_1$ fit. As one can see, it is smaller
than the NLO$_1$ ${\cal F}_L(x,Q^2)$, but becomes steeper at very small 
$x$. The NLO$_2$ fit gives a very similar form of ${\cal F}_L(x,Q^2)$ to the 
NLO$_1$ fit, while the NLO--in--$\alpha_s$ fit where momentum is not conserved 
leads to a similar form of ${\cal F}_L(x,Q^2)$ at high $x$ but
an even larger value than the NLO$_1$ fit at $x<0.1$. 
The LORSC prediction for ${\cal F}_L(x,Q^2)$ is weakly dependent on the 
value of $Q_0^2$ chosen: the value at $Q^2=5\Gev^2$ and $x=10^{-4}$
varies by $\pm 10\%$ within our range of 
$Q_0^2$ (increasing with $Q_0^2$), and by
less than this for higher $x$ and $Q^2$.\foot{The author has submitted 
two conference proceedings on the topic of the current 
paper \ref\rstproc{R.S. Thorne, Proc. of the International Workshop 
on Deep Inelastic Scattering, Rome, April, 1996, in print; 
Proc. of the 1996 HERA Physics Workshop, eds. G. Ingelman,
R. Klanner, and A. De Roeck, DESY, Hamburg, 1996, Vol. 1, p. 107.}  
and should point out that these are both incomplete. In the former 
the expression 
\fullsolfin\ was not used for the input for ${\cal F}_2(x,Q^2)$. 
Thus, the fit imposed far less constraint on $Q_0^2$ than
the full procedure, and a value of $5\Gev^2$ was used, which resulted in a
prediction of ${\cal F}_L(x,Q^2)$ that is much too small. The latter claimed 
that an input for ${\cal F}_L(x,Q^2)$ a little smaller than that consistent 
with the full set of expressions was needed for the best fit,
even at the optimum $Q_0^2$. This was due to 
there being no account whatever taken of the $b$-quark threshold. The
correction makes very little difference to the phenomenological results
in this latter case.} The very recent results on ${\cal F}_L(x,Q^2)$ 
for $0.01\gsim x \gsim 0.1$ from NMC \nmc\ are matched 
far better by the LO(x) ${\cal F}_L(x,Q^2)$ than the NLO$_1$ 
${\cal F}_L(x,Q^2)$ (the latter being 
rather large). However, it is fair to say that any problems with the 
NLO$_1$ ${\cal F}_L(x,Q^2)$ can almost certainly be attributed to the 
naive treatment of the charm 
quark threshold, i.e. the predicted ${\cal F}_L(x,Q^2)$ from the 
NLO--in--$\alpha_s$ calculation in the last of \grv\ 
matches the data well. Actually, it is rather uncertain whether in this 
region of 
parameter space ${\cal F}_L(x,Q^2)$ would be better described by the 
LORSC calculation, which ignores some ${\cal O}(\alpha_s^2)$ effects,
or the NLO--in--$\alpha_s$ calculation, which ignores the leading--$\ln(1/x)$
effects beyond second order in $\alpha_s$. 

Measurements of ${\cal F}_L(x,Q^2)$ at $x<10^{-2}$
would be a better discriminant between fixed--order--in--$\alpha_s$ 
calculations and those involving leading--$\ln (1/x)$ terms. 
However, the sort of ``determination'' of ${\cal F}_L(x,Q^2)$
already performed by H1 \ref\fldet{H1 collaboration, {\tt hep-ex/9611017}, 
preprint DESY 96--236.} is really 
only a consistency check for a particular NLO--in--$\alpha_s$ 
fit. It is by no means a true measurement 
of ${\cal F}_L(x,Q^2)$. In essence, all it proves is that the 
measurements of the cross--section are consistent with a particular 
NLO--in--$\alpha_s$
fit to ${\cal F}_2(x,Q^2)$ when the relationship between the cross--section
and the values of ${\cal F}_2(x,Q^2)$ is determined assuming 
the correctness of the NLO--in--$\alpha_s$ expressions for both structure 
functions. This is a perfectly correct procedure, and should be adopted for
any fit to ${\cal F}_2(x,Q^2)$ data, as it has been for the LO(x) fit. 
It would cause concern about the validity of 
the particular fit being used if it 
were to fail, but in itself says nothing 
whatever about the validity of a different 
approach, or about the actual value of ${\cal F}_L(x,Q^2)$.
Hence, real measurements of ${\cal F}_L(x,Q^2)$ 
are needed in order to find the real values of ${\cal F}_L(x,Q^2)$.
From \fl\ it is clear that such measurements at HERA would be an important 
(and probably essential) way of determining the validity of the 
approach in this paper, and the genuine importance of leading--$\ln(1/x)$ 
terms in structure functions.

Another potentially important discriminant between different methods of 
calculating structure functions is the measurement of the charm structure 
function. Both the LO(x) fit and the NLO$_1$ fit can provide good fits to 
the currently available charm data, as already seen, and the value of $m_c^2$ 
in the NLO$_1$ fit is rather more satisfactory than that in the LO(x) 
fit. However, these calculations involve incorrect treatments of the 
charm quark threshold. A correct treatment of this threshold at NLO in 
$\alpha_s$ \grs\ shows that the value of $m_c^2$ must be at least halved in
order to produce the same sort of value of the charm structure function as 
the approach used in this paper. Indeed, as seen in \ref\ballroeck{R.D. Ball
and A. De Roeck, {\tt hep-ph/9609309},
Proc. of the International Workshop on Deep 
Inelastic Scattering, Rome, April, 1996, in print.}, correct 
NLO--in--$\alpha_s$ 
calculations with $m_c^2=2.25\Gev^2$ undershoot the small--$x$ 
data (other than the H1, calculation which has a form of the gluon already
criticized and which would certainly badly undershoot the larger $x$ 
EMC data). In fact, standard NLO--in--$\alpha_s$ calculations using the 
coefficient functions for charm production in \grs\ only produce a 
sufficiently large charm structure function if $m_c^2\approx 1\Gev^2$
\ref\dickcharm{R.G. Roberts, private communication.}, i.e. 
somewhat lower than expected.   
 
A correct treatment of the charm quark threshold within the framework 
advocated in this paper has not been completely worked out (although, as 
already mentioned, work is in progress), and has certainly not been tested.
However, using the naive treatment in this paper, the quark mass required 
in the LORSC approach is somewhat higher than for the NLO--in--$\alpha_s$ 
approach. If the same sort of conclusion is true when using a correct 
treatment of quark thresholds, as we might expect (but cannot guarantee), then 
this will offer further support for the LORSC calculation. More generally, a 
comparison of the theoretical calculations with the ever--improving data on 
the charm structure function seems potentially to be a very useful way 
of discriminating between different methods of calculation.\foot{We also note
that as seen in \grs\ the longitudinal structure function is more
sensitive to the treatment of the charm quark threshold than 
${\cal F}_2(x,Q^2)$, with there being a sizeable discontinuity in 
${\cal F}_L(x,Q^2)$ at $Q^2=m_c^2$ when using the approach in this paper. 
Thus, for accurate predictions of ${\cal F}_L(x,Q^2)$ at values of $Q^2$ in 
the vicinity of $m_c^2$ a smooth treatment of the charm quark 
threshold is really needed in practice. The problem is not as acute
at the bottom quark threshold because the total squared charge of the quarks 
changes proportionally much less at this threshold than at the charm quark
threshold.}

In principle there are many other quantities which could be calculated 
within the LORSC framework and compared to those calculated using the 
NLO--in--$\alpha_s$ approach (or any other method) and to experimental data. 
Particularly obvious examples are the distribution of the transverse
energy flow in the final state in lepton--hadron scattering and the 
the cross--section for forward jet production, for both of which there exists 
some experimental data which does not seem to be terribly well described by 
the order--by--order--in--$\alpha_s$ approach, particularly in the latter case,
see e.g. \ref\transverse{H1 collaboration, paper pa02-073, 
submitted to ICHEP 1996, (Warsaw, July, 1996).} and 
\ref\fjets{H1 collaboration, paper pa03-049, 
submitted to ICHEP 1996, (Warsaw, July, 1996).}. We have not 
performed a LORSC calculation of any
such quantities, and it is very difficult to estimate the 
results, other than guess that they will probably lie somewhere 
between the fixed--order--in--$\alpha_s$ predictions and those obtained using 
BFKL physics naively. Such calculations are
obviously a priority, and work will begin soon. Only by comparing our
theoretical predictions with a wide variety of experimental data can we 
determine unambiguously which theoretical approach is correct. For the
moment we leave the theoretical arguments and the results of tables 1 and 
2 as the evidence supporting our particular approach.

\newsec{Conclusion and Summary.}

In this paper we have derived expressions for the structure functions 
${\cal F}_2(x,Q^2)$ and ${\cal F}_L(x,Q^2)$
in a theoretically correct, well--ordered manner. We have first done this 
for the particular expansions schemes which order the expressions strictly in
orders of $\alpha_s$, i.e. the standard loop expansion,
or in terms of the leading powers of $\ln (1/x)$ for given power of $\alpha_s$,
i.e. the leading--$\ln (1/x)$ expansion. In both cases we have demonstrated 
that a correct calculation automatically leads to  
factorization--scheme--invariant 
results which may be expressed in terms of Catani's physical 
anomalous dimensions \cat. Thus, we refute any suggestion that these physical 
anomalous dimensions are simply an example of anomalous dimensions in 
a convenient, physically motivated factorization scheme, 
but insist that they are fundamental 
pieces in the correct expressions for the structure functions. Likewise, we 
insist that no particular factorization scheme is fundamentally any 
more ``physical'' than 
any other. However, we also demonstrate that in the case of the 
leading--$\ln (1/x)$ expansion the correct expressions are 
more difficult to obtain than in the loop expansion, and 
a correct calculation requires more 
than just the knowledge of these physical anomalous dimensions. 

We have then argued that both the above expansion schemes are restrictive, and
lead to only part of the correct solution at any given order. We have 
shown that the only calculational method which is truly consistent with 
working to a given order in $\alpha_s$ within a given renormalization scheme 
is the renormalization--scheme--consistent expansion, in which one 
works to a given order in both $\alpha_s$ and in $\ln(1/x)$ for given power
of $\alpha_s$. After presenting our argument that this is indeed 
the only correct
expansion scheme, we have made use of Catani's physical anomalous 
dimensions (though this is not necessary, merely very convenient) to derive  
the leading--order, renormalization--scheme--consistent expressions for 
the structure functions ${\cal F}_2(x,Q^2)$ and ${\cal F}_L(x,Q^2)$. 

As part of our overall approach we have also taken a different 
view of the starting 
scale $Q_0^2$ from that normally taken. Rather than trying to guess some 
form for the input at some particular $Q_0^2$, or simply allowing the 
inputs at some arbitrary $Q_0^2$ to take any form they wish within a given 
parameterization, we have demanded firstly that our expressions should be as 
insensitive as possible to the choice of the input scale, and secondly that
any deviation from the flat Regge--type behaviour of the structure functions
must come from perturbative effects. Thus, our inputs take the form of  
nonperturbative functions, which are flat at small $x$, convoluted with 
functions of the physical anomalous dimensions which are evaluated at $Q_0^2$
and determined by the requirement of insensitivity to the 
value of $Q_0^2$. This leads to our inputs being determined entirely in 
terms of the flat nonperturbative inputs and one arbitrary scale $A_{LL}$ 
which roughly indicates the scale where perturbative physics should break 
down, i.e. $A_{LL}\lsim 1\Gev^2$. This gives us a great deal more predictive 
power than more usual approaches. We have some idea of the form of 
individual structure functions at small $x$: 
for a full range of sensible choices of 
$Q_0^2$ ($10\Gev^2 - 100\Gev^2$) and $A_{LL}$ we obtain 
${\cal F}_2(x,Q_0^2)$ roughly $\propto x^{-(0.25-0.33)}$
for $0.01\geq x\geq 0.00001$, which is clearly in good 
qualitative agreement with the data. However, we also have a much stronger 
prediction for the relationship between the small--$x$ inputs for 
${\cal F}_2(x,Q^2)$, ${\cal F}_L(x,Q^2)$ and for 
$d\,{\cal F}_2(x,Q^2)/d\ln Q^2$. Therefore, as well as some predictive 
(or at the very 
least, explanatory) power for individual structure functions, we 
have a consistency condition between the form of the different 
inputs, i.e. once we choose the input for ${\cal F}_2(x,Q^2)$ we have 
determined, up to a small amount of freedom, the small--$x$ inputs for
$d\,{\cal F}_2(x,Q^2)/d\ln Q^2$ and ${\cal F}_L(x,Q^2)$.
This is a feature unique to the approach in this paper. 

Not only are the features of the LORSC calculation unique and
compelling, but they also 
work rather well in practice. The LORSC expressions, including this 
constraint on the small--$x$ inputs, lead to very good fits to the data.
To qualify this statement, the $\chi^2$ for the LORSC fit to 1099 data on
${\cal F}_2(x,Q^2)$ ranging from
from $0.75\geq x\geq 0.000032$ and $1.5\Gev^2 \leq Q^2\leq
5000\Gev^2$ is better by at least 60 than any NLO--in--$\alpha_s$ fit, even 
though the NLO--in--$\alpha_s$ fit is allowed arbitrary, unjustified 
powerlike behaviour at small $x$ (i.e. the sea--quark distribution $\propto 
x^{-1-0.2}$), and the small--$x$ inputs for ${\cal F}_2(x,Q^2)$ and 
$d{\cal F}_2(x,Q^2)/d\ln Q^2$ are allowed to vary with respect to each other 
a great deal more than in the LORSC fit. In fact, as we would hope, all
of this superiority comes from the 
fit to the data with $x<0.1$. Moreover, much the same 
quality of fit is obtained for a range of $Q_0^2$ from $20\Gev^2 -
80\Gev^2$. Let us also put the difference in the quality of the LO(x) and the 
NLO$_1$ fits, i.e. a difference of $\chi^2$ of 64, into context:
if one obtains fits to the data using the NLO--in--$\alpha_s$ approach and 
allows the fit to become up to $64$ worse than the 
(best) NLO$_1$ fit, then values of 
$\Lambda_{\overline{\rm MS}}$ from $190\hbox{\rm MeV}$ to $380\hbox{\rm MeV}$
are allowed, i.e. $\alpha_s(M_Z^2)= 0.109\to 0.122$.   
As a caveat to the above, however, it is certainly true that
the calculations in this paper must be improved to take account of massive
quark thresholds in a better manner (as is the case with most 
NLO--in--$\alpha_s$ fits), and work towards this end is in progress.
Nevertheless, with the present treatment we feel
that the quality of the fit and the degree of explanatory (if not predictive)
power, not to mention the theoretical correctness, give strong
justification for the LORSC expressions. 

We do, however,  recognize that the quality of the fit alone, 
although impressive, is not 
such a substantial improvement on  more standard approaches that it 
necessarily convinces one that this approach has to be correct. In 
order to obtain verification we must compare with more and different 
experimental data. Hence, we have presented a LORSC prediction for 
${\cal F}_L(x,Q^2)$, comparing it to that obtained using the 
NLO--in--$\alpha_s$ approach. Hopefully there will be true measurements of 
${\cal F}_L(x,Q^2)$ at HERA some time in the future with which we can compare 
these predictions. We stress the 
importance of such measurements to the understanding of the physics 
which really underlies hadron interactions. In the near future we will also 
have predictions of the charm contribution to the structure function 
within the 
framework of a correct treatment of the massive quark, and comparison with
the ever--improving data on the charm structure function should also be a 
good discriminant between different theoretical approaches. Calculations and 
measurements of other, less inclusive quantities, such as forward jets, are 
another clear goal.

Finally, as we have noted, our calculation is at present only at leading 
order due to the lack of knowledge of 
next--to--leading--order--in--$\ln (1/x)$ 
coefficient functions and anomalous dimensions, or
equivalently of physical anomalous dimensions. This means that 
the NLO--in--$\alpha_s$ approach is still in principle superior to our 
approach at high $x$, where the leading--$\ln(1/x)$ terms at third 
order in $\alpha_s$ and beyond are not important but the 
${\cal O} (\alpha_s^2)$ terms are. Indeed, 
in practice the NLO--in--$\alpha_s$ fits, particularly
the NLO$_2$ fit, are slightly better than the LO(x) fit for data at
$x\geq 0.1$. We might therefore expect any predictions coming from the 
NLO--in--$\alpha_s$ approach to be more accurate than those coming from 
the LORSC approach down to values of $x$ somewhere in the region of $0.05$. 
The lack of the NLORSC expressions also means that a true determination of 
$\alpha_s(M_Z^2)$ from a global 
fit to structure function data is not yet possible, but that the best 
determination from fits to structure function data should 
at present come from using the NLO--in--$\alpha_s$
approach, with a correct treatment of heavy quark thresholds, using only 
large $x$ data. For a really fair comparison between the 
renormalization--scheme--consistent method and the conventional
order--by--order--in--$\alpha_s$ approach we really need the full 
next--to--leading--order, renormalization--scheme--consistent calculation.
Hopefully the required NLO in $\ln(1/x)$ quantities will soon become 
available \nlobfkl, and with a little work we will be able to 
make such a comparison. 
We would expect that when using the NLORSC expressions
the fit to the large $x$ data would become at least as good as for the 
NLO--in--$\alpha_s$ approach, and that the fit to the small--$x$ data 
would be of at least the same quality as for the LORSC fit. 
We wait expectantly to discover if this is indeed the case.  

\bigskip

{\bf Acknowledgements.}
\medskip
I would like to thank R.G. Roberts for continual help during the period of 
this work and for the use of the MRS fit program. I would also like to thank 
Mandy Cooper--Sarkar, Robin Devenish, Jeff Forshaw and Werner Vogelsang  
for helpful discussions.

\vfill 
\eject

\noindent {\bf Table 1}\hfil\break
\noindent Comparison of quality of fits using the full 
leading--order (including 
leading--$\ln (1/x)$ terms) renormalization--scheme--consistent expression, 
LO(x), and the two--loop fits MRSR$_1$, MRSR$_2$, NLO$_1$ and NLO$_2$.
For the LO(x) fit the H1 data chooses a
normalization  of $1.00$, the ZEUS data of $1.015$, and the BCDMS data
of $0.975$. The CCFR data is fixed at a normalization of $0.95$, and the 
rest is fixed at $1.00$. 
Similarly, for the NLO$_1$ fit the H1 data is fixed at a 
normalization of $0.985$, the ZEUS chooses a normalization of $0.99$, 
and the BCDMS data of $0.975$.
Again the CCFR data is fixed at a normalization of $0.95$, 
and the rest fixed at $1.00$.
Also, for the NLO$_2$ fit the H1 data is fixed at a 
normalization of $0.985$, the ZEUS chooses a normalization of $0.985$, 
and the BCDMS data of $0.97$.
Again the CCFR data is fixed at a normalization of $0.95$, 
and the rest fixed at $1.00$.  
In the R$_{1}$ and R$_2$ fits the BCDMS data has a fixed normalization of
$0.98$, the CCFR data of $0.95$ and the rest of $1.00$. 
 
\medskip

\hfil\vtop{{\offinterlineskip
\halign{ \strut\tabskip=0.6pc
\vrule#&  #\hfil&  \vrule#&  \hfil#& \vrule#& \hfil#&
\vrule#& \hfil#& \vrule#& \hfil#&
\vrule#& \hfil#& \vrule#& \hfil#& \vrule#\tabskip=0pt\cr
\noalign{\hrule}
& Experiment && data   && \omit &\omit& &\omit&
$\chi^2$&\omit& \omit &\omit&\omit&\cr
&\omit       && points && LO(x) &\omit& NLO$_1$ &\omit& NLO$_2$ &\omit& R$_1$ 
&\omit& R$_2$ &\cr
\noalign{\hrule}
& H1 ${\cal F}^{ep}_2$ && 193 && 123 && 145 && 145 && 158 && 149 &\cr
& ZEUS ${\cal F}^{ep}_2$ && 204 && 253 && 281 && 296 && 326 && 308 &\cr
\noalign{\hrule}
& BCDMS ${\cal F}^{\mu p}_2$ && 174 && 181 && 218 && 192 && 265 && 320 &\cr
& NMC ${\cal F}^{\mu p}_2$ && 129 && 122 && 131 && 148 && 163 && 135 &\cr
& NMC ${\cal F}^{\mu d}_2$ && 129 && 114 && 107 && 125 && 134 && 99 &\cr
& NMC ${\cal F}^{\mu n}_2/{\cal F}^{\mu p}_2$ && 85 && 142 && 137 && 138 
&& 136 && 132 &\cr
& E665 ${\cal F}^{\mu p}_2$ && 53 && 63 && 63 && 63 && 62 && 63 &\cr
\noalign{\hrule}
& CCFR ${\cal F}^{\nu N}_2$ && 66 && 59 && 48 && 40 && 41 && 56 &\cr
& CCFR ${\cal F}^{\nu N}_2$ && 66 && 48 && 39 && 36 && 51 && 47 &\cr
\noalign{\hrule}}}}\hfil

\bigskip

\noindent {\bf Table 2}\hfil\break
\noindent Comparison of quality of fits using the full leading--order 
(including 
leading--$\ln (1/x)$ terms) renormalization--scheme--consistent expression, 
LO(x), and the two--loop fits NLO$_1$ and NLO$_2$. The fits are identical to 
above, but the data are 
presented in terms of whether $x$ is less than $0.1$ or not.
 
\medskip

\hfil\vtop{{\offinterlineskip
\halign{ \strut\tabskip=0.6pc
\vrule#&  #\hfil&  \vrule#&  \hfil#& \vrule#& \hfil#&
\vrule#& \hfil#& \vrule#& \hfil#& \vrule#\tabskip=0pt\cr
\noalign{\hrule}
& \omit && data   && \omit &\omit&$\chi^2$& \omit&\omit&\cr
&\omit       && points && LO(x) &\omit& NLO$_1$& \omit &NLO$_2$&\cr
\noalign{\hrule}
& $x\geq 0.1$&& 551 && 622 && 615 && 595 &\cr
& $x<0.1$ && 548 && 483 && 554 && 589 &\cr
\noalign{\hrule}
& total && 1099 && 1105 && 1169 && 1184 &\cr
\noalign{\hrule}}}}\hfil

\footatend\vfill\supereject\immediate\closeout\rfile\writestoppt
\baselineskip=14pt\centerline{{\bf References}}\bigskip{\frenchspacing%
\parindent=20pt\escapechar=` \input refs.tmp\vfill\eject}\nonfrenchspacing

\vfill\eject\immediate\closeout\ffile{\parindent40pt
\baselineskip14pt\centerline{{\bf Figure Captions}}\nobreak\medskip
\escapechar=` \input figs.tmp\vfill\eject}

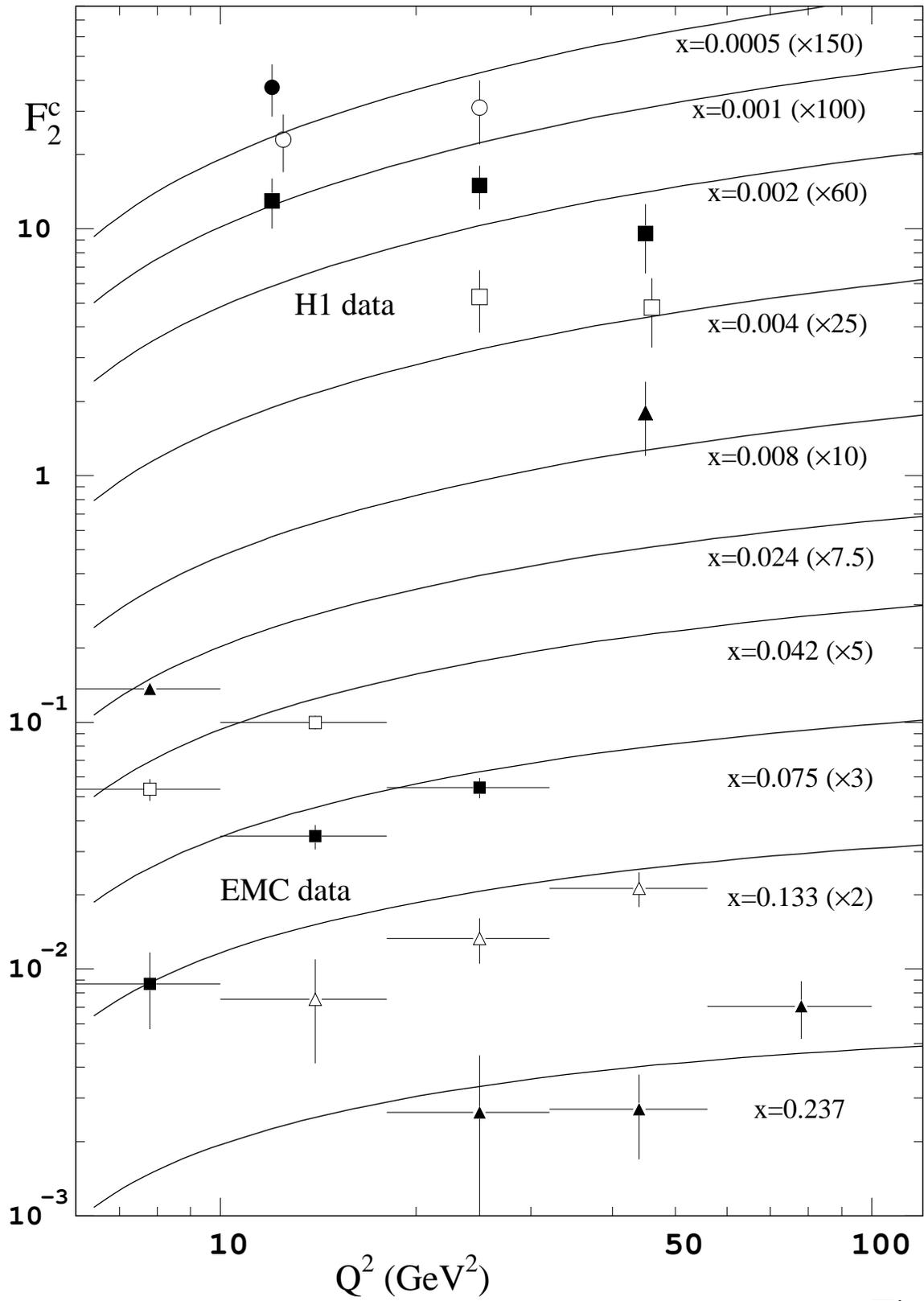
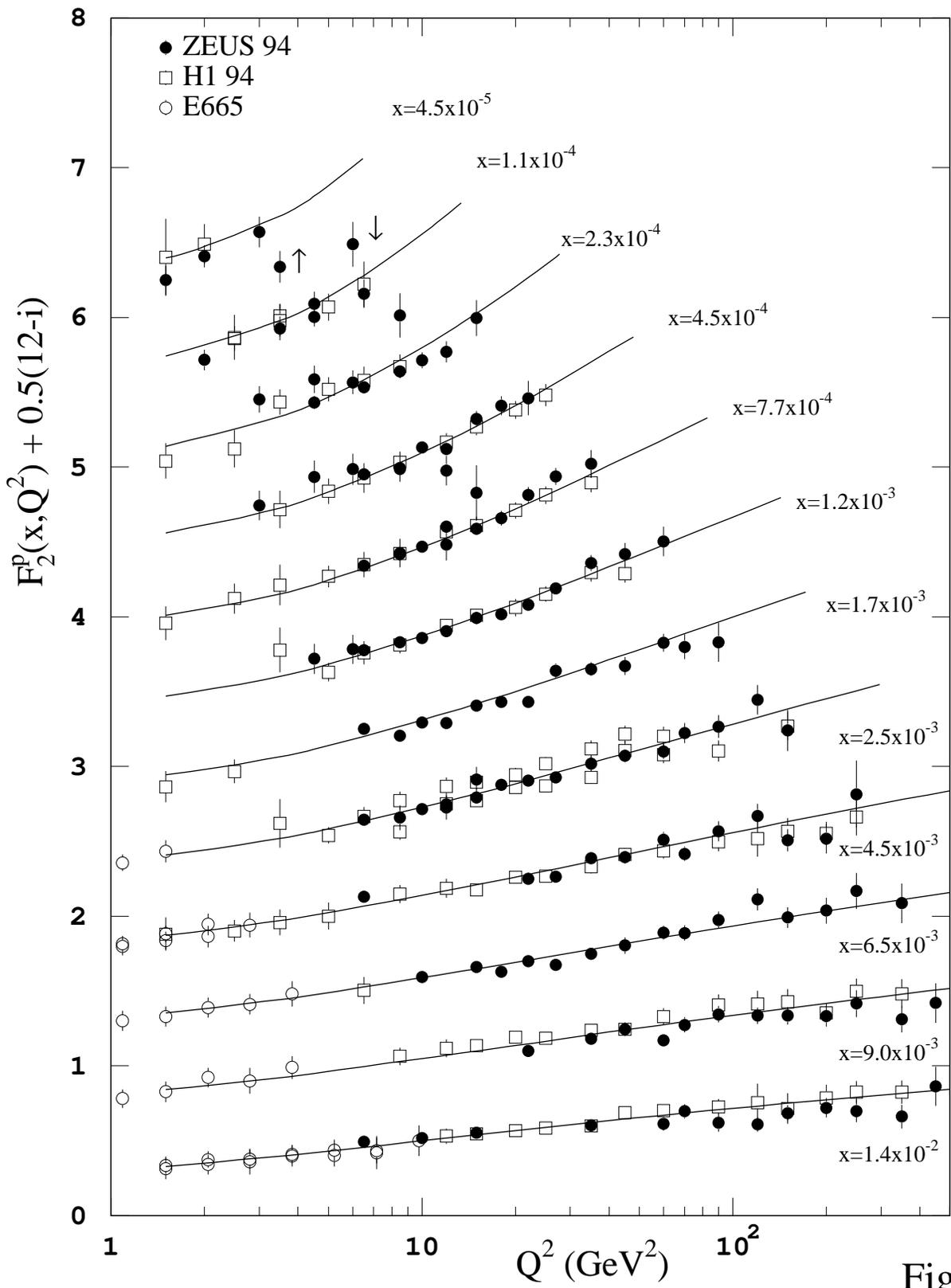
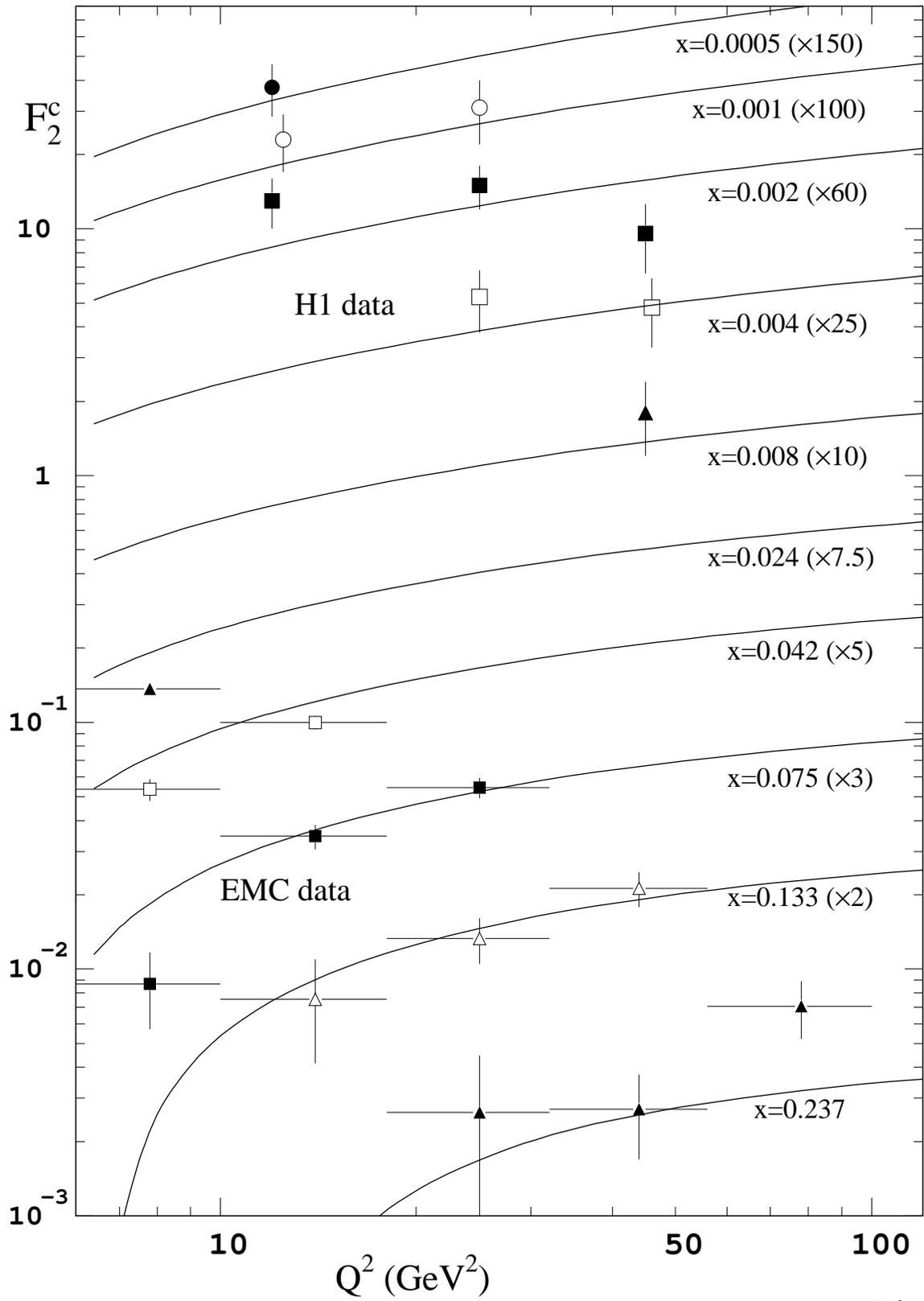
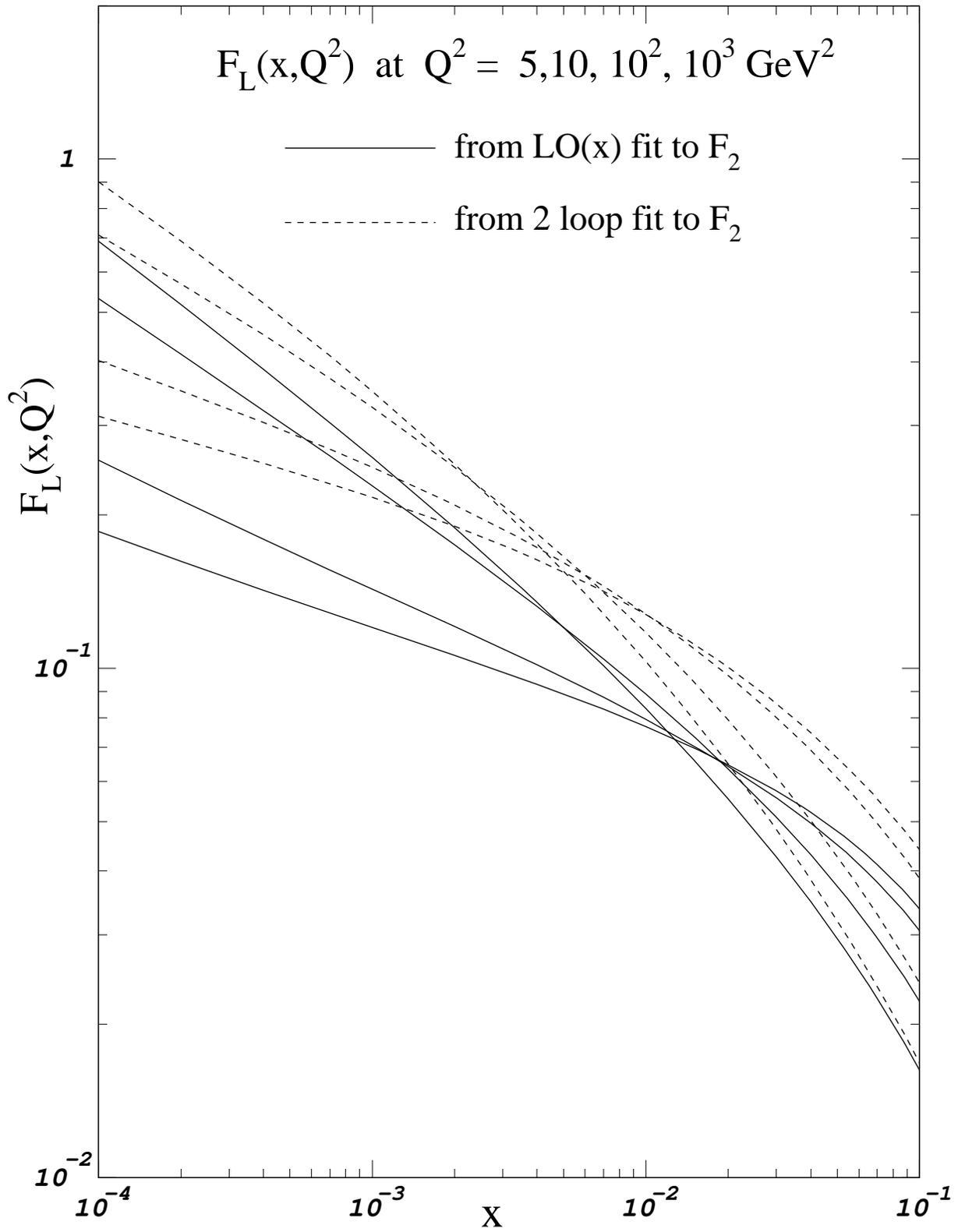
\end